\documentclass[rmp,twocolumn,amsmath,amssymb]{revtex4}

\usepackage[dvips]{color}
\usepackage{graphicx}

\begin{document}
\title{One dimensional Bosons: \\ From Condensed Matter Systems to Ultracold  Gases}
\author{M. A. Cazalilla}
\affiliation{Centro de Fisica de Materiales, Paseo Manuel de Lardizabal, 5.
 E-20018 San Sebastian, Spain}
\affiliation{Donostia International Physics Center, Paseo Manuel de Lardizabal, 4.
E-20018 San Sebastian, Spain} 
\author{R. Citro}
\affiliation{Dipartimento di Fisica
``E. R. Caianiello'', Universit\`a di Salerno, and Spin-CNR  I-84084 Fisciano (Sa), Italy}
\author{T. Giamarchi}
\affiliation{DPMC-MaNEP, University of Geneva, CH1211 Geneve, Switzerland}
\author{E. Orignac}
\affiliation{Laboratoire de Physique, CNRS UMR5672 and \'Ecole Normale Sup\'erieure de Lyon,
F-69364 Lyon Cedex 7, France}
\author{M. Rigol}
\affiliation{Department of Physics, Georgetown University, Washington, DC 20057, USA}
\date{\today}

\begin{abstract}
We review the physics of one-dimensional interacting bosonic
systems. Beginning with results from exactly solvable models and
computational approaches, we introduce the concept of bosonic
Tomonaga-Luttinger Liquids relevant for one-dimension, and compare it with
Bose-Einstein condensates existing in dimensions higher than one. We discuss the effects
of various perturbations on the Tomonaga-Luttinger liquid state as well as extensions to
multicomponent and out of equilibrium situations. Finally, we review
the experimental systems that can be described in terms of models of interacting
bosons in one dimension.
\end{abstract}
\maketitle

\tableofcontents

\section{Introduction} \label{sec:intro}

While still holding many surprises, one-dimensional (1D) quantum
many-body systems have fascinated physicists and mathematicians alike
for nearly a century. Indeed, shortly after the inception of Quantum
Mechanics, \citet{bethe_betheansatz_1931} found an exact solution to
the 1D Heisenberg model using an \emph{ansatz} for the wave function
that nowadays bears his name. This early exact solution of a 1D
quantum many-body problem of interacting (but fixed)
spin-$\frac{1}{2}$ particles was to be followed by a multitude of
others. Other simple 1D models, which could not be solved exactly,
were thoroughly studied by powerful methods especially suited for 1D.
For long, many researchers regarded these solutions as mere
mathematical curiosities that were, in general, of rather limited
interest for the real three-dimensional (3D) world.  Subsequent
technological developments in the 20th and 21st centuries led to the
discovery, chemical synthesis, and more recently fabrication of a wide
range of (quasi-)1D materials and physical systems. Interestingly
enough, the properties of these systems are sometimes fairly well
captured by the `toy models' of the past.  In this article, we shall
review the physics of some of these models as well as their
experimental realizations. However, unlike earlier reviews which have
mainly focused on 1D systems of
fermions~\cite{gogolin_1dbook,giamarchi_book_1d}, we shall only deal
with 1D systems where the constituent particles (or the relevant
excitations) obey Bose statistics.

As the reader will learn below, compared to higher dimensions, the
quantum statistics of the constituents plays a much less determinant
role in 1D. Nevertheless, when computing physical properties, quantum
statistics dictates the type of observables that may be experimentally
accessible.  Thus, because of the long standing interest in the
electronic properties of quasi-1D materials and nanostructures,
physicists have mainly focused their attention on 1D electron
systems. Much less attention has been given to 1D bosons, with the
important exception of spin systems.  We will show in
Sec.~\ref{sec:mapping} that spin-1/2 systems are mathematically equivalent
to a lattice gas of hard-core bosons, and thus are also reviewed in
this article. In fact, many quasi-1D materials exhibit
(Mott-)insulating phases whose magnetic properties can be modeled by
assuming that they consist of weakly coupled spin chains, similar to
those analyzed by Bethe in his groundbreaking work.

 Besides the spin chains, interest in other bosonic
systems  is rather recent and was initially spurred by the fabrication of long 1D arrays of
Josephson junctions. In these systems,  Cooper pairs (which, to a first
approximation, behave like bosons) can hop around in 1D. Even more
recently, further experimental stimulus has come from the 
studies of the behavior at low temperatures of liquid $^{4}$He
confined in elongated mesoscopic pores, as well as from availability of
ultracold atomic gases loaded in highly anisotropic traps and optical
lattices. Indeed, the strong confinement that can be achieved in these
systems has made it possible to realize \emph{tunable} low-dimensional
quantum gases in the strongly correlated regime.

As often happens in physics, the availability of new experimental
systems and methods has created an outburst of theoretical activity,
thus leading to a fascinating interplay between theory and experiment.
In this article, we shall attempt to survey the developments
concerning 1D systems of interacting bosons. Even with this
constraint, it is certainly impossible to provide a comprehensive
review of all aspects of this rapidly evolving field.

The plan of this article is as follows. In Sec.~\ref{sec:models} the
basic models are derived starting from the most general Hamiltonian of
bosons interacting through a two-body potential in the presence of an
arbitrary external potential. Models both in the continuum and on the
lattice are discussed. This section also introduces some important
mappings allowing to establish the mathematical relationship of some
of these boson models to other models describing $S=\frac{1}{2}$ spins 
or spinless fermions on 1D lattices. Next, in Sec.~\ref{sec:exactsol}, exact 
solutions of integrable models along with
with results on their correlation functions  are
reviewed. Some of these results are important by themselves and not
just merely academic models, as there currently exist fairly
faithful experimental realizations of them. In
Sec.~\ref{sec:numerical-results}, we describe some of the
computational approaches that can be used to tackle both integrable
and non-integrable models. We also discuss their application to some
1D models of  much current interest. The following section,
Sec.~\ref{sec:bosonize}, reviews the basic field-theoretic tools
that describe the universal low-energy phenomena in a broad class of
1D interacting boson models, the so-called Tomonaga-Luttinger
liquid (TLL) phase. For these systems, the method of bosonization and its
relationship to the hydrodynamic description of superfluids (which is
briefly reviewed in Sec.~\ref{sec:models}), are described.

The classification of phases and phase transitions exhibited by the
models introduced in Sec.~\ref{sec:models} is presented in
Sec.~\ref{sec:pertlut}, where besides the Mott transition between the
TLL phase and the Mott insulating phase, other kinds of instabilities
arising when the bosons move in the presence of a disorder potential
will be described. The low-energy picture is often unable to provide
the quantitative details that other methods such as exact solutions
(Sec.~\ref{sec:exactsol}) or computational approaches
(\ref{sec:numerical-results}) do. Thus, when results from any (or
both) of the latter methods are available, they complement the picture
provided by bosonization. Further extensions are considered in
Sec.~\ref{sec:coupled1dbosons}, where studies of coupled 1D systems
are reviewed. These include multicomponent systems (binary mixtures of
bosons or bosons and fermions).  Next we turn to the experimental
realizations of systems of interacting bosons in 1D, which include
spin ladders, superconducting wires, Josephson junctions, liquid $^4$He in nanopores, 
as well as the most recent ultracold atomic systems.  Finally, in
Sec.~\ref{sec:outlook} we shall provide a brief outlook for the field,
focusing on non-equilibrium dynamics and other topics that may become
important research topics in the future.

\section{Models for interacting bosons} \label{sec:models}

In this section, we introduce the basic models that describe
interacting bosons in 1D both in the continuum and on a lattice. These
models will be analyzed using various techniques in the rest of the
review. However, before embarking on the study of 1D physics per se,
let us briefly recall the main general results for bosons in
dimensions higher than one. This will serve as a reference with which
we can compare the results in 1D.

\subsection{Bosons in dimensions higher than one} \label{sec:bosons_larged}

As Einstein discovered shortly after Bose introduced a new type of
quantum statistics, when the temperature of a system of $N$
non-interacting bosons is lowered, a phase transition occurs. Below
the transition temperature, the occupation ($N_0$) of the lowest
available energy state becomes macroscopic, i.e., $N_0/N$ tends to a
constant for $N$ large. The transition was therefore named
`Bose-Einstein condensation', and the macroscopically occupied quantum
state, `Bose-Einstein condensate'. The acronym BEC is used
indistinctly for both concepts.  Mathematically, a BEC can be
described by a coherent state of matter, i.e., an eigenstate of the
boson field operator: $\hat{\Psi}(\mathbf{r})|\psi_{0}(\tau)\rangle =
\Psi_0(\mathbf{r},\tau)| \psi_0(\tau) \rangle$, hence $|\psi_0(\tau)
\rangle = e^{\hat{a}^{\dag}_0(\tau)} | 0\rangle$, where $|0\rangle$ is
the zero-particle state, $\hat{a}_0(\tau) =\int d\mathbf{r}\,
\Psi^*_0(\mathbf{r},\tau) \hat{\Psi}(\mathbf{r})$, and
$\Psi_0(\mathbf{r},\tau)$ is a complex function of space
($\mathbf{r}$) and time ($\tau$).

The above definition implies that the $U(1)$ symmetry group related to
particle conservation must be spontaneously broken. However, this is
sometimes problematic~\cite{leggett01,leggett_book_2006} because the
use of coherent states for massive particles violates the
super-selection rule forbidding the quantum superposition of states
with different particle numbers. The definition of BEC due
to~\citet{yang_odlro} circumvents this problem by relating the
existence of a BEC to the case in which the one-particle density
matrix of the system, $g_1(\mathbf{r},\mathbf{r}^{\prime},\tau) =
\langle \hat{\Psi}^{\dagger}(\mathbf{r},\tau)
\hat{\Psi}(\mathbf{r}^{\prime},\tau) \rangle$, behaves as
$g_1(\mathbf{r},\mathbf{r}^{\prime},\tau) \to
\Psi^*_0(\mathbf{r},\tau) \Psi_0(\mathbf{r}^{\prime},\tau)$
($\Psi_0(\mathbf{r},\tau) \neq 0$) for
$|\mathbf{r}-\mathbf{r}^{\prime}|\to +\infty$.  This behavior is
referred to as \emph{off-diagonal} long-range order (ODLRO) and
$\Psi_0(\mathbf{r},\tau)$ is called the \emph{order parameter} of the
BEC phase.

However, Yang's definition is not applicable to finite
systems. Following \citet{penrose56}, a more general criterion is obtained 
by diagonalizing the one-particle density matrix\footnote{This is
  mathematically always possible because
  $g_1(\mathbf{r},\mathbf{r}^{\prime},\tau) = \langle
  \hat{\Psi}^{\dag}(\mathbf{r},\tau)\hat{\Psi}(\mathbf{r}^{\prime},\tau)
  \rangle$, is a positive-definite Hermitian matrix provided
  $\mathbf{{r,r}^{\prime}}$ are regarded as matrix indices.} as
$g_1(\mathbf{r},\mathbf{r}^{\prime}) = \sum_{\alpha} N_\alpha(\tau) \:
\phi^{*}_{\alpha}(\mathbf{r},\tau) \phi_{\alpha}(\mathbf{r}',\tau)$,
where the \emph{natural orbitals} $\phi_{\alpha}$ are normalized such
that $\int d\mathbf{r}\, |\phi_{\alpha}(\mathbf{r},\tau)|^2 = 1$. The
existence of a BEC depends on the magnitude of the $N_\alpha$ compared
to $N$. When there is only one eigenvalue of $O(N)$ (say, $N_0\sim
O(N)$), the system is a BEC described by $\Psi_0(\mathbf{r},\tau) =
\sqrt{N_0(\tau)} \phi_0(\mathbf{r},\tau)$. When there are several such
eigenvalues one speaks of a `fragmented' BEC
(see~\citet{mueller_fragmented_BEC_2006} and references therein).
Furthermore, provided the depletion of the BEC is small, that is, for
$N-N_0\ll N$ (indeed, $N_0(T) \to N$ as $T \to 0$ for the
non-interacting gas), it is possible to describe the state of the
system using $|\psi^N_0(\tau)\rangle = \frac{1}{\sqrt{N!}} \left[
  \hat{a}^{\dag}_0(t) \right]^{N} |0\rangle$. Note that, differently
from the coherent state $|\Psi_0(\tau)\rangle$,
$|\psi^N_0(\tau)\rangle$ has a well defined particle number and can be
obtained from $|\psi_0(\tau)\rangle$ by projecting it onto a state
with total particle number equal to $N$.

The definitions of BEC due to Yang, Penrose and Onsager are equivalent
in the thermodynamic limit and also apply to interacting systems. In
general, the ratio $N_0/N$ is called the condensate fraction.  It is
important to not to confuse the condensate fraction, which for a BEC
is finite, with the superfluid fraction. The latter is a thermodynamic
property that is physically related to the fraction of the system mass
that is not dragged along by the walls of the container when the
latter rotates at constant angular frequency.  In charged systems, it is related
to Meissner effect, i.e., the expulsion of an externally applied
magnetic field. Mathematically, the superfluid fraction can be
obtained as the thermodynamic response to a change in the boundary
conditions of the system (see Sec.~\ref{sec:bosonize} for a discussion
concerning 1D systems). Indeed, the non-interacting Bose gas below the
condensation temperature is the canonical example of a BEC that is not
a superfluid. On the other hand, the 1D models discussed here do not
exhibit BEC (even at $T=0$).  Yet, they are good superfluids
(cf.  Sec.~\ref{sec:bosonization-method})

The above discussion does not provide any insights into how to compute
the BEC wavefunction, $\Psi_0(\mathbf{r},\tau)$, for a general system
of mass $m$ bosons interacting through a potential
$V_\mathrm{int}(\mathbf{r})$ and moving in an external potential
$V_\mathrm{ext}(\mathbf{r},\tau)$, which is described by the
Hamiltonian:
\begin{eqnarray}\label{eq:ham-sec}
  \hat{H}&=&\int d\mathbf{r}\,  \hat{\Psi}^\dagger(\mathbf{r}) \left[  -\frac{\hbar^2}{2m} \nabla^2
  +V_{ext}(\mathbf{r},\tau)\right] \hat{\Psi}(\mathbf{r})\nonumber\\
  &&+\int d\mathbf{r}\:  d\mathbf{r'}\,
  \hat{\Psi}^\dagger(\mathbf{r}) \hat{\Psi}^\dagger (\mathbf{r'}) V_\mathrm{int}(\mathbf{r'-r})
  \hat{\Psi}(\mathbf{r}) \hat{\Psi}(\mathbf{r'}).\quad
\end{eqnarray}
For such a system, in the spirit of a
mean-field theory \citet{gross-equation} and
\citet{pitaevskii-equation} independently derived an equation 
for the condensate wavefunction by
approximating $\hat{\Psi}(\mathbf{r})$ in the equation of motion,
$i\hbar \partial_{\tau} \hat{\Psi}(\mathbf{r},\tau)
=[\hat{\Psi}(\mathbf{r},\tau),\hat{H} - \mu \hat{N}]$, by its
expectation value $\langle \hat{\Psi}(\mathbf{r},\tau) \rangle =
\Psi_0(\mathbf{r},\tau)$ over a coherent state, where $\mu$ is the
chemical potential and $\hat{N}$ the number operator. The
Gross-Pitaevskii (GP) equation reads:
\begin{eqnarray}\label{eq:gross-pitaevskii}
  i\hbar \partial_{\tau} \Psi_0(\mathbf{r},\tau) =\left[-\frac{\hbar^2}{2m}
  \nabla^2 - \mu + V_{ext}(\mathbf{r},\tau)+ \right. \nonumber  \\
  + \left.  \int d\mathbf{r}^{\prime} \: V_\mathrm{int}(\mathbf{r-r}^{\prime}) |\Psi_0(\mathbf{r}^{\prime},\tau)|^2 \right]
  \Psi_0(\mathbf{r},\tau).
\end{eqnarray}
For an alternative derivation, which does not assume the spontaneous
break-down of the $U(1)$ symmetry, one can look for the extrema of the
functional $\langle\psi(\tau)
|i\hbar\partial_{\tau}-(\hat{H}-\mu\hat{N})|\psi(\tau)\rangle$, using
the (time-dependent Hartree) ansatz
$|\psi(\tau)\rangle=\frac{1}{\sqrt{N_0!}}\left[\hat{a}^{\dag}_0(\tau)\right]^{N_0}|0\rangle$,
where $\hat{a}_0(\tau) = \int d\mathbf{r} \, \Psi^*_0(\mathbf{r},
\tau) \: \hat{\Psi}(\mathbf{r}) |0\rangle$ and $\mu$ is a Lagrange
multiplier to ensure that $\int
d\mathbf{r}\:|\Psi_0(\mathbf{r},\tau)|^2 = N_0$.

The use of the Gross-Pitaevskii equation (\ref{eq:gross-pitaevskii})
to describe the interacting boson system assumes a small BEC
depletion, i.e., $N_0 \simeq N$. This is a good approximation in the
absence of strong correlations and at low temperatures. Thus, it is
particularly suitable for dilute ultracold
gases~\cite{pitaevskii1991,leggett01,Stringa_Pita_book}, for which the
range of inter-particle potential is much smaller than the
inter-particle distance $d$. In these systems, interactions are well
described by the Lee-Huang-Yang
pseudo-potential~\cite{huang_yang_pseudopot_1957,lee_yang_pseudopot_1957,lee_huang_yang_pseudopot_1957}
$V(\mathbf{r})\simeq\frac{4\pi \hbar^2 a_s}{m}\delta
(\mathbf{r}) \partial_{r}\left(r \cdot \right)$, where $a_s$ is the
$s$-wave scattering length. Typically, $a_s\ll d$, except near a
$s$-wave Feshbach
resonance~\cite{Stringa_Pita_book,leggett_book_2006}. For trapped gases, 
it has been rigorously established that in the limit of $N\to \infty$ and $Na_s$
fixed, the Gross-Pitaevskii approximation becomes exact for the ground
state energy and particle density~\cite{lieb2000}. Furthermore, in this
limit, the gas is both 100\% Bose condensed\cite{lieb2002} 
and 100\% superfluid~\cite{lieb2002b}.     

In a uniform system, the above definitions imply that the momentum
distribution:
\begin{equation}
 n(\mathbf{k}) = \int d\mathbf{r} \, e^{-i \mathbf{k}\cdot \mathbf{r}} g_1(\mathbf{r}) = N_0 \:
 \delta(\mathbf{k}) + \tilde{n}(\mathbf{k}),
\end{equation}
where $\tilde{n}(\mathbf{k})$ is a regular function of
$\mathbf{k}$. The Dirac delta function is the hallmark of the BEC in
condensed matter systems such like liquid $^{4}$He below the
$\lambda$-transition. On top of the condensate, interacting boson
systems support excitations with a dispersion that strongly deviates
from the free particle dispersion $\epsilon_0(\mathbf{k}) =
\frac{\hbar^2 \mathbf{k}^2}{2m}$.  A way to compute the excitation
spectrum is to regard the GP equation as a time-dependent Hartree
equation. Thus, its linearized form describes the condensate
excitations, which, for a uniform dilute gas, have an a dispersion of the
form $\epsilon(\mathbf{k})=\sqrt{\epsilon_0(\mathbf{k})\left[\epsilon_0(\mathbf{k})+2g\rho_0\right]}$,
where $\rho_0 = N/V$, and $g$ is the strength of the interaction. This
was first obtained by Bogoliubov, which proceeded differently by
deriving a quadratic Hamiltonian from (\ref{eq:ham-sec}) in terms of
$\delta\hat{\Psi}(\mathbf{r})=\hat{\Psi}(\mathbf{r})-\sqrt{N_0}$ and
$\delta\hat{\Psi}^{\dag}(\mathbf{r})=\hat{\Psi}^{\dag}(\mathbf{r})-\sqrt{N_0}$
and keeping only the (leading) terms up to $O(N_0)$. Note that, in the
$|\mathbf{k}| \to 0$ limit, the excitations above the BEC state are
linearly dispersing phonons: $\epsilon(\mathbf{k}) \simeq \hbar v_s
|\mathbf{k}|$, where $v_s = \sqrt{\frac{g\rho_0}{2m}}$. From the point
of view of the spontaneous break-down of the $U(1)$ symmetry, the
phonons are the Goldstone modes of the broken-symmetry phase.

We conclude this subsection by reviewing the hydrodynamic
approach. Although we derive it from the GP equation for a dilute gas,
its validity extends beyond the assumptions of this theory to
arbitrarily interacting superfluid systems in \emph{any} dimension. We
begin by setting
$\Psi_0(\mathbf{r},\tau)=\sqrt{\rho(\mathbf{r},\tau)}\:e^{i\theta(\mathbf{r},\tau)}$
in~(\ref{eq:gross-pitaevskii}).  Hence,
\begin{eqnarray}\label{eq:conservation}
\partial_\tau \rho&+& \frac{\hbar}{m} \nabla\cdot \left(\rho \nabla
  \theta \right) = 0, \\
\label{eq:eq-mot-velocity}
 \partial_\tau \theta&+&\left(\frac
   {\hbar  (\nabla \theta)^2}{2m}
 +\frac{1}{\hbar} \left[ V_{ext}+ g \rho \right]-
   \frac{\hbar}{2m} \frac{\nabla^2
  \sqrt{\rho}}{\sqrt{\rho}}\right)=0.\quad\quad
\end{eqnarray}
The first equation, (\ref{eq:conservation}), is the just continuity
equation where the particle current
$\mathbf{j}(\mathbf{r},\tau)=\rho(\mathbf{r},\tau)\mathbf{v}(\mathbf{r},\tau)=\frac{\hbar}{m}\rho(\mathbf{r},\tau)\nabla\theta(\mathbf{r},\tau)$.
The second equation describes a potential flow with velocity potential
$\frac{\hbar}{m} \theta(\mathbf{r},\tau)$.  The last term in
Eq.~(\ref{eq:eq-mot-velocity}) is called the quantum pressure. If we
call $\ell$ the typical distance characterizing the density
variations, the quantum pressure term scales as
$\sim\frac{\hbar^2}{m\ell^2}$ and therefore it is negligible compared
to the classical pressure term $g\rho$ when
$\ell\gg\xi=\hbar/\sqrt{m\rho g}$ ($\xi$ is known as the healing
length). In the limit of a slowly varying density profile, the quantum pressure
term in (\ref{eq:eq-mot-velocity}) can be dropped and the equations
can be written as:
\begin{eqnarray}
\label{eq:hydro-eqs}
\partial_{\tau} \rho +\mathbf{\nabla}\cdot(\rho \mathbf{v})=0, \\
\partial_{\tau} \mathbf{v}
+\mathbf{\nabla} \left( \frac{V_{ext}} m +\frac 1 2
\mathbf{v}^2\right)=-\frac{\mathbf{\nabla} P}{\rho m}.
\end{eqnarray}
These equations coincide with the classical Euler equation describing
the flow of a non-viscous fluid with an equation of state where $P(\rho)=
g\rho^2/2$. This result is in agreement with simple thermodynamic
considerations from which the classical fluid of
Eqs.~(\ref{eq:hydro-eqs}) has, at zero temperature, an energy
per unit volume $e(\rho)=g \rho^2/2$ and chemical potential
$\mu(\rho)=g \rho$. In the literature on ultracold gases, this
approximation is known as the Thomas-Fermi approximation. In the
static case ($\mathbf{v} =0$ and $\partial_{\tau} \rho = 0$), it leads
to the equation $\nabla\left[
  V_{ext}(\mathbf{r})+P(\rho(\mathbf{r}))\right]=0$, which allows to
determine the BEC density profile $\rho(\mathbf{r})$ for a given
trapping potential $V_\mathrm{ext}(\mathbf{r})$.

As mentioned above, the validity of the hydrodynamic equations is very
general, and applies to strongly-interacting boson as well as fermion
superfluids, for which the dependence of the chemical potential on the
density is very different. However, in 1D the assumptions that
underlie GP (as well as Bogoliubov) theory break down.  Quantum and
thermal fluctuations are strong enough to prevent the existence of
BEC, or equivalently, the spontaneous break-down of  the $U(1)$ symmetry in
the thermodynamic limit. This result is a consequence of the
Mermin-Wagner-Hohenberg (MWH) theorem
\cite{mermin_wagner_theorem,hohenberg67_theorem,mermin_theorem}, which
states that, at finite temperature, there cannot be a spontaneous
break-down of continuous symmetry groups (like $U(1)$) in
two-dimensional (2D) classical systems with short range
interactions. Indeed, at $T=0$, a 1D quantum system with a spectrum of
long wave-length excitations with linear dispersion can be mapped,
using functional integral methods, to a 2D classical system (see
discussion in Sec.~\ref{sec:mapping} for an example).  Therefore, the
MWH theorem also rules out the existence of a BEC in those 1D systems.
For the 1D non-interacting Bose gas, the excitations have a quadratic
dispersion, and thus the mapping and the theorem do not
apply. However, a direct proof of the absence of a BEC also in this
case is not difficult (see \emph{e.g.}~\citet{Stringa_Pita_book}).

In the absence of a BEC, the assumptions of the GP theory and the
closely related Bogoliubov method break down.  When applied to 1D
interacting boson systems in the thermodynamic limit, these methods
are plagued by infrared divergences. The latter are a manifestation of
the dominant effect of long wavelength thermal and quantum
fluctuations, as described by the MWH theorem. By decoupling the
density and phase fluctuations, \citet{popov72,popov_functional_book}
was able to deal with these infrared divergences within the functional
integral formalism. However, Popov's method relies on integrating the
short-wave length fluctuations perturbatively, which is a controlled
approximation only for a weakly interacting system. However, for
arbitrary interaction strength, it is necessary to resort to a
different set of tools to tackle the 1D world. Before going into their
discussion, we first need to define the various models that we will
use to describe the bosons interacting in 1D, and whose solutions are
reviewed in the following sections.

\subsection{Bosons in the continuum in 1D} \label{sec:bosons_conti}

Let us first consider the case of bosons moving in the continuum along one direction (henceforth denoted by $x$).
Thus, we  assume that a very strong confinement is applied in the transverse directions [denoted
$\mathbf{r}_{\perp}=(y,z)$] so that only the lowest energy transverse quantum state $\phi_0(\mathbf{r}_{\perp})$
needs to be considered. Hence, the many-body wavefunction reads:
\begin{equation} \label{eq:reductionto1D}
 \psi_B(\mathbf{r}_{1}, \ldots, \mathbf{r}_{N}) =
 \psi_B(x_1, \ldots, x_N) \: \prod_{i=1}^N \phi_0(\mathbf{r}_{i\perp}). \quad
\end{equation}
In what follows, we shall focus on the degrees of freedom described by $\psi_B(x_1, \ldots, x_N)$.
The most general Hamiltonian for a system of $N$  bosons interacting through a two-body potential $V_\mathrm{int}(x)$
while moving in an external potential $V_\mathrm{ext}(x)$ reads:
\begin{equation}
  \label{eq:basicmodel1d}
  \hat{H} = \sum_{i=1}^N \left[\frac{ \hat{p}^2_i}{2m} +
    V_\mathrm{ext}(\hat{x}_i) \right]
 +  \sum_{i<j=1}^N V_\mathrm{int}(\hat{x}_i-\hat{x}_j),
\end{equation}
where $m$ is the atom mass and $\hat{p}_i = -i\hbar \frac{\partial}{\partial x_i}$ the $i$-th particle
momentum operator ($\left[ \hat{x}_i,\hat{p}_j \right] = i \hbar \delta_{ij}$).

The simplest non-trivial model of interacting bosons in the continuum is the one introduced by \citet{lieb_bosons_1D,lieb_excit},
which is obtained from~(\ref{eq:basicmodel1d}) upon setting $V_\mathrm{ext}(x) = 0$
and considering a Dirac-delta interaction, $V_\mathrm{int}(x) = g\: \delta(x)$.
When a transverse external confinement is considered, an effective one-dimensional model with Dirac-delta interaction can still be obtained within a pseudopotential approximation by properly summing over the virtual excitations of the high-energy transverse modes~\cite{olshanii98} (cf. Sec.~\ref{subsec:LLgas}). For the Lieb-Liniger model the Hamiltonian reads:
\begin{equation}
  \label{eq:lieb-liniger-model}
  \hat{H} =-  \frac{\hbar^2}{2m} \sum_{i=1}^N \frac{\partial^2}{\partial x^2_i} + g
  \sum_{i<j=1}^N \delta(x_i-x_j).
\end{equation}
It is convenient to parametrize the interaction strength in this model using the parameter
$c =\frac{mg}{\hbar^2}$ which has the dimension of an inverse length. Thus, $c=0$ corresponds to free
bosons while $c\to +\infty$ is the hard-core or Tonks-Girardeau limit~\cite{girardeau_bosons1d}.
As it will be shown in Sec.~\ref{sec:tonks}, in this limit the model can be solved exactly by mapping it
onto a system of non-interacting fermions. Moreover, as shown by Lieb and Liniger, the model can be solved
for all values of $c$ using the Bethe ansatz. This solution is reviewed in Sec.~\ref{sec:lieb}.
\textcite{gurarie_resonant1DBosegas} has studied a generalization of the Lieb-Liniger model
with Feshbach resonant interactions, which
can be approximately solved by the Bethe ansatz. However, we shall not review this solution here and
refer the reader to the original paper~\cite{gurarie_resonant1DBosegas}.

The Lieb-Liniger model is not the only interacting boson
model that can be solved exactly. The following model
\begin{equation}
  \label{eq:calogero-sutherland}
  \hat{H}=-\frac{\hbar^2}{2 m} \sum_{i=1}^N  \frac{\partial^2}{\partial x_i^2} +
  \sum_{i<j=1}^N \frac{g}{(x_i-x_j)^2}\,,
\end{equation}
which was introduced by~\citet{calogero69_model1} has also this property. Moreover, Calogero showed that
the model is also solvable when a harmonic potential, i.e., for
$\sum_{i=1}^N V_\mathrm{ext}(\hat{x}_i) = \frac{1}{2} m \omega^2 \sum_{i=1}^L x_i^2$, is added to
Eq.~(\ref{eq:calogero-sutherland}). The interaction strength in the model~(\ref{eq:calogero-sutherland})
is characterized by the dimensionless parameter $\lambda$, where $\lambda(\lambda-1)=\frac{2mg}{\hbar^2}$.
Note that the interaction potential in this model is singular for $x_i = x_j$. To deal with this behavior,
we must require that the many-body wavefunction vanishes when $x_i = x_j$, that is, the model describes a
system of hard-core bosons with $1/x^2$ interactions.  For  $\lambda = 0,1$, one recovers the TG gas.
This model will be further discussed in Sec.~\ref{sec:calogero}.

Furthermore, besides the two types of interactions leading to the exactly solvable models of Lieb and Liniger and
Calogero, it is also possible to consider other kinds of interactions that are experimentally relevant.
Among them, three types of interactions play an especially important
role: i) $V_{\mathrm{int}}(x) \sim 1/|x|^3$ describes a system of (polarized) dipoles in 1D. Such
interaction is relevant for systems of dipolar ultracold atoms~\cite{pfau_chromium_bec} or bosonic molecules.
Note that, contrarily to 3D, this potential is integrable in 1D (i.e.,
$\int^{+x}_{-x} dx'\: V_{\mathrm{int}}(x')$ converges for large $x$) and thus it is essentially not different
from a short but finite range interaction potential; ii) The unscreened Coulomb potential
$V_{\mathrm{int}}(x) \propto 1/|x|$ is relevant for charged systems such as superconducting wires, Josephson
junction arrays, and trapped ion systems. This potential is nonintegrable even in 1D,  and this
leads to some modifications of the  properties of the
system at the long wavelengths compared to the case of short range interactions.

Furthermore, other models that are worth studying deal with the effect of different kinds of external
potentials such as periodic, disorder, or trapping potentials. The effect of weak periodic
potential will be considered in Sec.~\ref{sec:mott}, whereas the effect of a strong periodic potential is best discussed by using
the lattice models that will be introduced in the following section. In both cases, the potential leads to
the existence of bosonic Mott insulators. Disorder potentials are another type  that is
 relevant for both condensed matter systems, where it appears naturally, and  currently for ultracold atomic systems,
where it is introduced artificially to study its effects in a much more controlled way. In 1D Bose systems, as
it will be discussed in Sec.~\ref{sec:disorder}, it can lead to glassy phases. In addition, in ultracold atomic
systems, the atoms are in the gaseous phase and therefore need to be contained. As it will briefly be described in
Sec.~\ref{sec:coldatoms}, the confining potential can be either made using laser light or (time-dependent)
magnetic fields, and for small gas clouds, it can be well approximated by a harmonic potential,
$V_\mathrm{ext}(x) = \frac{1}{2} m \omega^2 x^2$.  Such a potential makes the system intrinsically
inhomogeneous, and thus can lead to a quite complicated physics, where different phases coexist in the same
trap. The effect of the trap will be discussed in several sections below.

\subsection{Bosons on the lattice in 1D}\label{se:josephson-th}

In the previous section, we have introduced several models of interacting bosons in the continuum.
However, when the bosons move in a deep periodic potential, \emph{e.g.} for
 $V_\mathrm{ext}(x) = V_0 \cos(G x)$ where $V_0$ is the largest energy scale in the problem, a more convenient
starting point is to project the Bose field operator on 
the basis of  Wannier orbitals $w_0(x)$ belonging to the lowest Bloch
band of $V_\mathrm{ext}(x)$.  Thus,
\begin{equation}
\Psi(x) \simeq \sum_{i=1}^L w_0(x-i a) \: \hat{b}_i, \label{eq:projfieldop}
\end{equation}
where $a = \frac{2\pi}{G}$ is the lattice parameter. In doing the above approximation, we have neglected the
projection of $\Psi(x)$ on higher bands. Thus, if $V_0$ is decreased, taking into account the
Wannier orbitals of higher bands may become necessary.  Upon inserting (\ref{eq:projfieldop}) into the second
quantized version of Eq.~(\ref{eq:basicmodel1d}), the following Hamiltonian is obtained:
\begin{eqnarray}
 \hat{H} = \sum_{i,j=1}^L \left[ -t_{ij} \hat{b}^{\dag}_{i} \hat{b}_{j} +
 \sum_{k,l=1}^L V^\mathrm{int}_{ik,jl} \hat{b}^{\dag}_{i} \hat{b}^{\dag}_k
 \hat{b}_{j} \hat{b}_{l} \right],
\end{eqnarray}
where $t_{ij} = -\int dx\,  w^*_0(x-i a) \hat{H}_0(x) w_0(x-j a)$, and
$\hat{H}_0 = \frac{\hbar^2}{2m} \partial^2_x + V_\mathrm{ext}(x)$, and
$V^\mathrm{int}_{ik,jl} = \int dx dx^{\prime} \: w^{*}_0(x- i a) w^{*}(x^{\prime}- k a) V^\mathrm{int}_\mathrm{int}(x-x^{\prime})
w_0(x^{\prime}-j a) w_0(x-l a)$. When the boundary conditions are periodic, we must further require that
$\hat{b}^\dagger_{L+1} =  \hat{b}^\dagger_{1}$ and $\hat{b}_{L+1} =  \hat{b}_{1}$. For a deep lattice potential,
it is possible to further simplify this model because in this limit, the orbitals $w_0(x-i a)$ are strongly
localized about $x = i a$ and it is sufficient to retain only the  diagonal as well as nearest neighbor terms,
which leads to the extended Hubbard model:
\begin{eqnarray}
 \hat{H}_{EBHM} = \sum_{i=1}^L \left[ -t \, \left( \hat{b}^{\dag}_{i} \hat{b}_{i+1}+ \mathrm{H.c.} \right)
 - \mu \hat{n}_i \right.  \nonumber \\ \left.
 +\, \frac{U}{2} \hat{b}^{\dagger}_i \hat{b}^{\dagger}_i \hat{b}_i \hat{b}_i   + V \hat{n}_i
 \hat{n}_{i+1} \right], \label{eq:EBHM}
\end{eqnarray}
where $\hat{n}_i = \hat{b}^{\dag}_i \hat{b}_i$ is the site occupation operator and $\mu$ is the chemical potential. We shall deal with
this model in Sec.~\ref{sec:mott}. The fist term is the kinetic energy of the bosons in the tight binding
approximation, while the last two terms describe an onsite interaction of strength $U$
and a nearest neighbor interaction of strength $V$.

When the range of the interaction is small compared to the lattice parameter $a$, it is possible to further neglect
the nearest neighbor interaction compared to the onsite interaction $U$ because the overlap of the Wannier
orbitals in a deep lattice potential makes $V$ small. The resulting model is known as the Bose-Hubbard model:
\begin{equation}
 \hat{H}_{BHM} = \sum_{i=1}^L \left[ -t\, \left( \hat{b}^{\dag}_{i} \hat{b}_{i+1}+ \mathrm{H.c.} \right)
 - \mu \hat{n}_i
  + \frac{U}{2} \hat{b}^{\dagger}_i \hat{b}^{\dagger}_i \hat{b}_i \hat{b}_i   \right].\label{eq:bose-hubbard}
\end{equation}
In condensed matter systems, such as Josephson junction arrays, it is usually difficult to compute
$t$ and $U$ from first principles. However, in cold atomic systems
the forms of the external potential and the atom-atom interaction are accurately known so that it is possible
to compute $t$ and $U$ from first principles~\cite{bloch_dalibard_zwerger_review_2008}.

For finite values of $U/t$, the Bose-Hubbard model \emph{is not} exactly solvable. However, for
$U/t\to +\infty$, the sectors of the Hilbert space where $n_i = 0, 1$ and $n_i > 1$ decouple and the model
describes a gas of hard-core bosons, which is the lattice analogue of the TG gas. For this system, the
Hamiltonian reduces to the kinetic energy:
\begin{equation}
  \label{eq:lattice-hcb}
  \hat{H}_{LTG}= \sum_{i=1}^L \left[ -t\, (\hat{b}^\dagger_i \hat{b}_{i+1} + \text{H.c.}) - \mu  \hat{n}_i \right],
\end{equation}
supplemented with the constraint that $(\hat{b}^{\dag}_i)^{2} | \Psi_\mathrm{phys} \rangle
= (\hat{b}_i)^{2} | \Psi_\mathrm{phys} \rangle  = 0$ on all physical states $|\Psi_\mathrm{phys}\rangle$.
Moreover, as in the continuum case, the model remains exactly solvable when an external potential of the form
$\hat{V}_\mathrm{ext}=\sum_{i} v_\mathrm{ext}(i) \hat{n}_i$ is added to the Hamiltonian.  The properties of the lattice hard-core
bosons both for $\hat{V}_\textrm{ext} = 0$, and for the experimentally relevant case of a harmonic trap
are discussed in  Sec.~\ref{sec:corr-latt}. We note in passing that in
dimension higher than one, the lattice gas of hard core bosons on
hypercubic lattice is
known rigorously to possess a BEC condensed ground
state~\cite{kennedy88}.  

When the interactions are long range, such as for Cooper pairs in Josephson junction arrays, which interact
via the Coulomb interaction, or for dipolar ultracold atoms and molecules, neglecting the nearest neighbor
interaction $ V\sum_i \hat{n}_i \hat{n}_{i+1}$ in (\ref{eq:EBHM}) is not entirely justified. In some cases,
however, $U$ may be sufficiently large, so that the $U \to +\infty$ limit is a reasonable approximation.
Under those conditions, the extended Bose-Hubbard model, Eq.~(\ref{eq:EBHM}), reduces to the so-called
$t$-$V$ model, whose Hamiltonian reads:
\begin{equation}
 \hat{H}_{t-V}= \sum_{i=1}^L \left[ -t\, (\hat{b}^\dagger_i \hat{b}_{i+1} + \text{H.c.}) - \mu \hat{n}_i
 + V  \hat{n}_i \hat{n}_{i+1}  \right], \label{eq:tvmodel}
\end{equation}
This model is  Bethe-Ansatz  solvable and is sometimes also  called the quantum lattice gas model
\cite{yang1966a,yang1966b,yang1966c}. As we will see in the next subsection, this model is equivalent to an
anisotropic spin-$\frac{1}{2}$ model called the XXZ chain~\cite{orbach1959,walker1959} and to the
6 vertex model of statistical mechanics~\cite{lieb1967}. It is convenient to introduce the dimensionless
parameter $\Delta=V/(2t)$ to measure the strength of the repulsion in units of the hopping.

\subsection{Mappings and various relationships} \label{sec:mapping}

In 1D, several transformations allow the various models introduced to be related to one other as well as to
others. We shall explore them in this section. The first mapping relates a system of hard-core bosons to a
spin-$\frac{1}{2}$ chain~\cite{holstein_primakoff,matsubara1956,fisher_xxz}. The latter is described by the
set of Pauli matrices $\{\hat{\sigma}_j^x,\hat{\sigma}_j^y,\hat{\sigma}_j^z\}_{j=1}^L$ (below we shall use
$\hat{\sigma}_j^\pm=\hat{\sigma}_j^x\pm i\hat{\sigma}_j^y$), where $j$ is the site index. The transformation due to
\citet{holstein_primakoff} reads:\footnote{The Holstein-Primakoff transformation is actually more general
as it can be used for higher $S$ spins and in any dimension.}
\begin{eqnarray}
 \label{eq:holstein_primakoff}
\hat{\sigma}_j^+&=&\hat{b}^{\dagger}_{j}\ \sqrt{1-\hat{n}_{j}}, \quad
 \hat{\sigma}_j^-=\sqrt{1-\hat{n}_j} \ \hat{b}_{j},\nonumber\\
 \hat{\sigma}_j^z&=&\hat{n}_j-1/2.
\end{eqnarray}
Hence, Eq.~(\ref{eq:lattice-hcb}) maps onto the XX spin-chain model:
\begin{equation}
\label{eq:xy-model}
  \hat{H}_0 = \sum_{j=1}^L \left[ -2t (\hat{\sigma}_j^x \hat{\sigma}_{j+1}^x + \hat{\sigma}_j^y \hat{\sigma}_{j+1}^y) - \mu \left(
  \sigma_z + \frac{1}{2} \right) \right]
\end{equation}
Furthermore, the nearest neighbor interaction in the $t$-$V$ model [cf. Eq.~\eqref{eq:tvmodel}]
becomes an Ising interaction:
\begin{equation}
 \hat{H}_\mathrm{int} = V \sum_{j=1}^L \left(\hat{\sigma}_j^z + \frac12\right) \left(\hat{\sigma}_{j+1}^z + \frac12\right).
\end{equation}
Therefore, the $t$-$V$ model maps onto the XXZ or Heisenberg-Ising spin-chain model in a magnetic field, and
spin systems can be seen as faithful experimental realizations of hard-core boson systems.

Another useful transformation relates $S=\frac{1}{2}$ spins (and hence hard-core bosons) to spinless fermions.
This is a fairly convenient way of circumventing the hard-core constraint by means of the
Pauli principle. Following \citet{jordan_transformation}, we introduce the transformation by relating
fermions to spins~\cite{jordan_transformation,lsm_61,katsura1962,niemeijer1967}:
\begin{eqnarray}
  \label{eq:jw-transf}
  \hat{\sigma}_j^+ &=& \hat{c}^{\dagger}_j\: e^{-i \pi \sum_{m<j} \hat{c}^\dagger_m \hat{c}_m},\quad
 \hat{\sigma}_j^- = e^{i \pi \sum_{m<j} \hat{c}^\dagger_m \hat{c}_m}\ \hat{c}_j, \nonumber\\
  \hat{\sigma}_j^z&=&\hat{c}^\dagger_j \hat{c}_j -1/2,
\end{eqnarray}
where $\{\hat{c}^{}_i, \hat{c}^\dagger_j\} = \delta_{ij}$, and otherwise anti-commute, as corresponds to  fermions.
By combining the transformations~\eqref{eq:jw-transf} and \eqref{eq:holstein_primakoff},
Eq.~(\ref{eq:lattice-hcb}) is mapped onto a non-interacting fermion Hamiltonian:
\begin{equation}
  H_0=\sum_{j=1}^L \left[ -t \left(\hat{c}^\dagger_j \hat{c}^{}_{j+1} + \text{H.c.} \right)  -\mu c^{\dag}_j c_j \right].  \label{eq:nonintferm}
\end{equation}
Nevertheless,  as given above, the mapping assumes open boundary conditions.
For periodic boundary conditions
\begin{equation}
\label{eq:hcb_bc}
\hat{b}^\dagger_{1} \hat{b}^{}_{L}=-\hat{c}^\dagger_{1}\hat{c}^{}_{L}\
e^{ i\pi \sum^L_{j=1}\hat{c}^\dagger_j \hat{c}_j},
\end{equation}
which means that when the total number of particles $N=\langle\hat{N}\rangle$, where
$\hat{N} = \sum^L_{m=1}\hat{c}^\dagger_m \hat{c}_m$ commutes with $\hat{H}$, in the chain
is odd, the fermions obey periodic boundary conditions, while if $N$ is even, they obey anti-periodic boundary
conditions.

A consequence of the mapping is that, similar to the situation encountered with TG gas in the continuum, the
spectra and thermodynamics of the lattice TG gas and the non-interacting fermion model (\ref{eq:nonintferm})
are identical.  In particular, the low energy excitations of the TG gas have linear dispersion, which
contribute a term linear in the temperature $T$ to the specific heat at low $T$ (see Sec.~\ref{sec:exactsol}
for an extended discussion). In addition, on the lattice it is also possible to show  (see \emph{e.g.}
the appendix in \citet{cazalilla_tonks_gases}) the equivalence of the Jordan-Wigner
approximation and  Girardeau's Bose-Fermi mapping, which
applies to the many-body wavefunctions of both models and is reviewed in Sec.~\ref{sec:tonks}.

At low lattice filling, that is, when the number of particles, $N$, is very small compared to the number of
lattice sites, $L$, single particle dispersion of the lattice model is well approximated by a parabola,
i.e., $\epsilon(k) = -2 t \cos k a \simeq - 2 t + \frac{\hbar^2 k^2}{2m^{*}}$ where $m^{*} = \hbar ^2/ (2 a^2 t)$
is the lattice effective mass. The TG gas of hard-core bosons in the continuum is thus recovered. Indeed,
by carefully taking the low-filling limit of the Bose-Hubbard model (\ref{eq:bose-hubbard}), it is also possible
to recover the Lieb-Liniger model (\ref{eq:lieb-liniger-model}) and hence, in the limit $U/t\gg 1$, the leading
corrections of $O(t/U)$ to the lattice TG Hamiltonian. These corrections can be
then used to compute various thermodynamic quantities of the  lattice and continuum TG
gases~\cite{cazalilla_hcbose_gases}. In the low-filling limit, the corrections yield an attractive interaction
between the Jordan-Wigner fermions, which is a particular case of the mapping found by~\citet{cheon1999_bosefermi}
between the Lieb-Liniger model and a model of spinless fermions with (momentum-dependent) attractive interactions.

 In the opposite limit of large lattice filling (i.e., $n_0 = N/L \gg 1$), the following
representation:
\begin{equation}
  \label{eq:phase-rep}
  \hat{b}^\dagger_j=\sqrt{\hat{n}_j} e^{-i\hat{\theta}_j},\quad\text{and}\quad
  \hat{b}_j=e^{i\hat{\theta}_j} \sqrt{\hat{n}_j}.
\end{equation}
in terms of the onsite particle-number ($\hat{n}_j$) and  phase ($\hat{\theta}_j$) operators becomes useful.
The phase operator $\hat{\theta}_j$ is defined as the canonical conjugate of $\hat{n}_j$, that is, we assume
that $\left[ \hat{n}_j,\hat{\theta}_j\right]=i\delta_{ij}$. Hence, $\hat{n}_j=i\frac{\partial}{\partial\theta_j}$
and the many-body wavefunctions become functions of the set of angles $\{ \theta_j \}_{j=1}^L$ where
$\theta_i \in [0, 2\pi)$. The angle nature of the $\theta_j$ stems from the discreteness of the eigenvalues of
$\hat{n}_j$. The operator $e^{i\hat{\theta}_j}$ decreases the eigenvalue of $\hat{n}_j$ by one, just like
$\hat{b}_j$ does, but if repeatedly applied to a eigenstate with small eigenvalue, it will yield unphysical
states with negative eigenvalues of $\hat{n}_j$. Indeed, this is one of the problems usually encountered when
working with $\hat{\theta}_j$. However, when the mean lattice filling $\langle\hat{n}_j\rangle=n_0\gg1$
these problems become less serious.

Using~(\ref{eq:phase-rep}), the kinetic energy term in the Bose-Hubbard model (\ref{eq:bose-hubbard})
becomes $-t\sum_{j}(\sqrt{\hat{n}_{j+1}} e^{i(\hat{\theta}_j-\hat{\theta}_{j+1})} \sqrt{\hat{n}_j}+\text{H.c.})$.
In addition, the interaction  energy and chemical potential  can be written as:
$ \sum_j \left[  \frac{U}{2} \hat{b}^{\dag}_j \hat{b}^{\dag}_j  \hat{b}_j \hat{b}_j  - \mu \hat{n}_j \right]
= \sum_j \left[ \frac{U}{2}  \hat{N}^2_j -  \delta  \mu \hat{N}_j \right]$, where $\hat{N}_j
= \hat{n}_j - \langle \hat{n}_j \rangle =  \hat{n}_j - n_0 = i \frac{\partial}{\partial \theta_j}$ and
$\delta \mu$ is the deviation of the chemical potential from the value for which $\langle\hat{n}_j\rangle=n_0$.
If the fluctuations of $\hat{N}_j$ are small, which is the case provided $U \gg t n_0$, $\hat{n}_j$ can be replaced
by $n_0$ and thus the following Hamiltonian is obtained:
\begin{equation}\label{eq:phasemodel}
 \hat{H}  =  \sum_{j=1}^L \left[ - E_J \cos (\hat{\theta}_j-\hat{\theta}_{j+1}) +
 \frac{E_C}{2} \hat{N}^2_j - \delta \mu \hat{N}_j \right],
\end{equation}
where $E_J = 2 n_0 t$, $E_C = U$. In the case of arrays of Josephson junctions (cf. Sec.~\ref{sec:experiments}),
the interaction is the long-range Coulomb potential, and a more realistic model is obtained by replacing
$\frac{E_C}{2}\sum_{j}\hat{N}^2_j$ by $\sum_{ij}V_{ij}\hat{N}_i\hat{N}_j$, where $V_{ij}$ is a function of $|i-j|$.

Interestingly, when formulated in the language of the Feynman path integral,  the
model in Eq.~(\ref{eq:phasemodel}) with $\delta \mu =0$ provides a particular example of a general relationship
between some  models of quantum bosons in  1D and some classical systems in 2D. As pointed
out by~\citet{feynman_statmech}, the partition function $Z =  \mathrm{Tr} \, e^{-\beta \hat{H}}$
($\beta = \frac{1}{k_BT}$, with $k_B$ being the Boltzmann constant) of a quantum system can be written as a
sum over all system configurations  weighted by $e^{-S}$,  where $S$ is the analytic
continuation of the classical action to imaginary time $\tau_E$. For the model in Eq.~(\ref{eq:phasemodel}),
$Z=\int \prod_{j=1}^L \left[d \theta_j\right] e^{-S[\theta_j]}$, where $\left[ d\theta_j \right]$ stands for
the integration measure over all possible configurations of the angle $\theta_j$, and $S[\theta_j]$ is
given by:
\begin{equation}\label{eq:actionphasemodel}
S[\theta_j] = \sum_{j=1}^L \int^{\hbar\beta}_0 d\tau_E \left[ \frac{\hbar \dot{\theta}^2_j}{2E_C} -
\frac{E_J}{\hbar} \cos \left(\theta_{j+1} - \theta_j \right)\right].
\end{equation}
where $\theta_j  = \frac{d\theta_j}{d\tau_E}$. The term  proportional to $\dot{\theta}^2_j$ in (\ref{eq:actionphasemodel})
plays the role of the kinetic energy, and stems from the operator
$E_C \: \hat{N}^2_j/2$  in the Hamiltonian, where  $\hat{N}_j = i\frac{\partial}{\partial \theta_j}$ plays the
role of the momentum.  Indeed, in
certain applications, such as \emph{e.g.}  the computational methods to be discussed in
Sec.~\ref{sec:numerical-results}, it is sometimes useful to discretize the integral over $\tau$.
A convenient discretization of the term $\propto \sum_j \dot{\theta}^2_j/2$ is provided by $-\sum_{j,\tau_T} \cos
\left( \theta_{j,\tau_E+\Delta \tau_E} - \theta_{j,\tau} \right)$. By suitably choosing the units of the
`lattice parameter' $\Delta\tau$, the  partition function corresponding to (\ref{eq:actionphasemodel}) can be related to that of
classical XY model. The latter describes a set of planar \emph{classical} spins $\mathbf{S}_\mathbf{r} = \left( \cos \theta_\mathbf{r},
\sin \theta_\mathbf{r}\right)$ interacting ferromagnetically on a  square lattice
and its \emph{classical} Hamiltonian reads:
\begin{equation}
H_{XY} = -\bar{J} \sum_{\langle \mathbf{r},\mathbf{r}^{\prime}\rangle} \mathbf{S}_{\mathbf{r}} \cdot
\mathbf{S}_{\mathbf{r}^{\prime}} = - \bar{J}  \sum_{\langle \mathbf{r},\mathbf{r}^{\prime}\rangle}
\cos \left( \theta_\mathbf{r} - \theta_{\mathbf{r}^{\prime}} \right).
\end{equation}
where $\mathbf{r} = (j, \frac{\tau_E}{\Delta \tau_E})$, and
$\langle \mathbf{r},\mathbf{r}^{\prime}\rangle$ means that the sum runs over nearest neighbor sites only.
Assuming that $\theta_{\mathbf{r}}$ varies slowly  with $\mathbf{r}$,
we may be tempted to take a continuum limit of the $XY$ model, where
$-\cos\left(\theta_\mathbf{r}-\theta_{\mathbf{r}^{\prime}}\right)$ is replaced by
$\frac{1}{2}\sum_{\langle \mathbf{r},\mathbf{r}^{\prime}\rangle}
\left(  \theta_\mathbf{r} - \theta_{\mathbf{r}^{\prime}}  \right)^2
= \frac{1}{2} \int d\mathbf{r}\, \left(\nabla_{\mathbf{r}} \theta \right)^2$.
Indeed, this procedure seems to imply that this model (and
the equivalent quantum model of Eq.~\eqref{eq:actionphasemodel})  are always superfluids
with linearly dispersing excitations in the quantum case. However, this
conclusion is incorrect as the  continuum limit neglects the existence of non-smooth
configurations of $\theta_\mathbf{r}$, which
are topological excitations corresponding to vortices of the classical $XY$ model and to
quantum phase slips in 1D  quantum models. The correct way of taking the continuum limit of models like
(\ref{eq:phasemodel}) or its ancestor, the Bose-Hubbard model, will be described in Sec.~\ref{sec:bosonize},
where the bosonization method is reviewed.

To sum up, we have seen that, equipped with the knowledge about a handful of 1D models, it is possible
to analyze a wide range of phenomena, which extend even  beyond 1D to 2D Classical Statistical Mechanics. In the following,
we begin our tour of 1D systems by studying the exactly solvable models, for which a rather comprehensive picture can
be obtained.

\section{Exact solutions}\label{sec:1dbosons-luttinger-homog} \label{sec:exactsol}

\subsection{The Tonks-Girardeau (hard-core boson) gas} \label{sec:tonks}

Let us first consider the Tonks-Girardeau (TG) model corresponding to the limit $c\to +\infty$ of
Eq. (\ref{eq:lieb-liniger-model}).
In this limit, the exact ground state energy was first obtained in \textcite{bijl_1937} and explicitly derived in \textcite{nagamiya_1940}.
The infinitely strong contact repulsion between the bosons
imposes a constraint in the form that any  many-body wavefunction of the TG gas
must vanish every time two particle meet at the same point. As first pointed out by \textcite{girardeau_bosons1d},
this constraint can be implemented by writing the wavefunction $\psi_B(x_1,\ldots)$ as follows:
\begin{equation}
  \label{eq:girardeau-transformation}
  \psi_B(x_1,\ldots,x_N)=S(x_1, \ldots, x_N)\:
\psi_F(x_1,\ldots,x_N),
\end{equation}
where $S(x_1, \ldots, x_N)= \prod_ {i>j=1}^N \mathrm{sign}(x_i-x_j)$ and $\psi_F(x_1,\ldots)$
is the many-body wavefunction of a (fictitious) gas of spinless fermions. Note that
the function $S(x_1,\ldots)$  compensates the sign change  of $\psi_F(x_1, \ldots)$
when any two particles are  exchanged and thus yields a wavefunction obeying Bose statistics.
Furthermore, eigenstates must  satisfy
the non-interacting Schr\"odinger equation when all $N$ coordinates are different. Hence,
in the absence of an external potential, and on a ring of circumference $L$ with periodic boundary conditions
(i.e., $\psi_{B}(x_1,\ldots,x_j+L,\ldots,x_N)=\psi_{B}(x_1,\ldots,x_j,\ldots,x_N)$), the (unnormalized)  ground state
wavefunction has the Bijl-Jastrow form:
\begin{equation}\label{eq:girardeau-gs-pbc}
  \psi_B^{0}(x_1,\ldots,x_N) \propto \prod_{i<j} \sin
  \frac{\pi}{L}|x_i-x_j|.
\end{equation}
This form of the ground state wavefunction is generic of various 1D models in the limit of infinitely strong repulsion and of the Calogero-Sutherland model as we will see later.
The ground state energy of the TG gas is then $E=\frac{\hbar^2 (\pi \rho_0)^2}{6\pi m}$ where $\rho_0=N/L$ is the mean
particle density; and the energy of the lowest excited state is: $E(q)\sim \frac{\hbar^2\pi \rho_0 }{m} |q|$ for
$|q|\ll \pi \rho_0$ suggesting a linear phonon spectrum. Like the free Fermi gas, the TG gas has a finite
compressibility at $T=0$ and a specific heat linear in temperature. As we will see  in Sec.~\ref{sec:bosonize},
these properties are  not limited to the case of infinite contact repulsion but are generic features of
interacting 1D bosons.

\subsubsection{Correlation functions in the continuum}\label{sec:corr-cont}

Besides thermodynamic properties, the mapping of (\ref{eq:girardeau-transformation}) also allows the calculation
of the correlation functions of the TG gas. Since Eq.~(\ref{eq:girardeau-transformation}) implies that the
probability of finding particles at given positions  is the same in the TG and in the free
spinless Fermi gases, all the density correlation functions of  both gases are
identical. In particular, the  pair correlation function, $D_2(x) = \langle \rho(x) \rho(0) \rangle/\rho^2_0$
is given:
\begin{equation}
\label{eq:paircorrelations_TG}
  D_2(x)=1-\left[\frac{\sin (\pi \rho_0 x)}{\pi \rho_0 x}\right]^2.
\end{equation}

 However, computing the one-particle density matrix,
\begin{multline} \label{eq:girardeau_2points}
  g_1(x,y) = \int \prod_{i=2}^N d x_i \,  \psi^*_B(x,\ldots x_N) \psi_B(y,\ldots x_N),
\end{multline}
 is considerably more involved. Upon inserting~(\ref{eq:girardeau-transformation}) into Eq.~(\ref{eq:girardeau_2points})
 we see that the  $\mathrm{sign}$ functions do not cancel out and therefore the bosonic and fermionic correlations are not
anymore identical.

 Nevertheless, it can be shown~\cite{schultz_1dbose} that $g_1(x,y)$ can be expressed as the Fredholm determinant of
a linear integral equation~\cite{tricomi_integeq}. Another representation~\cite{lenard64_bose1d} of $g_1(x,y)$
is in terms of a Toeplitz determinant. Such determinants have been thoroughly studied in relation to
the 2D Ising model~\cite{wu_determ,szego_determ}. The asymptotic long-distance behavior of the
Toeplitz determinant can be obtained~\cite{ovchinnikov2009} from the Fisher-Hartwig theorem~\cite{fisher-hartwig68},
which yields:
\begin{equation}\label{eq:tonks-ovchinnikov}
 g_1(x)=\rho_0 G^4(3/2) \left[\frac 1 {2 \rho_0 L |\sin (\pi x/L)|}\right]^{1/2},
\end{equation}
where $G(z)$ is Barnes' $G$ function; $G(3/2)=A^{-3/2} \pi^{1/4} e^{1/8} 2^{1/24}$;
$A=1.28242712\ldots$ is Glaisher's constant. Thus, $G^4(3/2)=1.306991\ldots$.  In the thermodynamic limit, a more
complete asymptotic expression is available~\cite{vaidya_bose1d_long,vaidya79,Gangardt_correlations}:
\begin{widetext}
\begin{equation}
  \label{eq:vaidya-girardeau}
  g_1(x)=\frac{\rho_\infty}{|\pi \rho_0 x|^{1/2}}
 \left[1- \frac 1 8 \left(\cos(2\pi
    \rho_0 x) +\frac 1 4\right) \frac 1 {(\pi \rho_0 x)^2}- \frac 3 {16}
    \frac{\sin (2\pi \rho_0 x)}{(\pi \rho_0 x)^3} + \frac{33}{2048}
    \frac{1}{(\pi \rho_0 x)^4} + \frac{93}{256} \frac{\cos (2\pi
      \rho_0 x)}{256 (\pi \rho_0 x)^4} +\ldots \right],
\end{equation}
\end{widetext}
where $\rho_\infty=G(3/2)^4/\sqrt{2}=\pi e^{1/2}2^{-1/3}A^{-6}$. The leading term of (\ref{eq:vaidya-girardeau})
agrees  with~(\ref{eq:tonks-ovchinnikov})  for $L\to \infty$.
The slow power law
decay of $g_1(x)$ at long distances leads to a divergence in
the momentum distribution: $n(k) \sim |k|^{-1/2}$ for $k\to 0$. The lack of a Delta function at $k = 0$
in $n(k)$ implies the absence of BEC. However, the $\sim k^{-1/2}$ divergence  can be viewed as a remnant of the
tendency of the system to form a BEC.  The power-law behavior of $g_1(x)$  as $|x|\to \infty$ (or $n(k)$ as $k\to 0$)  
is often referred to as \emph{quasi-long range} order.

Alternative ways of deriving the $g_1(x)$ follow from the analogy~\cite{lenard64_bose1d}
between the ground state wavefunction of the TG gas and the distribution of eigenvalues of random
matrices~\cite{mehta2004} from the circular unitary ensemble (CUE). It is also possible to show
\cite{jimbo1980,forrester_painleve} that the density matrix of the TG gas  satisfies
the Painlev\'e V nonlinear second order differential equation~\cite{ince1956}.  Furthermore, the Fredholm
determinant representation can be generalized to finite temperature~\cite{lenard66}.

The asymptotic expansion of the one-particle density matrix function at finite temperature in the grand-canonical
ensemble has been derived~\cite{its_tfinite_1991}. Asymptotically with distance, it decays exponentially, and, for
$\mu>0$,  the dominant term reads:
\begin{equation}\label{eq:its-girardeau-thermal}
g_1(x,0,\beta)\sim \frac{\sqrt{2m \beta^{-1}}} {\pi\hbar} \rho_\infty e^{-2|x|/r_c}F\left(\beta\mu\right),
\end{equation}
where $\beta^{-1} = k_B T$ and
 $\rho_\infty$ is the same constant as in Eq.~(\ref{eq:vaidya-girardeau}),
 the correlation radius $r_c$ is given  by
\begin{equation}\label{eq:rc}
r_c^{-1}=\frac{\sqrt{2m \beta^{-1}}} {2\pi \hbar}
\int_{-\infty}^{\infty} d\lambda \ln
\left|\frac{e^{\lambda^2-\beta \mu}+1}{e^{\lambda^2-\beta \mu}-1}\right|,\
\end{equation}
whereas $F(u)$ is a regular function of $u = \beta \mu$ given by:
\begin{eqnarray}
F(u)&=&\exp\left[-\frac 1 2 \int_u^\infty d\sigma
\left(\frac{dc}{d\sigma}\right)^2 \right],\\
\quad c(\sigma)&=& \frac 1 \pi\int_{-\infty}^{\infty} d\lambda \ln
\left|\frac{e^{\lambda^2-\sigma}+1}{e^{\lambda^2-\sigma}-1}\right|\nonumber.
\end{eqnarray}
Two regimes can be distinguished: For $0< \beta\mu \ll 1$, the correlation length $r_c$ is just proportional to
the De Broglie thermal length $\frac{\sqrt{2m \beta^{-1}}} {\hbar}$. For $1 \ll \beta \mu$, the integral in (\ref{eq:rc})
is dominated by $\lambda$ in the vicinity of $\pm \sqrt{\beta\mu}$. By linearizing $\lambda^2$ in the
vicinity of these points, the integral can be shown to be proportional to $(\beta\mu)^{-1/2}$. As a result, one
obtains $r_c \sim \hbar v_F \beta$ (where $v_F = \frac{\hbar\pi \rho_0}{m}$ is the Fermi velocity)
for a degenerate TG gas. For the one-particle Green's function,
$G^{<}_B(x,\tau;\beta) = \langle \hat{\Psi}^{\dag}(x,\tau)\hat{\Psi}(0,0) \rangle$
\textcite{korepin_book} showed that:
\begin{equation}
G^{<}_B(x,\tau;\beta) \propto \exp\left[\int_{-\infty}^{\infty} \frac{dk}{2\pi} \left|x
+ t \frac{\hbar k}{m}\right| \ln
\left|\frac{e^{\beta\frac{\hbar^2 k^2}{2m}-\beta\mu}-1}{e^{\beta\frac{\hbar^2 k^2}{2m}-\beta\mu}+1}\right|\right].
\end{equation}
For $1 \ll \beta\mu$, one can use the previous expansion of the integrand around the points
$k= \pm k_F =\pm \sqrt{2m \mu}/\hbar$, which gives
$G^{<}_B((x,\tau;\beta) \propto \exp[-\frac{\pi}{4 \beta \hbar v_F}(|x-v_F t|+|x+v_F t|)]$.
These results can  be also recovered by using the field theoretic approach reviewed  in Sec.~\ref{sec:bosonize}.


So far, we have dealt with the homogeneous TG gas, which can be relevant for condensed matter systems.
However, ultracold atomic gases are confined by inhomogeneous potentials. Fortunately, as mentioned in
Sec.~\ref{sec:bosons_conti}, the TG gas remains solvable even in the presence of an external potential.

The Bose-Fermi mapping, Eq.~\eqref{eq:girardeau-transformation}, is still valid in the presence of any external
confining potential, $V_\mathrm{ext}(x)$. However, the eigenstates are constructed from the Slater determinants:
\begin{equation}
 \label{eq:tgwf_general}
 \psi_F(x_1,\ldots,x_N) = \frac{1}{\sqrt{N!}}\det_{n=0,j=1}^{n=N-1,j=N} \varphi_n(x_j),
\end{equation}
where $\varphi_n(x_j)$ are the eigenfunctions  of the non-interacting  Schr\"odinger
equation in the  presence of $V_\mathrm{ext}(x)$. Considerable simplification results
from the fact that the confining potential  in experiments is, to a good  approximation,
harmonic:  $V_\mathrm{ext}(x)  = \frac{1}{2} m \omega^2 x^2$.
Using \eqref{eq:tgwf_general} and \eqref{eq:girardeau-transformation},
the  eigenfunctions of  many-body Hamiltonian are constructed from the harmonic oscillator orbitals:
$\varphi_n(x)=\frac{\mathcal{H}_n(x/\ell_{HO})}{\sqrt{2^nn! \ell_{HO} \sqrt{\pi}}}e^{-x^2/2\ell^2_{HO}}$,
where $\mathcal{H}_n(z)$ are the Hermite polynomials and $\ell_{HO} = \sqrt{\hbar/(m\omega)}$ the oscillator length
(in what follows and unless otherwise stated, all lengths are measured in units where $\ell_{HO} = 1$), which yields
the  following ground state wavefunction:
\begin{equation}
 \label{eq:tgwf-harmonic}
 \psi^0_B(x_1,\ldots,x_N) =C_N^{-1/2} \prod_{k=1}^N e^{-x_k^2/2}
 \prod_{i>j=1}^{N} |x_i-x_j|,
\end{equation}
where $C_N= N!\prod_{n=0}^{N-1} 2^{-n}\sqrt{\pi}\,n!$. Also in this case,
the special form of the mapping
\eqref{eq:girardeau-transformation} implies that the density profile and density correlations
of the harmonically trapped TG  gas are identical to those of a  harmonically trapped
gas of non-interacting spinless fermions.
The latter have been studied in detail~\cite{vignolo00,brack01} and will not be review here.
Instead, we focus on the off-diagonal  correlations and the momentum distribution which, as in the homogeneous
case, are different for the Fermi and TG gases.

The one-particle density matrix, $g_1(x,y)$, can be obtained evaluating the ($N$-$1$)-dimensional integral in
\eqref{eq:girardeau_2points} using \eqref{eq:tgwf-harmonic}. From this  result,
we can obtain the natural orbitals (cf.~section~\ref{sec:bosons_larged}), which obey
$ \int dy \, g_1(x,y)\, \phi_{\alpha}(y)= N_\alpha \,\phi_\alpha(x)$, as well as the momentum distribution:
$n(k)=\frac{1}{2\pi}\int dx\,dy\, g_1(x,y)\, e^{ik(x-y)}$. In the ground state, $N_{0}$ is the largest eigenvalue,
followed by $N_1$, etc. As discussed in Sec.~\ref{sec:bosons_larged}, when the system exhibits BEC, $N_0$ is of
$O(N)$ \cite{penrose56,leggett01}. Furthermore, note that, because the trapped system is not translationally invariant,
$N_{\alpha}$  and $n(k)$ are not proportional to each other. In other words, the natural orbitals $\phi_{\alpha}(x)$ are not
plane waves.

For small harmonically trapped systems, $g_1(x,y)$ was first computed by \citet{girardeau01,lapeyre02}.
To obtain the thermodynamic limit scaling of $n(k=0)$ and $N_0$ with $N$, \textcite{papenbrock03} studied $g_1(x,y)$ for
larger systems by writing the integrals in (\ref{eq:girardeau_2points}) in terms of an integration measure
that is identical to the joint probability density
for eigenvalues of $(N-1)$-dimensional random matrices  from the Gaussian unitary ensemble (GUE), and
expressed the measure in terms of harmonic oscillator orbitals. This enabled the computation of $g_1(x,y)$
in terms of determinants of the $(N-1)$-dimensional matrices:
\begin{eqnarray}
 \label{eq:opdm-harmonic1}
 g_1(x,y)&=&\frac{2^{N-1}e^{-(x^2+y^2)/2}}{\sqrt{\pi}(N-1)!}\,
 \det_{m,n=0}^{m,n=N-2}[B_{m,n}(x,y)], \nonumber\\
 B_{m,n}(x,y)&=&\int_{-\infty}^\infty dz|z-x||z-y|\varphi_m(z)\varphi_n(z).
\end{eqnarray}
The form of $g_1(x,y)$ above and its relation to the GUE had earlier been discussed by \citet{forrester_painleve}.
Equation \eqref{eq:opdm-harmonic1} allowed \textcite{papenbrock03} to study $g_1(x,y)$ and $n(k)$ for up to $N=160$.
The leading $N$ behavior of $n(k=0)$ was found to be $n(k=0)\propto N$. The behavior of $\lambda_0$ was then
inferred from the result for $n(k=0)$ and a scaling argument, which resulted in $N_0\propto\sqrt{N}$.

A detailed study of the lowest natural orbitals and their occupations $N_\alpha$ in a harmonic trap was given by
\citet{forrester_trapped_bosons}.  Using a numerical approach based on an expression similar to
Eq.~\eqref{eq:opdm-harmonic1}, they computed $g_1(x,y)$ and obtained the natural orbitals  by a quadrature
method  up to $N = 30$. By fitting the results of the two lowest natural orbitals ($\alpha=0,1$) to a law
$N_\alpha=aN^p+b+cN^{-q}$, they found that
\begin{eqnarray}
 \label{eq:NOForrester}
 N_0&=&1.43\sqrt{N}-0.56+0.12N^{-2/3}\nonumber\\
 N_1&=&0.61\sqrt{N}-0.56+0.12N^{-4/3}.
\end{eqnarray}
Furthermore, a mapping to  a classical Coulomb gas allowed \citet{forrester_trapped_bosons} to obtain an
asymptotic expression  for  $g_1(x,y)$ of the harmonically trapped TG gas in the limit of large $N$. Their
result reads:
\begin{equation}
 \label{eq:OPDMForrester}
 g_1(x,y)=N^{1/2} \frac{G^4(3/2)}{\pi} \frac{(1-x^2)^{1/8}(1-y^2)^{1/8}}{|x-y|^{1/2}}.
\end{equation}
This expression shows that $g_1(x,y)$ in the trap exhibits a power-law decay similar to the one found in
homogeneous systems. Using a scaling argument the behavior of $N_{\alpha=0,1}$ can be obtained from
Eq.~\eqref{eq:OPDMForrester}, which reproduces the leading behavior obtained numerically
[Eq.~\eqref{eq:NOForrester}]. In addition, the one-particle density matrix in harmonic traps was also
studied by~\textcite{Gangardt_correlations} using a modification of the replica trick. The leading order
in $N$ obtained by Gangardt agrees with~\eqref{eq:OPDMForrester}, and his
method further allows to obtain the finite-size corrections to Eq.~\eqref{eq:OPDMForrester}. Indeed,
the leading corrections in the trap and  homogeneous systems (cf. Eq.~\ref{eq:vaidya-girardeau})
are identical.

Another quantity of interest is the momentum distribution, $n(k)$. Whereas the small $k$ behavior can be
obtained from the asymptotic formula~(\ref{eq:OPDMForrester}), the large $k$ asymptotics gives information
about  short distance one-particle correlations not captured by  Eq.~\eqref{eq:OPDMForrester}.
For two hard-core bosons in a harmonic trap, \textcite{minguzzi02}
found that $n(k\to +\infty)\sim k^{-4}$. Since the singularities
arising in the integrals involved also appear in the many-body case, this behavior was believed to hold for arbitrary
$N$. Similar results were obtained numerically for up to eight particles in harmonic traps \cite{lapeyre02}
and analytically, using asymptotic expansions, for homogeneous systems \cite{forrester_trapped_bosons}.

\begin{figure}[!h]
\includegraphics[width=0.4\textwidth]{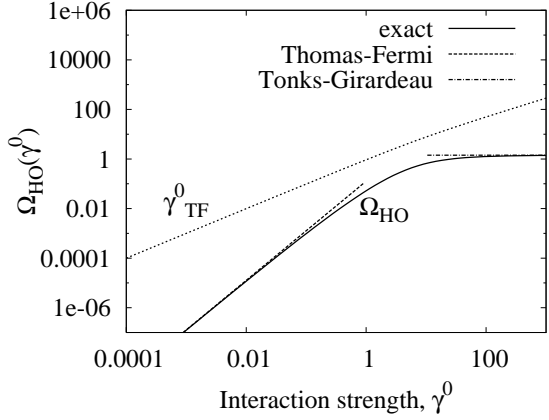}
\caption{\label{fig:p4_Olshanii} Dimensionless coefficient $\Omega_{HO}(\gamma^{0})$ (see text) as a function of
the interaction strength $\gamma^{0}$ at the center of the trap. The dotted line shows the Thomas-Fermi estimate
$\gamma^{0}_{\rm TF} = (8/3^{2/3}) (N m a_{\rm 1D}^2 \omega / \hbar)^{-2/3}$ for the interaction strength in the
center of the system as a function of $\gamma^{0}$. The numerical results are compared with the asymptotic
expressions in the Thomas-Fermi and Tonks-Girardeau regimes \cite{olshanii03}.}
\end{figure}

\textcite{olshanii03} later showed that the tail $\sim k^{-4}$ is a generic feature of delta-function interactions,
i.e., it applies to the Lieb-Liniger model at all values of the dimensionless parameter
$\gamma = c/\rho_0$. The behavior can be traced back to the kink
in the first derivative of the exact eigenstates at the point where two particles meet.
For homogeneous systems~\cite{olshanii03}:
\begin{equation}
n(k \to \infty)= \frac{1}{\hbar\rho_0}\frac{\gamma^2e'(\gamma)}{2\pi}
\left(\frac{\hbar\rho_0}{k}\right)^4,
\end{equation}
where the calculation of  $e(\gamma)$  is discussed in Sec.~\ref{sec:lieb}, and $n(k)$
is normalized so that $\int dk\, n(k)=1$. Further results for harmonically
trapped systems were obtained by means of the local density approximation (LDA).
In trapped systems, $n(k \to \infty)\sim \Omega_{HO} (\hbar \rho^0_0)^3/k^4$,
where $\Omega_{HO}$ is a dimensionless quantity and $\rho_0(x=0)$ is the density at the trap center.
Numerical results for $\Omega_{HO}$ for different values of $\gamma_0$ (where $\gamma_0$ is the $\gamma$ parameter
in the center of the trap) are shown in Fig.~\ref{fig:p4_Olshanii}. Based on the observed behavior of
$\Omega_{HO}$, \citet{olshanii03} proposed that measuring the high-$k$ tail of $n(k)$ allows one to
identify the transition between the weakly interacting Thomas-Fermi and the strongly interacting
Tonks-Girardeau regimes.

 A straightforward and efficient approach to computing the one-particle density matrix of hard-core bosons in generic
potentials in and out of equilibrium was introduced by \citet{pezerexacttonks}. Indeed, $g_1(x,y,\tau) = \langle \Psi^{\dag}(x,\tau) \Psi(y,\tau)\rangle$ (notice the  addition of the dependence on time $\tau$) can be written in terms of the
solutions   ($\varphi_i(x,\tau)$) of the single-particle  time-dependent Sch\"ordinger equation relevant to the problem:
\begin{equation}
 \label{eq:buljanini}
 g_1(x,y,\tau)=\sum_{i,j=0}^{N-1} \varphi^*_i(x,\tau)A_{ij}(x,y,\tau)\varphi_j(y,\tau),
\end{equation}
where, from the general definition of the many-body Tonks-Girardeau wave-function [Eq.~\eqref{eq:tgwf_general}]
and of the one-particle density matrix [Eq.~\eqref{eq:girardeau_2points}], one can write
\begin{equation}
 \label{eq:buljanfin}
 {\bf A}(x,y,\tau)=(\det{\bf M})\, ({\bf M}^{-1})^T,
\end{equation}
where $M_{ij}(x,y,\tau)=\delta_{ij}-2\int_x^y dx' \varphi^*_i(x',\tau)\varphi_j(x',\tau)$ (notice that $x<y$ without loss
of generality). This approach has allowed the study of hard-core boson systems out of equilibrium
\cite{pezerexacttonks,buljan07} and was generalized to study hard-core anyons by \citet{campo08}.

\subsubsection{Correlation functions on the lattice}\label{sec:corr-latt}

As it was briefly mentioned in Sec.~\ref{sec:mapping}, for hard-core bosons on the lattice, the Jordan-Wigner
transformation, Eq.~(\ref{eq:jw-transf}), plays the role of Girardeau's Bose-Fermi mapping
(\ref{eq:girardeau-transformation}). Thus, as in continuum case, the spectrum,
thermodynamic functions and the correlation function of the operator $\hat{n}_j$  are identical to those of the non-interacting
spinless lattice Fermi gas. However, the calculation of the one-particle density matrix is still a non-trivial problem requiring
similar  methods to those reviewed in Sec.~\ref{sec:corr-cont}.

 Using the Holstein-Primakoff transformation~\eqref{eq:holstein_primakoff}, the one-particle
 density matrix $g_1(m-n) =  \langle \hat{b}^\dagger_m  \hat{b}_{n}\rangle$
can be expressed in terms of spin correlation functions~\cite{barouch1970,barouch1971a,barouch1971b,johnson1971,mccoy1971,ovchinnikov2002_asymp_xx,ovchinnikov2004_ff_xx}: $g_1(m-n)= \langle \hat{\sigma}^+_m \hat{\sigma}^-_{n}\rangle = \langle \hat{\sigma}^x_m \hat{\sigma}^x_{n} \rangle +
\langle \hat{\sigma}^y_m \hat{\sigma}^y_{n} \rangle =  2  \langle \hat{\sigma}^x_m \hat{\sigma}^x_{n} \rangle$
(note that  $\langle \hat{\sigma}^x_n \hat{\sigma}^y_m\rangle =0$ and  $\langle \hat{\sigma}^x_n \hat{\sigma}^x_m\rangle=
\langle \hat{\sigma}^y_n \hat{\sigma}^y_m\rangle$ by the  $U(1)$ symmetry of the $XX$ model).  Thus, let us consider the following set of correlation functions: $S_{\nu\nu}(n-m,\tau)=\langle  \hat{\sigma}_m^\nu(\tau) \hat{\sigma}_{n}^\nu(\tau)\rangle$ where
$\nu=x,y,z$. We first note that~\cite{lsm_61}:
\begin{equation}
\label{eq:jwexp}
e^{i\pi \hat{c}^\dagger_m \hat{c}_m}=1-2 \hat{c}^\dagger_m \hat{c}^{}_m = \hat{A}_m \hat{B}_m,
\end{equation}
where $\hat{A}_m=\hat{c}^\dagger_m+\hat{c}_m$ and $\hat{B}_m=\hat{c}^\dagger_m-\hat{c}_m$.  Hence,
\begin{eqnarray}\label{eq:lsm}
S_{xx}(l-m)&=& \frac 1 4 \langle \hat{B}_l \hat{A}_{l+1} \hat{B}_{l+1}
\ldots \hat{A}_{m-1} \hat{B}_{m-1} \hat{A}_m \rangle \nonumber \\
S_{yy}(l-m)&=& \frac{1}{4} (-1)^{l-m}\langle \hat{A}_l \hat{B}_{l+1} \hat{A}_{l+1}
\ldots \hat{B}_{m-1} \hat{A}_{m-1} \hat{B}_m\rangle \nonumber \\
S_{zz}(l-m) &=&\frac 1 4 \langle \hat{A}_l \hat{B}_{l} \hat{A}_{m} \hat{B}_{m}\rangle.
\end{eqnarray}
Using Wick's theorem \cite{caianiello_52,barouch1971a}, these expectation values are reduced to
Pfaffians~\cite{itzykson-zuber}. At zero temperature, the correlators over the partially filled Fermi sea have
the following form $\langle \Psi_F |\hat{A}_l \hat{A}_m |\Psi_F \rangle = 0$, $\langle \Psi_F |\hat{B}_l \hat{B}_m |\Psi_F \rangle = 0$, and
$\langle \Psi_F |\hat{B}_l\hat{A}_m|\Psi_F \rangle =2G_0(l-m)$, where
$G_0(R) = \langle \Psi_F |\hat{c}^\dagger_{m+R}\hat{c}_m |\Psi_F
\rangle$ is the free-fermion one-particle correlation function on a finite chain.
The Pfaffians in Eq.~(\ref{eq:toeplitz}) then reduce to the Toeplitz determinant~\cite{lsm_61} of a $R\times R$
matrix: $G(R) = \text{det}_{l m} [2G_0(l-m-1)],\, l, m = 1, \ldots, R$. Thus,
\begin{equation}\label{eq:toeplitz}
S_{xx}(R)=\frac{1}{4}\, \vline
\begin{array}{cccc}
  G_{-1} & G_{-2} & \ldots & G_{-R} \\
  G_{0} & G_{-1} & \ldots  & G_{-R+1} \\
  . & . & \ldots & . \\
  . & . & \ldots & . \\
  . & . & \ldots & . \\
  G_{R-2} & G_{R-3}  & .. & G_{-1}
\end{array}
\vline\, .
\end{equation}
By taking the continuum limit of Eq.~(\ref{eq:toeplitz}) as explained in Sec.\ref{se:josephson-th},
the Toeplitz determinant representation of the continuum TG gas
is recovered. For a half-filled lattice with an even number of sites $L$ and a number of particles
$N=L/2$ odd~\cite{ovchinnikov2004_ff_xx}, the free fermion density-matrix
$G_0(l)= \frac{\sin( \pi l/2)}{L \sin(\pi l/L)}$, so that  $G_0(l) = 0$ for even $l$, and the Toeplitz determinant
(\ref{eq:toeplitz}) can be further simplified to yield $S_{xx}(R) = \frac 1 2 (C_{N/2})^2$
(for even $R$), $S_{xx}(R) = -\frac 1 2 C_{(R-1)/2} G_{(R+1)/2}$ (for odd $N$), where $C_R$ is the determinant of
the $R \times R$ matrix: $C_R = \textrm{det}_{l m} ((-1)^{l-m}G_0(2l - 2 m - 1)),\, l,m = 1,\ldots,R$.

On a finite chain and for odd $N$, $C_R$ is a Cauchy determinant, showed by
\textcite{ovchinnikov2002_asymp_xx} to yield:
\begin{equation}
\label{eq:cauchy_det} C_R = \left(\frac{2}{\pi}\right)^R \prod_{k=1}^{R-1} \left[
\frac{\sin\left(\frac{\pi(2k)}{L}\right)^2}{\sin\left(\frac{\pi(2k + 1)}{L}\right)
\sin\left(\frac{\pi(2k - 1)}{L}\right)}\right]^{R-k}.
\end{equation}
In the thermodynamic limit, Eq.~(\ref{eq:cauchy_det}) reduces to
\begin{equation}
  C_R= (2/\pi)^R \prod_{k=1}^{R-1} \left(\frac{4 k^2}{4k^2-1}\right)^{R-k},
\end{equation}
and hence the one-body density matrix,
\begin{equation}
  g_1(R) = 2 S_{xx}(R)\sim \frac{2C_0}{\sqrt{\pi}} \left(\frac{1}{R^{1/2}}
      -\frac{(-1)^R}{8R^{5/2}} \right) \label{eq:latticetgg1}
\end{equation}
for large $R$, where $2C_0/\sqrt{\pi}=0.588352\ldots$ can be expressed in terms of Glaisher's
constant~\cite{wu_determ,mccoy68,ovchinnikov2004_ff_xx}. Thus, also in this case,  the one-particle
density matrix of the bosons decays as a power law, indicating the absence
of a BEC at $T=0$. For positive temperature, the power law behavior is
replaced by an exponential decay~\cite{its1993}.

 Next, we turn to the dynamical correlations,
  $G^{<}_{\nu \nu}(R,\tau)  =\langle e^{i \hbar{H}\tau/\hbar} \hat{\sigma}_m^\nu e^{-i\hat{H}\tau/\hbar}
 \hat{\sigma}_{m+R}^\nu\rangle, \nu=x,y$. To obtain these objects, it  is convenient to consider
 the four spin correlator~\cite{vaidya78,mccoy1971}:
\begin{equation}
\label{eq:four-spin-corr} C_{\nu \nu}(R,\tau,N)=\langle
\hat{\sigma}_{1+\frac{N}{2}}^\nu(\tau)\hat{\sigma}_{1-R+N}^\nu(\tau)\hat{\sigma}_{1}^\nu(0)
\hat{\sigma}_{1-R+\frac{N}{2}}^\nu(0)\rangle,
\end{equation}
which is obtained by evaluating a Pfaffian~\cite{lsm_61,mccoy1971}. From the
cluster property,
{\setlength\arraycolsep{0.5pt}
\begin{eqnarray}
\label{eq:four-spin-corr-exp}
\lim_{N \to \infty} C_{\nu \nu}(R,\tau,N)= \left[G^{<}_{\nu \nu}(R,\tau)\right]^2.
\end{eqnarray}}
 $G^{<}_{\nu \nu}(R,\tau)$ can be obtained.  The long-time behavior of the
 one-particle Green's function  $G^{<}_B(R,\tau) =  \langle \hat{b}^{\dagger}_{n+R}(\tau) \hat{b}_n(0)\rangle$
was computed in~\cite{muller1983,muller1984}  for $R=0$ following a method due to
\citet{mccoy83a,mccoy83b}. These authors found that $G_B^{<}(0,\tau) \sim (i\tau)^{-1/2}$.
When comparing this result with Eq.~(\ref{eq:latticetgg1}),
we see that the leading asymptotic behavior is controlled by the same exponent ($=\frac{1}{2}$)
both in space and time.  This is a  consequence of the conformal invariance of the underlying field theory
(see Sect.~\ref{sec:bosonize}   and~\cite{gogolin_1dbook,giamarchi_book_1d} for an
extended discussion)

 At finite temperature, the asymptotic behavior of the  Green's function
has been obtained by~\textcite{its1993}:
\begin{equation}
  \label{eq:its-spacelike}
 G^{<}_B(R,\tau)
  \sim  \exp\left[|R|
   \int_{-\pi}^\pi \frac{dk}{2\pi} \ln \left|\tanh\beta\left(\mu-2 t \cos
          k\right)\right|\right],
\end{equation}
in the space-like regime (for $|R| > \frac{4\tau t}{\hbar}, \beta = \frac{1}{k_B T}$), and
{\setlength\arraycolsep{1pt}
\begin{eqnarray}
\label{eq:its-timelike}
   G^{<}_B(R,\tau)  &\sim&   \tau^{2(\nu_+^2+\nu_-^2)}
    \exp\Big[ \int_{-\pi}^\pi \frac{dk}{2\pi}\, |R -4\tau \sin k  | \nonumber\\
    &&\times \ln \left|\tanh\beta\left(\mu - 2t \cos k \right)\right|\Big],
\end{eqnarray}
in the time-like regime (for $|R|<4\tau t/\hbar$), where
\begin{equation}
  \label{eq:its-exponents}
  \nu_\pm=\frac 1 {2\pi} \ln \left|\tanh\beta \left(-\mu \mp 2 \cos
          p_0\right)\right|,
\end{equation}
where $p_0$ is defined by the equation $\sin p_0 = |R| \hbar /4\tau t $. For $\beta \mu \gg 1$ and $|\mu/t|<2$, one can expand in
the vicinity of $\pm k_F = \pi n_0$ such that $\mu=2t\cos k_F$, and obtain that the
correlation functions (\ref{eq:its-spacelike}) and (\ref{eq:its-timelike})
decay exponentially with $x$ on a lengthscale $\frac{\pi}{4\beta \hbar v_F}$,
as found previously in the continuum case. For infinite
temperature, the correlation functions are local and
gaussian~\cite{sur1975}.


Finally, we shall consider the effects of a trapping potential.
As it will be discussed Sec.~\ref{sec:experiments}, using a deep periodic potential generated by an optical lattice
provides us with a powerful tool to reach the strongly correlated regime where the system  behaves essentially
as a lattice TG gas in the presence of a trap~\cite{paredes04}.

Like in the continuum case, the lattice TG gas remains exactly solvable when
an external potential $\hat{V}_{\mathrm{ext}}$
is added to $\hat{H}_{LTG}$ (Eq.~\ref{eq:lattice-hcb}). For the experimentally relevant
harmonic potential, the Jordan-Wigner transformation maps the
model described by $\hat{H}_{LTG}+\hat{V}_\mathrm{ext}$ to the following fermionic
Hamiltonian:
\begin{equation}
\label{eq:HamFreeFtrap}
\hat{H}_F = \sum_{j}\left[ -t \left( \hat{c}^\dagger_{j} \hat{c}_{j+1}
+ \text{H.c.} \right) + (V x_i^2- \mu) \hat{n}_j \right],
\end{equation}
where $x_i = i a$, $a$ being the lattice parameter and $V$ the curvature of the trapping potential.
The above Hamiltonian can be easily diagonalized since it is quadratic.
Using the single-particle eigenstates of \eqref{eq:HamFreeFtrap}, the momentum distribution $n(k)$
can  be computed using Toeplitz determinants~ \cite{paredes04},
as explained before. In addition, an alternative and computationally more efficient way of
calculating one-particle correlations in the lattice, for arbitrary external potentials, was introduced by
\citet{rigol_muramatsu_04HCBa,rigol_muramatsu_05HCBb}.  As above, the one-particle density matrix
is expressed in term of spin operators
$g_1(i,j) = \langle \hat{b}^\dagger_i\hat{b}_j\rangle=\langle \hat{\sigma}^+_i\hat{\sigma}^-_j\rangle=
\delta_{ij}+(-1)^{\delta_{ij}} \langle \hat{\sigma}^-_j\hat{\sigma}^+_i\rangle$. Hence, in order
to determine $g_1(i,j)$ it suffices to calculate:
\begin{eqnarray}
\label{eq:greensHCB}
 G(i,j) &=&\langle \hat{\sigma}^{-}_i\hat{\sigma}^+_j\rangle=\langle\Psi_{F}|\prod^{i-1}_{\beta=1}
e^{i\pi \hat{c}^{\dag}_{\beta}\hat{c}^{}_{\beta}} \hat{c}^{}_i \hat{c}^{\dag}_j
\prod^{j-1}_{\gamma=1} e^{-i\pi \hat{c}^{\dag}_{\gamma}\hat{c}^{}_{\gamma}}
|\Psi_{F}\rangle \nonumber \\
&=&\det\left[ \left( {\bf P}^{i} \right)^{\dag}{\bf P}^{j}\right],
\end{eqnarray}
where $|\Psi_{F}\rangle=\prod^{N}_{\kappa=1} \sum^L_{\varrho=1} P_{\varrho \kappa}\hat{c}^{\dag}_{\varrho}\ |0 \rangle$
is the Slater determinant corresponding to the fermion ground state in the trap, and $({\bf P}^{\alpha})_{L,N+1}$, with
$\alpha=i,j$ is obtained using properties of Slater determinants and written as
\begin{eqnarray}
 P^{\alpha}_{\varrho \kappa}= \left\{ \begin{array}{rl}
 -P_{\varrho \kappa} & \text{for } \varrho<    \alpha,\,\kappa=1,\ldots,N \\
\,P_{\varrho \kappa} & \text{for } \varrho\geq \alpha,\,\kappa=1,\ldots,N \\
  \delta_{\alpha\varrho} & \text{for } \kappa=N + 1
\end{array}\right. .
\end{eqnarray}
From Eq.~\eqref{eq:greensHCB}, $G(i,j)$, and hence $g_1(i,j)$, are computed numerically. This method has the
additional advantage that it can be easily generalized to study off-diagonal correlations in systems out of equilibrium
\cite{rigol_muramatsu_05eHCBc}. It has also been generalized
to study hard-core anyons by \citet{Hao09}.

To discuss the properties of the trapped gas, it is convenient to define a the length scale determined by the
confining potential in Eq.~\eqref{eq:HamFreeFtrap}:
$\zeta=\left(V/t \right)^{-1/2}$. We also define the `characteristic' density $\tilde{\rho}=Na/\zeta$,
which is a dimensionless quantity ($a$ is the lattice parameter) that plays a similar role to the mean filling
$n_0$ in homogeneous systems \cite{rigol_muramatsu_04HCBa,rigol_muramatsu_05HCBb}.
For $\tilde{\rho}>2.68$ an incompressible plateau with $n_0=1$ is always present at the trap center.

\begin{figure}[!t]
\includegraphics[width=0.4\textwidth]{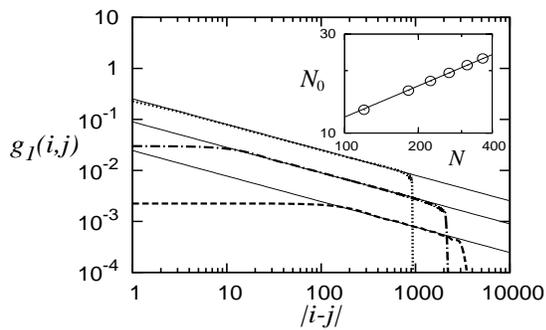}
\caption{One-particle density matrix of trapped hard-core bosons vs
  $|i-j|$ ($i$ located in the center of the
trap) for systems with $N=1000$, $\tilde{\rho}=2.0$, $n_0=0.75$ (dotted line), $N=100$,
$\tilde{\rho}=4.47\times 10^{-3}$, $n_0=0.03$ (dashed-dotted line), and $N=11$, $\tilde{\rho}=2.46\times 10^{-5}$,
$n_0=2.3\times 10^{-3}$ (dashed line), where $n_0$ is the filling of the central cite. The abrupt reduction of
$g_1(i,j)$ occurs when $n_j\rightarrow 0$. Thin continuous lines correspond to power laws $|i-j|^{-1/2}$. The inset
shows $N_0$ vs $N$ for systems with $\tilde{\rho}=1.0$ ($\bigcirc$). The straight line exhibits $\sqrt{N}$
behavior \cite{rigol_muramatsu_05HCBb}.}
\label{fig:lattice_opdm}
\end{figure}

A detailed study of $g_1(i,j)$, the lowest natural orbitals and their occupations, and the scaled momentum
distribution function, revealed that, in the regions where $n_i<1$, their behavior is very similar
to the one observed in the continuum trapped case \cite{rigol_muramatsu_04HCBa,rigol_muramatsu_05HCBb}.
One-particle correlations were found to decay as a power-law,
$g_1({i,j})\sim|i-j|^{-1/2}$,  at long distances
(see the main panel in Fig.~\ref{fig:lattice_opdm}), with a weak dependence on the density (discussed
in Sec.~\ref{sec:BHMTrap}). As a consequence of that power-law decay, the leading $N$ behavior of $N_0$
and $n'_{k=0}$ was found, both numerically and by scaling arguments, to be $N_0\propto \sqrt{N}$ (see
the inset in Fig.~\ref{fig:lattice_opdm}) and $n'_{k=0}\propto \sqrt{N}$, with proportionality constants
that are only functions of $\tilde{\rho}$. On the other hand, the high-$\alpha$ and high-$k$ asymptotics
of the occupation of the natural orbitals and of the momentum distribution function, respectively, are not
universal for arbitrary fillings. However, at very low filling, universal power-law decays
$N_\alpha \propto \alpha^{-4}$ and $n'(k)\propto k^{-4}$ were found in the lattice TG
\cite{rigol_muramatsu_04HCBa}, in agreement with what was discussed for the TG gas in the continuum.

Finite temperatures in experiments have dramatic effects in the long distance behavior of
correlations in 1D systems. For the trapped TG gas in a lattice, exact results for $g_1(i,j)$, the natural
orbitals and their occupations, as well as the momentum distribution function, can be obtained in the grand
canonical ensemble \cite{rigol_05}. Like in the homogeneous case, the power-law behavior
displayed in Fig.~\ref{fig:lattice_opdm} is replaced by an exponential decay, which implies that
$n(k=0)\sim O(1)$ and $N_0\sim O(1)$ at finite $T$. As a result, the behavior of $n(k)$ at low momenta
is very sensitive to the value of $T$. Hence, it can be used as a sensitive probe for thermometry~\cite{rigol_05}.
More recent studies of the  momentum distribution of the trapped lattice TG gas at finite $T$
have suggested that it can be well approximated by a L\'evy distribution \cite{ponomarev10}.

\subsection{The Lieb-Liniger model}\label{sec:lieb}

Let us now turn to the more general Lieb-Liniger model, Eq.~\eqref{eq:lieb-liniger-model}.
The model is integrable  \cite{lieb_bosons_1D,lieb_excit} by the Bethe Ansatz, i.e.,
its eigenfunctions are of the form:
\begin{equation}
  \label{eq:bethe-ansatz}
  \psi_B(x_1,\ldots, x_N)=\sum_P A(P) e^{i \sum_n k_{P(n)} x_n},
\end{equation}
for $x_1<x_2<\ldots<x_N$ where the $P$'s are the $N!$ possible permutations of the set $\{1,\ldots,N\}$. The
value of the wavefunction $\psi_B(x_1, \ldots)$ when the condition  $x_1<x_2<\ldots<x_N$ is not satisfied is obtained from the
symmetry of the wavefunction under permutation of the particle coordinates. The physical interpretation
of the Bethe Ansatz wavefunction is the following. When the particle coordinates are all distinct, the potential
energy term in (\ref{eq:lieb-liniger-model}) vanishes, and the Hamiltonian reduces to that of a
system of non-interacting particles. Thus, the eigenstates of the Hamiltonian can be written as a linear
combination of products of single particle plane waves. If we now consider a case with two particles $n$ and $m$
of respective momenta $k_n$ and $k_m$ having the same coordinate, a collision between these two particles occurs.
Due to the 1D nature of the system, the energy and momentum conservation laws imply that a particle
can only emerge out of the collision carrying the same momenta or exchanging it with the other particle.
Considering all possible  sequences of two-body collisions starting from a
particular set of momenta $k_1,\ldots,k_N$ leads to the form of
the wavefunction~(\ref{eq:bethe-ansatz}). When the permutations $P$ and $P'$ only differ by the transposition of
$1$ and $2$, the coefficients $A(P)$ and $A(P')$ are related by: $A(P)=\frac{k_1-k_2+ic}{k_1-k_2-ic} A(P')$ so
that the coefficients $A(P)$ are fully determined by two-body collisions. Considering three body collisions then
leads to a compatibility condition known as Yang-Baxter equation which amounts to require that any three-body
collision can be decomposed into sequences of three successive two-body collisions. The Bethe Ansatz
wavefunction can be seen as a generalization of the Girardeau wavefunction~(\ref{eq:girardeau-transformation}),
where the requirement of  a vanishing wavefunction for two particles meeting at the same point has been replaced
by a more complicated boundary condition.
As in the case of the TG gas, the total energy of the state (\ref{eq:bethe-ansatz}) is  a function of the $k_n$ by:
\begin{equation}
  \label{eq:energy-lieb}
  E=\sum_n \frac{\hbar^2 k_n^2}{2m}.
\end{equation}
However, the (pseudo-)momenta $k_n$ are determined by requiring that the wavefunction (\ref{eq:lieb-liniger-model})
obeys periodic boundary conditions, i.e.,
\begin{equation}
  \label{eq:bethe-pbc}
 e^{i k_n L} =\prod_{m=1\atop m\ne n}^N
 \frac{k_n-k_m+ic}{k_n-k_m-ic},
\end{equation}
for each $1\le n \le N$. Taking the logarithm, it is seen that the eigenstates are labeled by a set of integers
$\{I_n\}$, with:
\begin{equation}
  \label{eq:bethe-log}
  k_n = \frac{2\pi I_n}{L} + \frac 1 L \sum_m \log \left(
    \frac{k_n-k_m+ic}{k_n-k_m-ic}\right).
\end{equation}
The ground state is obtained by filling the pseudo-Fermi sea of the $I_n$ variables. In the continuum limit,
the $N$ equations (\ref{eq:bethe-pbc}) determining the $k_n$ pseudomomenta as a function of the $I_n$ quantum
numbers reduce to an integral equation for the density $\rho(k_n)=1/[L(k_{n+1}-k_n)]$ of pseudomomenta
that reads (for zero ground state total momentum):
\begin{equation}
  \label{eq:inteq-lieb}
  2\pi \rho(k)=1+2 \int_{-q_0}^{q_0}  \frac{c \rho(k')}{(k-k')^2+c^2},
\end{equation}
with $\rho(k)=0$ for $|k|>q_0$, while the ground state energy per unit length and the density become:
\begin{equation}
  \frac E L =\int_{-q_0}^{q_0} dk \frac{\hbar^2 k^2}{2m} \rho(k),\ \ \ \text{and}\ \ \
  \rho_0 =\int_{-q_0}^{q_0} dk \rho(k).
\end{equation}
By working with dimensionless variables and functions:
\begin{equation}
\label{eq:gamma} g(u)=\rho(q_0 x) \; ; \; \lambda=\frac c {q_0}\;
; \; \gamma = \frac c {\rho_0},
\end{equation}
the integral equation  can be recast as:
\begin{equation}\label{eq:lieb_dimensionless}
2\pi g(u)=1+ 2\lambda \int_{-1}^{1} \;\;  \frac{g(u') du'}{\lambda^2 +(u-u')^2},
\end{equation}
where $g(u)$ is normalized such that $\gamma \int_{-1}^1 g(u) du = \lambda$.

In the limit $c \to \infty$, $\lambda/(u^2+\lambda^2) \to 0$ so that $g(u)=1/(2\pi)$. Using the second equation
(\ref{eq:lieb_dimensionless}), one finds that $\lambda/\gamma \to 1/\pi$ i.e., $q_0=\pi \rho_0$. Thus, in this
limit, the function $\rho(k)=\theta(\pi \rho_0 -|k|)/(2\pi)$ and the ground state energy becomes that of the TG
gas. In the general case, the integral equations have to be solved numerically. The function $g(u)$ being fixed
by the ratio $\lambda=c/q_0$ and $\gamma$ being also fixed by $\lambda$, the physical properties of the Lieb-Liniger
gas depend only on the dimensionless ratio $\gamma=c/\rho_0$. An important consequence is that, in the Lieb-Liniger
model, low density corresponds to strong interaction, and high density corresponds to weak interaction, which is
the reverse of the 3D case. As a consequence of the above mentioned scaling property, the energy per unit length
is $E/L=\frac{\hbar^2 \rho_0^3}{2m}e(\gamma)$.  The Bogoliubov approximation gives a fair agreement with
$e(\gamma)$ for $0<\gamma<2$. The TG  regime (defined by $e(\gamma)= \pi^2/3$ with less than $10\%$ accuracy)
is reached for $\gamma>37$. Expansions of the ground state energy
to order $1/\gamma^3$ have been obtained\cite{guan2010} recently, and
are very accurate for $\gamma>3$, allowing to describe the crossover. \\
Until now, we have only considered the ground state energy with zero momentum. However, by considering the
equations~(\ref{eq:bethe-log}), it is seen that if we shift all the $I_n$ quantum numbers by the same integer
$r$ and at the same time we shift all the $k_n$ pseudomomenta by the quantity $2\pi r/L$, we obtain another
solution of the equations~(\ref{eq:bethe-log}). For such a solution, the wavefunction~(\ref{eq:bethe-ansatz}) is
multiplied by a factor $e^{i \frac{2\pi r}{L} \sum_n x_n}$ indicating a shift of the total momentum
$P= 2 \hbar \pi r N/L$ . At the same time, the ground state energy is shifted by a quantity
equal to $P^2/2N m$, in agreement with the Galilean invariance of (\ref{eq:lieb-liniger-model}).
The compressibility and the sound velocity can be derived from the expression of the energy per unit length.
Remarkably, the sound velocity obtained from the Bogoliubov approximation agrees with the exact sound velocity
derived from the Lieb-Liniger solution for $0<\gamma<10$ even though, as we have just seen, the range of
agreement for the energy densities is much narrower. For $\lambda \gg 1$, one can obtain an approximate solution
of the integral equations~(\ref{eq:lieb_dimensionless}) by replacing the kernel by $2/\lambda$. Then, one
finds~\cite{lieb_bosons_1D} $g(u)\simeq \lambda/(2\pi \lambda -2)$, which leads to $\pi
\lambda=(2+\gamma)$ and $E=\frac{\hbar^2 \pi^2 \rho_0^3}{6 m} \left(\frac{\gamma}{\gamma+2}\right)^2$. This
expression of the energy is accurate to $1\%$ for $\gamma>10$. For the non-interacting limit, $\lambda \to 0$,
the kernel of the integral equation~(\ref{eq:lieb_dimensionless}) becomes $2\pi \delta(u-u')$ and the solutions
become singular in that limit. This is an indication that 1D interacting bosons are not adiabatically
connected with non-interacting bosons, and present a non-generic physics.\\
Besides the ground state properties, the Bethe-Ansatz solution of the Lieb-Liniger model also allows the
study of excited states~\cite{lieb_excit}. Two types of excitations (that we will call I and II) are found.
Type I excitations are obtained by adding one particle of momentum $q>q_0$ to the $(N-1)$ particle ground state.
Because of the interaction, the $N-1$ pseudomomenta $k'_n$ ($1\le n \le N-1$) are shifted with respect to the
ground state pseudomomenta $k_n$ by $k'_n=k_n+\omega_n/L$, while the pseudomomentum $k_N=q$. Taking the continuum
limit of Eq.~(\ref{eq:bethe-log}), one finds the integral equation \cite{lieb_excit}:
\begin{equation}
  \label{eq:typeI-inteq}
  2\pi J(k)=2c \int_{-q_0}^{q_0} \frac{J(k')}{c^2+(k-k')^2} -\pi + 2
  \tan^{-1}\left(\frac{q-k}c\right),
\end{equation}
where $J(k)=\rho(k) \omega(k)$. The momentum and energy of type I excitations are obtained as a function of
$q$ and $J(k)$ as:
\begin{equation}
  \label{eq:typeI-dispersion}
  P=q + \int_{-q_0}^{q_0} J(k) dk;\quad
  \epsilon_I = -\mu + q^2 + 2 \int_{-q_0}^{q_0} k J(k) dk.
\end{equation}
The type I excitations are gapless, with a linear dispersion for $P\to 0$ and a velocity equal to the
thermodynamic velocity of sound. For weak coupling,  they reduce to the Bogoliubov excitations. In the TG limit,
they correspond to transferring one particle at the Fermi energy to a higher energy. Type II excitations are
obtained by removing one particle of momentum $0<k_h=q<q_0$ from the $(N+1)$ particle ground state. As in the
case of type I excitations, the interaction of the hole with the particles creates a shift of the pseudo-momenta,
with this time $k'_n=k_n+\omega_n/L$ for $n\le h$ and $k'_n=k_{n+1}+\omega_n/L$ for $n>h$. The corresponding
integral equation is \cite{lieb_excit}:
\begin{equation}
  \label{eq:typeII-inteq}
   2\pi J(k)=2c \int_{-q_0}^{q_0} \frac{J(k')}{c^2+(k-k')^2} +\pi - 2
  \tan^{-1}\left(\frac{q-k}c\right),
\end{equation}
with the same definition for $J(k)$. The momentum and energy of the type II excitation are:
\begin{equation}
  \label{eq:typeII-dispersion}
  P=-q + \int_{-q_0}^{q_0} J(k) dk;\quad
  \epsilon_{II} = \mu - q^2 + 2 \int_{-q_0}^{q_0} k J(k) dk.
\end{equation}
For $P=\pi \rho$, $\epsilon_{II}$ is maximum, while for $q=K$, $P=0$ and $\epsilon_{II}=0$. For $P\to 0$ the
dispersion of type II excitation vanishes linearly with $P$, with the same velocity as the type I excitations.
Type II excitations have no equivalent in the Bogoliubov theory. Type II and type I excitations are not independent
from each other~\cite{lieb_excit}: a type II excitation can be built from many type I excitations of vanishing
momentum. The Lieb-Liniger thermodynamic functions can be obtained using the Thermodynamic Bethe
Ansatz~\cite{yang1969_bosons1d}.

\subsection{The $t$-$V$ model}

Results similar to that above can be obtained for the $t$-$V$ model defined in Sec.~\ref{se:josephson-th}. Its
Bethe-Ansatz wavefunction, in second quantized form, reads:
\begin{eqnarray}
  \label{eq:ba-xxz}
  |\Psi_{B} \rangle&=&\sum_{1\le n_1<\ldots<n_N\le L} \sum_P e^{i
    \sum_{j=1}^N k_{P(j)} n_j} A(\mathcal{P}) \nonumber\\&&
    \qquad\qquad\qquad\qquad
    \times \hat{b}^\dagger_{n_N} \ldots  \hat{b}^\dagger_{n_1}
  |0\rangle,
\end{eqnarray}
where $|0\rangle$ is the vacuum state of the bosons, and $\mathcal{P}$ is a permutation. By imposing periodic boundary conditions,
the following set of equations for  the pseudomomenta $k_n$ is obtained:
\begin{equation}
  \label{eq:ba-xxz-eqs}
  e^{i k_j L} = (-)^{N-1} \prod_{l=1\atop l \ne j}^N \frac{1-2 \Delta e^{i k_j} +
    e^{i(k_j+k_l)}}{1-2 \Delta e^{i k_l} +  e^{i(k_j+k_l)}}. 
\end{equation}
By the same method as in the Lieb-Liniger gas, (\ref{eq:ba-xxz-eqs}) can be turned into
an integral equation for the density of pseudomomenta. The energy of the eigenstates is given by:
\begin{equation}
  \label{eq:ba-xxz-eigenstate}
  E=-2t \sum_{j=1}^N \cos k_j.
\end{equation}
For half-filling, solving the equations~(\ref{eq:ba-xxz-eqs}) shows that the quantum lattice gas model has
a period 2 density-wave ground state (with gapped excitations) for
$V>2t$. This state corresponds to the more general Mott phenomenon to be examined
in Sec.~\ref{sec:mott}. For $V<-2t$, the system has a collapsed ground state. Using the mappings described in
Sec.~\ref{sec:mapping},  in the spin language the collapsed state becomes a ferromagnetic ground state,
while the charge density wave state becomes an Ising like antiferromagnetic ground state. For $|V|<2t$, the ground state
has a uniform density and the excitations above the ground state are
gapless. For incommensurate filling, 
no density wave phase is obtained.

The Bethe Ansatz equations also allow the determination of the dispersion of excitations. In the half-filled case,
for $|V|<2t$, a continuum is obtained where the excitation energy, $\hbar \omega$ obeys:
\begin{eqnarray}
\label{eq:continuum-xxz}
&&\frac{\pi t}{2}  \frac{\sqrt{1-(V/2t)^2}} {\arccos(V/2t)} |\sin (qa)|
\le \hbar \omega \le \\&&\qquad\qquad\qquad\qquad\qquad
\pi 2  \frac{\sqrt{1-(V/2t)^2}} {\arccos(V/2t)}
|\sin (qa/2)|.\nonumber
\end{eqnarray}
For small momentum, the upper and lower bound of the continuum merge, yielding a linearly dispersion branch
$\hbar \omega(q)=\frac{\pi t \sqrt{1-(V/2t)^2}}{\arccos(V/2t)} |qa|$. In the case of attractive interactions
($-2t<V<0$) there are also bound states above the continuum with dispersion:
\begin{eqnarray}
  \label{eq:bound-states-xxz}
  \hbar \omega_n(q)&=& \pi t  \frac{\sqrt{1-(V/2t)^2}} {\arccos(V/2t)}
|\sin (qa/2)| \nonumber\\ &&\times\frac{\sqrt{1-\cos^2 y_n \cos^2 q/2 }}{\sin y_n},
\end{eqnarray}
where:
\begin{equation}
  y_n=\frac{n\pi}{2} \left[\frac{\pi}{\arccos (V/2t)} -1\right],
\end{equation}
and $y_n<\frac \pi 2$. For $V>2t$ and at half-filling, a density wave state is formed. The transition between
the liquid and the density wave was reviewed by \citet{shankar_spinless_conductivite}. The gap in the density-wave
state is:
\begin{equation}
  \label{eq:gap-xxz-af}
  E_G=-t \ln \left[ 2e^{-\lambda/2} \prod_1^\infty
    \left(\frac{1+e^{-4m\lambda}}{1+ e^{-(4m-2)\lambda}} \right)^2
  \right],
\end{equation}
where $\cosh \lambda = V/2t$. For $V \to 2t$, the gap behaves as: $E_G \sim 4t \exp(-\pi^2/\sqrt{8(1-V/2t)})$.
The expectation value of the particle number on site $n$ is
$\langle \hat{b}^\dagger_n \hat{b}^{}_n \rangle = (1+(-)^n P)$ where:
\begin{equation}
  \label{eq:op-xxz-af}
  P=\prod_1^\infty
  \left(\frac{1-e^{-2m\lambda}}{1+e^{-2m\lambda}}\right)^2.
\end{equation}
The order parameter $P$ of the density wave for $V-2t \to 0^{+}$ vanishes as
$P \sim \frac{\pi \sqrt{2}}{\sqrt{V/2t-1}} \exp\left[-\frac{\pi^2}{\sqrt{32(V/2t -1)}}\right]$ and goes to
$1$ for $V\gg 2$.

Besides the $t$-$V$ model, other integrable models of interacting lattice bosons have been constructed. These
models are reviewed by~\citet{amico04_boson_integrable_review}. They include correlated hopping and interactions
beyond next nearest neighbor, finely tuned in order to produce integrability. In the continuum
limit they reduce to the Lieb-Liniger model.

\subsection{Correlation functions of integrable models}\label{sec:lieb-corr}

\begin{figure}[!b]
\begin{tabular}{ll}
\includegraphics[width=4.4cm]{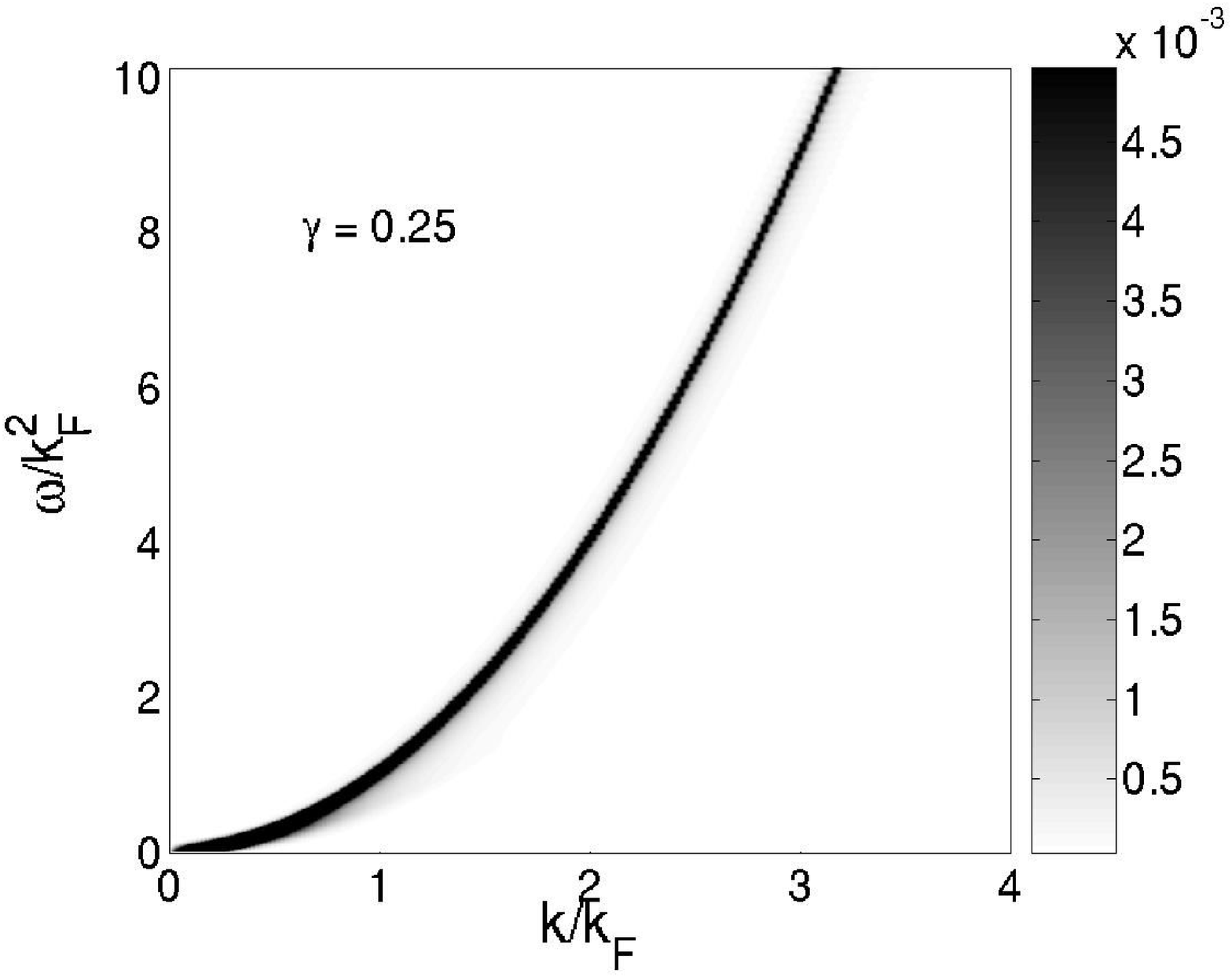}&
\includegraphics[width=4.4cm]{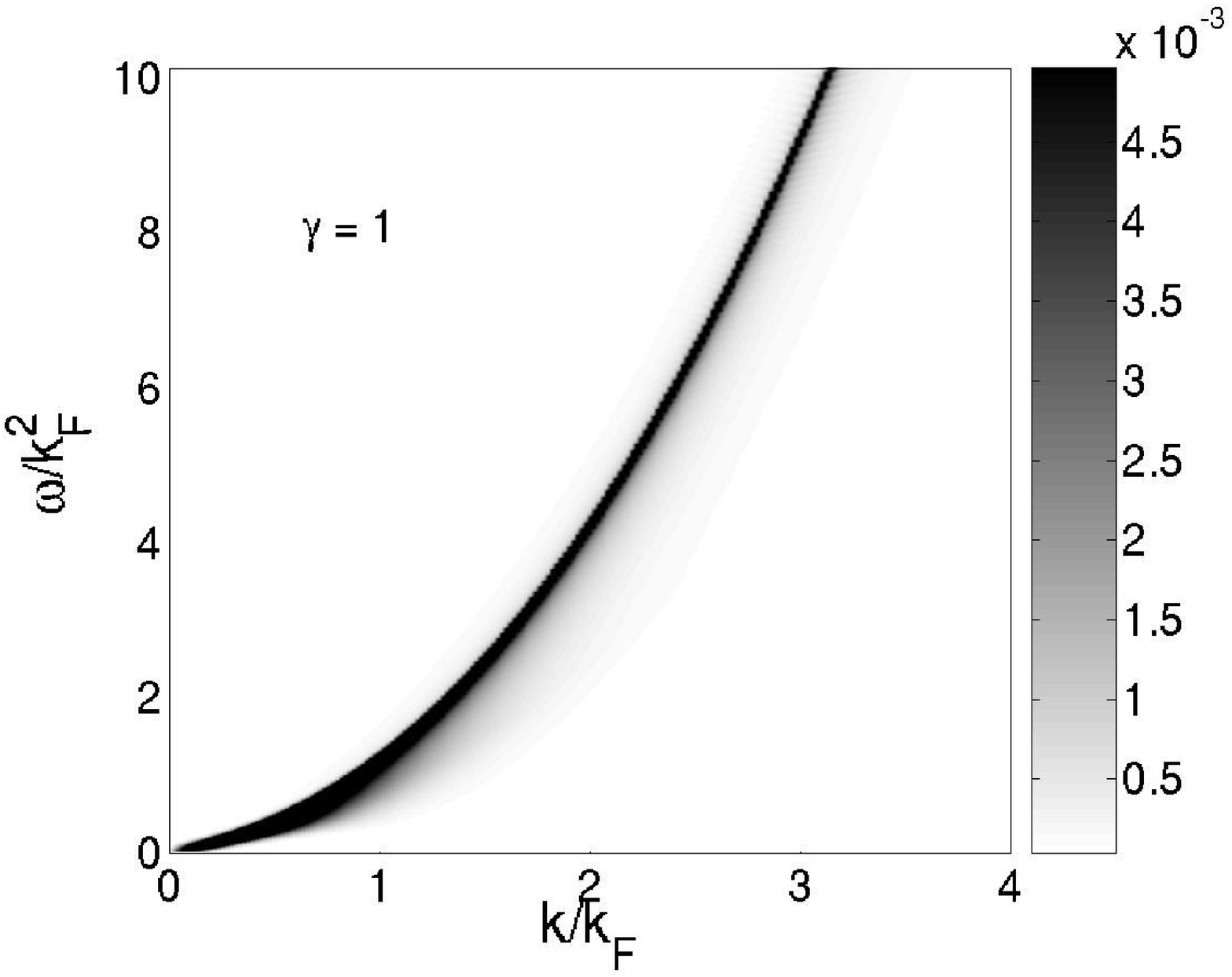} \\
\includegraphics[width=4.4cm]{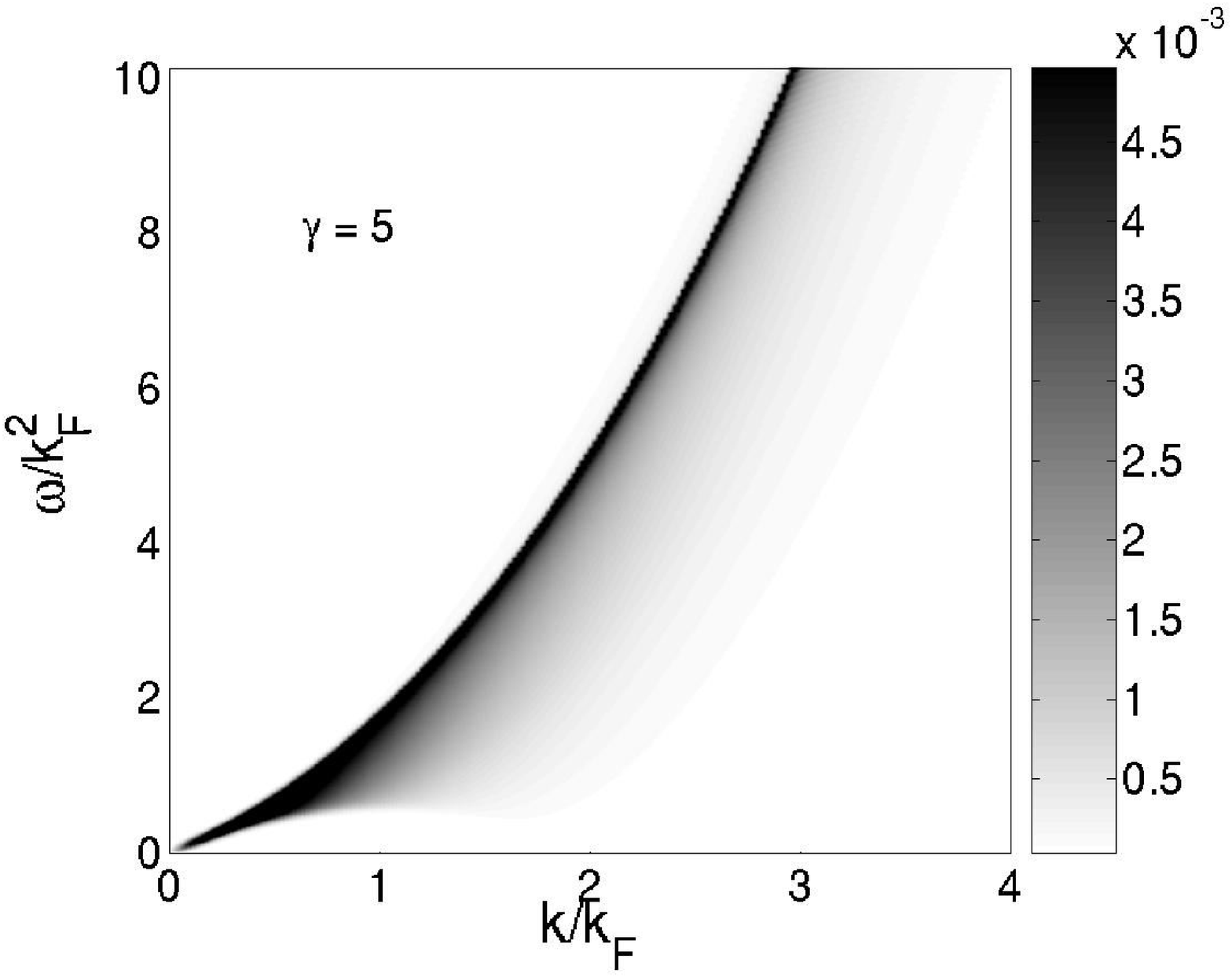}&
\includegraphics[width=4.4cm]{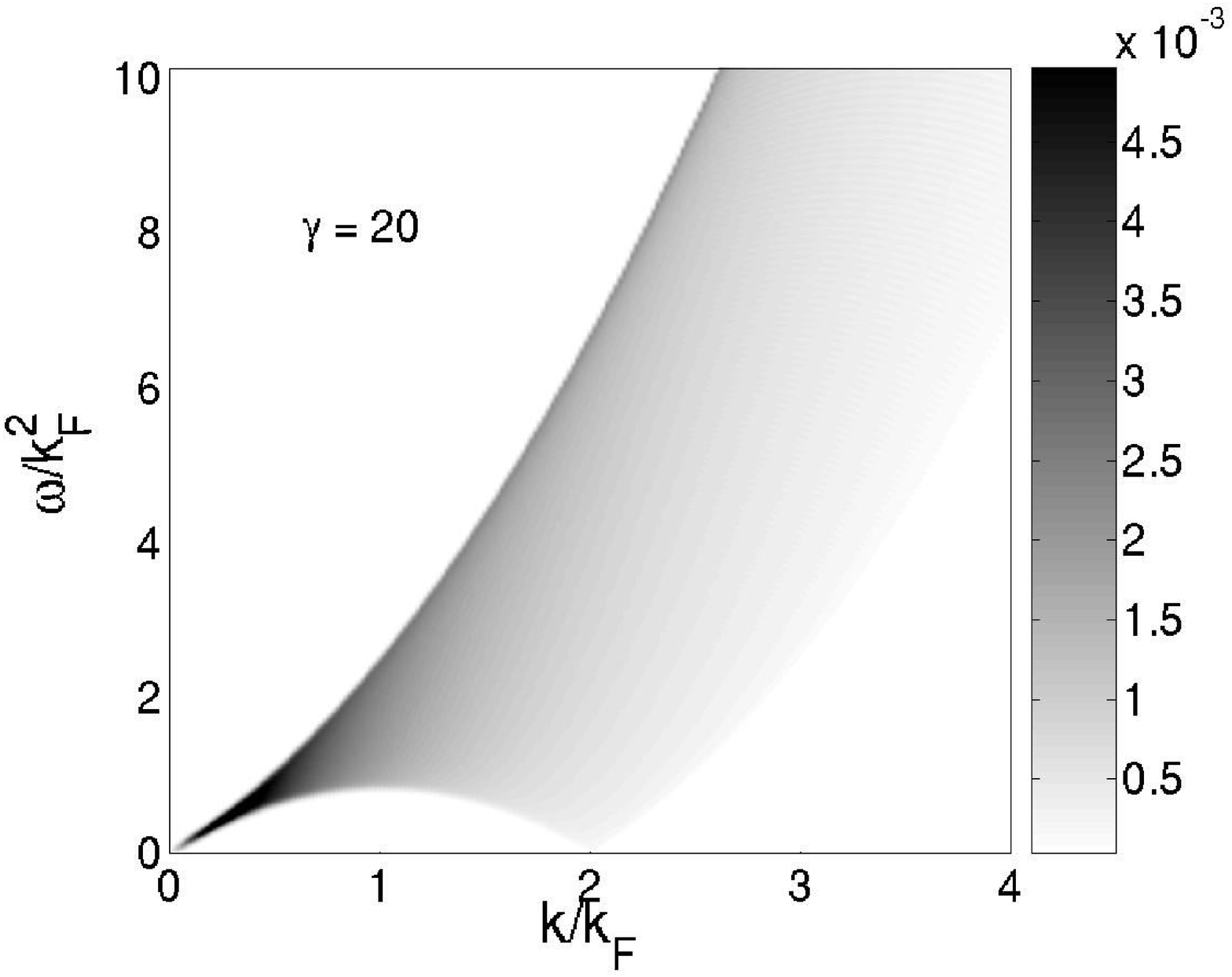}
\end{tabular}
\caption{\label{fig:caux-density}Intensity plots of the dynamical structure factor [$S(q,\omega)$]. Data
obtained from systems of length $L = 100$ at unit density, and $\gamma = 0.25, 1, 5,$ and $20$ \cite{caux_density}.}
\end{figure}

Recent progress in the theory of integrable systems has found that  the form factors (i.e., the matrix
elements between two Bethe-Ansatz eigenstates)  of a given operator can be obtained by
computing a determinant. These methods have been applied to the Lieb-Liniger
and $t$-$V$ models by \citet{caux_density,caux_one_particle,caux2005xxz,caux2005xxz_long}.

 For the Lieb-Liniger model,  \textcite{caux_density} and \textcite{caux_one_particle} computed the
 dynamic structure factor $S(k,\omega) = \int dx d\tau \, e^{i(\omega \tau - k x)} \, \langle \hat{\rho}(x,\tau) \hat{\rho}(0,0) \rangle$
 and the single-particle Green's function $G^{<}(k,\omega) = \int dx d\tau \, e^{i(\omega \tau - k x)} \,
  \langle \hat{\Psi}^{\dag}(x,\tau) \hat{\Psi}(0,0)\rangle$.  For $S(k,\omega)$, the matrix elements
  of $\hat{\rho}(k) = \int dx\,  e^{-i k x} \hat{\rho}(x)$ were calculated for up to $N=150$ particles. It was checked that the f-sum
  rule, $\int d\omega\:  \omega S(k,\omega) = 2\pi \rho_0 k^2$, was fulfilled to a few percent accuracy. Results for $S(k,\omega)$
are shown in Fig.~\ref{fig:caux-density}. There, one can see that for  $\gamma<1$, most of spectral weight of $S(k,\omega)$
is found  for $\hbar\omega$  in the vicinity of  the dispersion of the type I excitation $\epsilon_I(k)$
(cf. discussion in Sec.~\ref{sec:lieb})  so that $S(k,\omega) \simeq \frac{N k^2}{L\epsilon_I(k) }
\delta(\hbar\omega-\epsilon_I(k))$, i.e.,   Bogoliubov's
theory provides a good approximation for the dynamic structure factor. For larger $\gamma$, the most spectral weight of 
$S(k,\omega)$ is found when
$\epsilon_{II}(q) <\hbar \omega <\epsilon_I(q)$. Finally, at very large $\gamma$, the dynamical structure
factor of the TG gas,  with its characteristic particle-hole excitation spectrum becomes essentially identical to that of
the free-fermi gas.

In the case of the one-particle correlation functions, Fig.~\ref{fig:caux-spectral}, similar results were obtained.
For small $\gamma$, the spectral weight of the single particle correlation function peaks in the vicinity of the
Type I dispersion, in agreement with the Bogoliubov picture. For larger $\gamma$, the support of the spectral weight
broadened, with the type II dispersion forming the lower threshold.
For the $t$-$V$ model, similar calculations have been carried out by \citet{caux2005xxz,caux2005xxz_long}.
\begin{figure}[!h]
\begin{tabular}{ll}
\includegraphics[width=4.4cm]{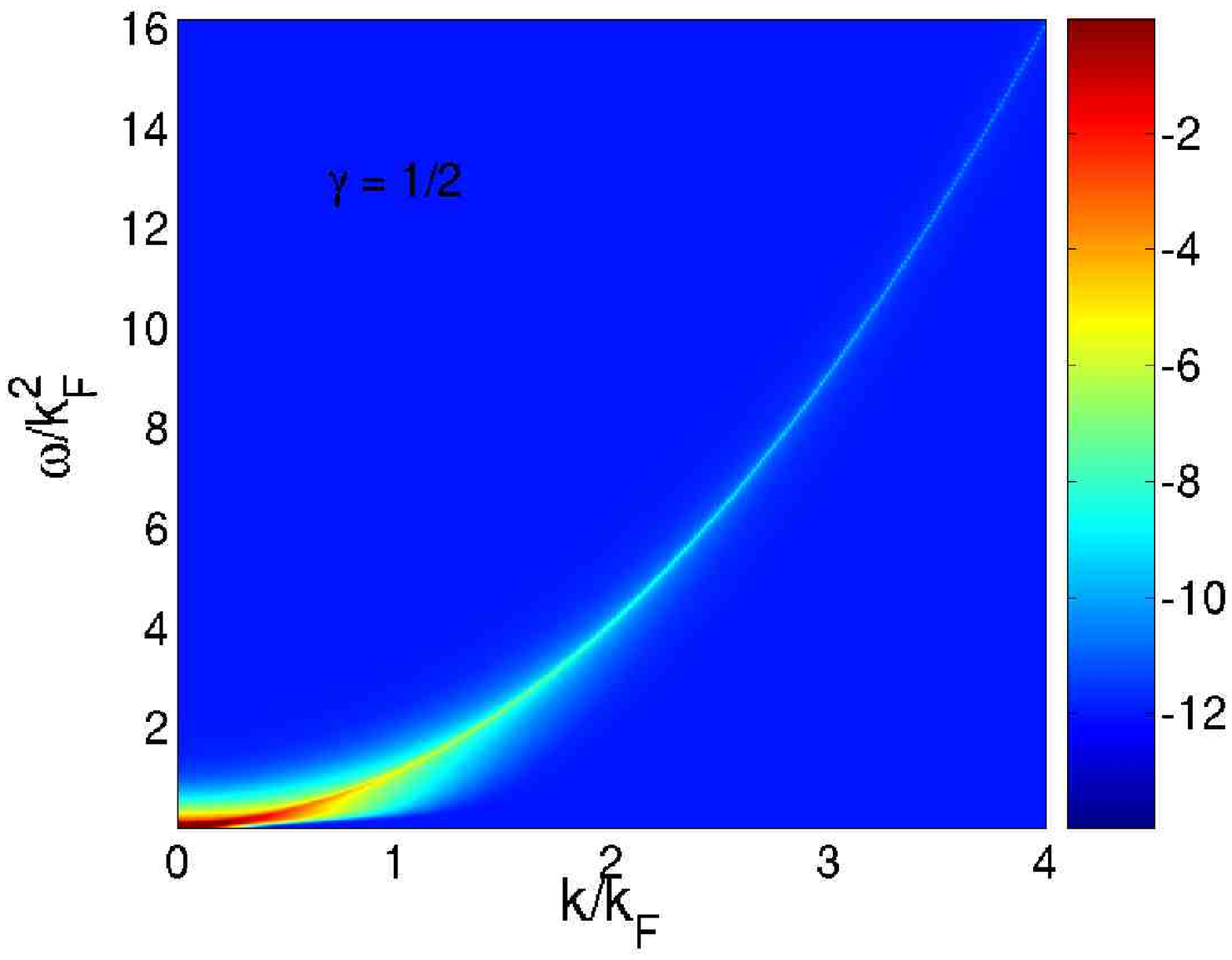} &
\includegraphics[width=4.4cm]{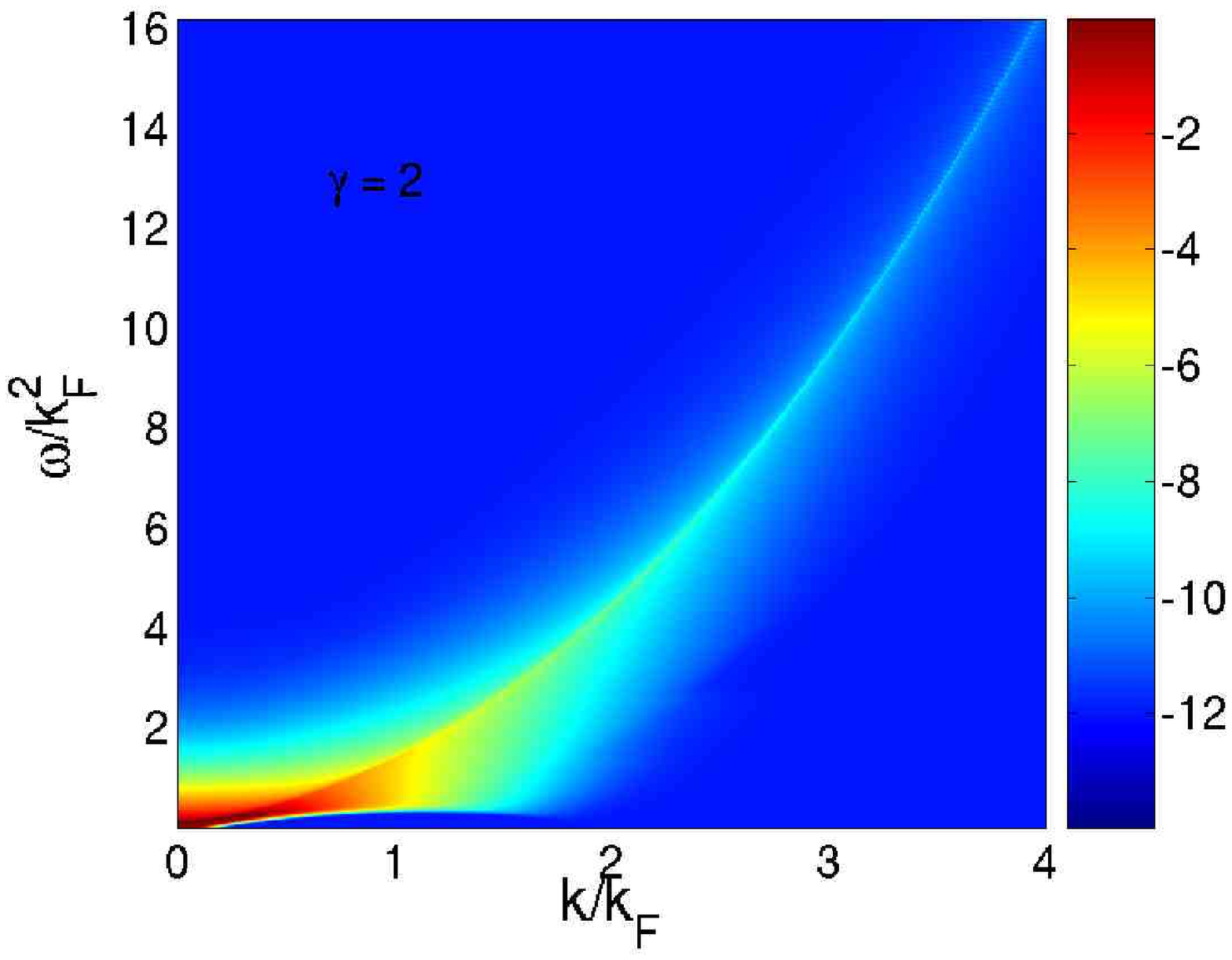} \\
\includegraphics[width=4.4cm]{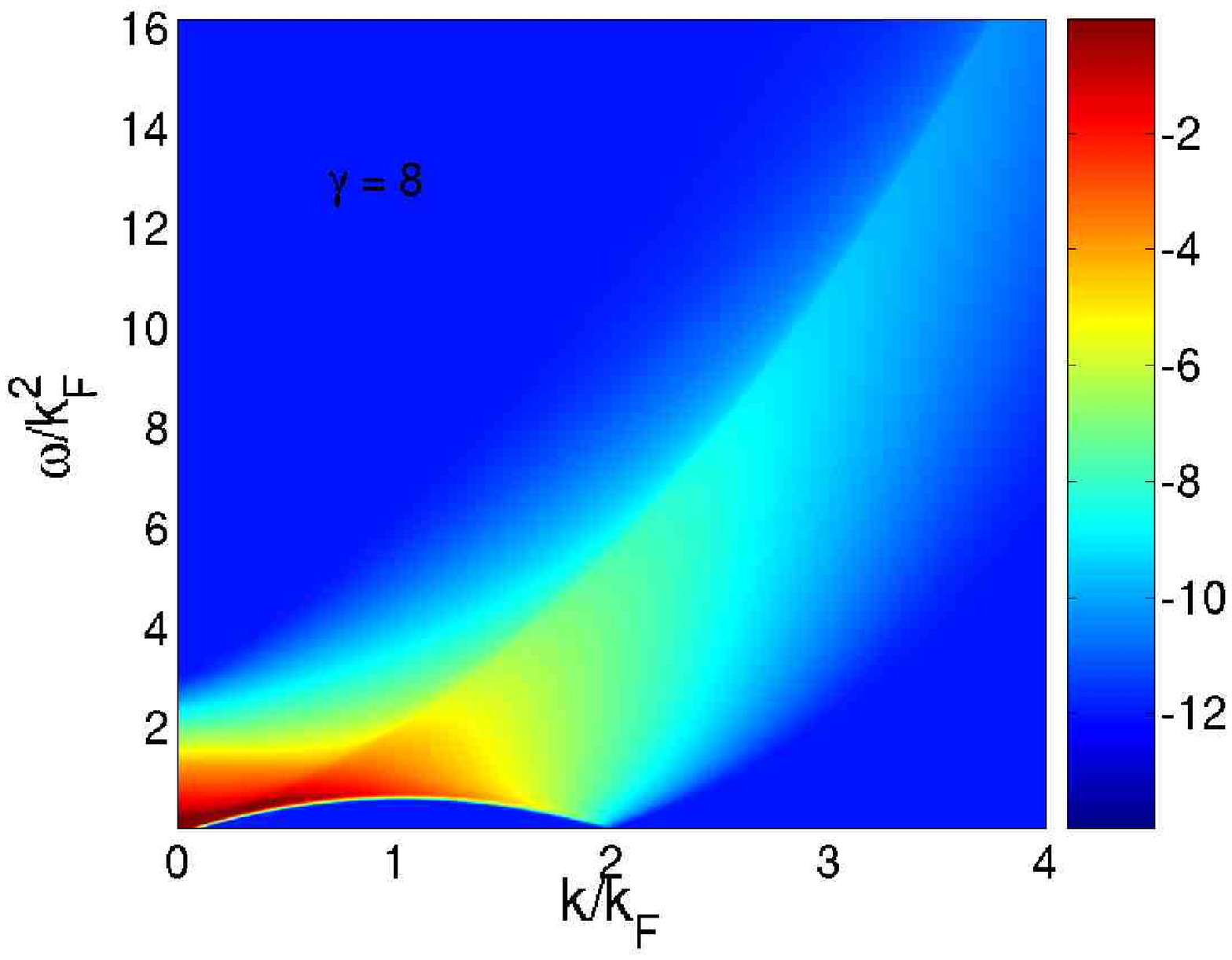} &
\includegraphics[width=4.4cm]{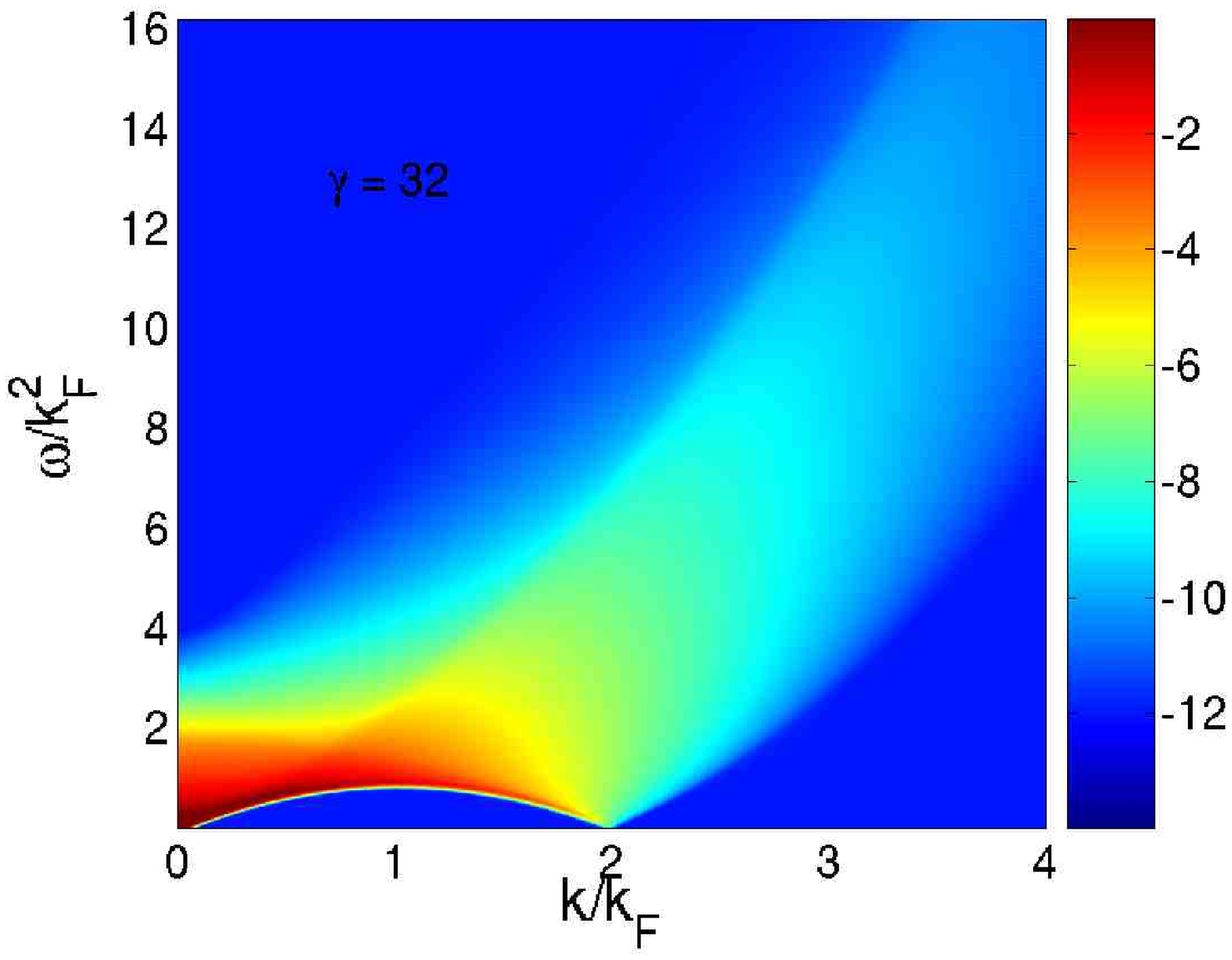}
\end{tabular}
\caption{\label{fig:caux-spectral} Intensity plots of the logarithm of the dynamical
one-particle correlation function of the Lieb-Liniger gas. Data obtained from systems of
length $L = 150$ at unit density, and $\gamma = 0.5, 2, 8,$ and $32$ \cite{caux_one_particle}.}
\end{figure}

In addition, local correlations of the form $g_n(\gamma,T) = \langle [ \hat{\Psi}^{\dag}(x)]^n
[\hat{\Psi}(x)]^n \rangle$ have been also investigated for the Lieb-Liniger model by relying on its integrability.
Indeed, $g_2(\gamma,T)$ follows, by virtue of the Hellman-Feynman theorem, from the free energy, which
can be computed using  the thermodynamic Bethe ansatz~\cite{kheruntsyan05}.
For $n>2$, one needs to resort to more  sophisticated methods. \textcite{gangardt2003_prl}
obtained the asymptotic behavior of $g_3$ at large and small
$\gamma$ for $T=0$. Later, \citet{cheianov_smith_zvonarev06} computed $g_3$ for all $\gamma$ at $T=0$ by relating
it  to the ground state expectation value of a conserved current of the Lieb-Liniger model. More recently,
\citet{Kormos_EV_LL_gas_2009,kormos_sinh_and_liebliniger_model} have obtained general
expressions for $g_n(\gamma,T)$ for arbitrary
$n$,  $\gamma$, and $T$  by matching the scattering matrices of the  Lieb-Liniger model and (the
non-relativistic limit of) the sinh-Gordon model, and then using the  form factors of the latter field theory
along with the thermodynamic Bethe-ansatz solution at finite $T$ and
chemical potential.  In Sec.~\ref{sec:experiments}
some of these results are reviewed in connection with the experiments.

\subsection{The Calogero-Sutherland model}\label{sec:calogero}

Let us finish our tour of integrable models with the Calogero model introduced in Sec.~\ref{sec:models}. This
model has the advantage of leading to relatively simple expressions for its correlation functions.
Moreover, the ground-state wavefunction of this model~(\ref{eq:calogero-sutherland}) in the presence of a
harmonic confinement potential $\frac 1 2 m \omega^2 \sum_n x_n^2$ is exactly known~\cite{sutherland71_model1}.
It takes the form:
\begin{equation}
  \label{eq:sutherland-gs}
  \psi^0_B(x_1,\ldots,x_N) \propto  e^{-\frac{m \omega}{2\hbar}
   \sum_{k=1} x_k^2} \prod_{j>i=1}^{N} |x_i-x_j|^\lambda,\nonumber
\end{equation}
where $\lambda$ is related to the dimensionless interaction parameter via $\lambda(\lambda-1)=\frac{mg}{\hbar^2}$.
For $\lambda=1/2,\,1,\,2$, the probability density of the particle coordinates $|\psi_B(x_1,\ldots,x_N)|^2$ can be
related to the joint probability density function of the eigenvalues of random matrices. More precisely,
$\lambda=1/2$ corresponds to the Gaussian Orthogonal Ensemble (GOE), $\lambda=1$ to the Gaussian Unitary Ensemble
(GUE), and $\lambda=2$ to the Gaussian Symplectic Ensemble (GSE). Many results for the random matrices
in the gaussian ensembles are available~\cite{mehta2004}, and translate into exact results for the correlation
functions of the Calogero-Sutherland models in the presence of a harmonic confinement. In particular, the
one-particle density is known to be exactly ($R^2 = \frac{2 N \hbar \lambda}{m \omega}$):
\begin{eqnarray}
  \label{eq:calogero-density-harmonic}
  \rho_0(x)&=&\frac{2 N}{\pi R^2}\theta(R^2 -x^2) \sqrt{R^2 -x^2}.
\end{eqnarray}

 In the homogeneous case with periodic boundary conditions,
 $\psi_B(x_1,\ldots,x_n+L,\ldots,x_N)=\psi_B(x_1,\ldots,x_n,\ldots,x_N)$, the Hamiltonian reads:
\begin{equation}
  \label{eq:sutherland-periodic}
  \hat{H}=-\frac{\hbar^2}{2 m} \sum_{i=1}^N  \frac{\partial^2}{\partial x_i^2} +
  \sum_{i > j=1}^N \frac{g\pi^2 }{L^2 \sin^2 \left[\frac{\pi
        (x_i-x_j)} L\right]}\, ,
\end{equation}
and the ground state is~\cite{sutherland71_model3}:
\begin{equation}
  \label{eq:sutherland-gs-periodic}
  \psi^0_B(x_1,\ldots,x_n) \propto \prod_{n<m} \sin \left|\frac{\pi
      \left(x_n-x_m\right)}{L}\right|^\lambda.
\end{equation}
The energy per unit length $E/L=\pi^2 \lambda^2 \rho_0^3/3$~\cite{sutherland71_model3}.
For $\lambda=0,1$ (i.e., $g=0$),  Eq.~(\ref{eq:sutherland-gs-periodic}) reduces to the
TG gas wavefunction (\ref{eq:girardeau-gs-pbc})
thus illustrating that the Bijl-Jastrow product form is generic of the ground-state of various 1D interacting models in the infinitely strong interaction limit.
For $\lambda=1/2$ or $\lambda=2$, the results for the respectively Circular Orthogonal Ensemble (COE) and
Circular Symplectic Ensemble (CSE) random matrices can be used \cite{mehta2004}. In particular, for
$\lambda=2$, the one-particle density matrix is
$g_1(x)=\frac{\mathrm{Si}(2\pi \rho_0x)}{2\pi x}$, where $\mathrm{Si}$ is the sine-integral function
\cite{abramowitz_math_functions}. Hence, the momentum distribution is
$n(k)=\theta(4\pi^2 \rho^2_0 - k^2) \ln(2\pi \rho_0/|k|)/(4\pi)$.
These results show that long range repulsive interactions (the case for $\lambda=2$) further weaken the
singularity at $k=0$ in the momentum distribution with respect to hard-core repulsion.  The above results
can also be used to obtain the static structure factor $S(k) = \int dx \: \langle \hat{\rho}(x) \hat{\rho}(0) \rangle$\cite{sutherland71_model3,mucciolo94_cs_dynamical}:
\begin{eqnarray}
    \label{eq:sutherland-s-k-half}
    S(k)&=&\frac{|k|}{\pi \rho_0}\left[1-\frac 1 2 \ln
      \left(1+\frac{|k|}{\pi \rho_0}\right)\right] \quad (|k|<2\pi \rho_0)
    \nonumber \\
    S(k)&=& 2- \frac{|k|}{2\pi \rho_0}\ln \left|\frac{|k|+\pi
        \rho_0}{|k|-\pi \rho_0}\right| \quad (|k|>2\pi \rho_0),
\end{eqnarray}
for $\lambda=1/2$,
\begin{eqnarray}
    \label{eq:sutherland-s-k-one}
    S(k)&=&\frac{|k|}{2 \pi \rho_0}  \quad (|k|<2\pi \rho_0)
    \nonumber\\
    S(k)&=& 1 \quad (|k|>2\pi \rho_0),
\end{eqnarray}
for $\lambda=1$, and
\begin{eqnarray}
    \label{eq:sutherland-s-k-two}
    S(k)&=&\frac{|k|}{4 \pi \rho_0}\left[1-\frac 1 2 \ln
      \left(1-\frac{|k|}{2\pi \rho_0}\right)\right] \quad
    (|k|<2\pi \rho_0)
    \nonumber\\
    S(k)&=& 1 \quad (|k|>2\pi \rho_0),
\end{eqnarray}
for $\lambda=2$. The Fourier transform of the density-density response function was also obtained
\cite{mucciolo94_cs_dynamical}. It was shown that for $\lambda=2$, the support of $S(q,\omega)$ touched
the axis $\omega=0$ for $q=0,\,2\pi\rho_0$ and $q=4\pi \rho_0$. The general case of rational $\lambda$
can be treated using Jack polynomials techniques~\cite{ha1994_cs_formfactors,Ha1995_cs_formfactors}.
The dynamical structure factor $S(k,\omega)$ is non-vanishing only for $\omega_-(k)<\omega<\omega_+(k)$ where
$\omega_\pm(k)=\frac{\hbar \pi \lambda \rho_0}{m} |k| \pm \frac{\hbar \lambda k^2}{2m}$ for
$|k|<2\pi \rho_0$ [$\omega_-(k)$ is a periodic function of $|k|$ with period $2\pi \rho_0$]. Near the
edges~\cite{pustilnik2006_cs},
$S(k,\omega \to \omega_-(k)) \propto (\omega-\omega_-(k))^{1/\lambda-1}$ and
$S(k,\omega \to \omega_+(k)) \propto (\omega_+(k)-\omega)^{\lambda-1}$. For $\lambda>1$ (repulsive interactions),
this implies a power law divergence of the structure factor for $\omega \to \omega_-(k)$ and a structure factor
vanishing as a power law for  $\omega \to \omega_+(k)$. For $\lambda<1$ (attraction), the behavior is reversed,
with a power law divergence only for $\omega \to \omega_+$ reminiscent of the results of \citet{caux_density} for the Lieb-Liniger model near the type I excitation.
Using replica methods~\cite{gangardt2001_cs_replicas}, the long distance behavior of the pair correlation function
 is obtained as:
\begin{equation}
  \label{eq:sutherland-density-corr}
  D_2(x)=1-\frac{1}{2\pi^2 \lambda (\rho_0 x)^2} +
  \sum_{m=1}^\infty \frac{2 d_m(\lambda)^2  \cos (2\pi \rho_0 m x),}{(2\pi \rho_0 |x|)^{2m^2/\lambda}},
\end{equation}
where
\begin{equation}
\label{eq:sutherland-coef-density}
  d_l(\lambda)=\frac{\prod_{a=1}^l
    \Gamma(1+a/\lambda)}{\prod_{a=1}^{l-1}
    \Gamma(1-a/\lambda)}\, .
\end{equation}
For $\lambda=p/q$ rational, the coefficients $d_l$ vanish for $l>p$. The expression
(\ref{eq:sutherland-density-corr}) then reduces to the one derived \cite{Ha1995_cs_formfactors}
for rational $\lambda$. The replica method can be generalized to time-dependent correlations
\cite{gangardt2001_cs_replicas}. 
For long distances, the one-body density matrix behaves as~\cite{astrakharchik05_cs}:
\begin{eqnarray}
  \label{eq:sutherland-1body}
  g_1(x) &=&\rho_0 \frac{A^2(\lambda)}{(2\pi \rho_0 |x|)^{\lambda/2}}
  \\&&\times\left[1 + \sum_{m=1}^\infty (-)^m \frac{D_m^2(\lambda) \cos (2\pi
      \rho_0 x)}{(2\pi \rho_0 |x|)^{2m^2/\lambda}} \right],\nonumber
\end{eqnarray}
where:
\begin{eqnarray}
  \label{eq:sutherland-coefs-1body}
  A(\lambda)&=&\frac{\Gamma(1+\lambda)^{1/2}}{\Gamma(1+\lambda/2)}\nonumber\\
  &&\times\exp \left[\int_0^\infty \frac{dt}{t} e^{-t}\left(\frac{\lambda}{4} -
      \frac{2 (\cosh(t/2)-1)}{(1-e^{-t})(e^{t/\lambda}-1)}
    \right)\right],\nonumber \\
  D_m(\lambda)&=&\prod_{a=1}^k
  \frac{\Gamma(1/2+a/\lambda)}{\Gamma(1/2+(1-a)/\lambda)}.
\end{eqnarray}
The asymptotic form (\ref{eq:sutherland-1body}) is similar to the one derived for the
TG gas, Eq.~(\ref{eq:vaidya-girardeau}) differing only by the (non-universal) coefficients $D_m(\lambda)$.

\section{Computational approaches} \label{sec:numerical-results}

The exactly solvable models discussed in the previous section allow one to obtain rather unique insights
into the physics of 1D systems. However, the results are restricted to these special models, and it is
difficult to ascertain how generic the physics of these special models can be. Besides, as we saw in the
previous section, it is still extremely difficult to extract correlation functions. One thus needs to
tackle various of the models in Sec.~\ref{sec:models} by generic methods that do not rely on integrability.
One such approach is to focus on the low-energy properties as will be discussed in Sec.~\ref{sec:bosonize}.
In order to go beyond low energies, one can use computational approaches. We thus present in this section
various computational techniques that are used for 1D interacting quantum problems, and discuss some of
the physical applications.

\subsection{Bosons in the continuum}\label{sec:QMCcontinuum}

\subsubsection{Methods}

Several methods have been used to tackle 3D and 1D systems in the continuum. Let us examine them briefly
before moving to the physical applications.

\paragraph{Variational Monte Carlo:}
Within this approach a variational trial wave function $\psi_T({\bf R},\alpha,\beta,\ldots)$ is introduced,
where ${\bf R}\equiv({\bf r}_1,\ldots,{\bf r}_N)$ are the particle coordinates, and $\alpha,\beta,\ldots$ are
variational parameters. The form of $\psi_T$ depends on the problem to be solved. One then minimizes the energy,
\begin{equation}
 E_{\text{VMC}}=\frac{\int d{\bf r}_1\ldots d{\bf r}_N\,\psi^*_{T}(\mathbf{R})\,\hat{H}\,\psi_{T}(\mathbf{R})}
                       {\int d{\bf r}_1\ldots d{\bf r}_N\,\psi^*_{T}(\mathbf{R})\,\psi_{T}(\mathbf{R})},
\end{equation}
with respect to the variational parameters, using the Metropolis Monte Carlo method of integration
\cite{umrigar99}. $E_{\text{VMC}}$ is an upper bound to the exact ground-state energy. Unfortunately the observables
computed within this approach are always
biased by the selection of $\psi_{T}$, so the method is only as good as the variational trial wavefunction itself.

\paragraph{Diffusion Monte Carlo:}
This is an exact method, within statistical errors, for computing ground-state properties of quantum systems
\cite{kalos74,reynolds82}. The starting point here is the many-body time-dependent Schr\"odinger equation
written in imaginary time $\tau_E$:
\begin{equation}
 \label{eq:imschr}
 \left[\hat{H}(\mathbf{R}) -\epsilon\right]\psi(\mathbf{R},\tau_E)=
 - \hbar \frac{\partial \psi(\mathbf{R},\tau)}{\partial{\tau_E}}
\end{equation}
where $\hat{H}= -\frac{\hbar^2}{2m}\sum_{i=1}^N\nabla^2_i + \hat{V}_\mathrm{int}(\mathbf{R}) + \hat{V}_\mathrm{ext}(\mathbf{R})$.
Upon expanding $\psi({\bf R},\tau_E)$ in terms of a complete set of eigenstates of the Hamiltonian
$\psi({\bf R},\tau_E)=\sum_n c_n \exp[-(\epsilon_n-\epsilon)\tau_E/\hbar]\, \psi^n({\bf R})$. Hence, for $\tau_E\to+\infty$,
the steady-state solution of Eq.~\eqref{eq:imschr} for $\epsilon$ close to the ground-state energy is the ground state
$\psi^0({\bf R})$. The observables are then computed from averages over $\psi({\bf R},\tau_E\to+\infty)$.

The term diffusion Monte Carlo stems from the similarity of Eq.~\eqref{eq:imschr} and the diffusion equation.
A direct simulation of~\eqref{eq:imschr} leads to large statistical fluctuations and a trial function
$\psi_{T}({\bf R})$ is required to guide the metropolis walk, i.e.,
$\psi({\bf R},\tau_E)\to \psi({\bf R},\tau_E) \psi_{T}({\bf R})$. $\psi_{T}({\bf R})$ is usually obtained
using variational Monte Carlo. $\psi_{T}({\bf R})$ can introduce a bias into the calculation of
observables that do not commute with $\hat{H}$, and corrective measures may need to be taken \cite{kalos74}.

\paragraph{Fixed-node diffusion Monte Carlo:}
The diffusion Monte Carlo method above cannot be used to compute excited states because
$\psi({\bf R},\tau_E) \psi_{T}({\bf R})$ is not always positive and cannot be interpreted as a probability density.
A solution to this problem is provided by the fixed-node diffusion Monte Carlo approach in which one enforces
the positive definiteness of $\psi({\bf R},\tau_E) \psi_T({\bf R})$ by imposing a nodal constraint so that
$\psi({\bf R},\tau_E)$ and $\psi_T({\bf R},\tau_E)$ change sign together. The trial wavefunction $\psi_T({\bf R})$ is
used for that purpose and the constraint is fixed throughout the calculation. The calculation of
$\psi({\bf R},\tau_E)$ then very much follows the approach used for the ground state. Here one just needs to
keep in mind that the asymptotic value of $\psi_B({\bf R},\tau_E\to+\infty)$ is only an approximation to the
exact excited state, and depends strongly on the parametrization of the nodal surface \cite{reynolds82}.

\subsubsection{The Lieb-Liniger gas}\label{subsec:LLgas}

Let us next discuss the application of the above methods to the Lieb-Liniger model, Eq.~(\ref{eq:lieb-liniger-model}).
As mentioned in Sec~\ref{sec:bosons_conti} (see also Sec.~\ref{sec:experiments} for a brief description
of the experimental methods), in order to realize such a system, a strong  transverse confinement must be
applied to an ultracold atomic gas. \textcite{olshanii98} pointed out that
doing so modifies the interaction potential between the atoms, from
the Lee-Huang-Yang pseudopotential  (cf. Sec.~\ref{sec:bosons_larged}) that describes their interactions
in the 3D gas in terms of the $s$-wave scattering length $a_s$, to a  delta-function interaction described
by the Lieb-Liniger model. The strength of the latter is given by the coupling $g$, which is related to
$a_s$ and the frequency of the transverse confinement $\omega_{\perp}$ by means of the expression~\cite{olshanii98}:

\begin{equation}
 \label{eq:1dcoupling}
 g=\frac{2\hbar^2 a_{s}}{m a_\perp^2}\frac{1}{1-C a_{s}/a_\perp},
\end{equation}
where $a_\perp=\sqrt{\hbar/m\omega_\perp}$, $C=|\zeta(1/2)|/\sqrt{2}=1.0326$, and $\zeta(\cdots)$ is the
Riemann zeta function. The  coupling $g$ can also be expressed in terms of an effective 1D
scattering length $a_{1D}$, $g=-2\hbar^2/{ma_{1D}}$, where $a_{1D}=-a_\perp(a_\perp/a_{s}-C)$.
Hence, $g$ increases as $\omega_\perp$ increases and the bosonic density profiles will start resembling
those of noninteracting fermions as correlations between bosons are enhanced and the 1D Tonks-Girardeau
regime is approached. These changes occur through a crossover that was studied theoretically, by means of
diffusion Monte Carlo simulations \cite{blume02,astrakharchik02}.

 In order to simulate the crossover from the 3D to 1D gas,  two different models for the
interatomic potential  $\hat{V}_\mathrm{int} = \sum_{i <j}v(r_{ij})$  ($r_{ij}=|{\bf r}_i-{\bf r}_j|$) in
the Hamiltonian ($\hat{H}$) of Eq.~(\ref{eq:imschr})  were considered in the numerical studies:
(i) a hard-core potential, $v(r_{ij})=\infty$  for $r_{ij}<a_{s}$ and $v(r_{ij})=0$ otherwise, where $a_{s}$
 corresponds to the 3D $s$-wave scattering length
a quantity that is experimentally measurable, and (ii) a soft-core potential, $v(r_{ij})=V_0$ for $r_{ij}<R$ and
$v(r_{ij})=0$ otherwise, where $R$ are $a_{s}$ are related by $a_{s}=R[1-\tanh(K_0R)/K_0R]$, with
$K_0^2=V_0m/\hbar^2$ and $V_0>0$ [note that for $V_0\rightarrow\infty$, the potential (ii)$\to$(i)]. The external
potential, whose shape drives the 3D to 1D crossover, was taken to be
$V_\mathrm{ext}({\bf r})=m(\omega^2_\perp \mathbf{r}^2_\perp+\omega^2 x^2)/2$ to closely resemble actual experimental traps.

For the Monte Carlo sampling, $\Psi_T({\bf R})$ was chosen to have a Bijl-Jastrow form
\cite{bijl40,jastrow55}
\begin{equation}\label{eq:bj_qmp}
 \Psi_T({\bf R})=\prod_{i=1}^N \varphi({\bf r}_i)\prod_{j<k,j=1}^N f(r_{jk}),
\end{equation}
which had been successfully used in a variational Monte Carlo study of the trapped bosonic gas in 3D
\cite{dubois01}. In Eq.~\eqref{eq:bj_qmp}, the single-particle orbital $\varphi({\bf r})$ accounts for the effect
of the external potential and was taken to be $\varphi({\bf r})=\exp(-\alpha_\perp r^2_\perp-\alpha x^2)$, which
is a harmonic-oscillator ground-state wave function with two variational parameters $\alpha_\perp$ and $\alpha$.
Those parameters were optimized using variational Monte Carlo simulations. The two-particle function $f(r_{jk})$
was selected to be the exact solution of the Schr\"odinger equation for two particles interacting via corresponding
two-body interatomic potential.

By changing the aspect ratio $\lambda=\omega/\omega_\perp$, the numerical simulations revealed the expected
crossover between the mean-field Gross-Pitaevskii regime (Sec.~\ref{sec:bosons_larged}) and the strongly interacting
Tonks-Girardeau limit (Sec.~\ref{sec:corr-cont}), both for the ground-state energy and for the density
profiles \cite{blume02,astrakharchik02}. For large anisotropies ($\lambda\ll 1$), a comparison of the full 3D
simulation with the Lieb-Liniger theory (Sec.~\ref{sec:lieb}) was also presented. The agreement between them was
remarkable and validated the analytical expressions for $a_{1D}$ and $g$ in terms of $a_{s}$ and $\omega_\perp$
\cite{astrakharchik02}.

 The ground state, one-particle, and two-particle correlation functions of the Lieb-Liniger model, for which no closed analytic
expressions are known, were calculated using the diffusion Monte Carlo method by \citet{astrakharchik03,astrakharchik06}.
For $x\rightarrow\infty$, the one-particle correlations were found to exhibit the power-law decay predicted by
the Tomonaga-Luttinger liquid theory described in Sec.~\ref{sec:bosonize}, while two-particle correlations were
seen to fermionize as the TG limit was approached. Local \cite{kheruntsyan03,kheruntsyan05} and
nonlocal \cite{sykes08,deuar09} two-particle correlations, as well as density profiles, have been studied at
finite temperatures by exact analytical methods and perturbation theory in some limits, and numerical calculations
using the stochastic gauge method \cite{drummond04}.

\subsubsection{The super-Tonks-Girardeau gas}

For negative values of $g$, the low energy eigenstates of the Hamiltonian \eqref{eq:lieb-liniger-model} are
cluster-like bound states \cite{mcguire64} and their energy is not proportional to $N$~\cite{lieb_bosons_1D}.
In order to access $g<0$ starting from a 3D gas, \citet{astrakharchik04a,astrakharchik04b} considered a short-range
two-body potential of the form $v(r_{ij})=-V_0/\cosh^2(r_{ij}/r_0)$, where $r_0$ was fixed to be much smaller
than $a_\perp$, and $V_0$ was varied. The inset in Fig.~\ref{fig:scatt1Dvs3D} shows how $a_s$ changes
with $V_0$ as a resonance is crossed. Following the effective theory of \citet{olshanii98}, $g$ and
$a_{1D}$ vs $a_s$ are depicted in Fig.~\ref{fig:scatt1Dvs3D}. That figure shows that there is a value
of $a_s$, $a^c_{s}$, at which $a_{1D}=0$ and $|g|\rightarrow\infty$. This is known as a confinement-induced
resonance. The Lieb-Liniger regime ($g>0$) is only accessible for $0<a_s<a^c_{s}$, where $a_{1D}<0$. On the
other hand, $g<0$ occurs for $a_s<0$ and $a_s>a^c_{s}$, where $a_{1D}>0$. It is also important to note that,
at the 3D resonance $a_s\rightarrow\infty$, $g$ and $a_{1D}$ reach an asymptotic value and  become
independent of the specific value of $a_s$.

\begin{figure}[!htb]
\includegraphics[width=0.39\textwidth]{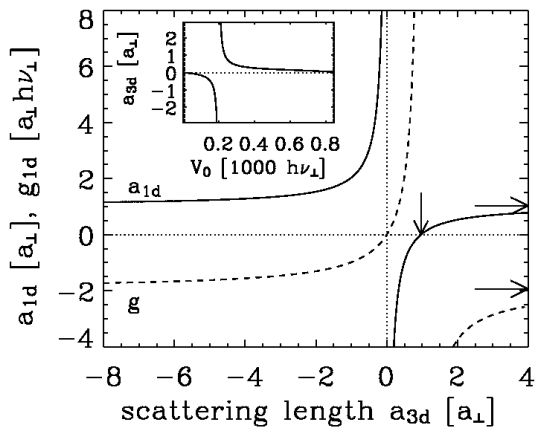}
\caption{$g$ [dashed line, Eq.~\eqref{eq:1dcoupling}] and $a_{1D}$ [solid line] as a function of
$a_{3d}=a_s$. The vertical arrow indicates the value of the $s$-wave scattering length $a_{s}$ where $g$
diverges, $a_{s}^{c}/a_{\perp}=0.9684$. Horizontal  arrows indicate the asymptotic values of $g$ and
$a_{1D}$, respectively, as $a_{s}\rightarrow \pm \infty$ ($g=-1.9368a_{\perp} \hbar \omega_{\perp}$
and $a_{1D}=1.0326a_{\perp}$). Inset: $a_{s}$ as a function of the well depth $V_0$ \cite{astrakharchik04a}.}
\label{fig:scatt1Dvs3D}
\end{figure}

Experimentally, for $g<0$, one is in general interested in the lowest energy solution without bound states (a
gas-like solution).  This solution describes a highly excited state of the system. A computational study of this
gas-like solution was done by \citet{astrakharchik04a,astrakharchik04b} using fixed-node diffusion Monte Carlo
simulations for both
3D and effective 1D systems. In that study, the result of the exact diagonalization of two particles was used to
construct the many-body nodal surface (expected to be a good approximation in the dilute limit). The many-body
energy was then found to be remarkably similar for the highly anisotropic 3D case and the effective 1D theory
using $g$ from Eq.~\eqref{eq:1dcoupling}. A variational Monte Carlo analysis of the stability of such a solution
suggested that it is stable if $\sqrt{N\lambda}a_{1D}/a_\perp\lesssim0.78$, i.e., the stability can be improved
by reducing the anisotropic parameter $\lambda$.

For small values of the gas parameter $\rho_0 a_{1D}$
($0<\rho_0 a_{1D}\lesssim0.2$), the energy of the gas-like solution is well described by a gas of hard rods,
i.e., particles interacting with the two-body potential $v(r_{ij})=\infty$ for $r_{ij}<a_{1D}$ and $V(r_{ij})=0$
otherwise \cite{astrakharchik05,girardeau_astrakharchik_10}. One- and two-particle correlation functions for $g<0$ display unique behavior. They
were computed both for the Hamiltonian \eqref{eq:lieb-liniger-model} and for the hard-rod model, and were shown
to behave similarly. The one-particle correlations decay faster than in the hard-core limit
(Sec.~\ref{sec:tonks}), for that reason, in this regime, the gas was called the `super-Tonks-Girardeau' gas. The
structure factor (the Fourier transform of the two-particle correlation matrix) exhibits a peak at
$\pi\rho_0$,\footnote{In the Tonks-Girardeau regime, for which the structure factor is identical
to the one of noninteracting fermions, only a kink is observed at $k_F$.} and, with increasing $\rho_0 a_{1D}$,
the speed of sound was shown to increase beyond the TG result \cite{astrakharchik05,mazzanti_astrakharchik_08}.
Analytically, it can be  seen that for $g\to\infty$ Eq.~(\ref{eq:girardeau-gs-pbc}) for the homogeneous gas and
Eq.~(\ref{eq:tgwf-harmonic}) for the trapped gas describes the ground state of the system.
For negative values of $g$, those states can be obtained as highly excited states of the
the model with attractive interactions~\cite{batchelor_supertonks,girardeau_astrakharchik_10},
which warrant their stability and suggest how to realize gas like states when $g<1$ in experiments
(see discussion in Sec.~\ref{sec:atomchips}). Like in the repulsive case, one possible method of detection
may rely upon the measurement of the local correlations. Kormos and coworkers~\cite{Kormos_supertonks} 
have obtained  the exact expression of the local correlators $g_n=\langle \left[\Psi^{\dag}(0)\right]^n \left[\Psi(0)\right]\rangle$ of 
the super-Tonks gas at finite temperature and at any value of the coupling using the same method employed 
for Lieb-Liniger gas~\citet{Kormos_EV_LL_gas_2009,kormos_sinh_and_liebliniger_model}.

\subsection{Bosons on a lattice}\label{sec:QMClattice}

\subsubsection{Methods} \label{sec:latticenum}

For lattice systems, in addition to the standard techniques valid in all dimensions, some specific and very
powerful techniques are applicable in 1D.

\paragraph{Exact diagonalization and density-matrix renormalization group:}
A natural way to gain insight into the properties of a quantum system using a computer is by means of exact
diagonalization. This approach is straightforward since one only needs to write the Hamiltonian for a finite
system in a convenient basis and diagonalize it. Once the eigenvalues and eigenvectors are computed, any physical
quantity can be calculated from them. A convenient basis to perform such diagonalizations for generic lattice
Hamiltonians, which may lack translational symmetry, is the site basis $\{|\alpha\rangle\}$. The number of states
in the site basis depends on the model, {e.g.}, two states for hard-core bosons ($|0\rangle$ and $|1\rangle$)
and infinite states for soft-core bosons.

Given the site basis, one can immediately generate the basis states for a finite lattice with $N$ sites as
\begin{equation}\label{eq:basisstates}
 |m\rangle\equiv|\alpha\rangle_1\otimes|\alpha\rangle_2\otimes\ldots\otimes|\alpha\rangle_N,
\end{equation}
from which the Hamiltonian matrix can be simply determined. Equation \eqref{eq:basisstates} reveals
the main limitation of exact diagonalization, namely, the size of the basis (the Hilbert space) increases
exponentially with the number of sites. If the site basis contains $n$ states, then the matrix that one needs
to diagonalize contains $n^N\times n^N$ elements.

Two general paths are usually followed to diagonalize those matrices, (i) full diagonalization using standard
dense matrix diagonalization approaches \cite{press_flannery_88}, which allow one to compute all eigenvalues
and eigenvectors required to study finite temperature properties and the exact time evolution of the system,
and (ii) iterative diagonalization techniques such as the Lanczos algorithm \cite{cullum_willoughby_85},
which give access the ground state and low energy excited states. The latter enable the study of larger system
sizes, but they are still restricted to few tens of lattice sites.

An alternative to such a brute force approach and an extremely accurate and efficient algorithm to study 1D lattice
systems is the density-matrix renormalization group (DMRG) proposed by \citet{white_92,white_93}. This approach
is similar in spirit to the numerical renormalization group (NRG) proposed by \citet{wilson_review_75} to study
the single-impurity Kondo and Anderson problems. The NRG is an iterative nonperturbative approach that allows
one to deal with the wide range of energy scales involved in those impurity problems by studying a sequence of
finite systems with varying size, where degrees of freedom are integrated out by properly modifying the original
Hamiltonian. However, it was early found by \citet{white_noack_92} that the NRG approach breaks down when
solving lattice problems, even for the very simple noninteracting tight-binding chain, a finding that motivated
the development of DMRG.

There are several reviews dedicated to the DMRG for studying equilibrium and nonequilibrium 1D systems
\cite{hallberg_03,schollwock_review_05,noack_manmana_05,manmana_muramatsu_05,dechiara_rizzi_08},
in which the reader can also find various justifications of the specific truncation procedure prescribed by DMRG.
A basic understanding of it can be gained using the Schmidt decomposition. Suppose we are interested in studying
the ground state (or an excited state) properties of a given Hamiltonian. Such a state $|\Psi\rangle$, with
density matrix $\hat{\rho}=|\Psi\rangle \langle\Psi|$, can in principle be divided into two parts $M$ and $N$ with
reduced density matrices $\hat{\rho}_M=\textrm{Tr}_N[\hat{\rho}]$ and $\hat{\rho}_N=\textrm{Tr}_M[\hat{\rho}]$,
where $\textrm{Tr}_M[\cdotp]$ and $\textrm{Tr}_N[\cdotp]$ mean tracing out the degrees of freedom of $M$ and $N$,
respectively. Remarkably, using the Schmidt decomposition \cite{schmidt_07} one can write $|\Psi\rangle$ in terms
of the eigenstates and eigenvalues of the reduced density matrices of each part
\begin{equation}
 |\Psi\rangle=\sum_\alpha \sqrt{w_\alpha} |m_\alpha\rangle |n_\alpha\rangle,
\end{equation}
where $\hat{\rho}_M|m_\alpha\rangle=w_\alpha|m_\alpha\rangle$ and $\hat{\rho}_N|n_\alpha\rangle=
w_\alpha|n_\alpha\rangle$, and the sum runs over the nonzero eigenvalues $w_\alpha$, which can be proven to be
identical for both parts. Hence, a convenient approximation to $|\Psi\rangle$ can be obtained by truncating the
sum above to the first $l$ eigenvalues, where $l$ can be much smaller than the dimension of the smallest of the
Hilbert spaces of $M$ and $N$ provided the $w_\alpha$'s decay sufficiently fast. This approximation was shown to be
optimal to minimize the difference between the exact $|\Psi\rangle$ and the approximated one \cite{white_93}.

The DMRG is a numerical implementation of the above truncation. We should stress that the DMRG is variational
and that two variants are usually used: (i) the infinite-system DMRG, and (ii) the finite-system DMRG. We will
explain them for the calculation of the ground state, but they can also be used to study excited states.

In the infinite-system DMRG, the idea is to start with the Hamiltonian ${\bf H}_{L}$ of a lattice with $L$ sites,
which can be diagonalized, and then: (1) Use an iterative approach to compute the ground state
$|\Psi\rangle$ of ${\bf H}_{L}$ (Hamiltonian of the superblock), and its energy. (2) Compute the reduced density
matrix of one half of the superblock (the ``system'' block). For definiteness, let us assume that the system block
is the left half of the superblock. (3) Use a dense matrix diagonalization approach to compute the eigenvectors
with the $l$ largest eigenvalues of the reduced density matrix from point 2. (4) Transform the Hamiltonian
(and all other operators of interest) of the system block to the reduced density matrix eigenbasis, i.e.,
${\bf H}'_{L/2}={\bf R}^\dagger{\bf H}_{L/2}{\bf R}$, where ${\bf R}$ is the rectangular matrix whose columns are
the $l$ eigenvectors of the reduced density matrix from point 3. (5) Construct a new system block from
${\bf H}'_{L/2}$ by adding a site to the right, ${\bf H}'_{L/2+1}$. Construct an ``environment'' block using the
system block and an added site to its left, ${\bf H}''_{L/2+1}$. Connect ${\bf H}'_{L/2+1}$ and ${\bf H}''_{L/2+1}$
to form the Hamiltonian of the new superblock, ${\bf H}_{L+2}$. (The same is done for all operators of interest.)
Here we note that: (i) we have assumed the original Hamiltonian is reflection symmetric, (ii) the superblock
has open boundary conditions, and (iii) the superblock Hamiltonian (and any other superblock operator) will have
dimension $2(l+n)$ ($n$ is the number basis states for a site) at most, as opposed to the actual lattice Hamiltonian
(or any other exact operator), which will have a dimension that scales exponentially with $L$. At this point, all
steps starting from 1 are repeated with $L\rightarrow L+2$. When convergence for the energy (for the ground state
expectation value of all operators of interest) has been reached at point 1, the iterative process is stopped.

There are many problems for which the infinite size algorithm exhibits poor convergence or no convergence at all.
The finite-system DMRG provides the means for studying such systems. The basic idea within the latter approach is
to reach convergence for the properties of interest in a finite-size chain. Results for the thermodynamic limit can
then be obtained by extrapolation or using scaling theory in the vicinity of a phase transition. The finite-system
DMRG utilizes the infinite-system DMRG in its first steps, namely, for building the finite size chain with the
desired length. Once the chain with the desired length has been constructed one needs to perform sweeps across the
chain by (i) increasing the system block size to the expense of the environment block size, up to a convenient
minimum size for the latter, and then (ii) reversing the process by increasing the environment block size to the
expense of the system block size, again up to a convenient minimum size for the latter. Those sweeps are repeated
until convergence is reached. In general, this approach yields excellent results for the ground state and low-lying
excited states properties. However, care should always be taken to check that the system is not trapped in some
metastable state. Problems with incommensurate fillings and disordered potentials are particularly challenging
in this respect. For the estimation of the errors as well as extensions of DMRG to deal with systems with periodic
boundary conditions, higher dimensions, finite temperatures, and nonequilibrium dynamics, we refer the reader to
the reviews previously mentioned.

\paragraph{World-line quantum Monte Carlo:}
Quantum Monte Carlo (QMC) approaches provide a different way of dealing with many-body systems. They can be used to
efficiently solve problems in one and higher dimensions. However, for fermionic and spin systems, QMC algorithms
can be severely limited by the sign problem \cite{loh_gubernatis_90,troyer_wiese_05}.

One of the early QMC algorithms devised to deal with lattice problems is the discrete-time world-line
algorithm, introduced by \citet{hirsch_sugar_82}. It is based in the path integral formulation of the
partition function in imaginary time. The goal is to compute observables within the canonical ensemble
$\langle\hat{O}\rangle=\mathrm{Tr}\lbrace\hat{O}e^{-\beta \hat{H}}\rbrace/Z$, where
$Z=\mathrm{Tr}\lbrace e^{-\beta \hat{H}}\rbrace$ is the partition function. For example, for a 1D
Hamiltonian that only couples nearest-neighbor sites, i.e.,
$\hat{H}=\sum_{i=1}^L \hat{H}_{i,i+1}$ and $[\hat{H}_{i,i+1},\hat{H}_{j,j+1}]=0$ if $j\geq i+2$, the
Hamiltonian can be then split as the sum of two terms $\hat{H}=\hat{H}_{\textrm{odd}}+\hat{H}_{\textrm{even}}$,
where $\hat{H}_{\textrm{odd}(\textrm{even})}=\sum_{i\,\text{odd(even)}}\hat{H}_{i,i+1}$. Since
$\hat{H}_{\textrm{odd}}$ and $\hat{H}_{\textrm{even}}$ do not commute, one can use the Trotter-Suzuki
decomposition \cite{trotter_59,suzuki_76} to write
\begin{equation}
e^{-\Delta\tau(\hat{H}_{\textrm{odd}}+ \hat{H}_{\textrm{even}})}=
e^{-\Delta \tau \hat{H}_{\textrm{odd}}} e^{-\Delta \tau \hat{H}_{\textrm{even}}}
+O[(\Delta\tau)^2],
\end{equation}
and $Z$ is approximated by $Z_{TS}$
{\setlength\arraycolsep{0.9pt}\begin{eqnarray}
&&Z_{TS}=\mathrm{Tr}\left\lbrace\prod^L_{m=1} e^{-\Delta \tau \hat{H}_{\textrm{odd}}}
e^{-\Delta \tau \hat{H}_{\textrm{even}}}\right\rbrace \nonumber \\&&=
\sum_{ m_1\cdots m_{2L}}
\left\langle m_1|e^{-\Delta\tau\hat{H}_{\textrm{odd}}}|m_{2L}\right\rangle
\left\langle m_{2L}|e^{-\Delta\tau\hat{H}_{\textrm{even}}}|m_{2L-1}\right\rangle
\nonumber \\ &&\qquad\quad\;
\cdots \left\langle m_3|e^{-\Delta\tau \hat{H}_{\textrm{odd}}}|m_{2}\right\rangle
\left\langle m_{2}|e^{-\Delta\tau \hat{H}_{\textrm{even}}}|m_{1}\right\rangle,
\label{eq:ME_WLQMC}
\end{eqnarray}
}where $L\Delta \tau=\beta$, and $\left\lbrace|m_{\ell}\rangle\right\rbrace$ are complete sets of states
introduced at each imaginary time slice. Since $\hat{H}_{\textrm{odd}}$ and $\hat{H}_{\textrm{even}}$
consist of a sum of mutually commuting pieces, the matrix elements in Eq.~\eqref{eq:ME_WLQMC} can be
reduced to products of the matrix elements of $e^{-\Delta\tau \hat{H}_{i,i+1}}$. A graphical representation
of each term of the sum in Eq.~\eqref{eq:ME_WLQMC} leads to a checkerboard picture of space-time,
where particles ``move'' along the so called world lines (the particle number
must be the same at each $\tau$, and periodic boundary conditions are applied to the imaginary
time axis). The systematic error introduced by the Trotter-Suzuki decomposition can be proven to be
$O[(\Delta\tau)^2]$ for the partition function and also for Hermitian observables
\cite{fye_86,assaad_02}.

The formulation above has been succeeded by a continuous time one with no discretization error
\cite{prokofev_svistunov_96,prokofev_svistunov_98}. The starting point for the latter is the operator identity
\begin{equation}
 e^{-\beta \hat{H}}=e^{-\beta \hat{H}_\textrm{d}}e^{-\int_0^\beta d\tau \hat{H}_\textrm{od}(\tau)}
\end{equation}
where the Hamiltonian $\hat{H}$ has been split as the sum of its diagonal $\hat{H}_\textrm{d}$, which
now can contain a term $-\mu\hat{N}$ so that one can work in the grand-canonical ensemble, and off-diagonal
$\hat{H}_\textrm{od}$ terms in the site basis $\{|\alpha\rangle\}$ introduced with Eq.~\eqref{eq:basisstates}.
In addition, $\hat{H}_\textrm{od}(\tau)=e^{\tau \hat{H}_\textrm{d}}\hat{H}_\textrm{od}e^{-\tau\hat{H}_\textrm{d}}$
and
\begin{eqnarray}
&&e^{-\int_0^\beta d\tau \hat{H}_\textrm{od}(\tau)}= 1-\int_0^\beta d\tau \hat{H}_\textrm{od}(\tau)
+\ldots+\\&&(-1)^n\int_0^\beta d\tau_n\ldots\int_0^{\tau_2}d\tau_1\,
\hat{H}_\textrm{od}(\tau_n)\ldots\hat{H}_\textrm{od}(\tau_1)+\ldots\nonumber
\end{eqnarray}
Once again, each term in the expansion of $e^{-\beta \hat{H}}$ has a graphical representation in terms of
world lines and the partition function $Z=\mathrm{Tr}\lbrace e^{-\beta \hat{H}}\rbrace$ can be written as a
sum over all possible paths of these world lines.

The Monte Carlo technique is then used to avoid the exponential sum over all the
possible world-line configurations by sampling them in such a way that accurate results for observables
of interest are obtained in polynomial time. The challenge is then to develop efficient update schemes
to perform such a sampling.

The updates within the discrete-time formulation are based on local deformations of the world lines
\cite{hirsch_sugar_82}. Observables that are diagonal in the occupation number (such as the
density and the density-density correlations), as well as observables that conserve the number of particles
in two contiguous sites (such as the kinetic energy and the current operator) can be easily computed while
other observables that do not conserve the fermion number locally, such as the one-particle density matrix
$g_1(i,j)=\langle \hat{b}_i \hat{b}_j^\dagger\rangle$, are almost impossible to calculate \cite{scalettar_99}.
This local update scheme may lead to long auto-correlation times \cite{kawashima_gubernatis_94}
in a way similar to classical simulations with local updates. However, those times can be reduced by
global moves in cluster (loop) algorithms, as proposed by \citet{swendsen_wang_87} for the classical Ising
model, and extended to quantum systems by \citet{evertz_lana_93,kawashima_gubernatis_94,evertz_03}.
In addition, other problems with this world-line formulation were removed by the loop algorithm, which
works in the grand canonical ensemble, and for which the time continuum limit $\Delta\tau\rightarrow 0$
in the Trotter-Suzuki decomposition was also implemented \cite{beard_wiese_96}.

Loop algorithms also have their limitations since they are difficult to construct for many
Hamiltonians of interest and may suffer from ``freezing''
when there is a high probability of a single cluster occupying the entire system
\cite{evertz_lana_93,kawashima_gubernatis_94,evertz_03}. These drawbacks
can be overcome by the continuous-time world-line approach with worm updates, which works
in an extended configuration space with open loops and for which all updates are local
\cite{prokofev_svistunov_96,prokofev_svistunov_98,prokofev_svistunov_09} and by the
stochastic series expansion (SSE) algorithm \cite{sandvik_99}.

\paragraph{Stochastic series expansions:}
The SSE algorithm is based on the power series expansion of the partition function
\begin{equation}
 Z=\mathrm{Tr}\left\lbrace e^{-\beta\hat{H}}\right\rbrace=\sum_m\sum_{n=0}^\infty
\frac{(-\beta)^n}{n!}\,\langle m|\hat{H}^n|m\rangle,
\end{equation}
where $\lbrace|m\rangle\rbrace$ is a convenient basis [such as the one in Eq.~\eqref{eq:basisstates}].
It is useful to write the Hamiltonian as the sum of symmetric bond operators $\hat{H}_{a_i,b_i}$, where $a_i$
denotes the operator type in the bond, $b_i\in\lbrace1,\ldots L_b\rbrace$ the bond index, and $L_b$ the number
of bonds. For example, if the basis in Eq.~\eqref{eq:basisstates} is used, then the operator
types in the bonds could be selected to be (i) diagonal, i.e., containing terms related to the density, and
(ii) off-diagonal, i.e., containing the hopping related terms.

The partition function can then be written as
\begin{equation}
 Z=\sum_m\sum_{n=0}^\infty\sum_{S_n}
\frac{(-\beta)^n}{n!}\,\langle m|\prod_{i=1}^n\hat{H}_{a_i,b_i}|m\rangle,
\end{equation}
where $S_n$ is the set of all concatenations of $n$ bond operators. The average expansion order can be shown
to be $\langle n\rangle\sim L\beta$, so that one can truncate the sum over $n$ at a finite cutoff $n_\textrm{max}$
without introducing systematic errors ($n_\textrm{max}$ can be adjusted during the warm up phase of
the simulation) \cite{sandvik_99}. By inserting $n_\textrm{max}-n$ unit operators $\hat{H}_{0,0}= 1$
one can rewrite $Z$ as
\begin{equation}
 Z=\sum_m\sum_{S_{n_\textrm{max}}}
\frac{(-\beta)^n (n_\textrm{max}-n)!}{n_\textrm{max}!}\,\langle m|\prod_{i=1}^{n_\textrm{max}}\hat{H}_{a_i,b_i}|m\rangle,
\end{equation}
where now $n$ is the number of nonunit operators in $S_{n_\textrm{max}}$. This last expression simplifies
the Monte-Carlo sampling, which once again is used to avoid summing over an exponentially large number
of terms. Various update schemes have been implemented within the SSE approach, such as the
``operator loop updates'' \cite{sandvik_99} and the ``directed loops'' \cite{syljuasen_sandvik_02,alet_wessel_05}.

In general, the SSE as well as the continuous-time world-line formulation with worm updates have been found
to be very efficient when dealing with spin and boson systems. We note that within these approaches:
(i) the superfluid density $\rho_s$ is calculated through
the fluctuations of the winding number $W$ in the world-line configurations generated during the simulations
$\rho_s=\langle W^2\rangle L/(2\beta t)$ \cite{pollock_ceperley_87}, where $W$ is the net number of times that
the world-lines wound around the periodic system, and (ii) the one-particle Green's function can also be
efficiently computed \cite{dorneich_troyer_01,prokofev_svistunov_09}.

\subsubsection{The Bose-Hubbard model and its phase diagram}\label{sec:BHMPeriodic}

Let us analyze the Bose-Hubbard model (\ref{eq:bose-hubbard}). As will be discussed in Sec.~\ref{sec:experiments}
this model is relevant in a wide variety of contexts such as ${}^4$He in various confined geometries (such as
in porous media or between surfaces), granular superconductors, Josephson arrays, and ultracold gases in optical
lattices. As with several models on a lattice, and in a way that will be examined in details in Sec.~\ref{sec:mott},
this model contains the necessary ingredients to describe a quantum phase transition between a superfluid and an
insulator. This transition, which takes place only at integer fillings and is driven by quantum fluctuations
rather than by thermal fluctuations, is the consequence of the competition between the delocalization effects of
the kinetic term, proportional to $t$, which reduce the phase fluctuations, and the localization effects of the
interaction term, proportional to $U$, which reduce the on-site particle number fluctuations.

The phase diagram of this model is qualitatively similar in all spatial dimensions. In 1D it can be calculated by
bosonization techniques \cite{haldane_bosons}, which we will introduce in Sec.~\ref{sec:bosonize} and
Sec.~\ref{sec:mott}. Its general properties in any dimension can be easily understood in terms of perturbative
arguments starting from the atomic limit and scaling arguments \cite{fisher_boson_loc}. For $t=0$, $\hat{n}_i$
commutes with the Hamiltonian and each site is occupied by an integer number of bosons $n$ (we drop the site index
because of the translational invariance). The site occupation can be computed minimizing the on-site energy
$E=-\mu n+ Un(n-1)/2$. One finds that for $n-1\leq\mu/U<n$ (where $n>0$) the occupation per site is $n$ and for
$\mu<0$ is zero. Now let us analyze what happens if one takes $\mu/U=n-1/2+\delta$, where $-1/2<\delta<1/2$ so
that the site occupation is $n$, and adds some small hopping $t$. If $t$ is smaller than the smallest of the
energies required to add a particle $E_p\sim(1/2-\delta)U$ or a hole $E_h\sim(1/2+\delta)U$, then one immediately
realizes that $n$ will not change. This is because the kinetic energy ($\sim t$) gained by the hopping of the
added particle (hole) will be smaller that the interaction energy that is needed to be overcome to add it. Hence,
there are finite regions in the plane $\mu/U$-$t/U$ in which $n$ is fixed at their values in the atomic limit.
Hopping in energetically unfavorable in those regions and the bosons remain localized, i.e., the system is insulating.
This insulating state, known as a Mott insulator, is characterized by an energy gap (the energy required to
add or remove a particle), which leads to a vanishing compressibility $\kappa=\partial n/\partial\mu=0$.
The lowest lying particle conserving excitations in the Mott insulator are particle-hole excitations.

By changing $\mu$ for any finite value of $t$ within the Mott insulating phase, one realizes at some point
$\mu_+(U,t)$ [$\mu_-(U,t)$] the kinetic energy gained by the added particle (hole) balances the interaction
energy required to add it. At that point, the added particle (hole) will be free to hop and a finite density
of those particles (holes) will led to BEC (in dimensions higher than one) and
superfluidity. Following the perturbative arguments given before, one can also realize that the gap
[$\Delta(U,t)=\mu_+(U,t)-\mu_-(U,t)$] in the Mott insulator should decrease as the hopping amplitude increases,
and this leads to a phase diagram with a lobe-like structure (see e.g., Fig.~\ref{fig:phasediag1Dclean}).

As will be discussed in details in Sec.~\ref{sec:mott}, two different universality classes exist for the Mott
transition: (i) A transition that occurs by changing the chemical potential (density), which is driven by density
fluctuations and belongs to the mean-field universality class. (This transition has $d_u=2$ as the upper critical
dimension.) (ii) A transition that occurs by changing $t/U$ at fixed density, which is driven by phase fluctuations
and belongs to the ($d+1$)-dimensional $XY$ universality class. (This transition has $d_u=3$ as the upper critical
dimension and $d_l=1$ as the lower critical dimension.)

Several analytical and computational approaches have been used to study the phase diagram of the Bose-Hubbard model
in one, two, three, and infinite dimensions. Among the most commonly used ones are mean-field theory, which is
exact in infinite dimensions but qualitatively correct in dimensions higher than one
\cite{fisher_boson_loc,rokhsar_kotliar_91}, quantum Monte-Carlo simulations
\cite{batrouni_scalettar_90,batrouni_scalettar_92,sansone_prokofev07,sansone_soyler_08}, strong coupling
expansions \cite{freericks_monien_96,elstner_monien_99,freericks_krishnamurthy_09}, and density matrix
renormalization group \cite{pai_pandit_96,kuhner_monien_98}.

In what follows, we restrict our discussion to the 1D case. It is interesting to note that despite the close
relation of the Bose-Hubbard model in 1D to the Lieb-Liniger model, the former is not Bethe-Ansatz solvable
\cite{haldane_80,choy_haldane_82}. The phase diagram for this model was first computed by
\citet{batrouni_scalettar_90} by means of the world-line quantum Monte Carlo approach. In that work,
the transition driven by changing the chemical potential was confirmed to have Gaussian exponents as
proposed by \citet{fisher_boson_loc}.

Studying the transition at constant density turned out to be a more challenging task. This transition,
as shown by \citet{haldane_bosons} (see also Sec.~\ref{sec:mott} and references therein) 
is in the same universality class as the two-dimensional
Berezinskii-Kosterlitz-Thouless (BKT) transition. As a consequence, the gap vanishes as
$\Delta\propto \exp[-\textrm{const}/\sqrt{(t/U)_\textrm{crit}-t/U}]$. This behavior makes the determination
of $(t/U)_\textrm{crit}$ in 1D much more difficult than in higher dimensions. Early attempts
for $n=1$ led to wide range of values, such as $(t/U)_\textrm{crit}=0.215\pm0.01$ from world-line quantum
Monte Carlo simulations \cite{batrouni_scalettar_90}, $(t/U)_\textrm{crit}=1/(2\sqrt{3})=0.289$
from a Bethe-Ansatz approximation \cite{krauth_91}, $(t/U)_\textrm{crit}=0.304\pm0.002$ from a combination
of exact diagonalization and renormalization group theory \cite{kashurnikov_svistunov_96},
$(t/U)_\textrm{crit}=0.300\pm0.005$ from world-line QMC simulations \cite{kashurnikov_krasavin_96}
and $(t/U)_\textrm{crit}=0.298$ from DMRG \cite{pai_pandit_96} after fitting the gap to the
exponential form given before, and $(t/U)_\textrm{crit}=0.265$ from a third order strong coupling
expansion after a constrained extrapolation method was used \cite{freericks_monien_96}.
\begin{figure}[!htb]
\centerline{\includegraphics[height=0.4\textwidth,angle=-90]{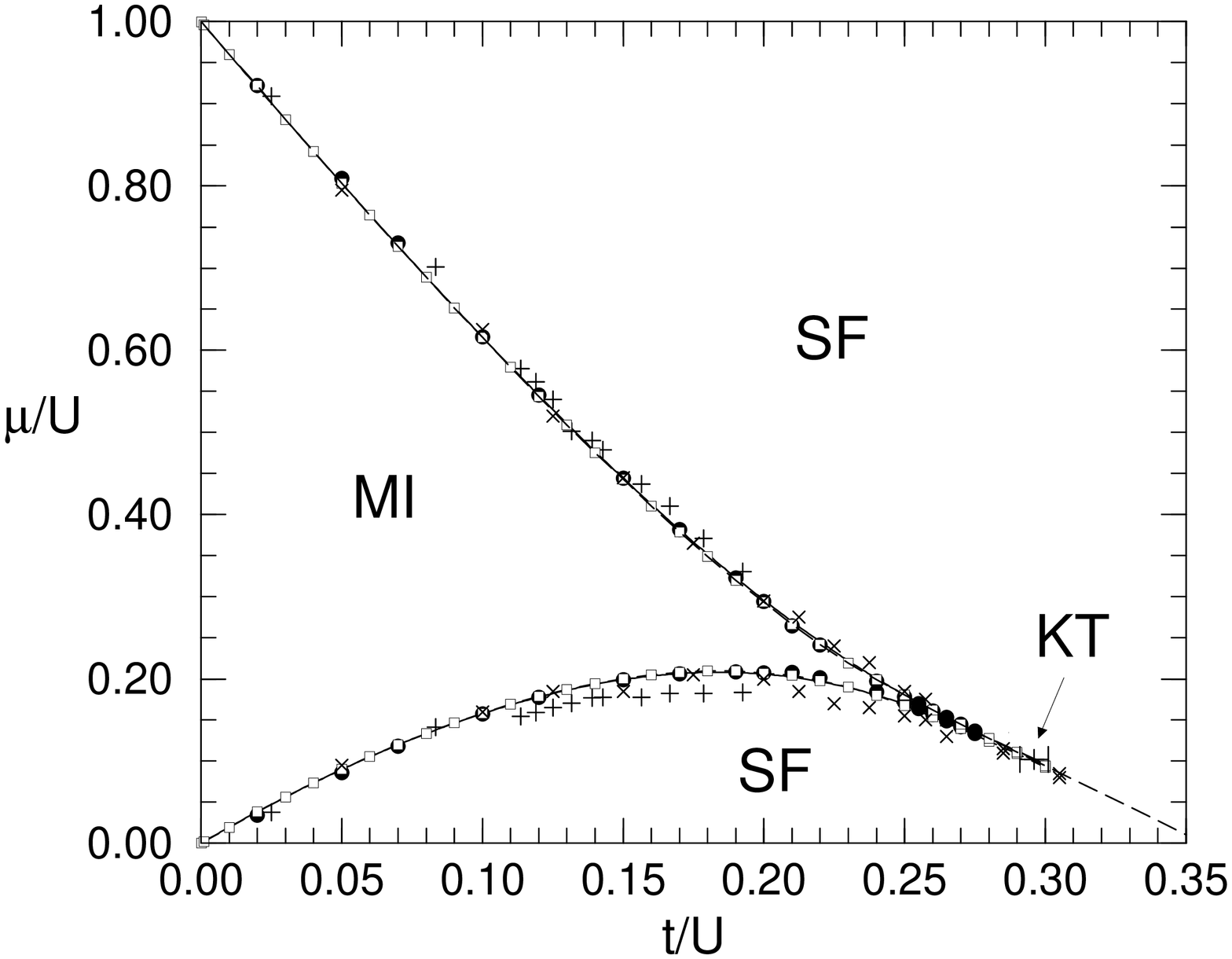}}
\caption{Phase diagram for the 1D Bose-Hubbard model obtained by means of:
two different sets of quantum Monte Carlo results [+] \cite{batrouni_scalettar_90} and
[$\times$] \cite{kashurnikov_krasavin_96}, [solid line] a Pad\'e analysis of $12^\textrm{th}$
order strong coupling expansions \cite{elstner_monien_99}, [filled circles]
earlier DMRG results \cite{kuhner_monien_98}, and [empty boxes] the DMRG results from
\citet{kuhner_white_00}. The dashed lines indicate the area with integer density.
The error bars in the $\mu/U$ direction are smaller than the circles, the error bar in the
$t/U$ direction is the error of $(t/U)_c$ for the BKT transition.
The Mott insulating phase is denoted by MI and the superfluid phase by SF \cite{kuhner_white_00}.}
\label{fig:phasediag1Dclean}
\end{figure}

A remarkable feature about the superfluid to Mott insulator transition that is unique to the phase diagram in
1D, and which was found in some of the previous studies, is that the Mott lobe exhibits a reentrant behavior
with increasing $t$. Later studies using DMRG \cite{kuhner_monien_98,kuhner_white_00}, and higher order strong
coupling expansions combined with Pade approximants \cite{elstner_monien_99}, provided some of the most accurate
results for the phase diagram. Those are depicted in Fig.~\ref{fig:phasediag1Dclean} and compared with the
results of quantum Monte Carlo simulations. The ratio $(t/U)_\textrm{crit}=0.297\pm0.010$ at the tip of the Mott
lobe was determined using DMRG  \cite{kuhner_white_00} by finding the value of $t/U$ at which the
Luttinger parameter $K = 2$ (see Sec.~\ref{sec:mott}).

\subsubsection{The Bose-Hubbard model in a trap}\label{sec:BHMTrap}

In relation to the experiments with ultracold bosons in optical lattices, a problem that has been
extensively studied recently is that of the Bose-Hubbard model in the presence of a confining potential.
The hamiltonian in this case is $\hat{H}_\textrm{trap}=\hat{H}_\textrm{BHM}+\sum_i V_i\, \hat{n}_i$,
where $V_i$ is usually taken to represent the nearly harmonic trap present in the experiments
$V_i=V (i-L/2)^2$.

The effect of the trapping potential is something that can be qualitatively, and in many
instances quantitatively, understood in terms of the local density approximation (LDA). Within the LDA,
one approximates local quantities in the inhomogeneous system by the corresponding quantities in a
homogeneous system with a local chemical potential equal to
\begin{equation}\label{eq:muLDA}
 \mu_\textrm{HOM}=\mu_i\equiv\mu_0-V\left(i-L/2\right)^2.
\end{equation}
Hence, density profiles in the trap can be constructed from vertical cuts across the homogeneous
phase diagram, which in 1D is depicted in Fig.~\ref{fig:phasediag1Dclean}, by starting a point
with $\mu=\mu_0$ and moving towards the point $\mu$ where the density vanishes. This leads to a wedding
cake structure in which superfluid and Mott insulating domains coexist space separated. Such a
structure was originally predicted by means of mean-field theory by \citet{jaksch_bose_hubbard}
and, for 1D systems, obtained by means of world-line QMC simulations by \citet{batrouni_rousseau_02}.

A consequence of the coexistence of superfluid and Mott insulating domains is that the total compressibility
in a trap never vanishes \cite{batrouni_rousseau_02}. This has motivated the definition and use of various
local quantities such as the local density fluctuations $\Delta_i=\langle n_i^2\rangle -\langle n_i\rangle^2$
and local compressibilities \cite{batrouni_rousseau_02,wessel_alet_04,rigol_batrouni_09}
\begin{equation}
 \kappa^{\textrm{l}_1}_i=\frac{\partial n_i}{\partial \mu_i} = \frac{1}{\beta}
\left[\left\langle  \left( \int_0^\beta d\tau \hat{n}_i(\tau)\right)^2\right\rangle -
\left\langle \int_0^\beta d\tau \hat{n}_i(\tau)\right\rangle^2\right],
\end{equation}
and
\begin{equation}\label{eq:compress}
\kappa^{\textrm{l}_2}_i=\frac{\partial N}{\partial \mu_i} = \int_0^\beta d\tau
\left[\langle\hat{n}_i(\tau)\hat{N}\rangle -
\langle\hat{n}_i(\tau)\rangle\langle\hat{N}\rangle\right],
\end{equation}
to characterize trapped systems. We note that $\kappa^{\textrm{l}_2}_i\equiv\partial n_i/\partial \mu$,
i.e., it can also be computed as the numerical derivative of experimentally measured density profiles.
This quantity vanishes in the Mott insulating domains and as such it can be taken as a local
order parameter \cite{wessel_alet_04}.

In Sec.~\ref{sec:corr-latt}, we used a dimensional argument to introduce the characteristic density
$\tilde{\rho}=Na/\zeta$ as the appropriate quantity to define the thermodynamic limit in a trapped system.
Hence, for each value of $U/t$, this quantity uniquely determines the local phases that are present in the
trap. A more rigorous derivation follows from the local density approximation
\cite{batrouni_krishnamurthy_08,roscilde_10}. Within the LDA $\langle n_i \rangle = n(\mu_i/t,U/t)$, where
$n(\mu_i/t,U/t)$ is the density in a homogeneous system with a chemical potential $\mu_i$ as defined in
Eq.~\eqref{eq:muLDA}, and $\mu_0$ is determined by the normalization condition
$N=\sum_i \langle n_i \rangle = \sum_i n(\mu_i/t,U/t)$. Approximating the sum by an integral, and taking
into account the reflection symmetry of the trapped system, $N=2\int_0^\infty dx\,n([\mu_0-V (x/a)^2]/t,U/t)$.
Finally, by making a change of variables $\mu_x=\mu_0-V (x/a)^2$, one gets\footnote{This result can be easily
generalized to higher dimensions where $\tilde{\rho}\equiv N a (V/t)^{d/2}$ \cite{batrouni_krishnamurthy_08}.}
\begin{equation}\label{eq:rhoLDA}
\tilde{\rho}\equiv N a \sqrt{V/t}=
\int_{-\infty}^{\mu_0} d\mu_x \dfrac{n(\mu_x/t,U/t)}{\sqrt{t(\mu_0-\mu_x)}}.
\end{equation}
As discussed before, within the LDA $\mu_0$ uniquely determines the density profiles, and Eq.~\eqref{eq:rhoLDA}
shows that $\mu_0$ is uniquely determined by $\tilde{\rho}$, so that indeed, $\tilde{\rho}$ uniquely determines
the local phases in the trap. This fact has been recently used to construct state diagrams for trapped systems
in 1D and 2D \cite{rigol_batrouni_09}, which are useful in order to compare experimental and theoretical results
in which different trap curvatures and system sizes are considered. Those state diagrams can be seen as the
equivalent to phase diagrams for homogeneous systems.

A word of caution is needed when using the LDA. For example, for a homogeneous
noninteracting bosonic gas in 2D there is no finite temperature transition to
a BEC, from which, within the LDA,  one could wrongly conclude that condensation
cannot take place in the two-dimensional trapped case. This is, of course,
incorrect as condensation does occur in the latter case
\cite{dalfovo_giorgini_review_99}. Equally, as one crosses the superfluid to
Mott insulator transition in a homogeneous system in any dimension, or any
other second order transition for that matter, the system becomes critical and
long-range correlation will preclude the local density approximation from
being valid in the finite size region where such a transition occurs in a
trap. A finite size scaling approach is required in those cases
\cite{campostrini_vicari_10a,campostrini_vicari_10b}. Finally, of
particular relevance to the 1D case, the superfluid phase in 1D is already a
critical phase with power-law decaying correlations, so that one needs to be
particularly careful when using the LDA in 1D
\cite{bergkvist_henelius_04,wessel_alet_04}.

With that in mind, theoretical studies have addressed how correlation behave in the 1D trapped
Bose-Hubbard model. Kollath {\it et al.}~\cite{kollath_schollwock_04} found that the same power-law
decay of the one-particle correlations that is known from homogeneous systems can still be observed
in the trap, in the regime of weak and intermediate interaction strength, after a proper rescaling
$\tilde{g}_1(i,j)=\langle \hat{b}^\dagger_i \hat{b}_j\rangle/\sqrt{n_i n^{}_j}$ is considered. In the infinite
repulsive limit (Sec.~\ref{sec:corr-latt}), on the other hand, \citet{rigol_muramatsu_05HCBb}
concluded that a different rescaling is required, and
$\tilde{g}_1(i,j)=\langle \hat{b}^\dagger_i \hat{b}_j\rangle/[n^{}_i(1-n_i)n^{}_j(1-n_j)]^{1/4}$ was proposed.
The latter rescaling is consistent with the results in the continuum (equivalent to the low density limit
in the lattice) presented in Sec.~\ref{sec:corr-cont}, where analytical expressions for the
one-particle density matrix in the trap are available.

In the early experiments, the momentum distribution function $n(k)$ of the trapped gas (measured after time of
flight) was one of the few probes available to extract information about those systems (see
Sec.~\ref{sec:experiments}). Hence, various numerical studies were devoted to identify how, (i) the peak high
and half width of $n(k)$ \cite{kollath_schollwock_04,wessel_alet_04}, (ii) the visibility
${\cal V} = \frac{n^\textrm{max}(k)-n^\textrm{min}(k)} {n^\textrm{max}(k)+n^\textrm{min}(k)}$
\cite{sengupta_rigol_05}, and (iii) other observables such as the total energy measured after time of
flight \cite{rigol_scalettar_06}, can be used to detect the formation of the Mott insulating domains
in 1D. More recently, local measurements have become available in trapped systems
\cite{gemelke_Zhang_09,bakr_gillen_09,bakr_peng_09} so that the local compressibility in Eq.~\eqref{eq:compress}
can be determined from the experiments and the local phases and transitions identified
\cite{kato_zhou_08,zhou_kato_09}. However, it has been argued \cite{pollet_prokofev_10} that $n(k)$ is
nevertheless the best quantity to determine the critical parameters for the formation of the Mott insulator
in the trap.

\section{Low energy universal description} \label{sec:bosonize}

At low temperatures, the models discussed in the previous sections exhibit a liquid phase in which no continuous
or discrete symmetry is broken. This phase has two salient features: i) The low energy excitations are collective
modes with linear dispersion. ii) At zero temperature, the correlation functions exhibit an algebraic decay
characterized by exponents that depend on the model parameters. These two properties are intimately related
to the MWH theorem, which rules  broken continuous symmetry in 1D.\footnote{Actually, the
Calogero-Sutherland models (see Sec.~\ref{sec:calogero}) escape one of the conditions of the theorem,
which require the interactions to be short-ranged. However, as the exact solution demonstrates, they also
exhibit these properties.} \citet{haldane81_luttinger_liquid} noticed that these features are quite
ubiquitous in 1D and thus defined  a \emph{universality class} of 1D systems that encompasses a large number
of models of interacting bosons (both integrable and non-integrable) in 1D \cite{giamarchi_book_1d,gogolin_1dbook}. This
universality class,  known as Tomonaga-Luttinger liquids (TLLs), is a  line of fixed points of the
renormalization group (RG)  characterized by a single parameter called the Tomonaga-Luttinger parameter.

\subsection{Bosonization method}\label{sec:bosonization-method}

The collective nature of the low-energy excitations in 1D can be understood as follows: In the presence of
interactions,  a particle must push its neighbors away in order to propagate. Thus, when confinement forces particles
to move on a line, any individual motion is quickly converted into a collective one. This collective character
motivates a field-theoretic description in terms of collective fields, which is known as `harmonic fluid approach'
(also called `bosonization'
\cite{mattis65_lieb_luttinger_model,luther74_peschel_correlation_functions,haldane81_luttinger_liquid,haldane_bosons,giamarchi_book_1d}
for historical reasons). For bosons, the collective fields are the density, $\hat{\rho}(x,\tau)$, and the
phase, $\hat{\theta}(x,\tau)$, out of which the boson field operator, $\hat{\Psi}^{\dagger}(x,\tau)$,
is written in polar form:
\begin{equation}
\hat{\Psi}^{\dag}(x,\tau)  = \left[\hat{\rho}(x,\tau)\right]^{1/2} \: e^{-i \hat{\theta}(x,\tau)}.
\label{eq:bosonfield}
\end{equation}
The above equation is the continuum version of~(\ref{eq:phase-rep}). The quantum mechanical nature of harmonic
fluid description requires that we also specify the commutation relations of the fields
$\hat{\rho}(x,\tau)$ and $\hat{\theta}(x,\tau)$:
\begin{equation}\label{eq:rho-theta-commutator}
\left[\hat{\rho}(x), \hat{\theta}(x') \right] =
i  \delta(x-x').
\end{equation}
Equation (\ref{eq:rho-theta-commutator}) is consistent with
$\left[ \hat{\rho}(x), e^{-i\hat{\theta}(x')}\right] = \delta(x-x') e^{-i\hat{\theta}(x')}$,
which is required by the canonical commutation relations of the field
operator~\cite{haldane_bosons,giamarchi_book_1d}. It expresses the well known fact in the theory of superfluids
that phase and density are canonically conjugated fields. The relation~(\ref{eq:rho-theta-commutator}) can be
derived from the well known commutation relation between  the momentum current,
$\hat{\jmath}_P(x) = \frac{1}{2} \sum_{j=1}^N \left[ \hat{p}_j \delta(x-\hat{x}_j) +
\delta(x-\hat{x}_j)\hat{p}_j \right]$, and the density operators:
\begin{equation}\label{eq:f-sum-rule}
\left[ \hat{\rho}(x), \hat{\jmath}_P(x') \right]  =  i\hbar
\hat{\rho}(x') \partial_{x'} \delta(x-x'),
\end{equation}
which follows directly from $\left[ \hat{x}_i, \hat{p}_j \right] = i \hbar \delta_{ij}$. Using
(\ref{eq:bosonized_rho}) and (\ref{eq:bosonized_psi}), and retaining only the slowly varying terms,
this equation leads to:
$\hat{\jmath}(x)=\frac \hbar m \sqrt{\hat{\rho}(x)} \partial_x \hat{\theta} \sqrt{\hat{\rho}(x)}$.
Using Eq.~(\ref{eq:f-sum-rule}), Eq.~(\ref{eq:rho-theta-commutator}) is recovered.

In a translationally invariant  system, the ground state density is a constant,
$\langle \hat{\rho}(x,\tau) \rangle = \rho_0 =  N/L$. At small excitation energies, we expect
that $\hat{\rho}(x,\tau)$
does dot deviate much from $\rho_0$. In a simple-minded approach, these small
deviations would be described by writing the density operator as follows:
\begin{equation}
\hat{\rho}(x,\tau) \simeq \rho_0 - \frac{1}{\pi} \partial_x \hat{\phi}(x,\tau), \label{eq:phidef}
\end{equation}
where $\hat{\phi}(x,\tau)$ is a slowly varying quantum field, which means that its Fourier components are
predominantly around $q\simeq 0$. Such description is however insufficient especially when the interactions
are strong. For example, in the case of the TG gas (cf. Sec.~\ref{sec:tonks}) the pair-correlation function,
Eq.~(\ref{eq:paircorrelations_TG}), is such that: $D_2(x)-1= (1-\cos (2\pi \rho_0 x))/(2x^2)$

The first term in the right hand side of this expression stems from the  $q\approx 0$ density fluctuations and
it can be recovered using the method of \citet{popov72, stoof_popov_02}, which is valid for weakly interacting
bosons. However, the second oscillating term stems from density fluctuations of wave number $q_0
\approx 2\pi \rho_0$ and cannot be recovered from (\ref{eq:phidef}). That oscillating contribution reflects
the discreteness of the constituent particles, which locally (but not globally) tend to develop a crystal-like
ordering with a lattice spacing of order $\rho^{-1}_0$. Similar oscillating terms involving harmonics of
$2 \pi \rho_0$ also appear in other correlations functions of the TG gas such as the one-particle density matrix,
cf. Eq.~(\ref{eq:vaidya-girardeau}), as well as in the correlations of other models of interacting bosons in
general. \citet{efetov_coupled_bosons}, argued that the density operator must contain, besides the
$\partial_x \hat{\phi}$ term, a term proportional to $\times \cos 2 \pi \int^x dx' \: \rho(x') =
\cos(2\pi \rho_0 x - 2 \hat{\phi}(x))$ in order to reproduce the oscillating contribution in $D_2(x)$.

A more complete derivation of the oscillating terms, which also automatically includes all higher harmonics of
$2\pi \rho_0 x$, was given by \citet{haldane_bosons}, and is reviewed in what follows. One enumerates the particles
on the line by assigning them an index $j = 1, \ldots N$, such that two consecutive values of $j$ correspond to
two neighboring particles. Next, a labeling field $\hat{\phi}_l(x,\tau)$ is introduced. This field is smooth on
the scale of $\rho^{-1}_0$ and such that $\hat{\phi}_l(x,\tau) = \pi j$ for $x = x_j(\tau)$, where $x_j(\tau)$ is the
position of the $j$-th particle. Hence,
{\setlength\arraycolsep{0.9pt}
\begin{eqnarray}
\hat{\rho}(x,\tau)  &=& \sum_{j=1}^N \delta[x - \hat{x}_j(\tau)] \simeq \partial_x \hat{\phi}_l(x,\tau)
\sum_j\delta[\hat{\phi}_l(x,\tau) - j \pi]\nonumber\\& =&  \frac{1}{\pi}\partial_x \phi_l(x,\tau)
\sum_{m=-\infty}^{+\infty} e^{2 i m \pi \phi_l(x,\tau)}. \label{eq:density_representation}
\end{eqnarray}
}
The last expression follows from Poisson's summation formula. If we imagine that the particle positions, $x_j(t)$,
fluctuate  about the 1D lattice defined by $x^0_j = j  \rho^{-1}_0$, it is possible to reintroduce
$\hat{\phi}(x,\tau)$ as $\hat{\phi}_l(x,\tau) = \pi \rho_0 x - \hat{\phi}(x,\tau)$. In terms of $\hat{\phi}(x,\tau)$,
\begin{equation}
\hat{\rho}(x,\tau)  \simeq  \left( \rho_0 - \frac{1}{\pi}\partial_x \hat{\phi}(x,\tau)\right) \:
\sum_{m=-\infty}^{+\infty} \alpha_m e^{2 i m \left( \pi \rho_0 x - \hat{\phi}(x,\tau) \right)},
\label{eq:bosonized_rho}
\end{equation}
which reduces to Eq.~(\ref{eq:phidef}) only when the $m =0$ term of the series is retained.

The harmonic fluid approach also  provides a representation for the boson field operator~\cite{haldane_bosons,didier_2009}
in terms of $\hat{\phi}(x,\tau)$ and $\hat{\theta}(x,\tau)$. From Eq.~(\ref{eq:bosonfield}) it can be seen that
this relies on finding a representation for the operator $\left[ \hat{\rho}(x,\tau)\right]^{1/2}$.
This can be achieved by noting that (by virtue of Fermi's trick) the square root of the sum of Dirac-delta functions
in Eq.~(\ref{eq:density_representation}) is also proportional to a sum of delta functions. Thus,
$\left[ \hat{\rho}(x,\tau)\right]^{1/2}$ in Eq.~(\ref{eq:bosonfield}) lends itself to the same treatment as
$\rho(x,\tau)$, which leads to the following result~\cite{haldane_bosons, cazalilla_correlations_1d,giamarchi_book_1d}:
\begin{eqnarray}
 \hat{\Psi}^{\dagger}(x,\tau) &\simeq& \left( \rho_0 - \frac{1}{\pi}\partial_x \hat{\phi}(x,\tau)\right)^{1/2}
\nonumber\\&&\times\left[\sum_{m=-\infty}^{+\infty} \beta_m e^{2 i m \left( \pi \rho_0 -
\hat{\phi}(x,\tau) \right)} \right] \: e^{-i\hat{\theta}(x,\tau)}.\quad\quad \label{eq:bosonized_psi}
\end{eqnarray}    
The expressions~(\ref{eq:bosonized_rho}) and~(\ref{eq:bosonized_psi}) must be understood as low-energy
representations of the density and field operators as an infinite series of harmonics of
$2(\pi \rho_0 x - \hat{\phi}(x))$. They can be used, for instance, to compute the asymptotic behavior of
correlation functions in 1D systems. However, it must be also emphasized that the coefficients $\alpha_m,\beta_m$
of each term in these series are not determined by the harmonic fluid approach and, in general, depend on the
microscopic details of model (i.e., they are \emph{non-universal}). In some integrable models, such as
the $t$-$V$ model it is possible to obtain these coefficients
analytically~\cite{lukyanov_spinchain_asymptotics,lukyanov_spinchain_correlations}. In the case of the TG gas,
they can be deduced from Eqs.~(\ref{eq:vaidya-girardeau}) and (\ref{eq:paircorrelations_TG}). Similarly, for the
CS models, Eqs.~(\ref{eq:sutherland-coefs-1body}) and (\ref{eq:sutherland-coef-density}) can be used to derive the
coefficients $\alpha_m$ and $\beta_m$. However, in the general case, one must resort to
semi-numerical~\cite{caux_density} or fully numerical
\cite{hikihara_correlations_xxz_magneticfield,bouillot_ladder_dmrg_dynamics} methods to compute them.

Finally, let us provide a representation of the Hamiltonian and total momentum operators describing the
(low-energy part of the) spectrum. For a system of bosons of mass $m$ interacting via a two-body potential
$V(x)$, such as (\ref{eq:ham-sec}) (with $V_{\rm ext}$) one can substitute the expressions
(\ref{eq:bosonized_rho}-\ref{eq:bosonized_psi}) to obtain the following effective Hamiltonian:
\begin{equation}\label{eq:effective_ham}
\hat{H} =  \frac{\hbar}{2\pi} \int dx \left[ v K \left(\partial_x \hat{\theta}(x) \right)^2
+ \frac{v}{K} \left( \partial_x \hat{\phi}(x) \right)^2 \right] + \cdots
\end{equation}
The ellipsis stands for an infinite series of \emph{irrelevant} operators in the RG sense, which yield subleading
corrections to the system properties. Similarly, applying the same identities, the momentum operator
($\hat{P} =  \int dx \:  \hat{\jmath}_P(x)$), becomes
\begin{equation}
\hat{P} = \frac{\hbar}{\pi} \int dx
\left(\rho_0 - \frac{1}{\pi} \partial_x \hat{\phi}(x) \right) \partial_x \hat{\theta}(x)
+ \cdots \label{eq:bosonized_momentum}
\end{equation}
The Hamiltonian (\ref{eq:effective_ham}) and the formulas (\ref{eq:bosonized_rho}-\ref{eq:bosonized_psi})
thus allow to compute the low energy behavior of \emph{all}
correlation function of the initial problem. 
Since (\ref{eq:effective_ham}) is a simple quadratic Hamiltonian, with
$[\hat{\theta}(x),\partial_{x'}\hat{\phi}(x')]=i\delta(x-x')$, this is a remarkable simplification.
The parameters $v_J = v K$ and $v_N = v/K$ in (\ref{eq:effective_ham}) are model dependent but can be related
to the ground state properties. All the interaction effects are then encoded into the two effective parameters
$v$ and $K$. Moreover, it can be shown
that the Hamiltonian (\ref{eq:effective_ham}) 
leads to ground state wavefunctions of the Bijl-Jastrow
form~\cite{fradkin93,cazalilla_correlations_1d}. 
As we will see in Sec.~\ref{sec:corr-funct-boso}, $K$ controls the behavior of correlation functions
at long distance. As to $v$, one can immediately see that (\ref{eq:effective_ham}) leads to a linear dispersion
$\omega = v |k|$ so that $v$ is the velocity of propagation of density disturbances. The bosonization technique
can reproduce the Bogoliubov spectrum simply by approximating the kinetic energy term as:
\begin{equation}
\int dx \frac{\hbar^2}{2m} (\partial_x \hat{\Psi}^\dagger)(\partial_x \hat{\Psi}) \simeq
\int dx \frac{\hbar^2}{2m} [ \rho_0 (\partial_x \hat{\theta})^2 +
(\frac{\partial_x \rho}{2 \sqrt{\rho_0}})^2 ],
\end{equation}
leading to the Hamiltonian:
\begin{eqnarray}
\hat{H} & \simeq & \frac{\hbar}{2\pi} \int dx \left[ \frac{\pi \hbar \rho_0}{
    m} \left(\partial_x \hat{\theta}(x) \right)^2 + \frac{g}{\pi
    \hbar}
  \left( \partial_x \hat{\phi}(x) \right)^2 \right. \nonumber \\
&& +\left.  \frac{\hbar }{4 m \pi \rho_0 }
\left( \partial_x^2 \hat{\phi}(x) \right)^2
 \right] + \cdots
\end{eqnarray}
and yielding a spectrum $\epsilon(k) = \sqrt{g \rho_0 \frac{ \hbar^2 k^2}{m} + \left(\frac{\hbar^2 k^2}{2m}\right)^2}$, which
is identical to the Bogoliubov spectrum discussed in Sec.~\ref{sec:bosons_larged}. Note however, that here no condensate fraction assumption was made. We will come back to that point in Sec.~\ref{sec:collective-exc}.

Since \emph{all} the low-energy properties only depend on $v$ and $K$, it is enough to determine these two
nonuniversal parameters to fully describe the system. There are very efficient ways to do such a calculation,
either based on analytical or computational approaches. First, in a Galilean invariant system such as
Eq.~(\ref{eq:lieb-liniger-model}), the value of $v_J$\cite{haldane_bosons}, is independent of interaction
as shown by the following argument. From Galilean invariance, the center of mass position
$\hat{X} = \frac{1}{\hat{N}} \int dx \: x \hat{\rho}(x)$ obeys the equation of motion:
\begin{equation}
\frac{d\hat{X}}{dt} = \frac{1}{i\hbar} \left[ \hat{X}, \hat{H} \right] = \frac{\hat{P}}{\hat{N}m}.
\label{eq:cm_motion}
\end{equation}
Using~(\ref{eq:bosonized_rho}), $\hat{X}=-\frac{1}{\hat{N} \pi}\int dx \,x\partial_x\hat{\phi}(x)+\cdots$,
along with Eq.~(\ref{eq:rho-theta-commutator})}), in order to compute $\left[ \hat{X}, \hat{H} \right]/(i\hbar)$,
and comparing the result with $\hat{P}/\hat{N}m$ where $\hat{P}$ is given by Eq.~(\ref{eq:bosonized_momentum}),
yields $v_J = \hbar \pi \rho_0/m$. Furthermore, this argument shows that Galilean invariance also requires the
existence of an irrelevant operator (contained in the ellipsis of~\ref{eq:effective_ham}) of the form
$-\frac{\hbar^2}{2\pi m} \int dx \partial_x \hat{\phi}(x) \left(\partial_x \hat{\theta}(x) \right)^2$,
which describes the curvature of the quadratic free-particle dispersion. Several authors
\cite{pereira2005xxz,Khodas_bose_liquid} have recently emphasized that 
such irrelevant operators lead to a damping rate $\gamma_q \sim q^2/m$ for the collective excitations in 1D.

If Galilean invariance is absent, as in lattice models like the Bose-Hubbard or the $t$-$V$ models, then $v_J$ can
be renormalized by the interactions. Yet, $v_J$  can be still related to the zero-temperature response of the system
to an infinitesimal phase twist $\Delta\chi$ in the boundary conditions
\cite{shastry_twist_1d,schulz_conductivite_1d,giamarchi_umklapp_1d,%
Maslov_josephson_LL,svistunov_sf_fraction,cazalilla_correlations_1d,giamarchi_book_1d}.
Physically, this corresponds to the existence of a persistent current flowing through the system at zero temperature 
upon connecting it to two large phase-coherent reservoirs, whose phase differs by  $\Delta \chi$.  Thus, it is very tempting to
think that $v_J$ is related to the superfluid fraction at zero temperature, $\rho_s(T=0) = \rho^0_s$ by means of
the relation $v_J = \frac{\hbar \pi \rho^0_s}{m}$. However, this can be misleading. To clarify this point,
let us define:
\begin{equation}
\rho_\mathrm{twist}(T,L)  = \frac{\pi L}{\hbar} \frac{\partial^2 F(T, L, \Delta\chi)}
{\partial \Delta \chi^2}\Bigg|_{\Delta \chi = 0},
\end{equation}
where $F(T, \Delta \chi)$ is the free energy of a system of length $L$ computed assuming twisted boundary conditions,
that is, $\hat{\Psi}(L) = e^{i\Delta\chi}\: \hat{\Psi}(0)$. The superfluid fraction is a thermodynamic property
obtained in the limit $\lim_{L\to +\infty}\rho_\mathrm{twist}(T,L) = \rho_s(T)$. However, it is important to stress
that the limits $L \to +\infty$ and $T \to 0$ of $\rho_\mathrm{twist}(T,L)$ do not commute
\cite{giamarchi_persistent_1d,svistunov_sf_fraction}. For example, for the TG gas, whose thermodynamics
is equivalent the 1D free Fermi gas thermodynamics, $\lim_{T\to 0} \lim_{L\to +\infty}\rho_s(T) = \rho_s(T= 0) = 0$
because the free Fermi gas is not superfluid. However, when the limits of the free Fermi gas
$\rho_\mathrm{twist}(T,L$) are taken in the reverse order,
$\lim_{L \to +\infty} \lim_{T\to 0} \rho_\mathrm{twist}(T,L) = \rho_0$. This corresponds to the result $v_J  = v_F$,
which follows from Galilean invariance. The product $v K$ is thus related to Kohn's stiffness \cite{kohn_stiffness}
which is the weight of the $\delta(\omega)$ in the conductivity, and is related to the capability of the system
to sustain persistent currents.

As to the other stiffness parameter, $v_N = v/K$, it can be related to the inverse of the macroscopic
compressibility at $T = 0$~\cite{haldane_bosons,giamarchi_book_1d}, which is defined as:
\begin{equation}
\kappa^{-1}_s = \rho^2_0 L \left[ \frac{\partial^2 E_0(N)}{\partial N^2}\right].
\end{equation}
where $E_0(N)$ is the ground state energy with $N$ particles. To relate $\kappa^{-1}_s$ and $v_N$, first note that
the operator $\delta \hat{N} = \hat{N} - N = -\int \frac{dx}{\pi} \partial_x \hat{\phi}(x) $, which means
that, by shifting $\hat{\phi}(x) \to \hat{\phi}(x) - \pi x \delta N  /L$ in Eq.~(\ref{eq:effective_ham}), and
taking the expectation value over the ground state, it is possible to obtain  $\partial^2 E_0(N)/\partial N^2$ as
the coefficient of the $O(\delta N^2_0)$ term. Hence, $\kappa^{-1}_s=\hbar\pi v_N \rho^2_0=\hbar\pi\rho^2_0 v /K$.
For Galilean invariant systems, this relationship can be written as
$\kappa_s = m K^2/(\hbar^2 \pi^2 \rho^3_0)$, which means that, as $K$ increases, the system becomes more and more
compressible. In particular, for a non-interacting boson system $\kappa_s = +\infty$, as the energy cost of adding
new particles in the ground state vanishes and therefore $K = +\infty$. Repulsive interactions reduce the value of
$K$. For instance, in the Lieb-Liniger model  $K\geq1$ (cf. Fig.~\ref{fig:LiebLiniger_TLL}), and $K=1$ for TG gas, which
stems from its equivalence to a free Fermi gas. In 1D boson systems with longer range interactions
(such as a dipolar interactions \cite{lozovik05,citro08}) regimes where $K<1$ are possible.

\subsection{Correlation functions: temperature, boundaries, and finite-size effects}\label{sec:corr-funct-boso}

Realization of 1D ultracold atomic systems (see Sec.~\ref{sec:experiments}), where the number of bosons confined
to 1D typically ranges from a few tens to a few hundred, has renewed the interest in understanding finite-size
systems~\cite{cazalilla_correlations_1d,batchelor_1Dbox_BA}. In a finite size system, the low-energy excitation
spectrum as well as the asymptotic behavior of correlation functions depend on the boundary conditions (BC's)
obeyed by the field operator. These can be either periodic (PBCs) or open (OBCs). In physical realizations, PBCs
are more convenient to describe systems near the thermodynamic limit, as they include translational invariance
from the start, whereas OBCs are more realistic for finite or semi-infinite systems. Moreover, numerical methods
such as the DMRG described in Sec.~\ref{sec:latticenum} work better with OBCs than with PBCs. The effects of
temperature can also be treated in the framework of finite-size effects. Indeed, in the Matsubara formalism,
a finite temperature $\beta = 1/(k_BT)$ corresponds to PBCs on the imaginary time of size $\hbar \beta$.
The harmonic fluid approach introduced in the previous section can be extended to deal with both finite-size
and boundary effects. The conformal invariance of the theory (\ref{eq:effective_ham}) greatly simplifies that
extension \cite{giamarchi_book_1d} by enabling the use of conformal transformations \cite{cardy_conformal_book}
to compute the finite size effects.

\subsubsection{Infinite systems and periodic boundary conditions}\label{sec:pbc-boso}

Since the temperature effects are already well documented in the literature \cite{giamarchi_book_1d},
we focus here on the spatial boundary conditions. Let us begin our discussion in the case of the field operator
obeying (twisted) periodic BC's, that is, $\hat{\Psi}^{\dagger}(x+L)=e^{-i \Delta \chi}\hat{\Psi}^{\dagger}(x)$.
Without the twist ($\Delta \chi=0$), this is simply the usual periodic boundary condition taken in many
thermodynamic systems. A physical realization of such PBCs, corresponds to loading an ultracold Bose gas in a
tight toroidal trap, which can be created by \emph{e.g.} combining a Gaussian laser beam with a magnetic harmonic
trap~\cite{ryu_toroidal_trap}.

In order to determine the appropriate boundary conditions for the fields $\hat{\phi}(x)$ and $\hat{\theta}(x)$,
we note that Eq.~(\ref{eq:bosonized_psi}) implies that the boson field operator obeys (twisted) periodic BC's
provided
\begin{eqnarray}
\hat{\phi}(x+L) &=& \hat{\phi}(x) - \pi (\hat{N}-N), \label{eq:phi_pbc}\\
\hat{\theta}(x+L) &=& \hat{\theta}(x) + \pi (\hat{J} + \frac{\Delta \chi}{\pi}), \label{eq:theta_pbc}
\end{eqnarray}
where the operator $N$ has integer eigenvalues whereas the eigenvalues of $J$ obey the selection rule
$(-1)^J$~\cite{haldane_bosons}. The conditions~(\ref{eq:phi_pbc}) and (\ref{eq:theta_pbc}) must be taken into
account when developing $\hat{\theta}(x)$ and $\hat{\phi}(x)$ in Fourier series, which for the present system read:
\begin{eqnarray}
\hat{\phi}(x) &=& \hat{\phi}_0 -  \delta \hat{N} \frac{x\pi}{L} - \frac{i}{2} \sum_{q\neq 0}
\left( \frac{2\pi K}{L |q|} \right)^{\frac{1}{2}} \mathrm{sgn}(q)\nonumber\\
&&\qquad\qquad\qquad\qquad\quad \times e^{-iq x}\left[ \hat{a}^{\dag}_q + \hat{a}_{-q}\right], \\
\hat{\theta}(x) &=& \hat{\theta}_0 + \hat{J} \frac{x\pi}{L}  + \frac{i}{2} \sum_{q \neq 0 }
\left( \frac{2\pi}{L|q| K} \right)^{\frac{1}{2}} e^{-i q x}
\left[ \hat{a}^{\dag}_{q} - \hat{a}_{-q} \right],\nonumber
\end{eqnarray}
where $q = 2\pi m/L$ ($m$ being an integer), $\left[ \hat{a}_{q}, \hat{a}^{\dagger}_{q'} \right] = \delta_{q,q'}$,
commuting otherwise, $\left[\hat{N} , e^{-i\hat{\theta}_0}\right] = e^{-i\hat{\theta}_0}$ and
$\left[ \hat{J}, e^{-i\hat{\phi}_0} \right] = e^{-i\hat{\phi}_0}$. The operator $\hat{\phi}_0$ is related to the
center of mass position of the liquid (and therefore $\pi^{-1} d\hat{\phi}_0/dt$ is the mean current
operator~\cite{haldane_bosons}), whilst $\hat{\theta}_0$ is related to the global phase.

Introducing these expressions into~(\ref{eq:effective_ham},\ref{eq:bosonized_momentum}) leads to:
\begin{eqnarray}
\hat{H} &=& \sum_{q\neq 0} \hbar v |q| \, \hat{a}^{\dag}_q \hat{a}_q  + \frac{\hbar \pi}{2 L K}
(\hat{N}-N)^2 \nonumber\\
&& \quad + \frac{\hbar\pi v K}{2L}  \left(\hat{J} + \frac{\Delta \chi}{\pi}\right)^2, \\
\hat{P} &=& \frac{\hbar \pi \hat{N} \hat{J}}{L} +  \sum_{q\neq 0} \hbar q \,  \hat{a}^{\dag}_q \hat{a}_q.
\end{eqnarray}
Thus, the elementary excitations of the system are running waves carrying momentum $\hbar q$ and energy
$\hbar v |q|$ ($v$ is the sound velocity). The system can also sustain persistent currents, quantized in values
proportional to the eigenvalue of $\hat{J}$, which for bosons must be even integers.

Furthermore, from Eqs.~(\ref{eq:phi_pbc}) and  (\ref{eq:theta_pbc}) the asymptotic behavior of correlation
functions obeying PBC's can be obtained. For the one-particle density matrix and the pair-correlation function,
it was obtained in~\cite{haldane_bosons}:
\begin{widetext}
\begin{eqnarray}\
g_1(x) &=& \langle \hat{\Psi}^{\dagger}(x) \hat{\Psi}(0) \rangle
=   \rho_{0} \left[  \frac{1}{\rho_{0} d(x|L)} \right]^{\frac{1}{2K}}
 \left\{  A_0 + \sum_{m= 1}^{+\infty} A_m
\left[  \frac{1}{\rho_{0} d(x|L)} \right]^{2m^2 K} \cos \left(2\pi m \rho_{0}
x \right) \right\}, \label{eq:g1pbc}\\
 D_2(x) &=& \rho^{-2}_{0} \langle \hat{\rho}(x) \hat{\rho}(0) \rangle =  \left\{ 1 - \frac{K}{2\pi^2}
\left[\frac{1}{\rho_{0} d(x|L)} \right]^2 +
\sum_{m > 0} B_{m}
\left[\frac{1}{\rho_{0} d(x| L)} \right]^{2m^2K}
\cos \left(2 \pi m  \rho_{0} x\right)  \right\}.\label{eq:densitypbc}
\end{eqnarray}
\end{widetext}
A more detailed treatment of the term $ (\rho_0 -
\partial_x \hat{\phi}/\pi)^{1/2}$ of Eq.~(\ref{eq:bosonized_psi})has
been developed in \cite{didier_2009}, allowing to obtain also the
subleading corrections to (\ref{eq:g1pbc}). 
As mentioned above, the coefficients $A_m=|\alpha_m|^2$ and $B_m=|\beta_m|^2$ are non-universal;
$d(x|L) = L |\sin(\pi x/L)|/\pi$ is the \emph{chord} function.

\begin{figure}
\includegraphics[width=0.45\textwidth]{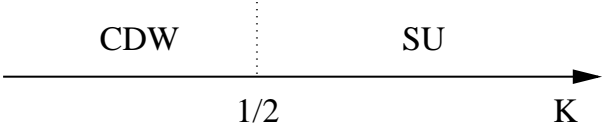}
\caption{\label{fig:diagbosons} ``Phase diagram'' indicating the slowest decaying fluctuation for
1D interacting bosons. The correlations are totally controlled by the Tomonaga-Luttinger liquid
parameter $K$. For $K>1/2$ the dominant order is superfluid (SU), while for $K<1/2$ the system exhibit dominant
charge density wave (CDW) fluctuations. Note that for a model with purely local interactions one has
$1 < K < \infty$ so the system is always dominated by superfluid order. Nearest neighbor repulsion on a lattice
can allow to reach the point $K = 1/2$.}
\end{figure}

Let us first examine these expressions for an infinite system. In the thermodynamic limit where
$L \to \infty$  $d(x|L) \to |x|$, and thus we retrieve the well-known power-law correlations
controlled by a single parameter, $K$. The Euclidean invariance of the theory ensures that a similar power-law
occurs in Matsubara time. We thus see that the system is in a critical state, with quasi-long range order both
in the one-particle correlation and in the density-density one. The later signals the tendency of the system
to form a charge density wave of bosons, while the first one is the tendency to BEC. Although there is no
long range order one can define a ``phase diagram'' by looking at the dominant correlation, i.e., the one
with the slowest decay. This is the one that should be realized as an ordered state should one couple
1D system, as will be discussed in Sec.~\ref{sec:coupled1dbosons}. Such a phase diagram is shown
in Fig.~\ref{fig:diagbosons}.
Upon comparison with, \emph{e.g.} Eqs.~(\ref{eq:vaidya-girardeau}) and (\ref{eq:paircorrelations_TG}), it can be
seen that the above series contains only the leading (i.e., the slowest decaying term) for each harmonic.
The remaining subleading corrections stem from less relevant operators, which have been neglected in the
derivation of (\ref{eq:g1pbc}) and (\ref{eq:densitypbc}).

\subsubsection{Open boundary conditions}

In the case of  a system of bosons trapped in a box with sufficiently high walls, OBCs of Dirichlet
type must be used, i.e., the field operator must obey $\hat{\Psi}(x) = 0$  and hence $\hat{\rho}(x) = 0$ for
$x = 0,L$. Hence, Eq.~(\ref{eq:density_representation}) implies that, for $x = 0, L$, the labeling field
$\hat{\phi}_l(x) \neq \pi j$, where $j$ is an integer. This is the minimum requirement to ensure that no particle
exist at $x =0, L$. Hence, $\hat{\phi}(x) = \pi \rho_0 x - \hat{\phi}_l(x) = \phi_0 \neq \pi j$ at $x = 0$. If
the number of particles (that is the eigenvalue of $\hat{N}$) is fixed (as  corresponds to a system of
trapped ultracold atoms), then it follows from $\delta \hat{N}  = \hat{N} - N =  -
\int^{L}_0 \frac{dx}{\pi}\, \partial_x \hat{\phi}(x) = \hat{\phi}(0) - \hat{\phi}(L)$ that
$\hat{\phi}(x = L) = \hat{\phi}(0) - \pi\hat{N}$.

When developed in a Fourier series of modes that obey the open BC's:
\begin{eqnarray}
\hat{\phi}(x) &=& \hat{\phi}_0 - \delta \hat{N} \frac{\pi x}{L} - \sum_{q > 0} \left( \frac{\pi K}{qL}
\right)^{\frac{1}{2}} \sin(q x) \left[ \hat{a}^{\dag}_{q} + \hat{a}_{q} \right], \nonumber\\
\hat{\theta}(x) &=& \hat{\theta}_0 - i  \sum_{q > 0}  \left( \frac{\pi K}{qL}\right)^{\frac{1}{2}}
\cos (q x) \left[ \hat{a}^{\dag}_q - \hat{a}_q  \right],
\end{eqnarray}
where $\left[ \delta \hat{N}, e^{-i\hat{\theta}_0} \right] = e^{-i\hat{\theta}_0}$, but $\hat{\phi}_0$
is a real (non-integer) number, i.e., it is \emph{not} an operator. The Hamiltonian operator becomes:
\begin{equation}
\hat{H} = \sum_{q > 0} \hbar v q \: \hat{a}^{\dag}_q \hat{a}_q + \frac{\hbar \pi v}{2L K} \left(\hat{N} - N\right)^2.
\end{equation}
Thus, the low-energy elementary excitations are standing waves characterized by a wavenumber $q > 0$ and an energy
$\hbar v q$. However, the total momentum operator, $P$, vanishes identically as the center of mass position of the
system is fixed (\emph{e.g.}, $\langle \hat{X} \rangle \propto  \phi_0 =$ const.) by the confinement. This means
that the operator $J$ undergoes large fluctuations and therefore its canonical conjugate operator,
$\phi_0$, acquires a constant expectation value.

The existence of boundaries leads to a dramatic change in the nature of the correlations as translational
invariance is  lost. Thus, for example, to leading order
{\setlength\arraycolsep{0.5pt}
\begin{eqnarray}
\rho_0(x) &=&  \rho_{0}  +
 C_1 \left[ \frac{1}{\rho_{0} d(2x|2L)} \right]^{K}
\cos (2 \pi \rho_{0} x + \delta_1),\quad\quad\\
g_1(x,y) &=& \rho_{0} B_0\:
\left[ \frac{\rho^{-1}_{0}
\sqrt{d(2x|2L) d(2y|2L)}}{d(x+y|2L) d(x-y|2L)}\right]^{\frac{1}{2K}},
 \label{eq: g1obc}
\end{eqnarray}
}where $C_{1}, \delta_1$, and $B_{0}$ are non-universal coefficients. For the full asymptotic series in harmonics of
$2\pi\rho_0$ of these and other correlation functions, we refer the interested reader to the
literature~\cite{Cazalilla_1DBG_02,cazalilla_correlations_1d}. Thus, because the system has boundaries, 
the ground state density $\rho_0(x) =  \langle \rho(x) \rangle$  is no longer uniform, and two-point correlation functions such as
$g_1(x,y) = \langle \hat{\Psi}^{\dag}(x) \hat{\Psi}(y) \rangle$ depend both on $x$ and $y$
\cite{eggert_openchains,hikihara_correlations_ladder_magneticfield,giamarchi_book_1d,%
Cazalilla_1DBG_02,cazalilla_correlations_1d}.
It is also important to notice that the boundaries do modify the asymptotic behavior of the correlation functions.
This can be seen in the previous expressions if we consider, for instance, the behavior of $g_1(x,y)$. Away from
the boundaries, (in the bulk) of the system, the leading behavior is the same as for PBC's, that is $~|x|^{-\frac{1}{2K}}$.
However, for $y \approx 0$ and $\rho^{-1}_0\ll x \ll L$, the decay of the asymptotic behavior is
$|x|^{-\frac{3}{4K}}$ controlled by the (\emph{boundary}) exponent  $\frac{3}{4K}$. The same remarks apply to the behavior
of other correlation functions near the boundaries at $x =0, L$.

Another type of OBCs (Neumann) corresponds to $\hat{\Psi}^{\dag}(x=0) = \Psi^*_0$ and
$\hat{\Psi}^{\dag}(x=L) =  \Psi^{*}_{L}$, where $\Psi_{0,L} = |\Psi_{0,L}| e^{-i \chi_{0,L}} \neq 0$
\cite{Maslov_josephson_LL}. This amounts to enforcing that  $\hat{\theta}(x = 0,L) = \chi_{0,L}$.
In this case, the results obtained for open BC's of  Dirichlet type can used upon interchanging the roles of
$\theta$ and $\phi$ and making the replacement $K\to K^{-1}$.

The form of the effective Hamiltonian, however, is different:
\begin{equation}
\hat{H} = \sum_{q > 0} \hbar v q \: \hat{a}^{\dag}_q \hat{a}_q + \frac{\hbar \pi v K}{2L} \left(\hat{J}
        + \frac{\Delta \chi}{\pi}\right)^2,
\end{equation}
where $\Delta \chi = \chi_L-\chi_0$ the phase difference across the system. Note that, in this case, the global
phase $\langle \hat{\theta}_0 \rangle = \Delta\chi$ is well defined, which means that the canonically conjugate
operator, $N$, undergoes large fluctuations about the ground state value $N = \langle \hat{N} \rangle$
that fixes the zero-temperature chemical  potential.

\subsubsection{The $t$-$V$, Lieb-Liniger, and Bose-Hubbard models as Tomonaga-Luttinger liquids}\label{sec:t-v-boso}

Let us now give the relation between some of the microscopic models that were examined in Sec.~\ref{sec:models}
and the TLL liquids. We examine in particular the $t$-$V$, Lieb-Liniger and Bose-Hubbard models. Indeed these
three models display TLL behavior at low energies, as can be seen from the exact results in
Sec.~\ref{sec:1dbosons-luttinger-homog} or the numerical ones in Sec.~\ref{sec:latticenum}.

\begin{figure}[!h]
\includegraphics[width=0.4\textwidth]{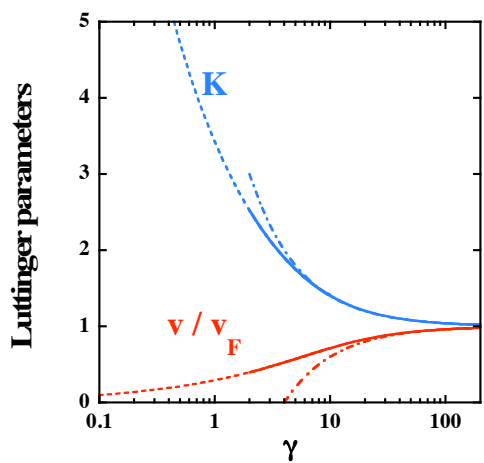}
\caption{\label{fig:LiebLiniger_TLL} Tomonaga-Luttinger liquid parameters for the
Lieb-Liniger model (cf. Sec.~\ref{sec:lieb}) as a function of
$\gamma = M g/\hbar^2 \rho_0$\cite{cazalilla_correlations_1d}.
$v_F  = \hbar v_F = \hbar \pi \rho_0/m$ is the Fermi velocity of a Fermi gas
of the same density $\rho_0$. The dashed lines correspond to the asymptotic
results following from Eq.~(\ref{eq:liebliniger_tllasymp}).}
\end{figure}

The phase diagram of the Lieb-Liniger model is the simplest, being described by the TLL fixed point for
\emph{all} values of  the single dimensionless parameter that characterizes the model, namely
$\gamma = m g/\hbar^2 \rho_0$. The existence of the Bethe-ansatz solution allows us to compute the ground state
energy as described in Sec.~\ref{sec:lieb}. From that result, the compressibility and hence the Tomonaga-Luttinger parameter,
$K = K(\gamma)$ are obtained. The sound velocity then follows from the Galilean invariance of the
model~\cite{haldane_bosons}: $v(\gamma) = \hbar \pi \rho_0/(mK(\gamma))$. These two functions are plotted in
Fig.~\ref{fig:LiebLiniger_TLL}. In general, analytical expressions are not available for arbitrary values of
$\gamma$.  However, in the limit large and small $\gamma$, the following asymptotic formulas are
known~\cite{buchler_cic_bec,cazalilla_correlations_1d}:
\begin{eqnarray}
K(\gamma) = \left\{ \begin{array}{cc}
1 + \frac{4}{\gamma^2} + O(\gamma^{-3}) & \text{for } \gamma \gg 1 \\
\frac{\pi}{\sqrt{\gamma}} \left( 1 - \frac{\sqrt{\gamma}}{2\pi}\right)
& \text{for } \gamma \ll 1
\end{array}\right. \label{eq:liebliniger_tllasymp}
\end{eqnarray}
The asymptotic formulas for the sound velocity $v(\gamma)$ follow from Galilean invariance.

The $t$-$V$ model is also integrable by the Bethe-ansatz and thus it is found to be a TLL when the lattice filling,
$n_0$, is away from $0, \frac{1}{2}$ and $1$. In the half-filled case ($n_0 = \frac{1}{2}$), for $2t < V$ the
system also belongs to the TLL universality class, and it is possible to obtain an analytic
expression for the Tomonaga-Luttinger parameter and the sound velocity:
\begin{equation}
K(\Delta) = \frac{1}{2 - \frac{2}{\pi} \cos^{-1} (\Delta)}, \quad
v(\Delta) = \frac{\pi v_F }{2} \frac{\sqrt{1-\Delta^2}}{\cos^{-1}\left(\Delta \right)},
\label{eq:KtV}
\end{equation}
where $\Delta = V/2t$ and $v_F = 2 t a/\hbar$, $a$ being the lattice parameter. Away from half-filling, one has
to resort in general to solving numerically the Bethe-ansatz equations to obtain $K$ and $v$ \cite{giamarchi_book_1d}.

Finally, the Bose-Hubbard model is a TLL for non-integer filling and also for sufficiently weak interactions at
integer filling ($U/t<0.297\pm 0.001$ for lattice filling $n_0=1$~\cite{kuhner_white_00}). Being a non-integrable
model, no analytic expressions are available for the Tomonaga-Luttinger parameter and the sound velocity. These parameters
must be therefore obtained from numerical calculations. Nevertheless, for $n_0 \leq 1$ and $U/t \gg 1$, the
following results have been obtained, to leading order in $U/t$ from a strong coupling expansion
\cite{cazalilla_tonks_gases}:
\begin{equation}
K \simeq 1 +  \frac{4t}{\pi U} \sin \pi n_0,\quad
v \simeq v_{B} \left(1 - \frac{4t}{U} n_0 \cos \pi n_0  \right),
\end{equation}
where $v_{B} = t a \sin(\pi n_0)/\hbar$.

The Tomonaga-Luttinger liquid parameters can be computed numerically for more complicated models (see e.g. the case of spin
ladders \cite{hikihara_correlations_ladder_magneticfield,bouillot_ladder_dmrg_dynamics}).
Typically one determines numerically the  compressibility. For DMRG, it is convenient to
determine $K$ directly from the static correlation functions,
but methods based on twisted boundary conditions, finite size effects, or direct computation of the spectrum
\cite{giamarchi_book_1d} are also possible. Proceeding in such a way one efficiently combines numerical
and analytical insights since thermodynamic quantities are less sensitive to finite size effects and can thus
be accurately obtained numerically. It is then possible to plug the obtained values of $v$ and $K$ in the analytic
expressions of the asymptotics of the correlation functions, which are more sensitive to finite size effects
and thus harder to calculate purely numerically. Such a combination has allowed for \emph{quantitative} tests
\cite{klanjsek_bpcp,thielemann_bpcp} of the Tomonaga-Luttinger liquid as we will see in Sec.~\ref{sec:experiments}

\subsection{The Thomas-Fermi approximation in 1D} \label{sec:GP}

In the presence of harmonic trapping along the longitudinal direction, 1D Bose gases become inhomogeneous,
finite-size systems. As a result of the finite-size, a crossover between the non-interacting BEC regime and
the quasi-long range ordered regime becomes observable. Meanwhile, inhomogeneity makes the excitation spectrum
different from the uniform case previously discussed. The equilibrium properties have been discussed in a
series of theoretical papers~\cite{petrov04_bec_review,olshanii_1d_bose_trap}, while the consequences of
harmonic trapping on the collective excitations were analyzed by an hydrodynamic
approach in~\cite{gangardt2006_bosons,petrov00,menotti02_bose_hydro1d,Ma_Ho99}.

As we saw in Sec.~\ref{sec:lieb} the dimensionless parameter,
$\gamma=c/\rho_0=\frac{mg}{\hbar^2 \rho_0}$ distinguishes
the weakly ($\gamma\ll 1$) and strongly ($\gamma \gg 1$) interacting regimes. In the presence of the axial
harmonic potential $V(x)=m\omega^2 x^2/2$, one can introduce another dimensionless quantity:
\begin{equation}\label{eq:alpha_par}
  \alpha=c \ell_{HO}=\frac{mg \ell_{HO}}{\hbar^2},
\end{equation}
where $\ell_{HO}=\sqrt{\hbar/m\omega}$ is oscillator length. This parameter can be regarded as the ratio between
the oscillator length and the interaction length $r_s=\hbar^2/mg$. The weakly interacting regime is characterized
by  $\alpha \ll 1$, which means that the relative motion of two particles approaching each other is governed by
the harmonic trapping rather than the interparticle distance.

In the weakly interacting limit ($\gamma \ll 1$), we expect the system to behave almost like a BEC, and thus the
Gross-Pitaevskii (GP) equation~(\ref{eq:gross-pitaevskii}) introduced in Sec.~\ref{sec:bosons_larged} can be
used to estimate (ground state) density profile as $\rho_0(x) = |\Psi_0(x)|^2$, where we write the order parameter
as $\Psi_0(\mathbf{r}) = \Psi_0(x) \phi_0(\mathbf{r}_{\perp})$ (cf. Eq.~\ref{eq:reductionto1D}). For $N\gg1$ and
$\gamma\ll1$ the Thomas-Fermi approximation becomes accurate (see discussion in Sec.~\ref{sec:bosons_larged} and
references therein).
Hence,
\begin{equation}
\label{eq:dens-prof-tf} \rho_0(x)=\frac \mu g \left(1-\frac{x^2}{R_{TF}^2}\right),
\end{equation}
where $R_{TF} = \sqrt{\frac{2\mu}{m\omega^2}}$ is the Thomas-Fermi radius. Imposing the normalization 
condition $\int_{-R_{TF}}^{R_{TF}} dx \rho(x)=N$, one obtains the expression of the
chemical potential $\mu$ as a function of the number of particles $N$:
\begin{equation}
\label{eq:chem-pot-N} \mu= \hbar \omega \left( \frac{3N\alpha}{4
\sqrt{2}}\right)^{2/3},
\end{equation}
where the parameter $\alpha=\sqrt{mg^2/(\hbar^3 \omega)}$. The condition for the validity of the approximation
$\gamma \ll 1$ then takes the form $(\alpha^2/N)^{2/3} \ll1$, i.e., either $\alpha \ll 1$ and any $N$ or
$\alpha \gg 1$ but $N\gg \alpha^2$. The latter condition reflects the fact that the weakly interacting regime
requires sufficiently large number of particles. Since $\rho_0(0)\ell_c=\sqrt{\hbar^2 \mu/(2m g^2)}$,
the regime of a weakly interacting gas also is characterized by the healing length $\ell_c$ being much larger
than the interparticle distance $1/\rho_0(0)$.

The limit $\alpha \gg 1$ and $N\ll \alpha^2$ corresponds to the case of the TG gas. In this case
the mapping to a free fermion hamiltonian gives a chemical potential $\mu=N\hbar \omega$ and the density
distribution \cite{mehta2004}:
\begin{equation}\label{eq:wigner-surmise}
\rho_0(x)=\frac{m \omega R_{TF}}{\pi \hbar} \left(1-\frac{x^2}{R_{TF}^2}\right)^{1/2}.
\end{equation}
Note that a similar density profile was also obtained in the case of Calogero-Sutherland model, as discussed in
Sec.~\ref{sec:calogero}. Finally, when the healing length $\xi = \sqrt{\frac{\hbar^2}{m\mu}} \gg R_{TF}$,
i.e., $\mu \ll \hbar \omega$, we can neglect the non-linear term in the GP equation and find the density profile
of the ground state in the form of a gaussian $\Psi_0(x)=\sqrt{\mu/g} e^{-x^2/(2\ell^2_{HO})}$. The chemical
potential is then $\mu=N \alpha \hbar \omega/\sqrt{\pi}$, and this regime is obtained for $\alpha \ll 1/N$.
This regime corresponds to interactions so weak that the system is forming a true BEC at zero temperature.

\begin{figure}[!h]
 \centering
 \includegraphics[width=0.4\textwidth]{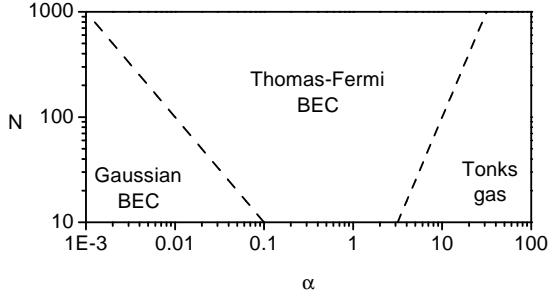}
 \caption{Zero temperature crossover diagram for a 1D Bose gas in a harmonic potential. For $\alpha \ll 1/N$,
 the ground state is a BEC. For $\alpha \gg \sqrt{N}$, the ground state is a TG gas. In between, the ground
 state is in the Thomas-Fermi regime \cite{petrov04_bec_review}.}
  \label{fig:petrov-zero-t}
\end{figure}
A more general approach to the interacting Bose gas in a trap can be developed \cite{olshanii_1d_bose_trap}.
If the trapping potential is sufficiently shallow, one can assume, in the spirit of the Thomas-Fermi approximation,
that the ground state energy for a given density profile $\rho_0(x)$ can be written as the integral
$P[\rho_0]=\int dx e(\rho_0(x))$, where $e(\rho_0)$ is the energy per unit length of a uniform interacting
Bose gas of mean density $\rho_0$. The ground state energy in the presence of the trapping potential
$V_{\text{ext}}$ and chemical potential $\mu$ can be approximated as:
\begin{equation}
\label{eq:lda_equation}
  P[\rho_0(x)]+ \int (V_{ext}(x)-\mu) \rho_0(x) dx,
\end{equation}
Varying the functional~(\ref{eq:lda_equation}) with respect to the particle density $\rho_0(x)$, one obtains
the equation:
\begin{equation}\label{eq:thomas-fermi}
  \left(\frac{\partial e(\rho_0)}{\partial \rho_0}\right)_{\rho_0=\rho_0(x)} + V_{\text{ext}}(x) = \mu,
\end{equation}
In the  weakly interacting (`mean field') limit $e(\rho_0)= g \rho_0^2/2$, and for a harmonic trap
Eq.~(\ref{eq:thomas-fermi}) reduces to (\ref{eq:dens-prof-tf}). In the opposite TG limit,
Eq.~(\ref{eq:wigner-surmise}) is recovered. Density profiles in the intermediate regime can be computed using
$e(\rho_0)$ derived from the numerical solution of the Bethe Ansatz equation \cite{olshanii_1d_bose_trap}.


\subsection{Trapped bosons at finite temperature}

Let us now turn to the properties of trapped 1D Bose gases at finite temperature and in particular the different
regimes of quantum degeneracy~\cite{petrov04_bec_review,petrov00}. In a non-interacting 1D Bose gas in a
harmonic trap, a sharp crossover from the classical regime to the trapped BEC is obtained \cite{ketterle_96}.
Its origin of the sharp crossover is related to the discrete nature of the levels in the trap. Interactions,
by smearing out the discrete nature of the levels can suppress this sharp crossover. The criterion for the
persistence of the sharp crossover is that the average interaction between particles is much smaller than the
level spacing $\hbar \omega$. With a gaussian profile for the density, the average interaction per particle
is $\sim g N (m \omega/\hbar)^{1/2}$. The criterion for persistence of the sharp crossover is then $\alpha\ll 1/N$.

In the Thomas-Fermi regime, it is convenient to expand the Hamiltonian to second order in terms of phase
and amplitude as was done in Eqs. (\ref{eq:conservation})-(\ref{eq:eq-mot-velocity}):
\begin{eqnarray}
  \label{eq:phase-amp-trap}
  \hat{H}&=&\int dx \Bigg[\frac{\hbar^2  \rho_0(x)}{2m}(\partial_x \hat{\theta}(x))^2
      \nonumber\\
       &&+\frac{\hbar^2(\partial_x \delta \hat{\rho}(x))^2}{8 m \rho_0(x)}
         + \frac g 2 (\delta \hat{\rho}(x))^2\Bigg].
\end{eqnarray}
In the limit of high temperature, we can treat the fluctuations of $\delta \hat{\rho} = \hat{\rho}(x)-\rho_0(x)$
and $\hat{\theta}$ as purely classical. Let us consider first the fluctuations of the operator $\delta \hat{\rho}$.
Since the Hamiltonian Eq.~(\ref{eq:phase-amp-trap}) is quadratic, we can easily compute the correlation functions
$\langle (\delta \hat{\rho}(x) - \delta \hat{\rho}(x'))^2\rangle$. In order to simplify further the calculation,
we make $\rho_0(x)=\rho_0(0)$ in Eq.~(\ref{eq:phase-amp-trap}). We then find that
$\langle (\delta \hat{\rho}(x) - \delta \hat{\rho}(x'))^2\rangle=
k_B T \sqrt{8m \rho_0(0)/(\hbar^2 g)} (1-e^{-\sqrt{8 m g \rho_0(0)}|x-x'|/\hbar})$.
The relative density fluctuation \cite{petrov00}, $\langle (\delta \hat{\rho}(x) -
\delta \hat{\rho}(x'))^2\rangle/\rho_0(0)^2\sim k_B T \sqrt{8m g^2 /(\hbar^2 \mu^3)}$, and is negligible provided
$T<T_d=\sqrt{\hbar^2 \mu^3 /(8m g^2)}$. With Eq.~(\ref{eq:chem-pot-N}), we have $T_d=\frac{3}{16} N \hbar \omega$,
and $T_d$ is just the degeneracy temperature. Below $T_d$ the density profile of the gas is given by
(\ref{eq:dens-prof-tf}) and above $T_d$ it reduces to the one of a classical gas. Let us now turn
to the phase fluctuations, using this time the Hamiltonian
$H_{\mathrm{phase}}=\int dx \: \hbar^2 \rho_0(x) (\partial_x \hat{\theta})^2/(2m)$. By the change of variable
$\rho_0(0) s =\int_0^x \rho_0(x') dx'$, we obtain the phase fluctuations as
$\langle (\hat{\theta}(x)-\hat{\theta}(x'))^2 \rangle =
\frac{\pi k_B T m}{2 \hbar^2 \rho_0(0)} |s(x)-s(x')|$. With the density profile
(\ref{eq:dens-prof-tf}), $s(x)=R_{TF} \ln [(R_{TF}+x)/(R_{TF}-x)]/2$
one has \cite{petrov00} $\langle (\hat{\theta}(x)-\hat{\theta}(x'))^2 \rangle =
\frac{\pi k_B T m}{4 \hbar^2 \rho_0(0) R_{TF}}  \ln [(R_{TF}+x)(R_{TF}-x')/((R_{TF}-x)(R_{TF}+x'))]$.
This time, the overall amplitude of the fluctuation is controlled by the parameter $T/T_{ph}$ where
$T_{ph}=\frac{\hbar \omega}{\mu} T_d \ll T_d$ since we are in the Thomas-Fermi regime. For $T<T_{ph}$,
both phase and density fluctuations are negligible, and the system behaves as a true BEC when $T_{ph}<T<T_d$,
there are no density fluctuations, but phase fluctuations are present. Such a regime is called
\emph{quasi-condensate}. For $T>T_d$, we have a classical gas. We note that
$T_{ph}\sim\hbar\omega(N/\alpha^2)^{1/3}$, so as $N$ is increased, the width of the quasicondensate regime broadens.

\begin{figure}[!h]
\includegraphics[width=0.4\textwidth]{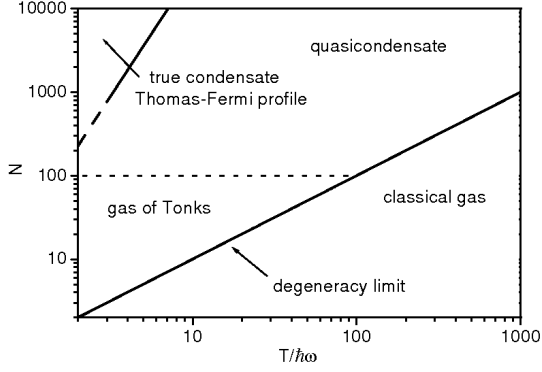}
\caption{\label{fig:1dPD_Petrov} Crossover diagram for a trapped 1D gas at finite temperature for $\alpha=10$.
For $N\ll\alpha^2$, the system is forming a TG gas below the degeneracy temperature $T_d\sim N \hbar \omega$.
For $N\gg \alpha^2$, the system is a quasicondensate when $T_{ph}\ll T\ll T_d$. For $T\ll T_{ph}\sim N^{1/3}$,
phase fluctuations are suppressed and the system is a true BEC. From \cite{petrov00}}
\end{figure}

The resulting crossover diagram at finite temperature, as computed by \cite{petrov00}, is shown in
Fig.\ \ref{fig:1dPD_Petrov}. It can be summarized as follows: for $N \gg \alpha^2$ the decrease of the
temperature to below $T_d$ leads to the appearance of a quasicondensate that becomes a true condensate
below $T_{ph}$. For $N<\alpha^2$ the system lies in the Tonks regime.

\subsection{Collective excitations}\label{sec:collective-exc}

Let us now turn to the collective excitations of the trapped gas. We will generalize the Tomonaga-Luttinger liquid
Hamiltonian and bosonization method to an inhomogeneous system \cite{petrov04_bec_review,citro08}. The
inhomogeneous Tomonaga-Luttinger liquid Hamiltonian is:
\begin{equation}\label{eq:hamiltonian}
  \hat{H}= \int_{-R}^R dx\,  \frac{\hbar v(x)}{2\pi} \Bigg[K(x) \left(\partial_x\hat{\theta}(x)\right)^2 +
  \frac{\left(\partial_x \hat{\phi}(x)\right)^2}{K(x)} \Bigg],
\end{equation}
where the Tomonaga-Luttinger parameters are given by
\begin{equation}
v(x) K(x)=\frac{\hbar\pi  \rho_0(x)}{m},\quad
\frac{v(x)}{K(x)}=\frac{1}{\pi}\partial_{\rho_0} \mu( \rho_0(x))  \label{lda-luttinger}.
\end{equation}
The field $\hat{\phi}(x,\tau)$ obeys the boundary conditions:
\begin{equation}
  \label{eq:bcs}
 \hat{\phi}(R,\tau)=\phi_1, \quad \text{and}\quad
 \hat{\phi}(-R,\tau)=\phi_0.
\end{equation}
which amount to require that no current is going in or out at the edges of the system. The fields
satisfy the following equations of motion\cite{safi_trans}:
\begin{eqnarray}
  \label{eq:eom}
  \partial_\tau \hat{\phi}(x,\tau)&=& v(x) K(x)  \partial_x\hat{\theta}(x,\tau), \nonumber \\
  \partial_\tau  \hat{\theta}(x,\tau)&=& \partial_x \left(\frac{v(x)}{K(x)}
  \partial_x  \hat{\phi}(x,\tau)\right).
\end{eqnarray}
Noting that density operator $\delta \hat{\rho} = -\partial_x \hat{\phi}/\pi$ and
and the $\hat{v} =\hbar \partial_x \hat{\theta}/m$, the above equations are the operator
1D equivalent of the linearized hydrodynamic equations discussed in Sec.~\ref{sec:bosons_larged} and
employed by \citet{menotti02_bose_hydro1d}.

Combining the two equations~(\ref{eq:eom}), using the boundary condition (\ref{eq:bcs}) and Fourier
transforming one obtains the eigenvalue equation for the normal models of the operator $\hat{\phi}(x)$:
\begin{equation} \label{eigen}
  -\omega_n^2 \varphi_n(x) = v(x) K(x) \partial_x \left(\frac{v(x)}{K(x)}
    \partial_x \varphi_{n}(x) \right).
\end{equation}
with boundary conditions $\varphi_n (\pm R)=0$. The eigenfunctions $\varphi_n$ are normalized as:
\begin{equation}
\label{eq:orthogonality}
 \int dx \frac{\varphi_n(x) \varphi_m(x)}{v(x) K(x)}= \delta_{n,m},
\end{equation}
The solution of the eigenvalue equation~(\ref{eigen}) gives access to the eigenmodes of the trapped gas.
The zero frequency solution is immediately obtained by substituting $\partial_x \varphi=K(x)/u(x)$ in
Eq.~(\ref{eigen}). With $\varphi(x)=\lambda  \rho_0(x)$, a solution with $\omega_n=\omega$, that
describes the harmonic oscillations of the center of mass of the cloud (the Kohn mode) is obtained.
For the particular case~\cite{petrov04_bec_review,citro08}:
\begin{equation}\label{eq:petrov-ansatz}
  v(x)=v_0 \sqrt{1-\frac {x^2}{R^2}},\quad
  K(x)=K_0 \left(1-\frac {x^2}{R^2}\right)^\alpha.
\end{equation}
the solutions $\varphi_n(x)$ of (\ref{eigen}) are obtained in terms of Gegenbauer (or ultraspherical)
polynomials~\cite{abramowitz_math_functions}:
\begin{eqnarray}
\label{eq:ultraspherical}
  \varphi_n(x) &=& A_n \left(1-\frac {x^2}{R^2}\right)^{\alpha+1/2}
  C_n^{(\alpha+1)}\left(\frac x R\right),  \\
\label{eq:eigen-trap}
  \omega_n^2 &=& \frac {v_0^2}{R^2} (n+1) (n+2\alpha+1),
\end{eqnarray}
where $A_n$ is a normalization factor.
The frequencies in (\ref{eq:eigen-trap}) are in agreement with the results of~\cite{menotti02_bose_hydro1d}.
In the case of $\alpha=1/2$, the dependence of $v(x)$ and $K(x)$ is the one predicted by the mean-field approach
of Sec.~\ref{sec:GP} [Eq.~\eqref{eq:eq-mot-velocity}]. The Gegenbauer polynomials in (\ref{eq:eigen-trap})
can then be expressed in terms of the simpler Legendre polynomials \cite{petrov00}. Another interesting
limit is $\alpha=0$, which corresponds to the case of the Tonks Girardeau gas, where the Gegenbauer polynomials
reduce to Chebyshev polynomials. Then, $\omega_n=(n+1)\omega$, as expected for fermions in a harmonic potential.

\section{Perturbations on 1D Superfluids} \label{sec:pertlut}

Since the Tomonaga-Luttinger liquid theory reviewed in Sec.~\ref{sec:bosonize} incorporates
all the interaction effects into a relatively simple quadratic  Hamiltonian,
it provides a convenient starting point to study the effects of various external potentials on 1D superfluid
systems. In this section, we consider the effect that three different kinds of external potentials have
on 1D superfluids. First, we discuss the effect of a periodic potential, i.e., a lattice. As already mentioned in
Sec.~\ref{sec:QMClattice}, this can lead to interaction driven insulating phases (Mott insulators). Then we
introduce the random potential and disorder driven insulating phases (Anderson insulators). We conclude with the
intermediate case of a quasiperiodic potential, for which recent cold atomic experiments have provided an
experimental realization \cite{roati08}.

\subsection{Mott transition} \label{sec:mott}

\subsubsection{Periodic potentials and the sine-Gordon model}

In this section we consider the effect of a periodic potential on an interacting boson system.
Its effect can be described in two complementary ways. If the potential is weak, it can be added as a
perturbation on the interacting bosons in the continuum. However, if the potential is strong, as described in Sec.~\ref{se:josephson-th},
is it better to start from a lattice model such as the extended Bose-Hubbard model (\ref{eq:EBHM}), or its descendants.
As we show below, both approaches lead to same low-energy field theory.

Let us begin by considering a weak periodic potential $V_{ext}(x) = V_0 \cos(Gx)$. The addition of this term
to the Hamiltonian (\ref{eq:basicmodel1d}) reduces its invariance under continuous space translations to
the discrete group of lattice translations $x\to x + \frac{2\pi}{G} m$, with integer $m$. As noted by
\citet{haldane_bosons}, using the expression (\ref{eq:bosonized_rho}) for the density leads to an additional term
$H_V$  (see e.g., \citet{giamarchi_book_1d} for a detailed derivation) in the TLL Hamiltonian, $H$
[cf. Eq.~\eqref{eq:effective_ham}].
\begin{equation}
 H_{V} =  \frac{\tilde{g}_u}{\pi} \int dx\, \cos \left(2p
   \hat{\phi}(x) +  \delta x \right).
\label{eq:sine_gordon_term}
\end{equation}
For a weak potential, $\tilde{g}_u \approx \pi \rho_0 V_0 \left(\frac{V_0}{\mu} \right)^{n-1}$  is the bare
coupling. In Eq.~(\ref{eq:sine_gordon_term}), we have retained the term with the smallest value of
$\delta = nG - 2 p \pi \rho_0$ ($n,p$ being integers), which measures the degree of incommensurability of the
potential; $\delta = 0$ corresponds to a commensurate number of bosons per site. For $p=1$ there is an integer number of
bosons per site since $\rho_0 = n G/(2\pi)$, while higher values of $p$, such as $p=2$, correspond (modulo
an integer) to one boson every two sites, etc. $\delta$ is thus a measure of the doping of the
system away from this commensurate value. There are some subtle issues depending on whether one works at constant
density or at constant chemical potential and we refer the reader to \citet{giamarchi_book_1d} for more details
on that point.

Remarkably, the \emph{same} effective model is obtained if one starts directly from a lattice model, such
as the Hubbard model. In that case the terms (\ref{eq:sine_gordon_term}) arise because, in a lattice system, the
quasi-momentum is only conserved modulo a reciprocal lattice vector. Terms in which the quasi-momentum
of the final state differs from the one of the initial state by a nonzero multiple of $G$,
nicknamed umklapp terms, can be obtained by using (\ref{eq:bosonized_rho}) and the condition that on a lattice
(e.g., for one particle per site) one has $2 \pi \rho_0 r_n = 2 \pi n$ and thus $e^{i 2 \pi \rho_0 r_n} = 1$.
For one particle per site the role of umklapp processes is to create a Mott insulator. The role of higher order
umklapps, allowing an interaction process to transfer several times a vector of the reciprocal lattice,
and its role in leading to Mott insulating phases was pointed out by
\citet{giamarchi_curvature,giamarchi_mott_review}. The final result is exactly (\ref{eq:sine_gordon_term}) with
the same condition for $\delta$. The main difference is in the amplitude of this term which now is proportional
to the interaction itself.

Hence, the model (\ref{eq:effective_ham}), and $H_\mathrm{sG} = H + H_V$ [after Eq.~(\ref{eq:sine_gordon_term})],
is thus able to give a complete description of the effects of periodic potential or lattice in a 1D interacting
bosonic system. This description depends on the value of the Tomonaga-Luttinger parameter, which is a measure of the
interactions, the lattice potential, and the doping $\delta$ away from commensurability. One immediately sees that
two different types of transition can be expected. In one case one remains commensurate $\delta = 0$. The term
$H_V$ is a simple cosine. The corresponding model is known as the ``sine-Gordon'' (sG) model for historical
reasons~\cite{coleman_book}. The transition can then occur as a function of the strength of the interactions, or
in our low energy model, as a function of the change of the Tomonaga-Luttinger parameter $K$. Alternatively we can fix the
interactions and change the degree of incommensurability $\delta$ inside the term $H_V$. Let us examine these two
type of transitions.

\subsubsection{Commensurate transition}

The first type of transition, which we denote Mott-$U$ transition, corresponds to keeping the density fixed, and
commensurate $\delta = 0$ and varying the Luttinger parameter. The physics of such a transition can be easily
understood by looking at the term $H_V$ to the Hamiltonian. 
When $H_{V}$ (rather than $H$) dominates the low-temperature properties
of the system, the fluctuations of the field $\hat{\phi}(x)$ are strongly suppressed, and $\hat{\phi}$ can order.
Since ordering $\hat{\phi}(x)$ means ordering the density, it follows that the bosons are localized at
the potential minima and the system becomes a Mott insulator (MI). To see this, let us consider the limit
$\tilde{g}_u\to -\infty$ for $\delta = 0$. In this limit, $\langle \cos 2 p \hat{\phi}(x) \rangle \to 1$ and the
ground state density following from~(\ref{eq:bosonized_rho}) is
$\langle \hat{\rho}(x) \rangle \simeq \rho_0 + \rho_0 \sum_{m>0} A_m \cos m\left( \frac{n}{p} G x \right)$.
This function is periodic for $x \to x+ \frac{2\pi p}{nG} l$ ($l$ being an integer). The insulating state is thus
a system where, on average, $n_0 = \frac{n}{p}$ bosons are localized per potential minimum. Note that there is no
violation of the Mermin-Wagne-Hohenberg theorem since, in the presence of the potential, the symmetry that is being broken
is the discrete invariance under lattice translations, which is possible at $T=0$ even in 1D.

To quantitatively study the effects of (\ref{eq:sine_gordon_term}) on the low-temperature properties of the
system, let us first consider a commensurate potential, and write renormalization group equations. The RG flow
maps one effective Hamiltonian $H_\mathrm{sG} = H+H_V$ characterized by a set of parameters
$\left(v,K,g_u\right)$ and a short-distance cutoff $a_0$ onto a new $H^{\prime}_\mathrm{sG}$ characterized by
$\left(v^{\prime},K^{\prime}, g^{\prime}_u\right)$ and a new cutoff scale $a^{\prime}_0 > a_0$ by progressively
integrating out high energy degrees of freedom. This method is well documented in the literature (see e.g.,
\citet{giamarchi_book_1d}) so we only quote the results. To lowest order in $g_u$, the RG flow is described by
the following set of differential equations for $[v(\ell), K(\ell), g_u(\ell]$ as functions of
$\ell = \ln \left( \frac{a^{\prime}_0}{a_0} \right)$,
\begin{equation}\label{eq:kt-flow}
\frac{dg_u}{d\ell} = (2-p^2 K) g_u,\quad
\frac{d K}{d\ell} = -g^2_u K^2.
\end{equation}
where $g_u = \tilde{g}_u a^2_0/\hbar v \ll 1$ is a dimensionless coupling. In addition to these equations,
$dv(\ell)/d\ell = 0$ to all orders due to the Lorentz invariance of the theory. Equations (\ref{eq:kt-flow})
describe a BKT flow \cite{kosterlitz_thouless,kosterlitz_renormalisation_xy}.
This flow has  two different regimes.  For weak coupling, The separatrix between them is given by $K = 2/p^2 + 2 g_u/p^3$ and
delimiters two phases: i) a phase in which $g_u(l) \to 0$. This corresponds to a Tomonaga-Luttinger liquid phase, in which
the decay of the correlations is algebraic, and the lattice plays asymptotically no role beyond a renormalization
of the parameters entering in the Tomonaga-Luttinger Hamiltonian; ii) a phase in which $g_u(l)$ scales up. This phase
corresponds to ordering of $\hat{\phi}$. As discussed before, this means that the particles are localized and
corresponds to a Mott insulator. It follows from the RG, or a variational treatment of the Hamiltonian
\cite{giamarchi_book_1d}, that a gap $G$ opens in the spectrum. Perturbative values are
$G/\mu \sim g^{1/(2-K)}_u$ for $2-K \gg 1$, but since the sine-Gordon model is integrable, exact values
are also available \cite{lukyanov_sinegordon_formfactors,gogolin_1dbook,giamarchi_book_1d}.

Hence,  generically for any commensurate filling, the system can undergo a transition between a
LL and a MI. At the transition point, the Luttinger parameter $K$ takes a universal value that depends \emph{only} on
the lattice filling: $K^* = 2/p^2$. Since $K/v$ is proportional to the compressibility and $v K$ to the Kohn
stiffness (cf. Sec.~\ref{sec:bosonization-method}), both compressibility and Kohn stiffness
jump discontinuously to zero when entering the MI phase. The exponents
of the various correlations have a \emph{universal} algebraic decay at the transition. In the MI phase, the
spectrum is gapped and the correlation functions decay exponentially.

Let us next examine some simple cases of the above transition. 
For $p=1$ which corresponds to an integer number of particles per site, relevant
the Bose-Hubbard model or the Lieb-Liniger gas in a periodic potential, the critical value is $K_1^* = 2$. The LL
phase is thus dominated by superfluid fluctuations since $K > 2$. One has a superfluid-MI transition. At the
transition the one-particle correlation function is thus $g_1(x) \sim |x|^{-1/4}$. For one particle every two
sites one has $p=2$ and thus $K_2^* = 1/2$. Note that this critical value is beyond the reach of a model with
only local interactions such as the Lieb-Liniger or the Bose-Hubbard (BH) models, as for such models
$1 < K < \infty$. A MI phase with one particle every two sites can thus never occur in those models. This is
quite natural since it would correspond to an ordered phase of the form $1010101010$ and  a purely local
interaction cannot hold the particles one site apart. In order to stabilize such MI phases one
needs longer range interactions, for example nearest neighbor ones as in the   $t-V$  or extended BH models,
 or dipolar interactions~\cite{citro08,burnell09}. In those cases   $K_2^* = 1/2$ can be reached, and the
 MI phase appears  Note that the critical behavior of such a transition is also  BKT-like but with 
 different values for the universal jump and for the correlation functions. For the transition with 
 one particle every two sites $g_1(x) \sim |x|^{-1}$. Moreover, logarithmic corrections to this scaling results exist 
 in some cases~\cite{giamarchi_logs,giamarchi_book_1d}. The transport properties can be also computed 
 near these phase transitions and we refer the reader to the literature~\cite{giamarchi_umklapp_1d,giamarchi_attract_1d,controzzi01,rosch_conservation_1d,giamarchi_book_1d}
for an extended discussion. Finally,  in systems with
a nearest neighbor interactions like the extended BH  model at unit lattice filling other interesting  phases,
besides the MI,  can also appear~\cite{dallatorre_haldane_insulators,berg_coupled_bosonic_haldane_ins}.

\begin{figure}[!h]
\includegraphics[width=0.4\textwidth]{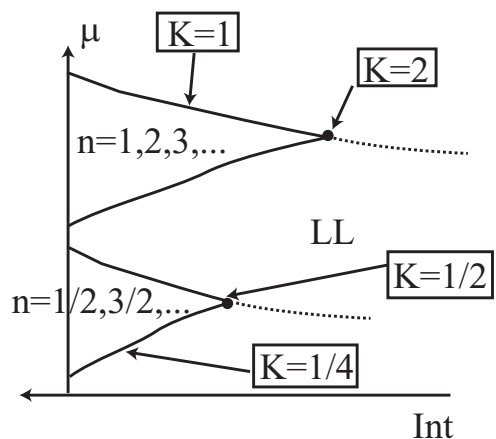}
\caption{\label{fig:criticalmott} Phase diagram for commensurate bosons as a function of the chemical
 potential $\mu$ and the interactions. Only the commensurabilities of one boson per site and one boson every
 two sites have been shown. The Mott phase for the bosons occurs for a commensurate filling and depends on the
 strength of the interactions. The Tomonaga-Luttinger liquid (LL) parameters take a universal value both for the Mott-$U$ and
 Mott-$\delta$ transitions. The value depends on the order of
 commensurability \cite{giamarchi_book_1d}.}
\end{figure}

The properties discussed above and summarized in Fig.~\ref{fig:criticalmott}
agree very well with the numerical results for the BH model of Sec.~\ref{sec:BHMPeriodic} and with
results on the extended BH model \cite{kuhner_white_00}, in which a value $K^* = 1/2$ was numerically found for
the transition from the TLL and the MI with one boson every two sites.

As a final remark, let us note that although for the BH model, which corresponds to the strong lattice
case, the phase diagram is more or less independent on the dimension, 1D systems in presence of a weak periodic
potential present a very interesting feature. Indeed from the scaling properties above we see that in 1D one can
realize a MI phase \emph{even} for a very weak periodic potential \emph{provided} that the interactions are
strong enough (e.g., $K < 2$ for one boson per site). This is a true quantum effect coming from the interferences
of the particles on the periodic potential that can lead to localization even if the periodic potential is much
weaker than the kinetic energy. In the TG limit ($K=1$) this would simply correspond to free fermions moving in a
periodic potential with an exact $2 k_F$ periodicity, where $k_F$ is the Fermi wavevector. In that case it is
well known that even an arbitrarily weak potential opens a gap. Such a feature would not be present in higher
dimensions were a strong lattice potential is needed to stabilize the Mott insulator phase, at least for the case
of local interactions.

\subsubsection{Incommensurate transition}

The situation is drastically different when the system is incommensurate. Let us now look at the case for which
the commensurate system $\delta=0$ would be a MI, i.e. $K < K^*$ and look at the effect of an increasing $\delta$
on the physics of the system. Naively, the presence of a $\delta x$ term inside the cosine means that the cosine
oscillates with space and thus is wiped out and its sole effect can be a renormalization of the LL parameters,
leading back to a purely quadratic Hamiltonian. This is traducing the fact that a \emph{doped} system is always
at large distance a LL with algebraic decay of the correlations. Of course, in order to have finite doping, since the
Mott phase has a gap, one needs to apply a chemical potential in excess of the gap $G$. One has thus the
possibility to have another MI-SF transition, driven by the change of density of the system, and sometimes
referred to as Mott-$\delta$ transition. Although it is not obvious in the original single particle language, the
LL representation makes it obvious that the doped Mott problem is directly related to one specially important
classical problem known as the commensurate-incommensurate phase transition
\cite{japaridze_cic_transition,pokrovsky_talapov_prl,schulz_cic2d}, describing particles adsorbed on a surface
with a different periodicity than the one of the adsorbed molecules. One can thus follow the same route than the
one used to determine the critical behavior of this transition to obtain the Mott-$\delta$ critical exponents.
For this reason the transition is often referred to as being in the universality class of the
commensurate-incommensurate (C-IC) phase transition.

In order to obtain the critical properties of such a transition two routes are possible. One is to use the
standard renormalization group procedure and establish equations involving the doping $\delta$
\cite{giamarchi_umklapp_1d,giamarchi_book_1d}. Although these equations are very useful in deriving some of the
physical quantities, specially away from the transition itself, they flow to strong coupling when one approaches
the transition line, making it difficult to extract the critical exponents. It is thus especially useful to use
a technique known as refermionization \cite{luther_exact,schulz_cic2d,giamarchi_book_1d} to deal with this issue.
The idea is simply to use the fact that in a similar way one can represent bosonic single particle operators by
collective variables as described by Sec.~\ref{sec:bosonize}, one also represent fermionic particles. It is thus
possible to map the sine-Gordon Hamiltonian to a fictitious Hamiltonian of interacting ``fermions''. We will not
detail here the relation between fermions and collective variables (the density operator is the same than for
bosons, but single particle operator is slightly different) since it has been well described in the
literature \cite{giamarchi_book_1d} and simply quote the final result \cite{schulz_cic2d}:
{\setlength\arraycolsep{0.5pt}
\begin{eqnarray} \label{eq:refermion}
&& \hat{H} = \sum_k \hbar v k (\hat{\psi}^\dagger_{k,R} \hat{\psi}_{k,R} - \hat{\psi}^\dagger_{k,L} \hat{\psi}_{k,L}) + \hat{H}_{\rm int} + \\
&&      g_u \sum_k (\hat{\psi}^\dagger_{k,R} \hat{\psi}_{k,L} + \textrm{H.c.})
      - \delta \sum_k (\hat{\psi}^\dagger_{k,R} \hat{\psi}_{k,R} + \hat{\psi}^\dagger_{k,L} \hat{\psi}_{k,L}) \nonumber
\end{eqnarray}
}where $\hat{H}_{\rm int}$ is an interaction term between the fermions. The above describes right and left moving
fermions ($R$ and $L$) indices, with a Dirac-like linear dispersion controlled by the velocity $v$. Those pseudo
fermions are in fact representing the kink excitations of the phase $\phi$ where the phase goes from one of the
minima of the cosine to the next, thereby changing by $\pm \pi/p^2$. The interaction term depends on the
commensurability of the system is proportional to $Kp^2- 1/(Kp^2)$. So while in general the fermions interact
making the fermionic theory as difficult to solve as the original sine-Gordon one, for a special value $K_0$ of
the LL parameter, this interaction disappear leading to a free fermionic theory. For $p=1$ one has $K_0 = 1$
that corresponds to the TG limit, in which case the fermions in (\ref{eq:refermion}) are nothing but the spinless
fermions of Eq.~(\ref{eq:girardeau-transformation}). This technique, introduced by \citet{luther_exact} has been
used in several contexts for 1D systems \cite{giamarchi_book_1d}. It provides a useful solution
when the non linear excitations of the sine-Gordon model are important to keep and when a simple quadratic
approximation of the cosine would be insufficient.

The former Mott term is now hybridizing the right and left movers and thus opens a gap at energy zero in the
initially gapless spectrum. The energy spectrum is $\epsilon_\pm(k) = \pm \sqrt{(v k)^2 + g_u^2}$. We thus
recover here the fact that this operator is opening a Mott gap. The refermionization allows thus to make quite
rigorous here the usually rather vague concept of the upper and lower Hubbard bands. The corresponding ``free''
particle are the kinks in the phase $\phi$ and thus describe the defects in the perfect arrangement of the Mott
insulator. For example starting from $10101010101010$ a configuration such as $10110101010010$ would contain
one kink and one anti-kink. The $\delta$ term corresponds to an average density for fermions. If $\delta$ is
nonzero, the chemical potential will sit either in the upper or lower band and one recovers that there are
gapless excitations given by the fermions. We thus see that the Mott-$\delta$ transition can be described
\cite{giamarchi_mott_review} as the \emph{doping} of a \emph{band} insulator.

Using this mapping on free fermions one can extract the critical behavior, which is markedly different from the
commensurate case. Right at the transition $\delta \to 0_+$, the dynamical exponent relating space and time
$\omega \sim k^z$ is $z=2$ contrarily to the commensurate case for which $z=1$ because of Lorentz
invariance. The effective velocity of excitations is  for $0<\delta \ll g_u/v$
\begin{equation}
 v^* = \left(\frac{d \epsilon(k)}{dk}\right)_{k=\pi \delta}  \simeq
 \frac{\pi v^2 \delta}{g_u},
\end{equation}
and goes to zero at the transition.  The compressibility which can be computed by $d N/d\mu$ \emph{diverges} at
the transition, before going to zero in the Mott phase. The Kohn stiffness, $D \propto v^* K^*$, vanishes
continuously with the doping $\delta$ since the Tomonaga-Luttinger parameter is a constant. Interestingly,
the exponents of the various correlation functions at the transition are universal and can be determined
by using the free fermion value. For example, the density correlator of the fermions decays as $|x|^{-2}$.
Since before the mapping the correlation would have corresponded to an exponent $2/K$ this means that at the
Mott-$\delta$ transition the Tomonaga-Luttinger parameter takes the universal value $K_0 = 1$. For a generic
commensurability $p$ it is easy to check that $K_0 = 1/p^2$. This is exactly half of the universal value
of the commensurate Mott-$U$ transition. A summary of is shown in Fig.~\ref{fig:criticalmott}. Interestingly,
several of these scalings both for the commensurate and incommensurate case are also valid and extendable to
higher dimensions \cite{fisher_boson_loc}.

These considerations also directly apply to the case of spin chains and ladders with a gapped phase and we refer
the reader to the literature \cite{chitra_spinchains_field,furusaki_dynamical_ladder,giamarchi_coupled_ladders}
and to Sec.~\ref{sec:experiments} for more details on this point.

\subsection{Disorder}\label{sec:disorder}

Another type of perturbation is provided by a disordered potential. For non-interacting particles, disorder can
give rise to Anderson localization: the single-particle eigenfunctions of the Hamiltonian decay exponentially
over a characteristic length $\xi_l$, called the {\it localization} length \cite{anderson_localisation}. Indeed, the
effect of disorder depends on dimensionality. In 1D, all eigenstates are localized for any nonzero disorder
strength \cite{mott_loc}. Exact solutions \cite{berezinskii-74,abrikosov-78,gogolin_disorder_revue,supersimmetry}
can be devised. For noninteracting particles, the localization length is of the order of the mean free path in 1D.
Higher dimensions \cite{scaling-dis,replicas-dis-2,supersimmetry} lead either to a full localization, albeit with
a potentially exponentially large localization length in 2D, or to the existence of a mobility edge
separating localized states (near band edges) from diffusive states (at band center) in 3D. Recently cold atomic
gases have allowed to directly observe the localization of noninteracting 1D particles
\cite{aspect-dis-2008,roati08}.

Although the non-interacting case is conceptually well understood, taking into account the combined effects
of disorder and interactions is a formidable problem. For fermions, the noninteracting case is a good starting
point and the effects of interactions can be at least tackled in a perturbative fashion
\cite{altshuler_aronov,diagrams-dis} or using renormalization group techniques
\cite{finkelstein_localization_interactions}.

Such an approach is impossible with bosons since the noninteracting disordered bosonic case is pathological. The
ground state in the non-interacting case is a highly inhomogeneous Bose condensate in which all particles are in
the lowest eigenstate of the Hamiltonian (which is necessarily localized). In the absence of repulsion, a
\emph{macroscopic} number of particles are thus in a \emph{finite} region of space. Such a state is clearly
unstable to the introduction of even the weakest interaction. The interactions should thus be included from the
start. Alternatively, the limit of very strong interactions can be considered starting from a crystal phase of
particles (fermions or bosons since statistics in that case does not matter)
\cite{giamarchi_varenna_wigner_review}.

\subsubsection{Incommensurate filling}

Let us next consider  the Hamiltonian of disordered interacting bosons. For convenience and generality, we will deal with 
the case of bosons on a lattice. Similar results and methods are of course applicable in the continuum. The Hamiltonian reads
\begin{equation}
\label{eq:ham-dis-bos}
  \hat{H}=\sum_{i=1}^L
  \left[ -t_i(\hat{b}^\dagger_i \hat{b}_{i+1} + \text{H.c.}) -\epsilon_i\hat{n}_i  + \frac{U}{2}
 \hat{b}^{\dag}_i \hat{b}^{\dag}_i  \hat{b}_i \hat{b}_i \right],\quad
\end{equation}
The $t_i$ describe random hopping from site to site. This type of disorder is particularly pertinent for spin
chains (see Sec.~\ref{sec:mapping}) since it that case it corresponds
to the case of random spin exchange \cite{hong2010}.
$\epsilon_i$ is a random chemical potential. Note that in the case of cold atomic gases $\epsilon_i$ also
contains in general the confining potential. Other types of disorder (random interactions, etc) can of course
also be treated, but we will confine our discussion here to the above two cases. Note that for the case of
hard-core bosons (or spins) these two types of disorder have an important difference. The first one respects
the particle hole symmetry of the problem (for soft-core bosons no such particle-hole symmetry exists),
while the second, being a random chemical potential, breaks it for each realization of the disorder, even if
it is still respected in average.

Just as for the case of a periodic potential, we use the bosonization method  introduced in
Sec.~\ref{sec:bosonize}. Let us start the with the on-site disorder. Similar results and equations can be derived
for the random hopping. Microscopic disorder is in general rarely Gaussian. For example impurities scattered in
random positions represent a Poissonian disorder. It might be important for practical cases to carefully take
into account the precise form of such correlations (see e.g., \citet{lugan_correlated_potentials}). However, the
Gaussian limit is generic if the length scales over which the properties of the system vary (e.g., the
localization length) are large compared to the microscopic scale of the disorder. In that case, the central limit
theorem applies. We thus for simplicity discuss the case of $\overline{\epsilon_i \epsilon_{i'}}=D \delta_{ii'}$,
where $D$ is the disorder strength.

For the case of incommensurate filling, or for bosons in the continuum, one can rewrite the coupling to a weak
disorder potential as
\begin{equation}
H_\textrm{dis}=\int dx V(x) \hat{\rho}(x),
\end{equation}
where, $\overline{V(x) V^\star(x')}=D \delta(x-x')$. Using the boson representation for the density of
Sec.~\ref{sec:bosonize} and keeping only the most relevant harmonics one has \cite{giamarchi_loc}:
\begin{equation} \label{eq:disorder-bosonized}
H_\textrm{dis}=\int dx V(x) \left[ -\frac 1 \pi \partial_x
 \hat{\phi}(x)+\rho_0 \left( e^{i2(\pi \rho_0
x-\hat{\phi}(x))}+\text{H.c.}\right)\right]
\end{equation}
The term proportional to $\partial_x \hat{\phi}$ describes ``forward'' scattering by the random potential. It
corresponds to a slowly (compared to the interparticle spacing $\rho_0^{-1}$) varying chemical potential. This
term can be absorbed, for incommensurate cases, in the quadratic part of the Hamiltonian. It is straightforward
to check that although it leads to exponential decay of density correlations, it cannot affect the superfluid
ones nor change the conductivity.

The main disorder effects are thus coming from the ``backscattering'' term, i.e., the term for which the momentum
exchanged with the impurities is of the order of $2\pi\rho_0$. This term can be treated by an RG procedure
\cite{giamarchi_loc}. The RG equations read:
\begin{equation}\label{eq:rg-back}
\frac{dK}{dl} =-\frac{K^2}{2} \tilde{D},\qquad
\frac{d\tilde{D}}{dl} = (3-2K) \tilde{D},
\end{equation}
where $\tilde{D}=D/(\pi^2u^2\rho_0)$. There is also an additional equation for the velocity showing that the
compressibility is not renormalized at this order of the flow. These equations indicate that both the disorder and
the interactions get renormalized when disorder (and interactions) is present. The equations have a BKT form.
There is a separatrix, depending on both $K$ and $\tilde{D}$, terminating at $K^* = 3/2$ separating a phase in
which $\tilde{D}$ scales to zero and a phase in which $\tilde{D}$ is relevant. The first one clearly corresponds to
a superfluid one. In the later one $\tilde{D}$ scales to strong coupling so the flow cannot be used beyond a certain
scale $l^*$ such that $\tilde{D}(l^*)\sim 1$. One can nevertheless infer the properties of such a phase from various
approximations. The simplest is to note that this phase contains the Tonks line $K = 1$ for which the bosons behave
as free fermions. Hence, it has the same long distance properties as free fermions in presence of disorder, and is
therefore a phase in which the particles are localized.
This Anderson localized phase has been discussed for the Tonks gas in \textcite{radic_2010}.

The critical properties of the transition can be extracted from the flow. Since the transition is of the BKT type,
$K$ jumps discontinuously at the transition. Note that contrarily to what happens for the periodic case, the system
remains \emph{compressible} in the localized phase, as one can see from the free fermion limit, or the RG flow.
The jump of $K$ indicates that the Kohn stiffness, which is finite in the superfluid, goes to zero in the
localized phase. The localization length, which can be extracted from the flow, diverges at the transition, in
the usual stretched exponential way characteristics of BKT transition. Note that if one defines a critical exponent
$\nu$ for the divergence of the localization length one has $\nu = \infty$ in 1D.  Disordered bosons in 1D provided
the first derivation of a superfluid-localized transition and the existence of a localized bosonic phase
\cite{giamarchi_loc}. This phase, nicknamed Bose glass, was surmised to exist in higher dimensions as well, and the
critical properties of the SF-BG transition were obtained by general scaling arguments \cite{fisher_boson_loc}.
In addition to 1D bosons, the same RG methodology has been applied to several systems using the mapping of
Sec.~\ref{sec:mapping}, such as spin chains and ladders or bosonic ladders. We refer the reader to the literature
for details on those points \cite{doty_xxz,orignac_2spinchains,orignac_2chain_bosonic,giamarchi_book_1d}.

\begin{figure}[!h]
\includegraphics[width=0.4\textwidth]{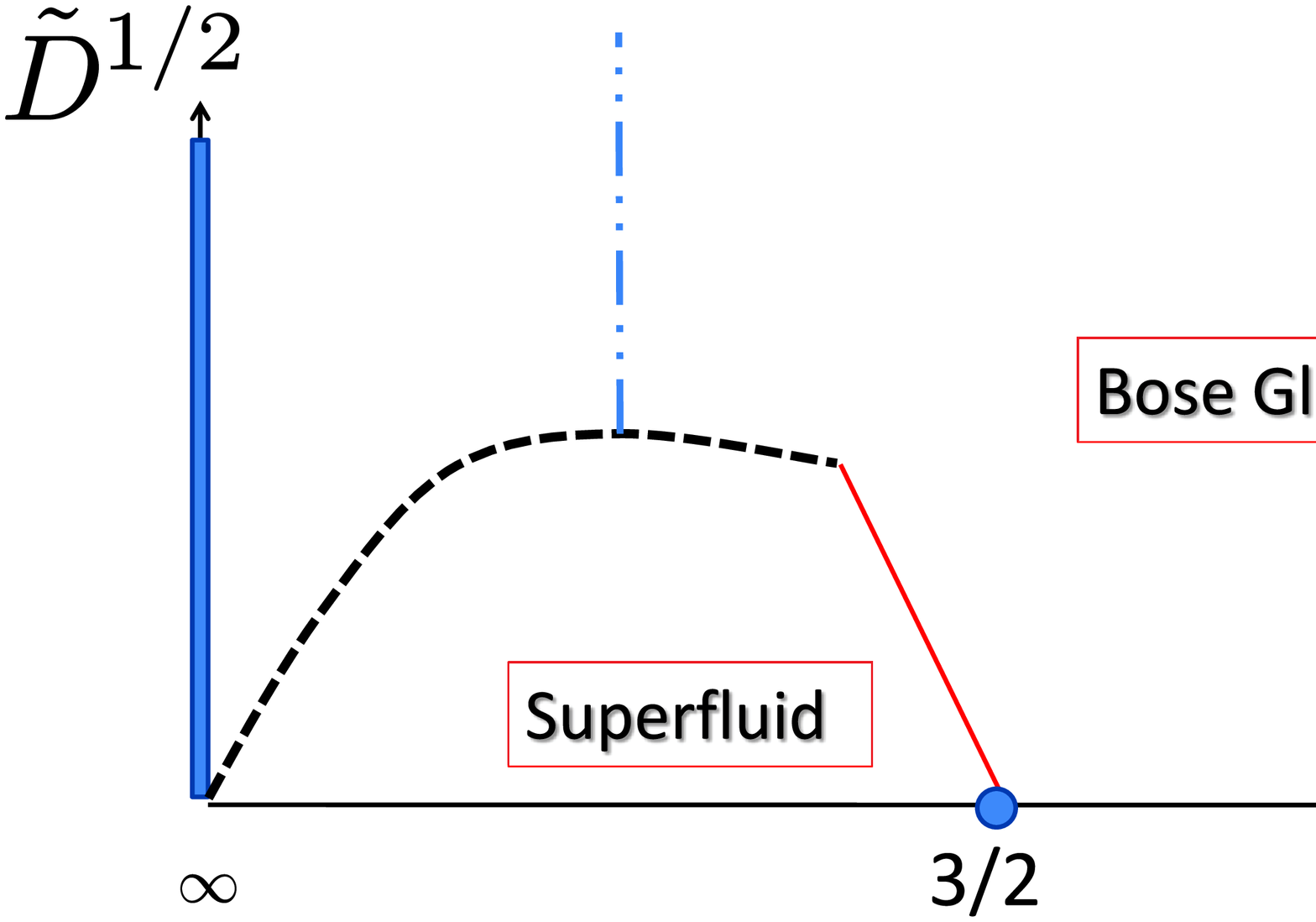}
\caption{\label{fig:phasediag_disorder} Phase diagram of interacting disordered bosons\cite{giamarchi_loc_lettre}, for incommensurate
filling, in 1D. $K$ is the Tomonaga-Luttinger parameter ($K=\infty$ for non interacting bosons, and $K$ decreases for
increasing repulsion). $\tilde{D}$ is the strength of the disorder. Noninteracting bosons are localized in a
finite region of space and have thus no thermodynamic limit. The solid line is the separatrix computed from the
RG (see text). The limit of the superfluid region must bend down (dashed line) for small interactions to be
compatible with the non-interacting limit, leading to a reentrant superconducting phase. The Bose glass phase
is a localized, compressible bosonic phase (see text). The question on whether to distinct localized
phases could exist (dash-dotted line) or whether there is only one single localized phase is still open.}
\end{figure}

If the value of $K$ becomes too large (i.e, if the interactions between the bosons become too small)
then the bosonization description becomes inadequate to describe the system since the chemical potential becomes
smaller than the disorder strength. Since the noninteracting line is localized irrespective of the strength of the
disorder it was suggested that the separatrix bend down to the point $K= \infty,\tilde{D}=0$, to lead to the
reentrant phase diagram of Fig.~\ref{fig:phasediag_disorder}.

These predictions were confirmed by different methods. On the analytic side, after bosonization the Hamiltonian
bears resemblance to the one describing the pinning of charge-density waves \cite{fukuyama_pinning}. This leads
to a physical interpretation of the Bose glass phase as a pinned density wave of bosons. The phase $\phi$ adjusts
in (\ref{eq:disorder-bosonized}) to the random phase $2\pi\rho_0 x_i$ where $x_i$ would be the position of the
impurities. For small disorder the boundary between the superfluid and the localized phase has recently been
reinvestigated \cite{falco_weakly_pinned_bosons,lugan_weakly_disordered_bosons,aleiner_altshuler_10},
confirming the general shape of Fig.~\ref{fig:phasediag_disorder}, and giving the precise position of the boundary.

The transition itself can be studied by two methods, slightly more approximate than the RG, since they amount to
neglecting the first RG equation and consider that the interactions are not renormalized by the disorder.
The first one is a self-consistent harmonic approximation, developed for the pinning of charge-density-waves
\cite{suzumura-loc}, the second a variational method based on replicas \cite{giamarchi_columnar_variat}.
Although they do not, of course, give the correct critical properties, they properly recover the transition at
$K=3/2$. The later method also allows the calculation of the correlation functions in the localized phase.
It, in particular, it shows that a second localization length, corresponding to the localization of the particles
in time exists. It also gives access to the frequency dependence of the conductivity at zero temperature.

The TG limit allows one to also obtain some of the properties in the localized phase.
The density correlations can directly be extracted and the single particle Green's function,
which can be represented as a Pfaffian \cite{klein1990} as discussed in Sec.~\ref{sec:tonks}. This allows to show
rigorously that for large distance the single particle Green's function of the bosons decays exponentially, and
thus that there is no off-diagonal quasi-long range order. For a fixed disorder realization ${h_i}$, one can
compute numerically the Pfaffian and then average over the different disorder realizations
\cite{young_random_ising,henelius_hc_bosons}. A similar numerical study of the effect of disorder has also been
done for a discrete binary probability distribution of random on-site energies \cite{krutitsky_binary}. The
probability distribution of on-site energies $\epsilon_i$ is given by
$p(\epsilon) = p_0\delta (\epsilon)+(1-p_0)\delta(\epsilon-U')$, where $1-p_0$ is the concentration of impurities,
$U'$ is the boson-impurity interaction term. This type of disorder can be obtained in experiments by the
interaction of the bosons with impurity atoms having a large effective mass \cite{gavish1995}. Hard-core bosons
also allow for an exact calculation of the superfluid fraction defined by
\citet{fisher1973,leggett1973,pollock_ceperley_87}.
The numerical results show that for small $t/U'$ the binary disorder destroys the superfluidity in the thermodynamic
limit in a similar manner as for Gaussian disorder. A study of binary disorder in case of finite but large $U$ has
been performed with strong coupling expansions and exact diagonalization by \citet{krutitsky_binary}.

Although the bosonization approach is very efficient to describe the moderately and strongly interacting bosons,
it is, as mentioned above, not applicable when the interactions become weak or alternatively if the disorder become
strong. If the disorder is very strong one has \emph{a priory} even to take into account broad distributions for
the various parameters. A very efficient real space renormalization group technique \cite{fisher_random_transverse}
has been developed to renormalize distribution of coupling constants. If the distributions are broad to start with
they become even broader, making the RG controlled. This RG has been applied with success to the study of strongly
disordered bosons \cite{altman_mottglass_rsrg,altman_bosons_rsrg_short}. This technique
gives a transition between a Bose glass and the SF phase. It is interesting that in the limit of strong disorder,
the transition is also of the BKT type but with some divergences of the distributions. Whether this indicates the
presence of two different localized phases, one for weak interaction and one for strong interactions (see
Fig.~\ref{fig:phasediag_disorder}) or whether the two can be smoothly connected is an interesting an still open
question. A proper order parameter separating the two phases would have to be defined.  Note that the computational
studies give conflicting results on that point \cite{batrouni_scalettar_92,rapsch_schollwock_99}

At finite temperature, the question of the conductivity of the Bose glass phase is a very interesting and still
open problem. If the system is in contact with a bath, an instanton calculation
\cite{nattermann_temperature_luttinger} leads back to Mott's variable range hopping \cite{mott_metal_insulator}
$\sigma(T) \sim e^{-(T_0/T)^{1/2}}$. This technique can also be used to compute the a.c. conductivity
\cite{rosenow_optical_instanton}. In the absence of such a bath the situation is more subtle
\cite{gornyi_manybody_localization,basko_manybody_localization,gornyi_manybody_localization_long,%
aleiner_altshuler_10}. In particular it has been suggested that the conductivity could be zero below
a certain temperature, and finite above, signaling a finite temperature many-body localization transition.
Consequences for bosonic systems have been investigated by \citet{aleiner_altshuler_10}.

\subsubsection{Commensurate filling}

Let us now turn to the case for which the system is at a commensurate filling. If the interactions is such that
the commensurate potential could open a Mott gap, then there will be a competition between Mott and Anderson
localization. To describe such a competition one must add to the Hamiltonian a term like
Eq.~(\ref{eq:sine_gordon_term}) at $\delta =0$, where $p$ describes the order of commensurability and is defined by
$p=1/(\rho_0 a)$.  In the case of a commensurability $p>1$, i.e., of an atom density wave competing with disorder,
it can be shown \cite{shankar_spinless_conductivite} that even a weak disorder turns the atom density wave into an
Anderson insulator, by breaking it into domains. Indeed, a density wave has a ground state degeneracy $p$, which
allows the formation of domain walls. Since the energy cost of a domain wall is a constant proportional to the
excitation gap of the pure atom density wave phase, while the typical energy gained by forming a domain of length
$L$ can be estimated to be of order $\propto -\sqrt{DL}$ by a random walk argument \cite{imry1975}, it is always
energetically advantageous to break the density wave phase into domains as soon as $D>0$. The resulting ground
state is gapless and thus a Bose glass.

In contrast, a Mott phase with each site occupied by an integer number of atoms has no ground state degeneracy, and
thus is stable in the presence of a weak (bounded) disorder. There will thus be several additional effects in that
case that have to be taken into account. First, at variance with the incommensurate case, it is not possible anymore
to eliminate the forward scattering by a simple shift of $\phi$. Indeed the forward scattering which is acting as a
slowly varying chemical potential is in competition with the commensurate term $\cos(2 \phi)$ just as was the case
for the doping in the Mott transition in Sec.~\ref{sec:mott}. The disorder will thus reduce the gap and ultimately
above a certain threshold destroy the commensurability. Generally, it will reduce the stability of the Mott region
compared to Fig~\ref{fig:criticalmott}. The forward scattering part of the disorder thus leads to a
\emph{delocalization} by a reduction of the Mott effects. This is quite general and extend to higher dimensions as
well. On the other hand, the effect of the backward potential will be to induce the Anderson localization and leads
to the Bose glass phase, as discussed in the previous section. It is easy to see that even if such a backward
scattering was not present to start with, the combination of the forward scattering and the commensurate potential
would always generate it. One can thus expect naively, when the disorder is increased, a transition between a Mott
insulator, which is \emph{incompressible} and localized to a Bose glass, which is \emph{compressible} and also
localized due to the backward scattering.

One interesting question \cite{fisher_boson_loc} is whether the Bose glass phase totally surrounds the shrunk Mott
lobes in Fig~\ref{fig:criticalmott}, or whether at commensurate filling, a direct MI-SF transition would be
possible. In 1D this question can be easily answered by looking at the combined renormalization of the commensurate
potential, and the one of the backward scattering. For small forward scattering, the commensurate
flow~(\ref{eq:kt-flow}) is essentially unperturbed. It reaches strong coupling before the forward disorder can cut
the RG, which signals the existence of the Mott gap. If the disorder is increased the flow will now be cut before
the scale at which the Mott gap develops. The Mott phase is thus destroyed by the disorder, which corresponds to
the shrinking of the Mott lobe. On the other hand, as seen in (\ref{eq:kt-flow}), $K$ has now being renormalized
to small values $K^*$.  Since when the commensurate potential flow is cut, backward scattering is present
(or generated) one should then start with the flow for the backward scatting disorder (\ref{eq:rg-back}).
Just at the destruction of the Mott lobe $K\sim 0$ one has $K < 3/2$ and the system is always in the
Bose glass phase. As the \emph{forward} scattering is increased (or alternatively as the initial value of $K$ would
be increased) the flow can be cut early enough such that $K^*>3/2$ and one can be in the superfluid phase where the
backward scattering is irrelevant. One thus sees that in 1D the RG predicts always a sliver of Bose glass between
the Mott and superfluid phase. Further analysis confirm these arguments
\cite{svistunov1996,herbut1998} and the topology of the phase diagram where the Mott lobes are surrounded by a Bose
glass phase applies to higher dimension as well \cite{pollet_prokofev_09}. The Mott insulator Bose glass phase
transition has been understood to be of the Griffiths type \cite{gurarie_pollet_09}. This is to be contrasted with
the transition between the Bose glass and the superfluid phase, which is a second order phase transition. The
dynamical critical exponent of the latter transition has been suggested to be $z=d$, with no upper critical
dimension $d_c=\infty$ \cite{fisher_boson_loc}. This relation is in agreement with the 1D result where the
Lorentz invariance of the bosonized theory implies indeed $z=1$ at the transition.

On the numerical side the combined effects of disorder and commensurability have been intensively studied.  One
of the first studies of the effect of bounded disorder on the Mott lobes in 1D
\cite{scalettar_batrouni_91,batrouni_scalettar_92} confirmed the shrinking of the Mott lobes, and the generation
of a compressible Bose glass phase in the weakly and strongly interacting regimes. The phase boundary between the
Mott insulator and the Bose glass was further studied by means of strong coupling expansions, both for finite
systems and in the thermodynamic limit \cite{freericks_monien_96}. While for finite system the results agreed with
the ones obtained in \citet{scalettar_batrouni_91,batrouni_scalettar_92}, \citet{freericks_monien_96} pointed out the difficulty in
obtaining the thermodynamic limit result from extrapolations of finite system calculations because of the effect of
rare regions imposed by the tails of the bounded disorder distribution.

\begin{figure}[!htb]
\centerline{\includegraphics[width=0.3\textwidth]{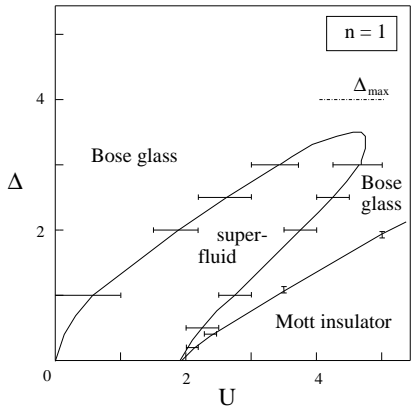}}
\caption{Phase diagram of the Bose-Hubbard model, for commensurate filling $n=1$, with an additional uniformly
distributed disorder in the interval $[-\Delta,\Delta]$. Notice the presence of the Bose glass phase between the
Mott insulator and the superfluid, and the reentrant behavior in the Bose glass phase when increasing the disorder
amplitude for $2\lesssim U\lesssim 5$ \cite{rapsch_schollwock_99}.}
\label{fig:phasediag1Ddisorder}
\end{figure}

In the thermodynamic limit, the effect of a
bounded and symmetric distribution with $|\epsilon_i|\leq\Delta$ in the grand-canonical phase diagram (depicted in
Fig.~\ref{fig:phasediag1Dclean}) is to shift the Mott insulating boundaries inward by $\Delta$. Later studies,
using QMC simulations \cite{prokofev_svistunov_98} and DMRG \cite{rapsch_schollwock_99}, mapped out the full phase
diagram for the Bose-Hubbard model in the presence of disorder [see also \cite{pai_pandit_96}]. The phase
diagram at fixed density $n=1$ is shown in Fig.~\ref{fig:phasediag1Ddisorder}. It exhibits, when starting from
the Mott insulating phase, a reentrant behavior into the Bose-glass phase with increasing disorder strength. It
also fully confirms the reentrance of the localized phase at small repulsion in agreement with
Fig.~\ref{fig:phasediag_disorder}.

Another very interesting class of effects occur when the filling is commensurate and the disorder respects
particle-hole symmetry. This is not the case of the random on site potential, and is traduced by the presence of
the forward scattering term. On the contrary for hard-core bosons this would be the case of the random hopping
term. This is a rather natural situation for spin chains, for which random exchange can be realized.  In that case,
there is an important difference with the case (\ref{eq:disorder-bosonized}). Because of the commensurability
$e^{i 2 \pi \rho_0 x} = 1$ and the disorder term is
\begin{equation} \label{eq:disorder-bosonized-comm}
H_\textrm{dis}=\int dx V(x) \rho_0 \cos(2 \hat{\phi}(x)).
\end{equation}
Although the two terms look superficially similar, there is thus no random phase on which $\phi(x)$ must pin.
$\phi(x)$ is thus oscillating between the two minima $\phi = 0$ and $\phi = \pi/2$, and the physics of the problem
is totally controlled by the kink between these two minima. Thus although the initial steps of the flow
(\ref{eq:rg-back}) are identical for the two problems, the strong coupling fixed points are different. In the
particle-hole symmetric case a delocalized state exists in the center of the band, leading to various singularities.
This is a situation which is well adapted to the above mentioned real space RG \cite{fisher_random_transverse}, and
the system is dominated by broad distributions (so called random singlet phase).

\subsubsection{Superlattices and quasiperiodic potentials} \label{sec:aubry}

As seen in the previous sections, the periodic and disordered potentials albeit leading to very different physical
phases, seem to share some common features. In particular, disorder can be viewed as a potential for which all
Fourier harmonics would be present. It is thus a natural question on whether one could generalize the case of a
simple periodic potential to potentials with several periodicities or even to quasiperiodic potentials. In
addition to the pure theoretical interest of this problem, it has become, thanks to optical lattices in cold atomic
gases, an extremely relevant question experimentally. Let us consider first the case of a superlattice, i.e., of
a potential with two commensurate periods $Q_1$ and $Q_2$. For simplicity we assume in what follows that the first
periodic potential is very large and thus defines the model on a lattice, and that the second potential is
superimposed on this lattice. This defines a variant of the Bose-Hubbard Hamiltonian
\begin{eqnarray} \label{eq:superlatt-bose}
   \hat{H}=\sum_{i=1}^L \left[ -t_i (\hat{b}^\dagger_i \hat{b}_{i+1} + \textrm{H.c.})+ \epsilon_i
  \hat{n}_i  + \frac{U}{2} \hat{b}^{\dag}_i \hat{b}^{\dag}_i \hat{b}_i \hat{b}_i \right], \quad
\end{eqnarray}
where $t_{i+l}=t_i$ and $\epsilon_i=\epsilon_{i+l}$ with $l\ge 2$.

The case of a commensurate superlattice (i.e., when $Q_1$ and $Q_2$ are commensurate) falls in the problems already
described in Sec.~\ref{sec:mott}. This situation allows one to obtain Mott insulating states with fractional filling
\cite{buonsante_04a,buonsante_04b,buonsante_05,rousseau_arovas_06}. The physics can be easily understood in the TG
limit. In that case one has free fermions in a periodic potential. The eigenstates of the non-interacting fermion
Hamiltonian form $l$ distinct bands separated by energy gaps. Within a given band, the eigenstates are indexed by a
quasi-momentum defined modulo $2\pi/l$, so that quasi-momenta can be taken in the interval $[-\pi/l,\pi/l]$ (called
the first Brillouin zone in solid state physics). The filling is determined by the condition
$n_0 = N/L=\sum_{n=1}^l \int_{-k_{F,n}}^{k_{F,n}}\frac{dk}{2\pi}$ where $n$ is the band index, and $k_{F,n}$ the Fermi
wavevector in the $n$-th band. As the density is increased, bands get progressively filled. From this picture, it is
clear that when the highest non-empty band is partially filled, the system will be in a gapless state, which as we
saw previously can be described as a Tomonaga-Luttinger liquid with exponent $K=1$. However, when the highest non-empty band
is completely filled, there is a gap to all excitations above the ground state and one obtains a Mott insulator.
This situation is obtained every time the filling is an integer multiples of $1/l$, thus allowing the observation
of a Mott insulating state for a filling with less than one-boson per site. In some cases, two energy bands may
cross and, since there will be no gap, the system may not be insulating for some integer multiples of $1/l$
\cite{rousseau_arovas_06}.

For finite $U$, one can use bosonization and is back to the situation described in Sec.~\ref{sec:mott} where one
has added a periodic potential leading to an operator with $p > 1$. These terms will be relevant for $K<2/p^2$, so
that they can only contribute in the case of systems with long-range interaction, as was discussed in
Sec.~\ref{sec:mott}.

\begin{figure*}[!htb]
\centerline{
\includegraphics[width=0.44\textwidth]{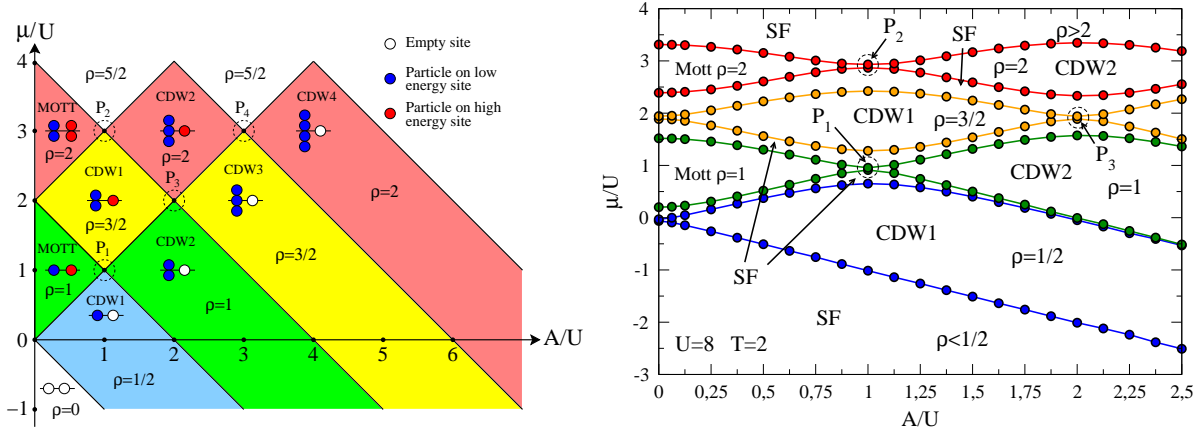}
\includegraphics[width=0.44\textwidth]{figure_phasediag1Dcomsuperlattice}}
\caption{(Left panel) Phase diagram of soft-core bosons in the atomic limit ($t=0$) in a superlattice with
$l=2$. $A$ is the amplitude of the periodic potential, $T$ is its period, $\mu$ is the
chemical potential, $\rho$ the number of particle per site. (Right panel) Phase diagram computed with QMC simulations for $U=8t$. In the latter, notice the appearance
of superfluid phases (SF in the figure) in between the insulating ones \cite{rousseau_arovas_06}.}
\label{fig:comsupPD}
\end{figure*}

An instructive way to understand the effect of the competition between $U$ and the superlattice potential is
to consider the atomic limit. For concreteness, let us assume that $\epsilon_i=A \cos(2\pi i/l)$ with $l=2$.
In this case, we have to deal with a two-site problem and $n=N/2$ ($N$ is the number of bosons). In order to
minimize the energy, the first particle should be added to the site with $\epsilon=-A$, i.e., the boundary
between $n=0$ and $n=1/2$ occurs  at $\mu\equiv E(n=1/2)-E(n=0)=-A$, and for $\mu<-A$ the system is
empty. The fate of the second particle added will depend on the relation between $A$ and $U$. If $A<U$
the energy is minimized by adding it to the site with $\epsilon=A$, i.e., the energy increases in $A$,
while if $A>U$ then the second particle should be added to the site with $\epsilon_i=-A$, i.e., the
energy increases in $U-A$. As a result, the boundary between $n=1/2$ and $n=1$ occurs at
$\mu=\min(A,U-A)$, so that for $-A<\mu<\min(A,U-A)$ the density in the system is
$n=1/2$ (this phase is also referred to in the literature as a charge-density wave), and so on (see left panel
in Fig.~\ref{fig:comsupPD}). As in the Bose-Hubbard model (Sec.~\ref{sec:BHMPeriodic}), starting from the atomic
limit result, one can perturbatively realize that the effect of a very small hopping amplitude will be to generate
compressible superfluid phases between the boundaries of the insulating phases. An exact phase diagram for finite
hopping from \citet{rousseau_arovas_06}, obtained using QMC simulations, is presented in the right panel in
Fig.~\ref{fig:comsupPD}.

Incommensurate superlattices exhibit a quite different behavior. The case of two incommensurate periodicities is
known as the Harper model \cite{hiramoto_Harper}, and constitutes one of the cases of quasiperiodic potentials. How
such quasiperiodic potentials lead to properties similar or different from disordered ones is a long standing
question. For noninteracting particles the question can be addressed analytically. In a lattice, the model is known
as the Aubry-Andr\'e model
\cite{aubry,almostMathieu,hofstadter}
\begin{equation}
\hat{H}=\sum_{i=1}^L \left[ -t(\hat{b}^\dagger_i \hat{b}_{i+1}+\textrm{H.c.})+V_2 \cos(2\pi\beta i) \hat{n}_i \right].
\end{equation}
Here $t$ is the tunneling rate and $V_2$ is the amplitude of the quasiperiodic modulation of the potential energy,
while $\beta$ is an irrational number. Such model exhibits a localization transition even in 1D with a critical
value $V_2/t=2$. A duality transformation maps the Aubry-Andr\'e model at $V_2/t>2$ on the same model at $V_2/t<2$
\cite{aubry}. Below criticality, all states are extended Bloch-like states characteristic of a periodic potential.
Above criticality, all states are exponentially localized and the spectrum is pointlike. At criticality the
spectrum is a Cantor set, and the gaps form a devil's staircase \cite{harper}. Some differences at the semiclassical
level between this type of potential and localization by disorder were pointed out
\cite{albert_semiclassical_quasiperiodic}.
A generalization of the Aubry-Andr\'e model to the presence of a non-linear term in the dynamics was discussed in \textcite{flach_2009}.
The crossover between extended to localized states in incommensurate
superlattices has been experimentally observed for a noninteracting ${}^{39}$K BEC where the effect of interactions
has been canceled by tuning a static magnetic field in proximity of a Feshbach resonance to set the scattering
length to zero \cite{roati08}.

What remains of such a transition in the presence of interactions is of course a very challenging question.
Here
the effects of interactions can be taken into account within a mean-field type of approach\cite{larcher_2009} or by bosonization. A unifying description of all types of
potentials including periodic, disordered, quasiperiodic was proposed by
\citet{vidal_quasi_interactions_short,vidal_quasiinter_long}, by generalizing the equations~(\ref{eq:kt-flow}) and
(\ref{eq:rg-back}) to a potential with arbitrary Fourier components $V(q)$. For quasiperiodic potentials, such as
the Fibonacci sequence, the transition between a superfluid phase and a phase dominated by the QP potential was
found. The critical point $K_c$ depends on an exponent characteristic of the QP potential itself. These conclusions
were confirmed by DMRG calculations on quasiperiodic spin and Hubbard chains
\cite{hida_quasi_spinless_DMRG,hida_precious,hida_quasi_spinfull}. The case of the Aubry-Andr\'e potential was
studied by exact diagonalization \cite{roth2003a,roth2003b} on $8$ and $12$ sites system and also for the case
of a specific choice of the height of the secondary lattice \cite{roscilde_08}.

\begin{figure*}[!t]
\centerline{\includegraphics[width=0.95\textwidth]{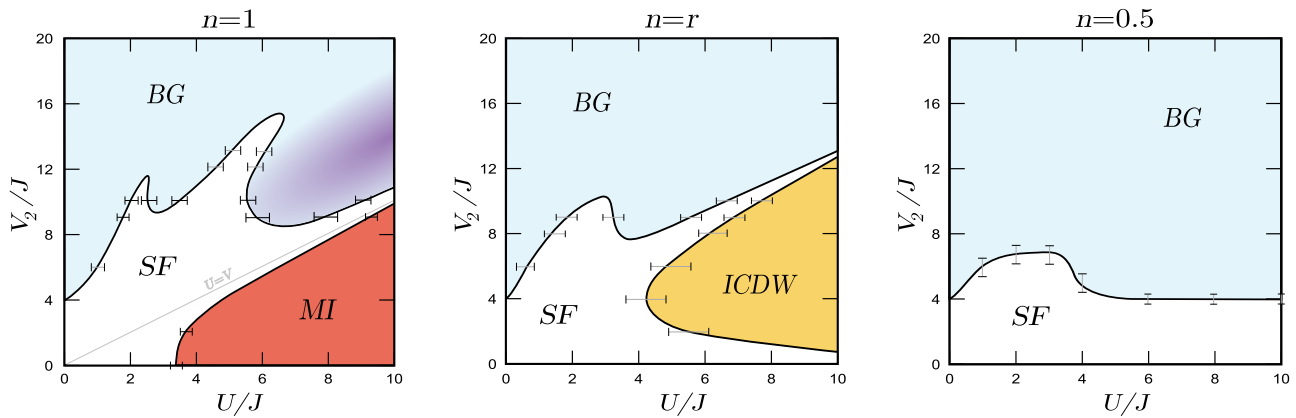}}
\caption{Phase diagrams of the bichromatic Bose-Hubbard model for densities $n=1$, $r$ (the ratio of the potential
wave lengths) and $n=0.5$. \emph{SF} stands for the superfluid phase, \emph{MI} for the Mott-insulating phase,
\emph{BG} for the ``Bose-glass'' phase (meaning localized but with zero one-particle gap) and \emph{ICDW} for
incommensurate charge-density wave phase. The $U=V_2$ line on the phase diagram with $n=1$ indicates the $J=0$
limit for which the gap of the one-particle excitation vanishes. In the phase diagram with density $n=1$, the
darker (violet) region in the BG phase is localized but could have a small gap which cannot be resolved numerically
\cite{roux_barthel_08}.}
\label{fig:aubry-phase-diag}
\end{figure*}

DMRG approaches \cite{roux_barthel_08,deng_disorder}, combined with the above analytical considerations, allowed for
a rather complete description of the physics of such quasi-periodic systems. As discussed above, for fillings
commensurate with either the primary or the secondary lattice, the periodic potential changes the simple quadratic
Hamiltonian of the Tomonaga-Luttinger liquid into a sine-Gordon Hamiltonian which describes the physics of the Mott
transition \cite{giamarchi_book_1d}. A Mott insulator is obtained in case of filling commensurate with the primary
lattice, and a pinned incommensurate density wave (ICDW) for filling commensurate with the secondary lattice. For
fillings incommensurate with both lattices, the potential is irrelevant under RG flow and one expects from
perturbative analysis a superfluid phase for all the values of the interaction strength. As a consequence, when one
of the potentials is commensurate, no Bose Glass phase can be created by the other potential in the vicinity of the
Mott Insulator superfluid transition in the regime where bosonization is applicable. In agreement with the
limiting cases of free and hard-core bosons described by an Aubry-Andr\'e problem, the transition towards the Bose
glass phase is found at $V_2/t\ge 2$, the critical value of $V_2$ being higher for bosons with finite interaction
strength. Another feature of an interacting Bose gas in quasiperiodic potential is the shrinking of
the Mott-lobes as a function of $V_2$. The computed phase diagram in the $(V_2/J,U/J)$ plane for both
commensurate and incommensurate fillings is reported in Fig. \ref{fig:aubry-phase-diag}.

The Bose glass phase has been shown to be localized by computing the expansion of the 1D cloud
\cite{roux_barthel_08}, as is relevant for the experimental situation. Although from the point of view of the phase
diagram, the localized nature of the phase in interacting quasiperiodic and disordered systems are found to be very
similar, some experimental probes have been found to show some marked differences between the two cases
\cite{orso_modulation_qp_disorder}. Interacting quasiperiodic potentials are a problem deserving more
experimental and theoretical investigations. We should add that the influence of a third harmonic as well as
confining potentials in such systems have already been investigated by \citet{roscilde_08}.

\section{Mixtures, coupled systems, and non-equilibrium}

In this Section we briefly examine several developments going beyond the simple case of 1D bosons in presence of an
external potential. We present three main directions where new physics is obtained by either: i) mixing bosonic
systems, or bosonic and fermionic systems; ii) going away from the 1D situation by coupling chains or structures.
This leads to quasi-1D systems which links the 1D world with its higher dimensional counterpart; and
iii) setting systems out of equilibrium.

\subsection{Multicomponent systems and mixtures}\label{sec:multi}

Compared to the case of a single component bosonic system, described in the previous sections, novel physics can be
obtained by mixing several components in a 1D situation or giving to the bosons an internal degree of freedom.

\subsubsection{Bose-Bose mixtures}

Internal degrees of freedom for bosons lead to a wider range of physical phenomena. Let us focus here on the
case of two internal degrees of freedom and refer the reader to the literature for the higher symmetry cases
\cite{boulat_2010,essler_2009,cao2007,lee2009}. For two components this is similar to giving a ``spin'' 1/2 to the bosons. The
corresponding case for fermions is well understood \cite{giamarchi_book_1d}. Because of the Pauli principle a
contact interaction in such a system can only exist between opposite spins $U = U_{\uparrow,\downarrow}$. The
ground state of the two component fermionic systems, is in the universality class of LL, with dominant
antiferromagnetic correlations. Quite remarkably because in the LL everything depends on collective excitations,
the Hilbert space separates into two independent sectors, one for the collective charge excitations, one for the
spins. This phenomenon, known as spin-charge separation is one of the hallmark of the 1D properties.

For the case of two component bosons the situation is much richer. First, even in the case of contact interactions,
interactions between the same species must be considered. The properties of the system thus depend on three
interaction constants $U_{\uparrow,\uparrow}$, $U_{\downarrow,\downarrow}$, $U_{\uparrow,\downarrow}$. The dominant
magnetic exchange will be crucially dependent on those interactions \cite{duan_spin_exchange_cold}, and can go for
dominantly antiferromagnetic if $U_{\uparrow,\uparrow}, U_{\downarrow,\downarrow} \gg U_{\uparrow,\downarrow}$
while they are dominantly \emph{ferromagnetic} in the opposite case. In particular, the isotropic case corresponds
to a ferromagnetic ground state \cite{eisenberg_ferro_bosons}. The phase diagram of such systems can be worked
out by using either LL description or numerical approaches such as DMRG or time evolving block decimation (TEBD)
\cite{kleine_2velocities_bosons,kleine_2velocities_bosons_long,mathey09}. Probes such as noise correlations, as
will be discussed in the next section, can be used to study the phases of such binary bosonic mixtures
\cite{mathey_noise_correlations}. When the exchange is dominantly antiferromagnetic, the physics is very similar to
the fermionic counterpart, with two collective modes for the charge and spin excitations, and can be described in
the LL framework. Such bosonic systems thus provide an alternative to probe for spin-charge separation
\cite{kleine_2velocities_bosons,kleine_2velocities_bosons_long} compared to the electronic case for which
this property could only be probed for quantum wires
\cite{auslaender_quantumwire_tunneling,tserkovnyak_quantumwire_tunneling}. Depending on the interactions, one of
the velocities can vanish, signaling an instability of the two component LL. This signals a phase separation that
corresponds to entering into the ferromagnetic regime which we now examine.

When the ground state of the system is ferromagnetic, a phenomenon unusual to 1D takes place. Indeed, the system
has now spin excitations that have a dispersion behaving as $k^2$ \cite{sutherland68,li2003,fuchs_effective_mass_ferro_bosons,guan2007}, which
cannot be described by a LL framework. This leads to the interesting question of the interplay between charge
and spin excitations, and it has been proposed \cite{zvonarev_ferro_cold} that such a system
belong to a new universality class, nicknamed ``ferromagnetic liquid''. This field is currently very active
\cite{akhanjee_ferro_liquid,guan2007,matveev_exponent_ferro_liquid,kamenev_exponents_ferro_liquid,zvonarev_bose_hubbard,%
zvonarev_gaudin_yang,caux2009}, and we refer the reader to the literature for further discussions on this subject and
references.

\subsubsection{Bose-Fermi mixtures}\label{sec:bosefermi}

Cold atomic systems have allowed for the very interesting possibility to realize Bose-Fermi mixtures.
Such systems can be described by the techniques described in the previous sections, both in the continuum case and
on a lattice. In the continuum, a Bose-Fermi mixture can be mapped, in the limit when the boson-boson interaction
becomes hard-core, to an integrable Gaudin-Yang model \cite{gaudin_fermions,yang_fermions}. For special symmetries
between the masses and couplings a mapping to an integrable model exists
\cite{lai1971,lai1974,batchelor2005,frahm05_bosefermi1d,guan2008,imambekov2006_mixture,imambekov05_bosefermi1d}.
In the case of a lattice, the system can be described by a generalized Bose-Hubbard description in which the
parameters can be obtained from the continuum description \cite{albus2003}. Similarly to the continuum case, if
the boson-boson repulsion becomes infinite, the hard-core bosons can be mapped onto fermions either by using the boson-fermion mapping\cite{girardeau-minguzzi_2007} or by using the
Jordan-Wigner transformation of Sec.~\ref{sec:mapping} leading \cite{sengupta2005} to a 1D
Fermi-Hubbard model that can be described be Bethe ansatz \cite{lieb_hubbard_exact} or by the various
low energy or numerical methods appropriate for fermionic systems \cite{giamarchi_book_1d}.

Quite generally the low energy of Bose-Fermi mixtures can be described by the LL description
\cite{cazalilla03_mixture,mathey04_mix_polarons}. Like for Bose-Bose mixtures, the resulting low energy Hamiltonian
has a quadratic form and can be diagonalized by a linear change of variables
\cite{muttalib1986,loss1994,cazalilla03_mixture}, which leads to two collective modes with two different velocities.
The elementary excitations are polaronic modes \cite{mathey04_mix_polarons,mathey07_bose_fermi}. In similar way,
the exponents of the correlation functions are given by the resulting TLL Hamiltonian
\cite{mathey07_bose_fermi,orignac2010}. Also similarly to the Bose-Bose mixtures, when the
velocity of one of the mode vanishes, the two component Tomonaga-Luttinger liquid shows an instability. This instability
can be, depending on the coupling constants, either a phase separation or a collapse of the system. This is the
analog of the Wentzel-Bardeen instability \cite{wentzel1951,bardeen1951} obtained in 1D electron-phonon systems
\cite{loss1994}.

Several interesting extensions to this physics exist when the components of the mixture are close in density
\cite{cazalilla03_mixture,mathey04_mix_polarons}. In particular, bound states of bosons and fermions can form
leading to a LL of composite particles \cite{burovski2009}. In the presence of a lattice, commensurability with the
lattice can give rise to partially gapped phases, in the similar way than the Mott physics described in the
previous section \cite{mathey07_bose_fermi,mathey09}, and, in general, to very rich phase diagrams
\cite{sengupta2005,pollet2006,hebert2007,hebert2008,pollet2008,rizzi2008,zujev2008}. Similarly, disorder effects
can be investigated for Bose-Fermi mixtures \cite{crepin_2010}.

The lattice Bose-Fermi Hubbard model 
has also been studied by world-line quantum Monte Carlo simulations \cite{takeuchi2005,takeuchi2005a}
and phase separation was found both as a function of the interparticle interaction and of the fermion density. The
mixed phase was instead studied using Stochastic Series Expansions by \citet{sengupta2005}. For weak coupling, a
two-component Tomonaga-Luttinger liquid behavior was found while at strong coupling, and commensurate filling, a phase with
total gapped density mode was recognized. Doping the gapped phase $n_F+n_B=1$ with fermions was shown to lead to
supersolid order \cite{hebert2008}. The ground state phase diagram was obtained using the canonical worm algorithm
for $n_F=n_B=1/2$ by \citet{pollet2006} and for other commensurate as well as incommensurate fillings by
\citet{zujev2008}.
The effect of harmonic trapping was instead considered using both DMRG and quantum Monte Carlo calculations
by \citet{pollet2008} and a complete analysis of the boson visibility appeared in \cite{varney2008}.

\subsection{Coupled systems}\label{sec:coupled1dbosons}

Although the extended Bose-Hubbard model [Eq.~(\ref{eq:EBHM})] 
describes a single chain system, we have seen that its bosonization description requires doubling the number of
degrees of freedom, which means that it effectively becomes two coupled (hard-core boson or spin-$\frac{1}{2}$)
systems. Here, we shall further pursue this issue and consider the rich physics that arises when coupling many such
1D systems. Reasons for studying coupled 1D boson systems are manifold. The main motivation is, of course, an
experimental one, since many experimental realizations of 1D systems are typically not created in isolation but in
arrays containing many of them. Considering the possible ways they can be coupled such as via quantum tunneling
and/or interactions leads to many interesting theoretical questions. These include the understanding of the way in
which the properties and excitations of system evolve, as a function of the temperature or the energy scale at
which the system is probed, from the fairly peculiar properties of a 1D to the more familiar ones of higher
dimensional systems \cite{giamarchi_book_carr}. Other important questions concern the competition between different
ordered phases which are stabilized either by interactions or tunneling between the different 1D systems.
We next review the work done on some of these questions.

One of the simplest situations allowing to transition from a low dimensional to a high dimensional system is
provided by coupling zero dimensional objects. This is for example the case in spin systems where pairs of spins
can form dimers. Such dimers form zero dimensional objects that can then be coupled into a higher dimensional
structure by additional magnetic exchanges. Since the dimers are normally in a singlet state, separated by a gap
from the triplet, a weak coupling between them is irrelevant and the ground state still consists in a collection
of uncoupled singlets. A quantum phase transition can thus be obtained in two ways: i) one can increase the
coupling up to a point at which the ground state changes and becomes an antiferromagnetic one
\cite{sachdev_nature_2008}, ii) one can apply a magnetic field leading to transitions between the singlet and
triplet states. As can be seen from the mappings of Sec.~\ref{sec:mapping}, this model corresponds then to
hard-core bosons (representing the triplet states) on a lattice and can lead to Bose-Einstein condensation and
various interesting phases \cite{giamarchi_nature_2008}, which will be further described in the next section
on experiments.

For bosonic chains, a typical coupling between the 1D systems is provided by a Josephson coupling
\begin{equation}
H_J =   -t_{\perp} \int \sum_{\langle \mathbf{R},\mathbf{R}'\rangle}
\left[ \hat{\Psi}^{\dag}_{\bf R}(x) \hat{\Psi}_{\bf R}(x) + \mathrm{H.c.} \right],
\end{equation}
Such a term will combine with the 1D physics and lead to novel phases. In the case for which the physics of the
1D systems is described by a Tomonaga-Luttinger liquid, the interchain coupling is strongly renormalized by the 1D
fluctuations \cite{efetov_coupled_bosons,ho_deconfinement_coldatoms,cazalilla_deconfinement_long}. Using the
various mappings of Sec.~\ref{sec:mapping}, such physics is relevant in various context ranging from spin chains
\cite{schulz_coupled_spinchains} to classically coupled $XY$ planes
\cite{benfatto_BKT_in_cuprates,cazalilla_BKT_cold,mathey08_phase_locking}. Although the system becomes essentially
an anisotropic 3D ordered system (superfluid or magnetically ordered), new modes (Higgs modes) appear
\cite{schulz_coupled_spinchains,cazalilla_deconfinement_long,huber07,huber08,menotti08} due to the quasi-1D nature of the system. Such modes
have also be observed in strongly correlated anisotropic 3D systems.

When the 1D physics is gapped, such as in a Mott insulator, an interesting competition occurs between the 1D gap
and the Josephson coupling. This leads to a quantum phase transition for which the system goes from a 1D Mott
insulator to a higher dimensional superfluid. Such a deconfinement \cite{giamarchi_book_carr} is relevant for a
variety of systems ranging from spins to fermions. In the case of coupled bosonic systems it can be studied
\cite{ho_deconfinement_coldatoms,cazalilla_deconfinement_long} by the low energy methods described in the previous
sections. Interesting results are also found in the case of a finite number of coupled chains (so called ladder
systems \cite{ladder}) or planes. We refer the reader to the literature
\cite{donohue_commensurate_bosonic_ladder,luthra_bosonic_twolegladder} for more on this subject.

\subsection{Nonequilibrium dynamics}
\label{sec:dynamics}

Another topic of much current interest is the nonequilibrium dynamics of 1D quantum systems
\cite{cazalilla_rigol_10,dziarmaga_10,polkovnikov_sengupta_10}. Research on this topic is mainly motivated by ultracold gases
experiments, where the unique degree of tunability, isolation, and long coherent times have enabled the
exploration of phenomena not previously accessible in condensed matter experiments. In the latter, a strong
coupling to the environment, combined with the short time scales associated to their microscopic properties,
usually lead to rapid decoherence.

As discussed in Sec.~\ref{sec:coldatoms}, driving cold gases with time-dependent potentials can be used to probe
several properties of interest, such as the excitation spectra. However, the physics out of equilibrium is far
richer than that. Optical lattices and Feshbach resonances can be used to transition between weakly and strongly
interacting regimes in controllable time scales, and to generate exotic out-of-equilibrium states
\cite{greiner_mandel_02b,winkler_thalhammer_2006,will_best_10,strohmaier_greif_10}. Those could lead to new
phenomena and phases not present in systems in thermal equilibrium. Studying the dynamics of cold gases can also
help us gain a better understanding of the inner working of statistical mechanics in isolated quantum systems
\cite{rigol_dunjko_08}. Furthermore, now that nearly integrable systems can be realized in experiments
\cite{paredes04,kinoshita04}, one can study their dynamics and address questions previously
considered purely academic, such as the effect of integrability in the properties of the gas after relaxation
\cite{kinoshita_wenger_06}.


Some of the early studies of the nonequilibrium dynamics in 1D geometries addressed the effect of
a lattice potential and the onset of the superfluid to Mott insulator transition in the transport
properties of a bosonic gas \cite{stoferle_moritz_04,fertig_ohara_05}. In those experiments, the
systems were loaded in deep two-dimensional optical lattices with a weaker lattice along the
1D tubes (see Sec.~\ref{sec:coldatoms}). The harmonic trap along the 1D tubes was then
suddenly displaced a few lattice sites and the dynamics of the center of mass studied by
TOF expansion. Surprisingly, it was found that even very weak optical lattices could produce large
damping rates,\footnote{In a harmonic trap, in the absence of a lattice, no damping should occur
as the center of mass follows the classical equation of motion of a harmonic oscillator.} and that, in some
regimes, over-damping took place with the center of mass staying away from the center
of the trap. The former effect was related to the large population of high momenta resulting from
the strong transverse confinement \cite{ruostekoski_isella_05}, while the latter was related to
the appearance of a Mott insulator in the trap \cite{rigol_rousseau_05,rey_pupillo_05,pupillo_rey_06Dip}.
More recently, time-dependent DMRG studies have accurately reproduced the experimental findings in the regime where
the one-band Bose-Hubbard model is a valid representation of the
experiments \cite{danshita_clark_09,montangero_fazio_09}.

Correlations in 1D systems can manifest themselves in surprising ways out of equilibrium.
As noted by \citet{sutherland_98}, if a gas of interacting bosons is allowed to expand under a
1D geometry, the resulting momentum distribution after long expansion times can be very different
from the initial momentum distribution of the trapped system, as opposed to what will happen
in the usual TOF expansion in three dimensions. As a matter of fact, in the
Tonks-Girardeau regime, it has been shown that no-matter whether there is a lattice
\cite{rigol_muramatsu_05eHCBb,rigol_muramatsu_05eHCBc} or not \cite{minguzzi_gangardt_05} along the 1D tubes,
the expansion of a TG gas results in the dynamical fermionization of the bosonic momentum distribution
function. This is something that only occurs out of equilibrium and leads to a bosonic momentum distribution
with a Fermi edge. The expansion of the more generic Lieb-Liniger gas has been studied by
\citet{jukic_pezer_08} using a Fermi-Bose transformation for time-dependent states \cite{buljan_pezer_08}.
Those authors found that, asymptotically, the wave functions acquire a Tonks-Girardeau
structure but the properties of the gas are still be very different from those of a TG gas
in equilibrium.

The ultimate effect of one-dimensionality in the dynamics of isolated systems may be the lack of
thermalization associated to integrability. In a remarkable experiment, \citet{kinoshita_wenger_06}
studied the relaxation dynamics of an array of 1D bosonic gases created by a deep 2D optical lattice.
They found that, as long as the system remained 1D, no relaxation occurred towards the
expected thermal result. In the TG limit, in which the 1D system is integrable, the lack of thermalization can be
understood to be a result of the constraints imposed by conserved quantities that make the TG gas integrable.
Interestingly, in that regime, few-body observables after relaxation can still be described by a
generalization of the Gibbs ensemble (the GGE) \cite{rigol_dunjko_07}. The GGE density matrix can be
constructed following \citet{jaynes_57a,jaynes_57b} principle of maximization of the many-body entropy
subject to constraints, which in this case are a complete set of integrals of motion.
The relevance of the GGE to different 1D integrable systems and few-body observables, as well as it limits
of applicability, have been the subject of several studies during the last years
\cite{rigol_dunjko_07,rigol_muramatsu_06,cazalilla_06,calabrese_cardy_07a,cramer_dawson_08,barthel_schollwock_08,
kollar_eckstein_08,rossini_silva_09,iucci_cazalilla_09,fioretto_mussardo_10,iucci_cazalilla_10,cassidy_clark_10}.


In the opposite limit of nonintegrable systems, i.e., far from any integrable point, the understanding of
thermalization in isolated quantum systems, whose dynamics is unitary, has remained controversial. Early
studies in 1D lattice models resulted in mixed results in which thermalization was reported in some regimes and
not in others \cite{kollath_lauchli_07,manmana_wessel_07,roux_09}. In 2D lattices, thermalization was observed
in rather small systems \cite{rigol_dunjko_08} and was explained in terms of the eigenstate thermalization
hypothesis (ETH) \cite{deutsch_91,srednicki_94}.
Later systematic studies in 1D lattices have shown that thermalization also occurs in finite nonintegrable
1D systems. However, thermalization and the eigenstate thermalization hypothesis break down as one approaches an
integrable point \cite{rigol_09a,rigol_09b,roux_10}. This may be the reason behind the lack of thermalization
observed by \citet{kinoshita_wenger_06} for all 1D regimes, while relaxation to thermal equilibrium was inferred
to occur in 1D experiments on atom ships \cite{hofferberth07}. While the former were not sufficiently
away from integrability the latter were \cite{mazets_schumm_08,mazets_schmiedmayer_10}.

Many questions remain open in this exciting area of research. We will mention some of them in the Outlook section
at the end.

\section{Experimental Systems}\label{sec:experiments}

Experimentally, 1D quantum liquids have been created in various forms. We shall review several of them
in this Section. In the first three subsections we examine condensed matter realizations, whereas the last
subsection is devoted to ultracold atom realizations. Some of the former, such as spin ladder compounds
submitted to strong magnetic fields, have yielded a wealth of experimental evidence
for exotic behavior related to the Tomonaga-Luttinger liquid. Ultracold atomic systems have properties
that are largely tunable and therefore hold many promises for reaching regimes that are hardly accessible in
condensed matter systems. However, their properties can only be tested by a still rather limited number of
probes. Hence, the discussion of these systems is intimately linked to the developments related to the
probes used to investigate their properties.

\subsection{Josephson junctions}

\paragraph{Theory:}
In a superconductor, electrons with opposite spins form Cooper pairs. Those pairs have bosonic statistics so
that a superconductor can be seen as a system of interacting bosons.
In a Josephson junction, two grains ($\mu$m sized) of superconducting metal are separated by a barrier of
insulating or non-superconducting material. Each grain is small enough to have a well defined superconducting
phase $\theta_j$. The operator measuring the number of Cooper pairs transferred to the $j$-th grain is
$\hat{N}_j=\hat{Q}_j/(2e)$, where $\hat{Q}_j$ measures the charge imbalance of the grain and $2e$ is the
Cooper-pair charge. The electrostatic energy of the charged grain is $E_C \hat{N}_j^2/2 -\delta \mu \hat{N}_j$,
where $E_C = 4 e^2/C$ and $\delta \mu = V_g/2e$, $C$ being the capacity of the grain and $V_g$ the gate potential.
The phase rigidity of the superconductor leads to a contribution
$-E_J\sum_{j} \cos \left(\hat{\theta}_j -\hat{\theta}_{j+1}-\frac{2e}{\hbar}
\int_{\mathbf{r}_j}^{\mathbf{r}_{j+1}} d\mathbf{r}\cdot \mathbf{A}(\mathbf{r})\right)$
in the case of a chain, where $\mathbf{A}$ is the electromagnetic vector potential. Since
$[\hat{N}_i,\hat{\theta}_j]=i \delta_{ij}$, the Josephson junction chain is a realization of the model
\eqref{eq:phasemodel}, where $V_g$ controls the chemical potential $\delta \mu$ of the Cooper pairs
\cite{bradley_josephson1d,fazio_josephson_junction_review}.

The  model defined by the Hamiltonian
\eqref{eq:phasemodel}, the so-called self-charging model, was shown to possess for $V_g=0$ (the particle-hole
symmetric case) a zero temperature superconductor-insulator phase transition~\cite{bradley_josephson1d} driven by
the ratio of Josephson to charging energy, analogous to the transition between superfluid order and bosonic Mott
insulator discussed in Sec.~\ref{sec:pertlut}. The insulating regime is called Coulomb blockade of Cooper pair
tunneling (CBPCT). Using an approximate mapping of the Hamiltonian~(\ref{eq:phasemodel}) onto the $t$-$V$
model~\cite{glazman_josephson_1d}, the non-particle-hole symmetric case ($V_g \ne 0$) can be analyzed. A
CBPCT phase is predicted for integer filling, recovering the result of~\citet{bradley_josephson1d} while for
fillings close to a half-integer number of particles per site a density wave phase is obtained. For incommensurate
filling or half-filling and weak/short-ranged electrostatic repulsion, a fluid phase is predicted. This model
has also been studied with quantum Monte Carlo simulations~\cite{baltin_qmc_josephson1d}.

\paragraph{Experiments:}
As we have seen previously, Josephson junction arrays provide an experimental realization of interacting boson
systems. Experiments to probe the phase transition between the bosonic Mott insulator and the Tomonaga-Luttinger liquid
have been reported. Chains of Josephson junctions are prepared by lithography. The capacities and influence coefficients
are determined by the geometry. Because of the finite dimension of the grains, the application of a magnetic field
can be used to create a dephasing that reduces the tunneling amplitude of the Cooper pairs
$\tilde{E}_J= E_J |\cos (\pi \Phi/\Phi_0)|$, where $\Phi$ is the
magnetic flux applied between the grains and $\Phi_0=h/(2e)$
is the flux quantum for a Cooper pair. Since the repulsion between Cooper pairs is not affected, this allows one to vary the ratio of Coulomb to Josephson energy and induce the
superconductor-insulator transition  by increasing the applied magnetic field.

Experiments with $\mathrm{Al/Al_2O_3/Al}$ junctions of $200\mathrm{nm}$ size show, as a function of magnetic field,
a transition from a regime of high electrical conductance to a regime with a voltage threshold for electrical
conduction \cite{chow_josephson1d,haviland_josephson1d,watanabe_josephson1d}, in qualitative agreement with the
theoretical predictions \cite{bradley_josephson1d}. However, a study of scaling near the quantum phase transition
\cite{kuo_josephson1d_scaling} showed  disagreement between the measured critical exponent $\nu$ and the theoretical
expectations \cite{fisher_boson_loc}. A possible explanation is the presence of random offset charges
on the junctions (which in the boson language correspond to a random potential) which might turn the transition into
an Anderson localization transition as discussed in Sec.~\ref{sec:disorder}. Another explanation could be related to
the dissipation which has been neglected in Sec.~\ref{se:josephson-th}.

Dissipation can be included by adding to the Matsubara action a term proportional to
$\frac 1 {\hbar\beta} \frac{R_Q}{R} \sum_n |\omega_n| |\hat{\theta}(\omega_n)^2|$, where $R_Q=h/(4e^2)$ is the
quantum of resistance, $R$ is the resistance of the junction, and $\theta$ is the phase \textit{difference} across
the junction \cite{bobbert_dissipative_josephson1d_a,bobbert_dissipative_josephson1d_b,%
korshunov_dissipative_josephson1d,korshunov_dissipative_josephson1d_long}. In the presence of such terms, a richer
phase diagram is obtained with new superconducting phases. However, in the vicinity of the superconductor-insulator
critical point, the dissipative contribution to the action could become relevant, and change the universality class
of the phase transition.

Although the problem of quantum phase transitions in dissipative systems is interesting in its own right, this
change of universality class would make Josephson junctions arrays not well-suited as an experimental realizations
of the model~(\ref{eq:phasemodel}). Another, more direct realization of interacting bosons using Josephson junctions
is based on the vortices in the Josephson junction arrays
\cite{vanoudenaarden_josephson_mott,oudenaarden_josephson_mott_long}. Here, the vortices induced in the array
by the magnetic field behave as bosonic quantum particles. The array acts on the vortices as a periodic potential,
and a Mott insulating state of the vortices is observable when the number of vortices per site is commensurate.
The advantage of this type of experiment is that the vortices are insensitive to the random offset charges.
The main inconvenient is that the quantum effects are weak \cite{bruder_josephson1d}.

\subsection{Superfluid helium in porous media}\label{sec:nanopores}

Mesoporous materials with pores of diameter of nanometer scale have
been available since the 1990s. Folded sheet mesoporous materials
(FSM-16), synthesized from layered silicate kanemite
possess pore forming long straight channels arranged in a honeycomb
lattice. The diameter of the pores can
be controlled between 1 and 5 nm using appropriate
surfactants during the synthesis process.
${}^4$He atoms adsorbed in these
pores are confined in the transverse direction, and the pores can only
interact with each other via the free surface of the materials.
For temperatures that are low compared to
the transverse confinement energy scale the ${}^4$He adatoms
are thus expected to exhibit the properties of a 1D
Bose fluid \cite{wada_nanopores_4he}.
With a typical phonon velocity $u$ of the order of $200
\mathrm{m.s^{-1}}$,  and pores of diameter $d=18\pm 2$ \AA\
the temperature scale below which the 1D physics
should appear is $T\sim \hbar u/(\pi k_B d)\sim 1 \mathrm{K}$.
With pores of a larger diameter,
one-dimensionalization occurs at lower temperatures.
Experiments have shown that the first layer
of ${}^4$He adatoms on the surface of nanopore was inert. For coverages $n<n_1=2.4\mathrm{mmol}$, the ${}^4$He atoms
occupy only the inert layer, and heat capacity measurements  gave $C/T \to 0$ as $T\to 0$ for these low
coverages. For coverage $n>1.15 n_1 \sim 2.7 \mathrm{mmol}$, the specific heat behaves as $C/T \sim \alpha(n) + bT$,
for $T\to 0$, with $b=1.4 \mathrm{mJ.K^{-3}}$ independent of $n$. The $bT$ contribution is attributed to the inert
layer, and the contribution $\alpha(n)$ to the fluid layer.

\begin{figure}[!htb]
  \centering
  \includegraphics[width=0.4\textwidth,angle=-90]{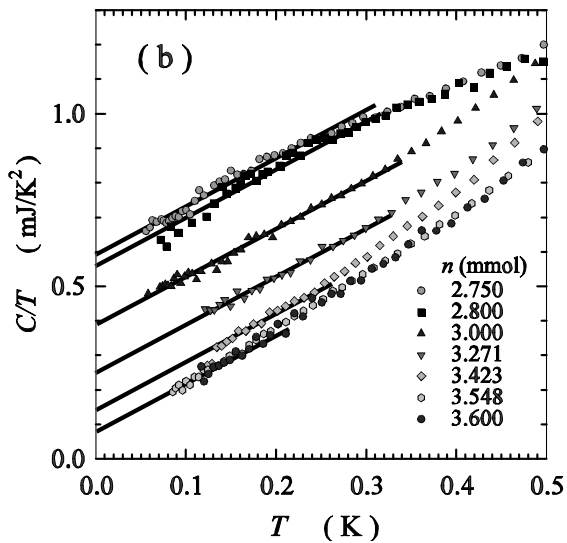}
  \caption{Specific heat over temperature $C_v/T$ of Helium 4 in nanopores of $18$ \AA\
    diameter. The $T$-linear contribution comes from the inert layer,
    and the constant contribution from the quasi-1D
    fluid \cite{wada_nanopores_4he}.}
  \label{fig:wada-nanopores}
\end{figure}

As shown in Fig.~\ref{fig:wada-nanopores}, the $T$-linear behavior of the specific
heat is the one expected in a 1D Tomonaga-Luttinger liquid.
By fitting $\alpha$ to the Tomonaga-Luttinger liquid expression, it was
possible to measure $u$ as a function of coverage.
Since the density of the 1D ${}^4$He fluid can be
obtained from the coverage knowing the pore diameters, the adsorption area $S$, and the first layer coverage, by
assuming a Lieb-Liniger interaction, one can obtain the Lieb-Liniger interaction constant. For
$n\sim 2.75\mathrm{mmol}$, \citet{wada_nanopores_4he} obtained an interaction $c\sim 0.7$\AA$^{-1}$. For higher densities,
$n> n_f\sim 1.4n_1$, the phonon velocity $u$ is found to diverge. This corresponds to the hard-core repulsion between
the ${}^4$He atoms turning the second layer into a 1D Mott insulator. The hard-core area was estimated to be
$A_0=(4.24 $\AA)$^2$.

\subsection{Two-leg ladders under an applied field}

Recently, quantum dimer spin systems in magnetic fields became of great interest because the experimental data
on ladder geometries showed the possibility of a quantum phase transition (QPT) driven by the magnetic field
\cite{sachdev-2000,giamarchi_nature_2008}.

In such systems, the application of a magnetic field induces a zero
temperature QPT between a spin-gap phase
with all dimers in the singlet state and zero magnetization and a
phase where a non-zero density of spin dimers are the triplet state
giving a nonzero magnetization. By increasing $H$, the spin gap
between the singlet ground state and the lowest triplet ($S_z=-1$)
excited states of the coupled dimers system decreases.  It closes at
$hc_1$, where the dimer system enters the gapless phase corresponding
to the partially magnetized state.  At $h>hc_2$ the system becomes
fully polarized and the gap reopens. This transition has been shown
\cite{giamarchi_coupled_ladders} to fall in the universality class of
BEC, where the dimers in the triplet state behave as hard-core bosons,
the magnetization is proportional to the density of hard-core bosons,
and the staggered magnetization in the plane orthogonal to the applied
field plays the role of the BEC order parameter. Basic experimental
evidence was recently obtained by magnetization
\cite{nikuni_bec_tlcucl} and neutron \cite{ruegg_bec_tlcucl}
measurements on the 3D material class XCuCl$_3$ (X = Tl,K,NH$_4$).  We
refer the reader to \citet{giamarchi_nature_2008} for more details and
  references.

As discussed in Sec.~\ref{sec:exactsol}, BEC cannot occur in 1D, but
quasi-long range order is expected in the magnetized phase. In 1D
systems, the continuous phase transition from the commensurate spin
gapped (C) zero uniform magnetization phase to an incommensurate phase
(IC) with nonzero magnetization has been studied theoretically
by mapping the spin-$1/2$ AF ladder in an external magnetic field $H$
onto a 1D system of interacting hard-core bosons
\cite{chitra_spinchains_field,furusaki_dynamical_ladder,
  giamarchi_coupled_ladders,hikihara_xxz}, where $H$ acts as a
chemical potential.  The interaction term in the hard-core boson
picture is determined by the exchange coupling constants, which can be
experimentally extracted, $J_\perp$ on the ladder rungs and
$J_\parallel$ on the ladder legs, and by $H$, which controls the
density of hard-core bosons. Thus the Hamiltonian is very well
controlled, allowing for a precise determination of the Tomonaga-Luttinger
parameters through the measurements of the magnetization and
transverse staggered spin-spin correlation function $\langle
\hat{S}^+(x)\hat{S}^-(x') \rangle$. By combining DMRG calculations
with bosonization, the LL parameters can be obtained numerically for
arbitrary $H$ \cite{hikihara_xxz,bouillot_ladder_dmrg_dynamics}.

These studies are relevant for several experimental compounds. One of
the first studied was Cu$_2$(C$_5$H$_{12}$N$_2$)$_2$Cl$_4$
\cite{chaboussant_cuhpcl}, but questions on whether its magnetic
structure is actually made of coupled ladders was
raised\cite{stone_ladderorganic_3d}.
Recently, CuBr$_4$(C$_5$H$_{12}$N)$_2$ (BPCB)
\cite{patyal_1990} was shown to be an excellent realization of
a ladder system. Namely,
the low-temperature magnetization data was well described by the $XXZ$
chain model \cite{watson_bpcb} in the strong-coupling limit
($J_\perp\gg J_\parallel$) of a ladder \cite{tachiki70,chaboussant_cuhpcl,
  mila_field,giamarchi_coupled_ladders}.

In connection with the magnon BEC, another important question is the dimensional crossover, between the 1D and
the 3D character in a system made of weakly coupled ladders as
discussed in Sec.~\ref{sec:coupled1dbosons}. At temperatures much larger than the inter-ladder
coupling, the system can be viewed as a collection of 1D ladders
\cite{sachdev_qaf_magfield,chitra_spinchains_field,giamarchi_coupled_ladders,furusaki_dynamical_ladder}
in the universality class of Tomonaga-Luttinger liquids at high field. At lower temperature,
interladder coupling cannot be ignored, and the system falls into the universality class of magnon BEC condensates.
 From the experimental point of view, such a change of regime
between the 1D and 3D limits is relevant for the above mentioned
compound  BPCB. Experimentally, to probe
these different regimes and LL behavior characterized by power-laws,
Nuclear Magnetic Resonance (NMR) Knight shift and relaxation time
$T^{-1}$ measurements have been performed and compared  with the theoretical predictions from bosonization and DMRG
\cite{giamarchi_coupled_ladders,klanjsek_bpcp,ruegg_magnetocaloric,thielemann_spinons,thielemann_bpcp}.

\begin{figure}[!htb]
\includegraphics[width=0.35\textwidth,angle=-90]{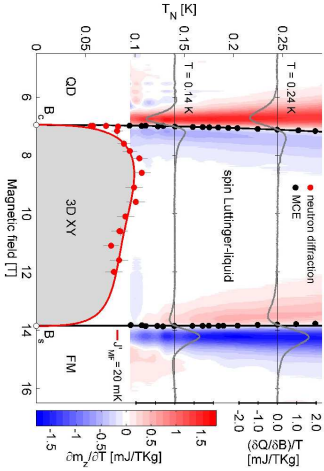}
\caption{Low $T$ phase diagram of (Hpip)$_2$CuBr4. The crossover temperature to the spin LL phase is derived from
magnetocaloric measurements and the phase transition to the BEC 3D-$XY$ magnetic order from neutron diffraction.
The red line is based on theoretical prediction \cite{thielemann_bpcp}.}
\label{fig:ladder-phase-diag}
\end{figure}

The above comparison allows for a quantitative check of the LL
theory \cite{klanjsek_bpcp}.
More generally phase diagram and order parameter have been found in
good agreement with the theoretical predictions
(Fig.~\ref{fig:ladder-phase-diag}). The neutron scattering spectrum
has been measured and compared with exact
solutions as described in Sec.~\ref{sec:lieb-corr} \cite{thielemann_spinons}
or DMRG calculations \cite{bouillot_ladder_dmrg_dynamics}.

Since they lead to a good realization of interacting bosons, these
ladder systems open interesting possibilities such as the study of
disordered bosons and the Bose glass phase (see
Sec.~\ref{sec:disorder}) \cite{hong2010}.

\subsection{Trapped atoms} \label{sec:coldatoms}

Atom trapping techniques allow for the realization of ultracold vapors
of bosonic atoms, thus offering another route for the experimental study of low-dimensional interacting bosons.
In experiments, Bose-Einstein condensates are usually created in 3D geometries.
However, by making the trap very anisotropic \cite{schreck01,gorlitz01,dettmer01}, or by loading the condensate
in 2D optical lattices \cite{greiner01,moritz03}, or by means of
atom chips \cite{folman00,amerongen2008_chip}, the quasi-1D regime can be accessed. In this section,
we first review the basic principles of atom
trapping techniques and the particular techniques developed for
probing the atomic clouds. Then, we review the experiments done on
quasi-1D bosonic systems.

\subsubsection{Atom trapping techniques}
\setcounter{paragraph}{0}

\paragraph{Optical trapping:}
In this method, ultracold atoms are trapped in standing light patterns
created by laser interference. The atoms respond to the electric field of the laser light by acquiring a small dipole
moment, which in turn yields a force on the atom proportional to the gradient of the electric field. The optical
potential thus created is proportional to the square of the electric field, that is, to the laser intensity. The
proportionality factor is the atom polarizability. The latter, for a two-level atom (a cartoon model of
alkali atoms such as $^{87}$Rb or $^{133}$Cs), is positive when the laser is \emph{blue-detuned} from a
characteristic atomic transition and negative when it is \emph{red-detuned}. Therefore, atoms accumulate in
regions of high light intensity when the laser is red detuned whereas they are attracted towards regions of low
light intensity when the laser is blue detuned. The two types of optical potentials are  indeed used in
the experiments to be described below.\cite{bloch_dalibard_zwerger_review_2008}

The simplest type of optical potential is created by a retro-reflected
laser beam (propagating, say, along the $z$-direction), which produces
a standing wave potential of the form $V_{\mathrm{opt}}(x) = V_{0z}
\sin^2 (k z)$, where $V_{0z}$ is proportional to the laser intensity
and $k = \frac{2\pi}{\lambda}$, $\lambda$ being the laser wavelength.
The strength of the optical potential, $V_{0z}$ is conventionally
measured in units of the recoil energy of the atom $E_{R} =
\frac{\hbar^2 k^2}{2M}$, where $M$ is the atom mass. Combining three
such mutually incoherent standing waves perpendicular to each other
yields the following three dimensional optical potential:
\begin{equation}
V_{\mathrm{opt}}(x,y,z) = \sum_{i=1}^{3}V_{0i} \sin^2 (k x_i). \label{eq:optpot}
\end{equation}
The minima of this potential occur at a cubic Bravais lattice. In
addition, in the experiments, a (harmonic) potential is needed in order to confine
the gas. The latter may or may not be generated by the same lasers
that create the lattice.

\paragraph{magnetic trapping and atoms on chip:}
Another technique for trapping atoms relies on magnetic fields.
The advantage of this technique is that the magnetic field can be
created by wires deposited on a surface (the so called atom chip
trap) by microfabrication techniques. Atoms
of total spin $\mathbf{F}$ interact with a magnetic field
$\mathbf{B}(\mathrm{r})$ by the Zeeman coupling
$H_{\mathrm{Zeeman}}=-g \mu_B \mathbf{F}\cdot\mathbf{B}$. When the
magnetic field is slowly varying in space, the spin follows
adiabatically the direction of the magnetic field resulting in a
potential proportional to $||\mathbf{B}||(r)$ \cite{folman2002}. When
the magnetic moment $g \mu_B \mathbf{F}$ is parallel to the magnetic
field, this potential is minimum where the strength of the magnetic
field is maximum. The atom is then said to be in a \textit{strong
  field seeking state}. Since Earnshaw's theorem \cite{ketterle1992}
prohibits maxima of the magnetic field in vacuum, in such situation
trapping is possible only if a source of magnetic field is located
inside the trap. With a magnetic moment antiparallel to the magnetic
field, the potential is minimum when the magnetic field intensity is
minimum. The atom is then said to be in a \textit{weak field seeking
  state}, which is metastable. Since Earnshaw's theorem does not
prohibit minima of the magnetic field in vacuum, no source of magnetic
field is needed to be present inside that trap.

Hence, most atom chip traps operate in the metastable weak field
seeking state. By superimposing the magnetic field created by a wire
and a uniform field $\mathbf{B}_{\text{bias}}$ orthogonal to the wire,
one obtains a line of vanishing magnetic field which traps the weak
field seeking atoms called a \textit{side guide} trap. However, in
this setup the adiabaticity condition $g\mu_B B /\hbar\gg dB/d\tau/B$ is
not satisfied and Majorana spin flips switch atoms to the strong field
seeking state causing losses from the trap. Thus, a second uniform
offset magnetic field $\mathbf{B}_{\text{o}}$ parallel to the wire is
superimposed to the trapping field to ensure that the adiabaticity
condition remains satisfied in the trap. The trapping in the
transverse direction is then harmonic \cite{folman2002,fortagh2004}.
With a Z-shaped wire, the atoms can also be confined along the central
part of the wire by a shallow harmonic confining potential.
Since magnetic trapping depends on the atoms remaining in the weak
field seeking state, Feshbach resonances cannot be used to reach a
regime of strong interactions. As a result, atom on chip trapping can
only be used to study the regime of weak interaction.

\subsubsection{Probes}\label{sec:coldatoms-probes}
\setcounter{paragraph}{0}

A usual probe in condensed matter physics is the measurement of the
structure factor:
\begin{eqnarray}
  S(q,\omega)=\int dx dt e^{-i(qx -\omega \tau)} \langle \hat{\rho}(x,\tau)
  \hat{\rho}(0,0)\rangle,
\end{eqnarray}
a quantity which can be predicted either via computational techniques
or analytical approaches.   Using linear response
theory, \citet{brunello2001}, showed that the structure factor of a
trapped gas could be measured with Bragg spectroscopy.
In Bragg spectroscopy measurements, \cite{stenger1999}, two laser beams of
different wavelengths are shone on an atomic clouds. This creates a
time dependent potential acting on the atoms:
\begin{eqnarray}
  H=V_0 \int dx \cos (qx-\omega \tau) \hat{\rho}(x,\tau)
\end{eqnarray}
After the perturbation has been applied, the total energy $E$ or the
total momentum $P$ of the system is measured. \citet{brunello2001}
showed that both quantities are proportional to the structure factor.
The effect of the trapping potential was considered by
\citet{minguzzi2009}. \citet{clement2009} have used Bragg spectroscopy
techniques to monitor the evolution of a trapped bosonic gas in a
1D lattice from the superfluid to the Mott insulator as
the potential is ramped up.

\paragraph{Time-of-flight measurements:}\label{sec:tof}
In a time-of-flight (TOF) measurement, the trapping potential containing the atomic gas is suddenly switched off
and, after expansion, the density of the atomic cloud is measured using absorption imaging techniques. Assuming
that interactions between atoms can be neglected and applying the stationary phase approximation to the solution of
the time-dependent Schrodinger equation, the annihilation operator of
a boson at position ${\bf r}$ and time $\tau$ is found to be
$\hat{\psi}({\bf r},\tau)\sim \hat{\psi}({\bf k}({\bf r}))/\tau^{d/2}$, where the momentum ${\bf k}({\bf r})=m{\bf r}/\hbar \tau$,
i.e., it provides a correspondence between the position of a boson after TOF and its initial momentum in the trap, and
the factor $\tau^{d/2}$ comes from the free evolution. After some straightforward manipulations, the density
distribution measured after a TOF $\tau$ at the position ${\bf r}$ is then found to be proportional to the initial
momentum distribution function 
\begin{equation}\label{eq:time-of-flight}
  \langle \hat{\rho}({\bf r},\tau)\rangle \propto \frac{1}{\tau}\langle
  \hat{\rho}({\bf k}({\bf r}))\rangle.
\end{equation}
The above expression means that BEC is reflected by a large peak in $\langle\hat{\rho}({\bf r}=0,\tau)\rangle$ after TOF
\cite{anderson_ensher_95,bradley_sackett_95,davis_mewes_95}. If a BEC is
loaded in the lowest band of an optical lattice, one needs to relate the momentum distribution function
$\hat{\rho}({\bf k})$ to the quasi-momentum distribution function $\hat{n}_{\bf k}$ in the lattice. The latter satisfies
$\hat{n}_{\bf k}=\hat{n}_{\bf k+Q}$, where ${\bf Q}$ is an arbitrary reciprocal lattice vector. Using the 3D
version of Eq.~\eqref{eq:projfieldop}, one finds that $\hat{\rho}({\bf k})=|w_0({\bf k})|^2 \hat{n}_{\bf k}$, where
$w_0({\bf k})$ is the Fourier transform of the Wannier orbitals. This relation implies that after TOF expansion, a BEC
released from an optical lattice will exhibit Bragg peaks at ${\bf k}({\bf r})={\bf Q}$ corresponding to reciprocal
lattice vectors \cite{greiner02}.

We should stress, however, that the density distribution after a TOF can be different from the in-trap momentum
distribution if interaction effects occur or when the TOF is not sufficiently long to neglect the
initial size of the cloud \cite{pedri2001,gerbier2008}. For optical lattice experiments with small filling factors,
the ballistic expansion is typically a good approximation. Moreover by applying the Feshbach resonance technique,
one can also always tune the collision interaction to a negligible magnitude at the beginning of the expansion to
satisfy the condition of the ballistic expansion. Another deviation from the true BEC behavior occurs when the
temperature is increased from zero, $T>T_\phi$, $T_\phi$ being the temperature at which the phase fluctuations become
of order unity. In this quasi-BEC regime the detection signal decreases and Bragg peaks becomes less visible. If
instead, the interaction is increased while the temperature is still very low, i.e., in the Mott state, coherence is
destroyed and there is no interference pattern in $\langle \hat{\rho}({\bf r},\tau)\rangle$.

For a Tomonaga-Luttinger liquid, the boson creation operator is represented by $\hat{\psi}^\dagger(x)\sim e^{-i\hat{\theta}(x)}$,
thus $ \langle \hat{\rho}(k)\rangle\propto k^{1/2K-1}$ displaying a characteristic power-law divergence at zero momentum
for $K>1/2$. In particular, in the Tonks-Girardeau limit, the Tomonaga-Luttinger parameter $K=1$, and the momentum distribution presents a
square-root divergence at $k=0$. During the past years experimental groups were able to increase the interactions
to reach this regime \cite{paredes04,kinoshita04}.

\paragraph{Noise correlations:}
Since atomic clouds are mesoscopic, the density fluctuates between
different time-of-flight
experiments. \citet{altman_noise_correlations}
thus proposed to use
time of flight spectroscopy to measure correlations
 between occupation numbers for different momenta:
\begin{eqnarray}
  \label{eq:momenta-correlations}
{\mathcal G}_{\mathbf{k},\mathbf{k}^\prime}=\langle \hat{n}_\mathbf{k} \hat{n}_{\mathbf{k}^\prime}\rangle - \langle \hat{n}_\mathbf{k}\rangle \langle \hat{n}_{\mathbf{k}^\prime}\rangle
\end{eqnarray}
where $\hat{n}_{\bf k}$ is the occupation number for momentum ${\bf k}$, and the average is taken over the initial
state. \citet{altman_noise_correlations} considered  Fermi superfluids
and bosonic Mott
insulators.

For 1D boson systems, the correlations
(\ref{eq:momenta-correlations})  were
studied by \citet{mathey_noise_correlations}.
For $K\gg 1$, the
correlation function ${\mathcal G}_{k,k^\prime}$ shows  a large peak for
$k=k'=0$, sharp power law
peak-like features for $k=\pm k'$, and power law dip-like features
for $k=0$ or $k'=0$.  The peak as $k=k'$ is the result of boson bunching
familiar from quantum optics. The location of the other features
 is predicted \cite{mathey_noise_correlations} by the Bogoliubov
approximation.
However, their power law character is a signature of the Tomonaga-Luttinger liquid
physics unique to one-dimension \cite{mathey_noise_correlations}.
For lattice hard-core bosons, noise correlations have been computed in the presence
of commensurate and incommensurate superlattices, and in the presence of a trap, by
\citet{rey_satija_06a,rey_satija_06b,rey_satija_06c} and their scaling has been studied by
\citet{he_rigol_10}, using the methods described in Sec.~\ref{sec:tonks}.

A different approach to probe static quantum correlations
with noise measurement has been proposed by
\citet{polkovnikov_interference_between_condensates}.
The idea is to start from two independent 1D condensates
parallel to the $x$-axis and separated by a distance $d$, both tightly confined in the radial direction $z$. At
time zero, both condensates are let to expand in the radial
direction.
Because the condensates were tightly confined,
expansion is much faster in the radial direction $z$, and one can neglect
expansion along the direction $x$.
Under this assumption, the total annihilation operator of the bosons reads:
\begin{align}
  \label{eq:two-condensates-expansion}
  \hat{\Psi}(z,x,\tau) &\simeq  \frac{-i m}{2\pi \hbar \tau}   \sum_{p=1,2} \hat{\Psi}_{p}(z)
  e^{\frac{im}{2\hbar \tau} (x+ s_p d/2)^2} ,
\end{align}
where $s_{p=1} = -1$ and $s_{p=2} = +1$  and $\hat{\psi}_{1,2}(z)$ is the annihilation operator of bosons initially
trapped in the condensate $1$ or $2$.
Absorption imaging techniques can then be used to measure the density profile integrated along the beam axis:
\begin{equation}
  \label{eq:integ-dens-prof}
  \hat{\rho}(z,\tau)=\int_0^L  dx \,  \hat{\Psi}^\dagger(x,z,\tau)  \hat{\Psi}(x,z,\tau),
\end{equation}
Using Eq.~(\ref{eq:two-condensates-expansion}), one can express $\hat{\rho}(z,\tau)$ as a function of the annihilation
operators of condensates $1$ or $2$,
which has an oscillatory contribution (that accounts for the interference fringes in the optical absorption images)
proportional to $\hat{A}_Q e^{i Q z} $,  where we have introduced $Q=\frac{md}{\hbar \tau}$ and
\begin{equation}
  \label{eq:demler-correlator}
 \hat{A}_Q= \left( \frac{m}{2\pi \hbar \tau}\right)^2  \int_0^L dz_1 \hat{\Psi}^\dagger_1
(z_1) \hat{\Psi}_2 (z_1).
\end{equation}

From one experimental realization (`shot', for short) to another, the quantity $A_Q$ varies randomly. As a
result, the amplitude of the fringes varies from experiment to
experiment, and one needs  the  moments of the correlation function $\langle (A_Q^\dagger)^n A_Q^n\rangle$ to characterize the full
probability distribution of $\hat{A}_Q$ \cite{glauber1963,Mandel1965}.
As we will see, the probability
distribution of $\hat{A}_Q$ is non-gaussian, and depends on the 
Luttinger parameter thus permitting its experimental determination.
The higher moments are given by the expression:
\begin{align}
  \label{eq:2condensates-higher-moments}
  \langle (\hat{A}_Q^\dagger)^n \hat{A}_Q^n\rangle = &\int_0^L dx_1 \ldots dx_n
  dx'_1 \ldots dx'_n \,\times  \\
& \left|\langle \hat{\Psi}^\dagger(x_1)\ldots \hat{\Psi}^\dagger(x_n)
 \hat{\Psi}^\dagger(x'_1)\ldots \hat{\Psi}^\dagger(x'_n)\rangle \right|^2.\nonumber
\end{align}
Using bosonization, this integral can be rewritten as:
\begin{eqnarray}
  \label{eq:2condensates-higher-moments-boso}
  \langle (\hat{A}_Q^\dagger)^n \hat{A}_Q^n\rangle& \sim&  L^{n(2-1/K)} \int_0^1 d\omega_1 \ldots d\omega_n
  d\omega'_1 \ldots d\omega'_n \times \nonumber\\&&\left|\frac{\prod_{1\le k<\ell\le n}
      |\omega_k-\omega_\ell||\omega'_k-\omega'_\ell|}{\prod_{1\le k,
        \ell \le n}|\omega_k-\omega'_\ell|}  \right|^{1/K},
\end{eqnarray}
where the change of variables $x_i=L\omega_i$,  $x'_i=L\omega'_i$ has been used.
Equation~(\ref{eq:2condensates-higher-moments-boso}) shows that the higher moments are of the form:
$\langle (\hat{A}_Q^\dagger)^n \hat{A}_Q^n\rangle \sim (\langle (\hat{A}_Q^\dagger) \hat{A}_Q\rangle)^n Z_{2n}(K)$, 
where $Z_{2n}(K)$ is a dimensionless factor given by the multidimensional integral in (\ref{eq:2condensates-higher-moments-boso})
that depends only on $n$ and $K$. The quantities $Z_{2n}(K)$ are the moments of the distribution function of the
normalized fringe interference contrast. The integrals in
(\ref{eq:2condensates-higher-moments-boso}) can be expressed using
Jack polynomials \cite{fendley_integrals},  contour integration
\cite{konik_integrals} or Baxter's Q operator
\cite{gritsev2006_contrast}  methods.
\citet{imambekov-random-surfaces},
related the distribution function  to the statistics of random
surfaces, allowing for  Monte-Carlo computation.
With periodic boundary conditions, the
normalized distribution $\tilde{W}(\alpha)$  such that $Z_{2n}(K)/Z_2(K)^n=\int d\alpha\,
\alpha^{n} \tilde{W}(\alpha)$  goes to a  Gumbell
distribution $W_G(\alpha)=K \exp(K(\alpha-1)-\gamma -e^{
  K(\alpha-1)-\gamma})$  in the limit $K\to \infty$.
 This is the result of the rare events \cite{imambekov-varenna} in the fluctuations of the equivalent random surface model. The prediction
of a Gumbel distribution of interference fringe contrasts was  checked experimentally \cite{hofferberth-noise}
using ${}^{87}$Rb atoms in radiofrequency microtraps on atom chips.

\subsubsection{One-dimensional bosons  with cold atoms} \label{sec:atomchips}
\setcounter{paragraph}{0}

On a chip, atoms can be trapped in a quasi-1D
geometry allowing one to access the weakly interacting regime.
A study of the density profile of atoms on chip \cite{trebbia2006} showed that the
Hartree-Fock approximation breaks down already in the weakly interaction regime for quasi-1D trapping. Later
experiments \cite{amerongen2008_chip} showed that the density profiles could be well fitted using the Yang-Yang
thermodynamics \cite{yang1969_bosons1d} of the Lieb-Liniger gas
\cite{lieb_bosons_1D}. To access the regime of strong interaction,
optical lattices are more suitable.

\begin{figure}
 \centering
 \includegraphics[width=8cm]{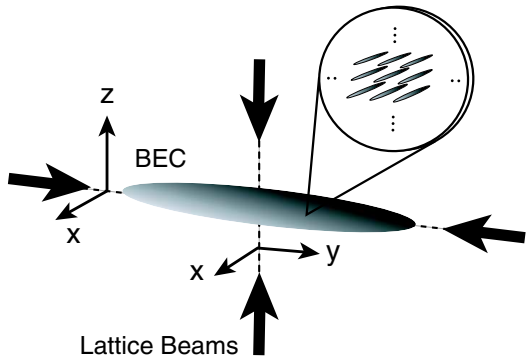}
 \caption{\label{fig:2DOL}
Experimental setup for the creation of a 2D Optical lattice \cite{greiner01}. The optical potential
resulting form the interference of a pair of retro-reflected, mutually perpendicular, lasers splits a BEC into
an array of several thousand 1D gas tubes.}
\end{figure}

In what follows, we shall consider ultracold quantum gases in strongly
anisotropic lattices. By `strongly anisotropic', we mean that the
optical potential is of the form of Eq.~(\ref{eq:optpot}) but such
that $|V_{0z}| \ll |V_{0x}| = |V_{0y}|$. Under this condition, it is
possible to confine atoms to 1D
\cite{greiner01,moritz03,stoferle04,kinoshita04,paredes04,laburthe04,kinoshita05,fertig05,haller_sine_Gordon_MI_2010}.
Indeed, this requires that the chemical potential of the 1D gas,
$\mu$, is much smaller than the trap frequency, which for a deep
lattice $\simeq k |\frac{V_{0z}}{M}|^{1/2}$. Thus, the atoms are
strongly localized about the minima of the potential in the transverse
directions (i.e., $y$ and $z$ in the previous example), forming
elongated clouds (referred to as ``tubes''). Since the potential along
the longitudinal ($x$) direction is much weaker, all the dynamics will
occur in that direction. The resulting system is an ensemble of a few
thousand (finite-size) 1D ultracold gas tubes, which is sometimes
called a 2D optical lattice (in the case where $V_{0x} =0$, see
Fig.~\ref{fig:2DOL}), or, as we shall do in what follows, a quasi-1D lattice.

Nevertheless, once the condition for one-dimensionality of the tubes
is achieved, two quasi-1D regimes are possible.
This is because, even if the atoms only undergo zero-point motion
transversally, the hopping amplitude between tubes $t_{\perp}$ is in
general non vanishing. Regime (i), if over typical duration of a
experiment, $\tau_\mathrm{exp} \lesssim 10^2$ s, the characteristic
hopping time $\tau_\mathrm{hop} \sim \frac{\hbar}{t_{\perp}} \gtrsim
\tau_\mathrm{exp}$, then hoping events will rarely occur and therefore
phase coherence between different tubes cannot be established. In the
RG language used in Sec.~\ref{sec:coupled1dbosons}, the energy scale
associated with the observation time $\sim \frac{h}{\tau_{\rm exp}}$
behaves as an infrared cut-off of the RG flow preventing the
renormalized Josephson coupling, $g_{J}$, from becoming of order one.
Regime (ii), if $\tau_\mathrm{hop}\ll \tau_\mathrm{exp}$, then the
establishment of long range phase coherence will be possible (but it may be
prevented by other terms of the Hamiltonian, see below and
Sec.~\ref{sec:coupled1dbosons}). Indeed, experimental groups have
explored the two regimes, as we shall review in more detail below.



The regime of a phase coherent ensemble of tubes and the transition to the phase incoherent
regime was explored in the early experiments by \citet{greiner01}, where the 1D regime was first reached in a
quasi-1D lattice, and later more thoroughly by \citet{moritz03,stoferle04}. However, other groups have
focused directly in the 1D (phase-incoherent) regime
\cite{kinoshita04,paredes04,kinoshita05,haller_sine_Gordon_MI_2010}. In what follows, these experiments are
reviewed and the theoretical background for the probes used in some of them will be also discussed. We begin with
the experiments that explored the phase incoherent 1D regime, and near the end of this subsection, discuss the
experiments where inter-tube hoping may become a relevant perturbation on the system.

\paragraph{Strongly Interacting 1D Bosons in optical lattices:}
For alkali atoms, which at ultracold temperatures (from $\sim 10$ nK to $\sim 1\,\, \mu$K) interact dominantly
via the $s$-wave channel (see Sec.~\ref{sec:models}), the interaction
is short ranged. When ultracold atoms are loaded into an anisotropic optical lattice, they occupy the lowest available (Bloch) band. This
system is thus amenable to a lattice description in terms of the
(anisotropic) Bose-Hubbard model. Furthermore,
if the experimental conditions are such that the hopping between tubes can be neglected (see discussion above), the
system can be described as an array of independent tubes. Each tube
is described by a 1D Bose-Hubbard model
Eq.~(\ref{eq:bose-hubbard}).  The particle number
within each tube is largest at the center and decreases towards the edges
of the lattice in response to the existence of the harmonic trap, which makes the system inhomogeneous.

\begin{figure}
\centering
  \includegraphics[width=8cm]{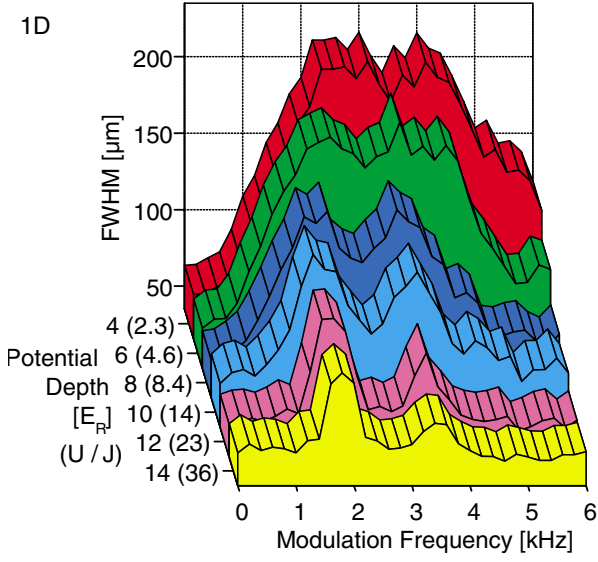}
  \caption{\label{fig:latticeshaking1D}
Energy absorption of a strongly interacting 1D Bose system measured as a function of the longitudinal potential
$V_{0x}/E_R$ ($E_R$ is th atom recoil energy). Notice the broad spectrum of the superfluid (small $V_{0x}/E_R$)
and the double peak structure characterizing the Mott insulator (large $V_{0x}/E_R$) \cite{stoferle04}.}
\end{figure}

Tuning the longitudinal potential ($\propto V_{0x}$), effectively changes the ratio $t/U$ of the 1D Bose-Hubbard
model. For deep potentials, it is thus possible to access the strongly interacting regime of the model and to
observe the transition from the superfluid (i.e., Tomonaga-Luttinger liquid) phase to the Mott
insulator. Indeed, shortly after this transition was observed in a 3D optical lattice~\cite{greiner02},
the 1D transition was observed by \citet{stoferle04}. More recently, the Mott transition in a weak lattice, as discussed in Sec.~\ref{sec:mott}, was also observed \cite{haller_sine_Gordon_MI_2010}. In order to probe the excitation spectra
of the phases realized by tuning $V_{0x}$ in the
quasi-1D optical lattice, \citet{stoferle04} followed a novel spectroscopic method which relied on modulating in time
 the amplitude of the longitudinal potential. In equations, this amounts to replacing
$V_{0x} \to V_{0x} + \delta V_{0x} \cos \omega \tau$ in Eq.~(\ref{eq:optpot}). The time dependent potential heats
up the system, and the transferred energy can be estimated in a time of flight experiment from the enhancement
of the width at half maximum of the peak around zero momentum. The measured spectra for different
values of the lattice depth ($V_{0x}/E_R$) are displayed in Fig.~\ref{fig:latticeshaking1D}. The superfluid phase
is characterized by a broad spectrum, whereas the emergence of the 1D Mott insulator is characterized
by the appearance of two peaks: A prominent one at $\hbar \omega \simeq U$ and a smaller one at
$\hbar \omega \simeq 2  U$.

In order to qualitatively understand these spectra, \citet{iucci_absorption} employed linear
response theory, and considered both the limit of small and large potential $V_{0x}$.  When $V_{0x}$ is smaller than
$\mu$, the lattice modulation couples to the density operator. Using the bosonization formula
Eq.~(~\ref{eq:bosonized_rho}), the $q\approx 0$ term  $\propto \partial_x \phi$  gives a vanishing contribution to
the energy absorption within linear response and, to leading order,
the modulation of the amplitude of the optical potential is described by the perturbation
$\hat{H}_1(\tau) =  B \rho_0 \delta V_{0x} \cos \omega \tau \int dx \cos (2 \hat{\phi}(x,\tau) + x \delta )$
where $\delta = 2G - 2\pi \rho_0$, that is, the modulation couples to the density operator for
$q \approx 2\pi \rho_0$. In the limit of large $V_{0x}$, it is most convenient to begin with the Bose-Hubbard
model. Within this framework, it can be shown that the lattice modulation can be expressed
as a modulation of the hopping amplitude,
$\hat{H}_1(\tau) = - \delta t \cos \omega \tau \sum_{m} \left[\hat{b}^{\dag}_{m+1}\hat{b}_m + \textrm{H.c.}\right]$, where
$\delta t = \frac{d t}{d V_{0x}} \delta V_{0x}$. Hence, the (time-averaged) energy absorption rate per particle reads:
\begin{equation}
\dot{\epsilon}(\omega) = \frac{\delta V^2_{0x}}{N} \omega \: \mathrm{Im} \left[-\chi_{\mathcal{O}}(\omega) \right],
\end{equation}
where $\chi_{\mathcal{O}}(\omega)$ is the Fourier transform of
$\chi(\tau) = -\frac{i}{\hbar} \langle \left[ \mathcal{\hat{O}}(\tau),\mathcal{\hat{O}}(0) \right] \rangle$.  The operator
$\mathcal{\hat{O}} = \int dx \: \cos(2\hat{\phi}(x) + x \delta )$, in the weak lattice regime, and
$\mathcal{\hat{O}} = -\delta J  \sum_{m} \left[ \hat{b}^{\dag}_{m+1}\hat{b}_m + \textrm{H.c.}\right]$,
in the strong lattice regime.

For a weak commensurate lattice, in the superfluid (Tomonaga-Luttinger liquid) phase,
$\dot{\epsilon}(\omega) \sim  \hbar \omega \left( \frac{\hbar \omega}{\mu} \right)^{2K-2}$, where $K$ is the
Tomonaga-Luttinger liquid parameter and $\mu$ the chemical potential. Since for the superfluid phase ($K\gtrsim 2$),
the calculated low frequency part contains very little spectral weight, using the $f$-sum rule,
\citet{iucci_absorption} argued that the absorption spectrum of the superfluid phase should be broad,
in agreement with the experiment and the calculations of \citet{caux_density} based on the Bethe-ansatz.
For the Mott insulator phase, in the weak coupling limit, a threshold behavior
of the form $\dot{\epsilon}(\omega) \sim  \hbar \omega F(\hbar \omega)
\theta(\hbar \omega - \Delta)$ exists,
where $\Delta$ is the Mott gap and $F(x)$ is a smooth function. This threshold behavior
was observed in the experiments of \cite{haller_sine_Gordon_MI_2010}, which recently explored the
phase diagram of the 1D Bose gas in a weak periodic potential.

\begin{figure}
  \centering
  \includegraphics[width=8cm]{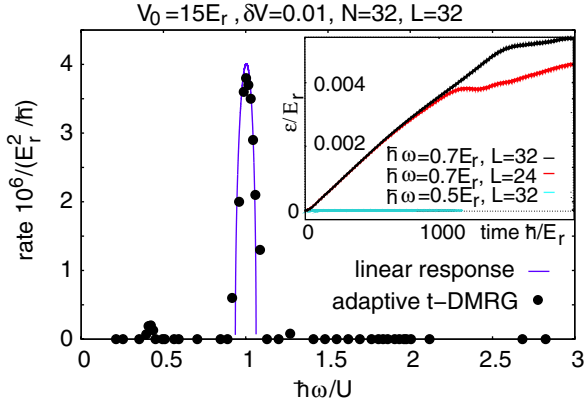}
  \caption{\label{fig:drmg_vs_linresponse}
 Comparison of the analytical results of \citet{iucci_absorption} (continuous curve)
 the time-dependent DMRG calculations of~\citet{kollath_latticeshaking06} (black dots) for
 the energy absorption rate due to a periodic modulation of the lattice and. The inset shows the behavior
 of the energy absorption with time, and how it deviates from the linear response for different system sizes.}
\end{figure}

In the strong coupling regime the perturbation reads, in an expansion
to $O(t^2/U)$, in the subspace of particle hole states:
\begin{equation}
\dot{\epsilon}(\omega) = \frac{(\delta t)^2}{2t} \hbar \omega \left( \frac{n_0+1}{2n_0+1}\right)
 \sqrt{1-\left[\frac{\hbar \omega - U}{(2n_0+1)t} \right]^2},
\end{equation}
and zero otherwise.  The accuracy of this expression was tested using time-dependent DMRG by
\citet{kollath_latticeshaking06}, and the results are shown in
Fig.~\ref{fig:drmg_vs_linresponse}. The later numerical technique
allows to go beyond linear response and to incorporate the effects of
the trap and thus to deal with the experiments of~\citet{stoferle04}
for which  $\delta V_{0x}/V_{0x} \sim 10\%$.
 Numerically investigating of the
inhomogeneity brought about by the trap, it was shown by \citet{kollath_latticeshaking06} that the
smaller peak near $\hbar \omega = 2U$ is indeed a measurement of the incommensurability in the 1D tubes.

An alternative interpretation of the broad resonance observed in the superfluid phase was given in terms of a parametric instability of Bogoliubov modes and their nonlinear dynamics\cite{tozzo_2005}. This interpretation assumes that the Bogoliugov approximation is also applicable  for strongly interacting bosons in 1D.


\paragraph{Approaching the Tonks-Girardeau limit:}
For bosonic atoms, there are essentially two routes into the strongly interacting regime, which
asymptotically approaches the Tonks-Girardeau limit where fermionization occurs. The two routes were
explored simultaneously by \citet{paredes04} and \citet{kinoshita04}. One of them~\cite{paredes04}
relies on reaching the strong interacting limit of the 1D Bose Hubbard model while ensuring that
the system remains incommensurate with the lattice potential (\emph{e.g}, at average filling $n_0 < 1$).
In the presence of a longitudinal harmonic confinement, this is by no means straightforward, because,
as discussed in Sec.~\ref{sec:BHMTrap}, the center of the cloud easily becomes Mott insulating as the
ratio $t/U$ decreases.  Thus, carefully controlling the trapping potential, \citet{paredes04}
entered the strongly interacting regime where the bosons undergo fermionization. The signatures of the
latter were observed in the momentum distribution as measured in time-of-light experiments
(cf. Sec.~\ref{sec:tof}). After averaging over the distribution of tubes resulting from the
inhomogeneous density profile, \citet{paredes04} were able to reproduce
the experimentally measured momentum distribution of the strongest interacting systems.
Computational studies of the approach to the Tonks-Girardeau regime were done by
\citet{pollet_rombouts_04,wessel_alet_05}.

%
%
The other route into the strongly correlated regime for bosons in 1D is to first load the atoms in a deep 2D optical
lattice ($V_{0x} = 0$ and $V_{0y}= V_{0z} \gtrsim 30\, E_R$ in Eq.~\ref{eq:optpot}). The resulting system is
described by the Lieb-Liniger model in a harmonic trap.  The strongly interacting regime of this model is reached
by making $\gamma = \frac{Mg}{\hbar \rho_0}$  large. This can be achieved either by decreasing the mean density,
$\rho_0$, which for small atom numbers $\lesssim 10$ per tube ultimately poses detection  problems, and/or by
increasing the coupling constant $g$ (cf. Eq.~\ref{eq:1dcoupling}). As predicted by \citet{olshanii98}, this is
possible either by making the transverse confinement tighter~\cite{kinoshita04,kinoshita05} or by increasing the
scattering length using Feshbach resonance~\cite{haller_sine_Gordon_MI_2010}. \citet{kinoshita04,kinoshita05}
used a blue-detuned 2D optical lattice potential (cf. Fig.~\ref{fig:2DOL}) to confine an ultracold $^{87}$Rb gas
into an array of 1D traps ($\sim 10^3$ tubes). In addition, the atoms were confined longitudinally by an optical
(red detuned) dipole trap. The blue-detuned lattice potential makes it possible to reach a tighter transverse
confinement without increasing the probability of spontaneous emission, which due to the atom recoil may result
in the latter being lost from the trap. In addition, as mentioned above, by virtue of Eq.(\ref{eq:1dcoupling}),
a tighter confinement allows for the increase of the interaction coupling $g$.

By measuring the mean atom energy after the expansion in 1D (turning off the optical dipole trap only),
\citet{kinoshita04} were able to detect
the incipient signatures of fermionization of the 1D gases. However, the most dramatic signatures of strong
interactions were observed in a later experiment~\cite{kinoshita05}, in which local pair correlations $g_2(x)$
were measured.

For the Lieb-Liniger model, $g_2$ (note that we drop $x$ assuming translational invariance) is a thermodynamic quantity
that can be obtained from the free-energy by using the Hellman-Feynman theorem:
\begin{equation}
g_2(\gamma, \vartheta) = \rho^{-2}_0 \langle  \left[ \hat{\Psi}^{\dag}(x)\right]^2  \left[ \hat{\Psi}(x)\right]^2 \rangle  =
-2\frac{T}{L} \frac{\partial F(T)}{\partial g},
\end{equation}
where $F = -T \ln \: \mathrm{Tr}\: e^{-H/T}$ is the free energy for the Lieb-Liniger model ($H$ is given
by Eq.~\ref{eq:lieb-liniger-model}), $\vartheta = T/T_d$, $T$ being the absolute temperature and
$T_d = \frac{\hbar^2 \rho_0}{2M}$ the characteristic temperature for quantum degeneracy. \citet{gangardt2003_prl}
first obtained $g_2(\gamma,\vartheta)$ for $\tau = 0$ for all $\gamma$ values using the Bethe ansatz
result for the ground state energy. \citet{kheruntsyan05} extended these results to finite temperatures. For large
and small $\gamma$ values, the following asymptotic limits have been derived~\cite{gangardt2003_prl,kheruntsyan05}:
\begin{align}
\label{eq:g2gammatau}
g_2(\gamma,\vartheta) \simeq \left\{ \begin{array}{c}
 \frac{4}{3} \left(\frac{\pi}{\gamma} \right)^2 \left[ 1 + \frac{\vartheta^2}{4\pi^2} \right]  \,
\gamma \gg 1 \, \mbox{and}\,  \vartheta \ll 1, \\
\frac{2\vartheta}{\gamma^2} \quad 1 \ll \vartheta \ll \gamma^2, \\
1 - \frac{2}{\pi}\sqrt{\gamma} + \frac{\pi\vartheta^2}{24 \gamma^{3/2}}\,\, \vartheta \ll \gamma \ll 1,\\
1 + \frac{\vartheta}{2\sqrt{\gamma}}\,\, \gamma\ll   \vartheta \ll \sqrt{\gamma}, \\
2 - \frac{4\gamma}{\vartheta^2} \quad \sqrt{\gamma}  \ll \vartheta \ll 1, \\
2 - \gamma \sqrt{\frac{2\pi}{\vartheta}} \,\, \vartheta \gg \max\{1,\gamma^2\}.
 \end{array}
 \right.
\end{align}
The results for large $\gamma$ can be also reproduced from the low-density limit of a strong coupling expansion
for the Bose-Hubbard model~\cite{cazalilla_tonks_gases}.

\begin{figure}
  \centering
  \includegraphics[width=9cm]{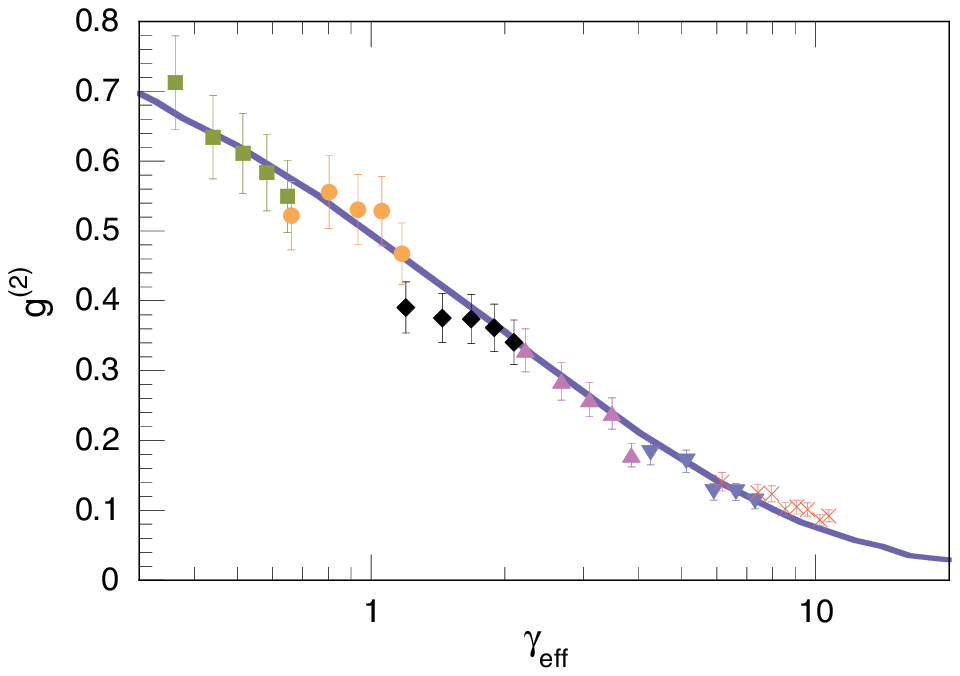}
  \caption{\label{fig:kinoshita05}
  Pair correlation function as determined from photo-association
  by~\citet{kinoshita05} versus dimensionless parameter $\gamma$.
The solid line is from the 1D interacting
  bosons theory of \citet{gangardt2003_prl}.The different symbols
  correspond to different dipole trap power intensities $P_{cd}$,
and thus different one-dimensional densities:
squares, $P_{cd} = 3.1$ W; circles, $P_{cd}
  = 1$ W; diamonds,
$P_{cd} = 320$ mW; up triangles, $P_{cd} = 110$ mW; and down triangles,
$P_{cd} = 36$ mW.}.
\end{figure}

The asymptotic expressions (\ref{eq:g2gammatau}) show that for $T \to 0$, $g_2 \to 0$ as
$\gamma \to +\infty$. The latter reflects the fermionization that becomes complete in the Tonks-Girardeau
limit, where $g_2 = 0$ at all temperatures. The dramatic decrease in $g_2$~\cite{kinoshita05} as it crosses over
from  the behavior for weakly interacting bosons  (i.e., $g_2 \approx 1$ for $\gamma \sim 0.1$) to strongly
interacting bosons (i.e.,  $g_2 \approx 0.1$ for $\gamma \simeq 10$) was observed by tuning the interaction
coupling $g$, as described above. The measurements are depicted in Fig.~\ref{fig:kinoshita05}, which also show
the excellent agreement with theory.

In order to measure $g_2$, atoms in the optical
lattice were photo-associated into molecules using a broad laser beam resonant with a well-defined molecular state
of the Rb-Rb dimer. The molecule formation was subsequently detected as a loss of atoms from the trap. One important
issue that must be addressed in order to compare theory and experiment is the inhomogeneity of the 1D systems due
to the harmonic trap. Fortunately, the LDA allows one to obtain $g_2(x)$ for the trapped system
from the $g_2(\gamma)$ calculated for the uniform system.

Another experimentally accessible local correlation is
$g_3 =  \rho^{-3}_0\langle \left[\hat{\Psi}^{\dag}(x) \right]^3\left[\hat{\Psi}(x) \right]^3 \rangle$. This local correlation
function is related to the rate of three-body recombination processes where molecules from out of three-atom
encounters. The calculation of $g_3$ is less straightforward. \citet{gangardt2003_prl} obtained
the following asymptotic expressions for the Lieb-Liniger model at large and small $\gamma$ and $T = 0$:
 \begin{align}
 g_3(\gamma) = \left\{
 \begin{array}{c}
 \frac{4\pi^2}{3\gamma^2}  \quad \gamma \gg 1, \\
 1 - \frac{6}{\pi} \sqrt{\gamma} \quad \gamma \ll 1.
 \end{array}
 \right.
 \end{align}
Subsequently, in a mathematical \emph{tour de force}, \citet{cheianov_smith_zvonarev06} computed the behavior of
$g_3(\gamma)$ at $T=0$ for the entire crossover from weakly to strongly interacting gas. More generally, at  any finite $T$ and for intermediate  of $\gamma$, one has to rely on the general expressions recently obtained by~\citet{Kormos_EV_LL_gas_2009}.
These results predict a strong reduction in $g_3(\gamma)$ at low $T$ as the 1D gas enters the strongly interacting regime. This behavior should manifest itself in a strong reduction of the atom losses due to three-body recombination. Indeed, this is consistent
with the experiments in optical lattices in the 1D regime~\cite{laburthe04}.

Finally, in a recent experiment, \citet{haller_gustavsson_09} have been able to access
the super-Tonks-Girardeau regime in a 2D optical lattice by starting with a low energy initial state in
the Tonks-Girardeau regime (with a large and positive value of $g$) and quickly changing $g$ across the
confinement induced resonance (see Sec.~\ref{sec:QMCcontinuum}) to a negative final value. In this
experiment, $g$ was changed by means of a Feshbach resonance. The ratio between the lowest compressional
and dipole modes of oscillation in the highly excited state (with $g<0$) created in such a way was found
to be larger than the one in the Tonks-Girardeau regime, a hallmark of the super-Tonks-Girardeau regime
\cite{astrakharchik05}.

\paragraph{Coupled condensates}

The behavior of the Bose gas in
the quasi-1D optical lattice as the transverse potential is tuned and phase coherence between the
tubes emerges has been explored by \citet{moritz03,stoferle04}.  A theoretical discussion of these experiments has been given by some of us in
\citet{cazalilla_deconfinement_longpaper} and we thus refer the interested reader to this work.

\begin{figure}[htbp]
  \centering
  \includegraphics[width=8cm]{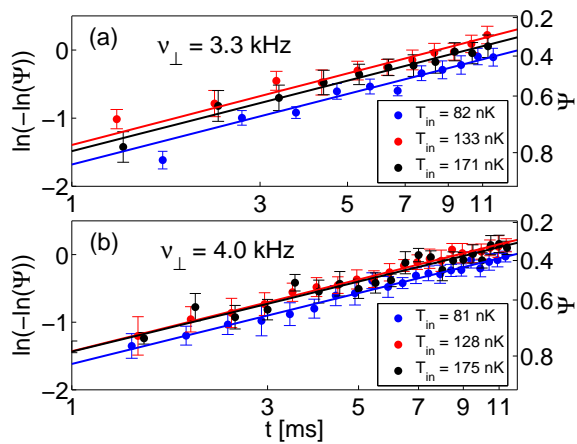}
  \caption{Double logarithmic plot of the coherence factor versus time
    for decoupled 1d condensates. Each point is the average of 15
    measurements, and error bars indicate the standard error. The
    slopes of the linear fits are in good agreement with a $2/3$
    exponent \cite{hofferberth07}.}
  \label{fig:schmiedmayer-split}
\end{figure}

Using atom chips, a 1D Bose gas with a few thousand atoms can be trapped in the 1D quasicondensate regime
$k_BT, \mu \ll \hbar \nu_\perp$ \cite{hofferberth07}. Applying a radiofrequency (rf) induced adiabatic potential,
the 1D gas can be split into two 1D quasicondensates. The height of the barrier between the two condensates can be
adjusted by controlling the amplitude of the applied rf field. This allows  one to achieve both Josephson coupled
and fully decoupled quasicondensates. The fluctuations of the relative phase of the two condensates are measured
by the quantity:
\begin{eqnarray}
  \label{eq:coherence-factor}
  \mathcal{\hat{A}}(\tau)=\frac 1 L \left|\int dx e^{i (\hat{\theta}_1(z,\tau)-\hat{\theta}_2(z,\tau))}
  \right|,
\end{eqnarray}
where $\hat{\theta}_1,\hat{\theta}_2$ are the respective phases of the two condensates obtained after the splitting. From the
theoretical point of view, split condensates can be analyzed within the Tomonaga-Luttinger liquid framework
\cite{bistritzer2007_interferometers,burkov2007}. One introduces $\hat{\theta}_+=(\hat{\theta}_1+\hat{\theta}_2)/2$ and
$\hat{\theta}_-=\hat{\theta}_1-\hat{\theta}_2$. The Hamiltonian $\hat{H}=\hat{H}[\hat{\theta}_1]+\hat{H}[\hat{\theta}_2]$ can be rewritten as
$\hat{H}[\hat{\theta}_+]+\hat{H}[\hat{\theta}_-]$. In the initial state, the wavefunction factorizes
$\hat{\Psi}[\hat{\theta}_1,\hat{\theta}_2]=\hat{\Psi}_+[\hat{\theta}_+]+\hat{\Psi}_-[\hat{\theta}_-]$, where $\hat{\Psi}_+$ is determined by the initial state
of the condensate and $\hat{\Psi}_-$ is initially localized near $\hat{\theta}_-(z)=0$. Assuming an initial state with
$\langle \hat{\theta}_-(x) \hat{\theta}_-(x')\rangle = \delta(x-x')/2\rho$, the quantum dynamics of the phase $\hat{\theta}_{-}$
under the Hamiltonian (\ref{eq:effective_ham}) gives
$\langle\mathcal{\hat{A}}\rangle=\mathcal{A}_0 e^{-\tau/\tau_Q}$ with $\tau_Q=2K^2/(\pi^2 v \rho)$ \cite{bistritzer2007_interferometers}.
For long times, $\tau\gg \hbar/(k_B T)$, the symmetric and antisymmetric modes interact with each other. The symmetric
modes remain in thermal equilibrium, and generate friction
$1/\tau_{\rm f}(k) \propto k^{3/2}$ and noise $\zeta(x,\tau)$
(satisfying the fluctuation dissipation relation) for the dynamics of the antisymmetric mode. The field $\hat{\theta}_-$
can then be treated as a classical variable, satisfying the Langevin equation:
\begin{eqnarray}
  \partial_\tau^2 \hat{\theta}_-(k,\tau) +\frac{ \partial_\tau \hat{\theta}_-(k,\tau)}{ \tau_{\rm f}(k)} + (v
  k)^2 \hat{\theta}_-(k,\tau) = \zeta (k,\tau). \quad
\end{eqnarray}
Solving this second order differential equation by the variation of the constant method with initial conditions
$\hat{\theta}(k,\tau)=\partial_\tau \hat{\theta}(k,\tau)$, one finds that $\langle\hat{\theta}_-(x,\tau)^2\rangle=2 (\tau/\tau_C)^{2/3}$,
and thus $\mathcal{A}\propto e^{-(\tau/\tau_C)^{2/3}}$. This behavior was indeed observed in experiments in the case of
a large potential barrier \cite{hofferberth07}, as shown in Fig.~\ref{fig:schmiedmayer-split}.

\section{Outlook}\label{sec:outlook} \setcounter{paragraph}{0}

Quantum interacting bosons in 1D provide the conjunction between solid
theoretical descriptions and a wide variety of experimental
situations. In the previous sections we have seen that the properties
of those systems can be captured by a variety of techniques ranging
from exact solutions and computational methods to asymptotic theories
such as Tomonaga-Luttinger liquids. Although we have now a firm
understanding of several of these properties, many challenges
remain. This is in particular prompted by many remarkable experiments,
various of which were discussed in Sec.~\ref{sec:experiments}. In the
present section, we discuss some of the open problems and recent
developments in the field and provide an outlook for future
research. Our list is of course non-exhaustive and many unexpected
developments will certainly occur as well. These are all the
ingredients of a very lively field that will surely continue to
provide exciting results in the years to come.

\paragraph{Beyond Tomonaga-Luttinger liquid theory:}
The low-energy, long wavelength description based on the bosonization
technique introduces a particle-hole symmetry that is not present in
the original models that it intends to describe. In the recent years,
there has been intensive theoretical efforts to improve bosonization
by including non-linear terms
\cite{bettelheim2006_hydodynamics,Khodas_bose_liquid,pereira2005xxz,imambekov09}. In
particular, damping rates and universal crossover functions for the
spectral functions have been derived. A related problem is the
existence of a quantum critical regime at sufficiently large
temperature which does not obey Luttinger liquid
scaling\cite{guan2010}. The search for experimental
signatures of these effects going beyond the simplest Tomonaga-Luttinger liquid
theory is one possible future development of interacting bosons in
1D. In a similar way, as discussed in Sec.~\ref{sec:multi}, systems
which do not obey the paradigm of Tomonaga-Luttinger liquids
\cite{zvonarev07_1d_bose_ferro} have been found and will lead to
interesting further developments.

\paragraph{Disorder and many-body localization:}

As discussed in Sec.~\ref{sec:disorder}, important developments have
occurred in the understanding of the interplay between disorder and
interactions. However, many questions still remain open. In
particular, the understanding of the finite temperature transport
properties of disordered quantum systems, a topic that should lead to a
wealth of future developments
\cite{aleiner_altshuler_10,oganesyan_huse_07,pal_huse_10,monthus_garel_10}.


Another interesting open problem is the far-from-equilibrium dynamics
of interacting bosons in the presence of a random potential. This
problem is relevant for the expansion of condensates in experiments in
which either a quasiperiodic or speckle potential is present
 \cite{roati08}. More generally, understanding how those systems
respond to quenches, and the steady states in the presence of
time-dependent \cite{dallatorre_noise}
or time-independent noise, are of very much current
interest. A related question is the one of aging in disordered and
glassy quantum systems. It is well known that in disordered systems,
below a certain temperature, the relaxation may become extremely slow
so that the system never equilibrates on any reasonable timescale.
In such regime, both correlation $C(\tau,\tau')$  and response
$R(\tau,\tau')$
 functions depend on both $\tau$ and $\tau'$ and not just $\tau-\tau'$
as in equilibrium. Some of these properties were understood for classical glasses
\cite{cugliandolo03} but remain largely to be understood for quantum glassy systems.

\paragraph{Dimensional crossover:}
In Sec.~\ref{sec:coupled1dbosons}, we discussed briefly the issue of
the dimensional crossover. A particularly interesting case is the one
of coupled chains in a single plane. In that case, the system possesses long
range order in the ground state but, for any strictly positive temperature,
long range order is replaced by quasi-long range order with a temperature
dependent exponent. At higher temperature, the system undergoes a BKT transition
and is fully disordered above the transition. By contrast, a single
chain has no long range order at any strictly positive temperature,
with a correlation length that diverges with temperature. There is
  thus an important issue of understanding the crossover from between
the 1D decoupled chain regime and the 2D regime. Since both regimes
are strongly fluctuating, this crossover cannot be tackled within a
mean field approximation, in contrast to the 1D-3D crossover.

Another important problem is the magnon BEC in coupled two-leg ladders when
the interladder coupling is frustrated. In that case, one has to
understand whether the formation of the magnon BEC is suppressed
leaving the system in a critical state
 \cite{emery_smectic,vishwanath_slide_LL}, or whether a more exotic
broken symmetry develops, making the system off-critical albeit with
an unconventional spin ordering.  A related problem exists in arrays
of Josephson junctions \cite{tewari05,tewari06} where dissipative
effects can also come into play.

\paragraph{Ergodicity, quantum chaos, and thermalization:}

In Sec.~\ref{sec:dynamics}, we briefly touched upon some of the recent
experimental and theoretical studies that have dealt with the
nonequilibrium dynamics of isolated quantum systems in 1D. This field
is in its infancy and many important questions are currently being
addressed both by experimentalists and theorists.

Most of the computational works so far have considered either small
systems (with less than 10 particles), for which all time scales are
accessible by means of full exact diagonalization techniques, or
systems with up to $\sim100$ particles, for which only the short time
scales ($\sim 10 \hbar/t$) can be addressed with time-dependent DMRG
and related techniques. It is becoming apparent that in order to gain
further understanding of the long time limit in large systems (or in
the thermodynamic limit) a close collaboration will be needed between
cold gases experiments, with their ``quantum analog simulators'', and
theory. Among the things we would like to learn from those studies is
what kind of memory of the initial conditions few-body observables
retain in isolated quantum systems after relaxation
\cite{olshanii_yurovsky_10}, what typicality
\cite{tasaki_98,goldstein_lebowitz_06,popescu_06,reimann_08} means for
1D systems (some of which are integrable or close to integrable
points), the relation between thermalization and quantum chaos
\cite{santos_rigol_10a,santos_rigol_10b}, the time-scales for thermalization
\cite{moeckel_kehrein_08,barmettler_punk_09,eckstein_kollar_09}, and
the proper definition of thermodynamic quantities
\cite{polkovnikov_08,polkovnikov_10}.

\begin{acknowledgments}
  
  We thank G. Astrakarchik, G. G. Batrouni, M. Batchelor, F. Dalfovo,
  S. Giorgini, X.-W. Guan, A. Muramatsu, G. Mussardo, A. Minguzzi,  H.-C.
  N\"agerl, 
 G. Orso, L. Pollet, N. Prokof'ev, R. T. Scalettar, B. 
Svistunov, and M. Troyer 
 for valuable comments on the manuscript.  MAC thanks
 the hospitality of M. Ueda  (Ueda ERATO Macroscopic Quantum Control Project 
 of JST, Japan) and M. Oshikawa (ISSP) at the Univesity of Tokyo, 
 where  parts of this review where written, and
the support of the Spanish MEC grant no. FIS2007-066711-C02-02.
  MR was supported by the US Office of Naval Research 
under Grant No.~N000140910966 and by the National Science Foundation under 
Grant No.~OCI-0904597. This work was supported in part by the Swiss
National Science Foundation under MaNEP and Division II.  
\end{acknowledgments}

\bibliographystyle{apsrmp}

\end{document}